\documentclass[aps,prd,preprint,nosuperscriptaddress,nofootinbib,tightenlines,titlepage,floatfix]{revtex4-1}
\usepackage{amsmath}
\usepackage{amsfonts}
\usepackage{amssymb}
\usepackage{color}
\usepackage{graphicx}
\usepackage{subfigure}
\usepackage{longtable}
\usepackage[bookmarks,pdfpagelabels]{hyperref}
\usepackage{bbm}
\usepackage{slashed}
\usepackage{xspace}
\usepackage{xcolor}
\usepackage[utf8]{inputenc}

\newcommand{\di}{\mathrm{d}}
\newcommand{\La}{\mathcal{L}}
\newcommand{\M}{\mathcal{M}}
\newcommand{\Mfi}{\mathcal{M}_{fi}}
\newcommand{\MSbar}{\ensuremath{\overline{\textrm{MS}}}}

\newcommand{\Order}[1]{\mathcal{O}\left(#1\right)}

\newcommand{\Id}{\mathbbm{1}}
\newcommand{\MeV}{\textrm{ MeV}}
\newcommand{\GeV}{\textrm{ GeV}}
\newcommand{\TeV}{\textrm{ TeV}}
\newcommand{\fb}{\textrm{ fb}}
\newcommand{\bra}[1]{\left\langle#1\right|}
\newcommand{\ket}[1]{\left|#1\right\rangle}
\newcommand{\braket}[2]{\left\langle#1\middle|#2\right\rangle}
\newcommand{\abs}[1]{\left|#1\right|}

\renewcommand{\Re}{\mathfrak{Re}}
\renewcommand{\Im}{\mathfrak{Im}}
\DeclareMathOperator{\Tr}{Tr}

\newcommand{\obs}{\ensuremath\mathcal{O}}

\newcommand{\eq}[1]{eq.~(\ref{#1})}
\newcommand{\fig}[1]{Fig.~\ref{#1}}
\newcommand{\tab}[1]{Table~\ref{#1}}
\makeatletter
\newcommand\eqs[1]{%
  eqs.~(\my@refs #1,\relax\noexpand\@eolst)}
\newcommand\figs[1]{%
  Figs.~(\my@refs #1,\relax\noexpand\@eolst)}
\newcommand\tabs[1]{%
  Tables~(\my@refs #1,\relax\noexpand\@eolst)}
\def\my@refs #1,#2\@eolst{%
   \ifx\relax#2\relax
      \ref{#1}%
   \else
      \ref{#1},
      \my@refs #2\@eolst%
   \fi}
\makeatother
\newcommand{\chap}[1]{section~\ref{#1}}
\newcommand{\secref}[1]{section~\ref{#1}}

\newcommand{\program}[1]{\mbox{#1}\xspace}

\newcommand{\VBFNLO}{\program{VBFNLO}}
\newcommand{\HAWK}{\program{HAWK}}

\newcommand{\ie}{i.e.\xspace}
\newcommand{\eg}{e.g.\xspace}

\begin{document}
\preprint{KA-TP-35-2016}
\title{Vector-Boson Fusion and Vector-Boson Scattering}
\author{Michael Rauch}
\affiliation{Institute for Theoretical Physics, Karlsruhe Institute of Technology (KIT), Germany\\}

\begin{abstract}
Vector-boson fusion and vector-boson scattering are an
important class of processes for the Large Hadron Collider at CERN. It
is characterized by two high-energetic jets in the forward regions of
the detector and reduced jet activity in the central region. The higher
center-of-mass energy during the current and subsequent runs strongly
boosts the sensitivity in these processes and allows to test the
predictions of the Standard Model to a high precision.

In this review, we first present the main phenomenological features of
vector-boson fusion and scattering processes. Then we discuss the
effects of higher-order corrections, which are available at NLO QCD for
all processes and up to N3LO QCD and NLO electro-weak for VBF-H
production. An additional refinement is the addition of parton-shower
effects, where recently a lot of progress has been made. The appearance
of triple and quartic gauge vertices in the production processes enables
us to probe anomalous gauge couplings. We introduce and compare the
different parametrizations used in the literature and also discuss the
issue of unitarity violation and common unitarization procedures.
Finally, we give a short overview of current and possible future
experimental searches.
\end{abstract}

\maketitle
\tableofcontents
\clearpage

\section{Introduction}
\label{chap:introduction}

With the start of the Large Hadron Collider (LHC) at CERN, a new era in
particle physics has opened. Its run-I phase with center-of-mass
energies of 7 and 8~TeV has lead to the discovery of the Higgs boson
with a mass of 125~GeV~\cite{Aad:2012tfa,Chatrchyan:2012xdj}. With this
measurement, all particles predicted by the Standard Model (SM) have
been observed and its parameters determined. In 2015, run~II has
started with an increased center-of-mass energy of 13~TeV. This and
subsequent runs, approved for two further decades including a
high-luminosity upgrade, will allow to test the predictions of the SM to
high precision. With a projected final integrated luminosity of
3~ab$^{-1}$, its great success will be further strengthened, or
deviations can lead us to a new theory of beyond-the-Standard-Model
physics. 

A class of processes which strongly benefits from the higher
center-of-mass energy of the LHC is vector-boson fusion (VBF) and
vector-boson scattering (VBS), which we will discuss in this report.
In these processes, a quark or anti-quark scatters with another quark or
anti-quark via a space-like exchange of an electroweak gauge boson,
namely a $W$ or $Z$ boson or a photon. Off this $t$-channel exchange, a
single Higgs or electroweak gauge boson can be emitted, which are VBF
processes in the stricter sense. The emission of two electroweak bosons
defines the VBS process class. For most of this article, the distinction
between vector-boson fusion and scattering is not necessary, and
so we will in general use the label VBF to refer to both and mark explicitly
when this is not the case. Also, we will mainly focus on VBF production
of electroweak gauge bosons. For VBF-Higgs production, we refer the
reader to the yellow reports of the LHC Higgs cross section working
group~\cite{Dittmaier:2011ti,Dittmaier:2012vm,Heinemeyer:2013tqa,deFlorian:2016spz}
for an overview of the current theoretical status.

This electroweak $t$-channel exchange is the defining feature of VBF
processes, which we will first discuss in \chap{chap:processclass}. 
Restricting ourselves to the leading-order (LO) Born approximation,
important distributions for VBF processes are then presented. This
includes a very characteristic feature, the two final-state
\mbox{(anti-)}quarks which appear as jets in the forward regions of the
detector. These also allow us to distinguish VBF processes from other
production modes with the same final state. Possible backgrounds are on
the one hand triboson production, where as in VBF no strong coupling
constant enters at LO. There, the two jets originate from the decay of a
gauge boson, \ie instead of $t$-channel exchange an $s$-channel
resonance occurs. On the other hand, the two quark lines can be
connected by a gluon, giving rise to QCD-induced $V(V)jj$ production. The
appearance of two powers of the strong coupling constant and additional
partonic subprocesses with two quarks and two gluons means that for
total cross sections, QCD-induced production will dominate. We will
discuss how to suppress these alternative production modes compared to
VBF and also possible interference contributions in
\secref{sec:qcdvbfint}.

In order to obtain precise predictions, the inclusion of higher-order
corrections is necessary. Contributions at next-to-leading order (NLO)
in the strong coupling constant (NLO QCD) are known for all VBF
processes and have been studied in detail. In \chap{chap:nlo}, we will
first discuss the QCD corrections to the $qqV$ vertex, as these are both
ultraviolet and infrared divergent. Explicit expressions are derived,
employing the Catani-Seymour subtraction scheme~\cite{Catani:1996vz}. 
A discussion on the phenomenological impact of these corrections is
given in \secref{sec:nloqcdpheno}. Further higher-order corrections for
VBF processes have been calculated so far only for VBF-Higgs production.
There, both NLO electroweak and next-to-next-to-leading order (NNLO) QCD
corrections are available for differential distributions, and even
next-to-next-to-next-to-leading order (N3LO) QCD for inclusive cross sections.
Their impact is discussed in sections~\ref{sec:nloew} and \ref{sec:nnloqcd},
respectively.

A further possible refinement on the description of VBF processes is the
inclusion of parton-shower effects. In \chap{chap:partonshower}, first a
short general introduction is given to present the main features of
parton showers and the relevant formulae. Subsequently, we apply parton
showers to VBF processes. Phenomenological results for the combination
with LO cross sections are presented in \secref{sec:partonshowerLO}.
Great progress has been made in recent years for matching NLO QCD
calculations and parton showers. A popular choice for VBF processes has
been to use the POWHEG-BOX
program~\cite{Nason:2004rx,Frixione:2007vw,Alioli:2010xd}. Results
obtained within this framework are discussed in
\secref{sec:partonshowerNLOPOWHEG}. In parton showers, a further scale
appears, the starting scale of the parton shower. This allows one to
obtain error estimates on higher-order effects by varying this scale,
similar to changing the factorization and renormalization scales in an
NLO calculation. The impact of varying all three scales is shown in
\secref{sec:partonshowerNLOunc}. There, we also discuss the agreement
between the two popular matching schemes,
POWHEG-type~\cite{Nason:2004rx,Frixione:2007vw} and
MC@NLO-type~\cite{Frixione:2002ik}. These are equivalent to the accuracy
required for combining with NLO QCD calculations, but differ in
higher-order terms.

The appearance of triple and quartic gauge couplings in VBF processes
makes them an ideal tool to study anomalous contributions to these
vertices. A convenient parametrization of the effects, based on Lorentz
and gauge invariance, are effective field theories. Two approaches
exist, which differ in the expansion parameter and the power counting of
the operators. One is based on the canonical dimension of the operators,
the other on their chiral dimension. In \chap{chap:anomcoupl}, we
introduce the two methods and denote the operators relevant for
anomalous gauge couplings and relations between different
parametrizations. 
The high-energy behavior of VBS processes is dominated by a delicate
cancellation between different Feynman diagrams. Considering the
equivalent $2\rightarrow2$ process, individual Feynman diagrams for
scattering of longitudinal gauge bosons show a leading behavior of
$E^4$, where $E$ denotes the center-of-mass energy of the process. Exact
cancellations due to the gauge structure of the SM then guarantee that
the total amplitude exhibits as leading term a constant dependance on
energy.  Anomalous couplings spoil this cancellation, and the
corresponding rise of the amplitude with energy eventually leads to a
violation of unitarity of the $S$ matrix. In \secref{sec:anomunitarity},
we first derive the (tree-level) unitarity bound through a partial-wave
analysis.  Then we present different methods to circumvent this bound
and obtain a unitarity-conserving amplitude for all energies accessible
at the LHC.  Methods discussed are the application of form factors, or
using the $K$-matrix method~\cite{Heitler1,Heitler:1947mca}, an inverse
stereographic projection onto the unitarity circle. Finally, we show the
impact of anomalous couplings and unitarization methods on differential
cross sections in \secref{sec:anomcs}.

Lastly, \chap{chap:experiment} presents an overview of the experimental
status on VBF processes. We show the status of current measurements and
observations obtained from run-I of the LHC, and discuss the prospects
for subsequent runs.

\section{Characteristics of VBF Processes}
\label{chap:processclass}

\begin{figure}
\begin{tabular}{c@{\hspace*{-4ex}}c@{\hspace*{-4ex}}c}
\includegraphics[width=0.35\textwidth]{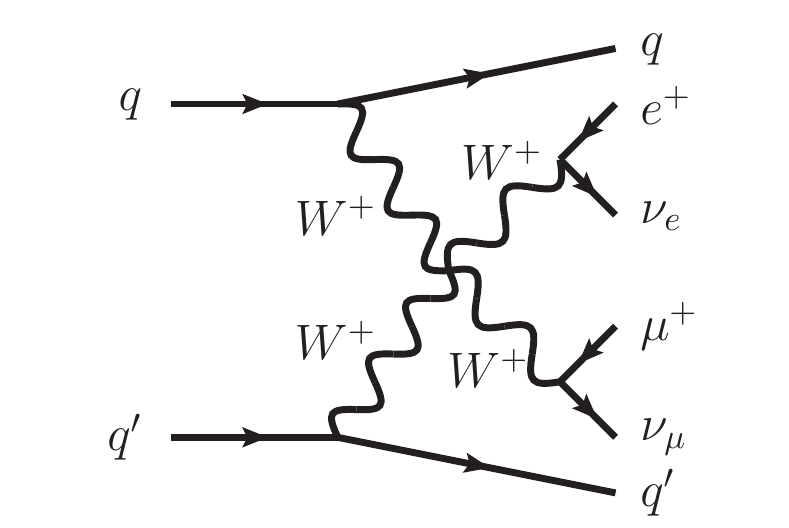} &
\includegraphics[width=0.35\textwidth]{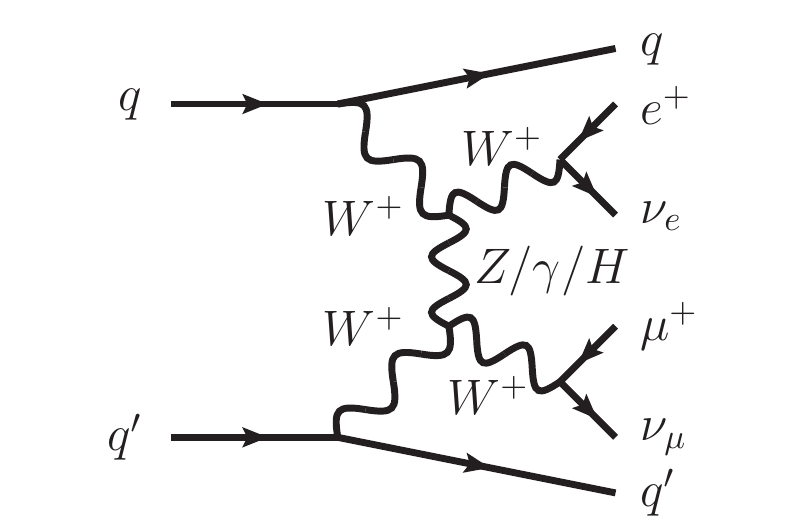} &
\includegraphics[width=0.35\textwidth]{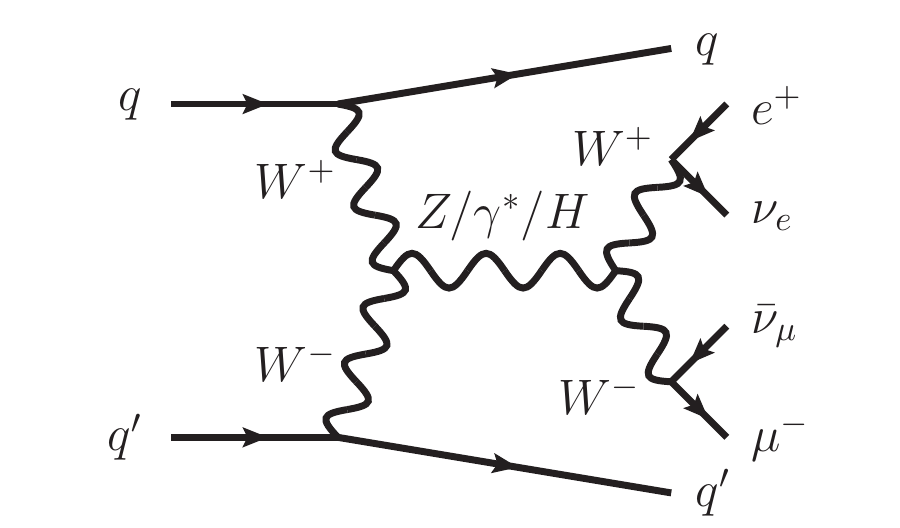} \\[1em]
\includegraphics[width=0.35\textwidth]{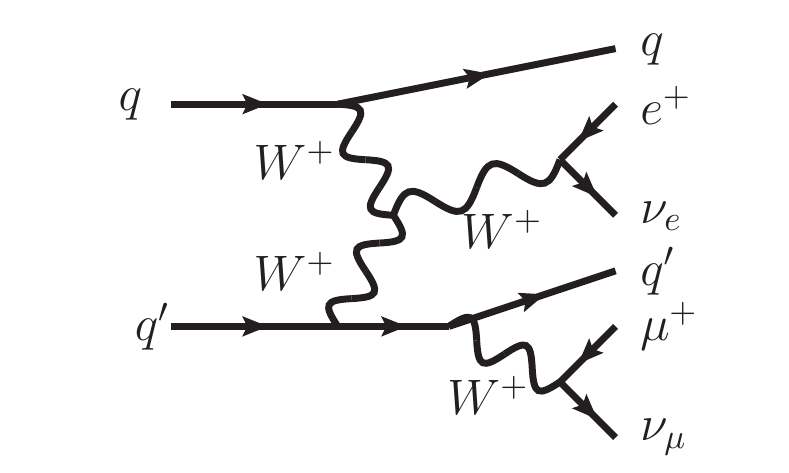} &
\includegraphics[width=0.35\textwidth]{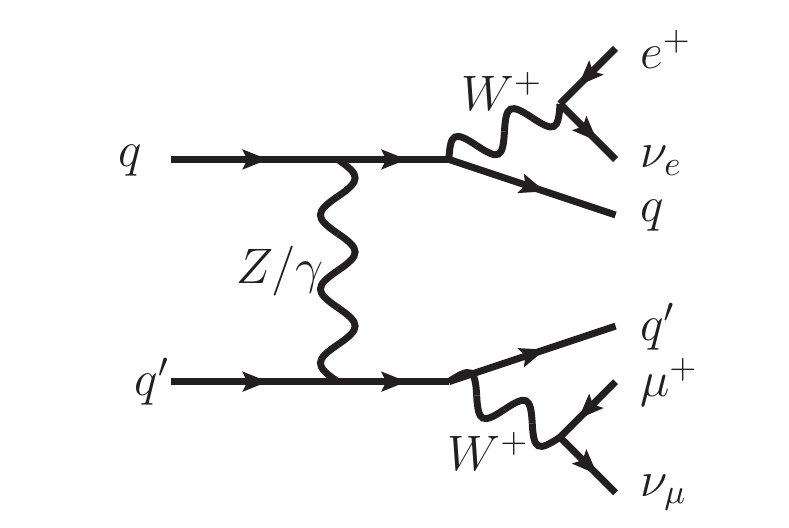} &
\includegraphics[width=0.35\textwidth]{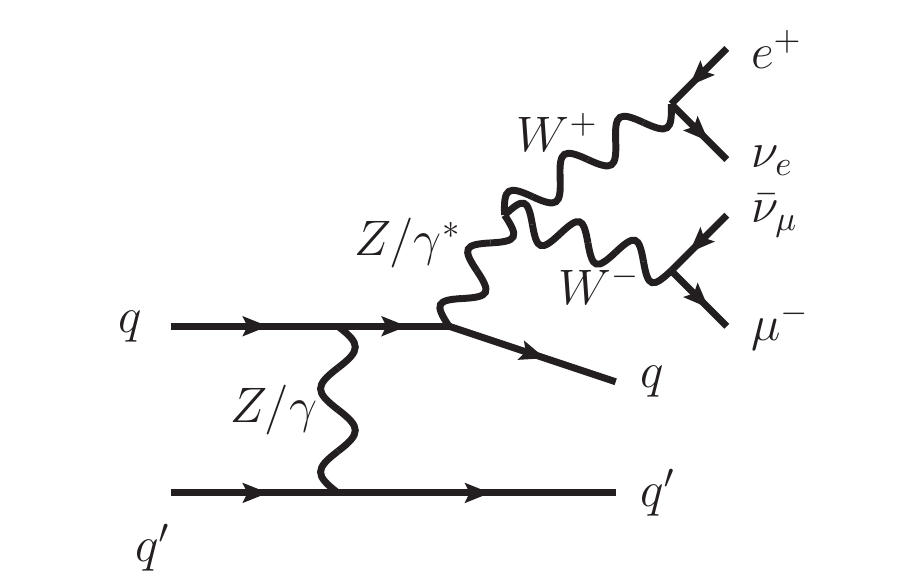} \\[1em]
\includegraphics[width=0.35\textwidth]{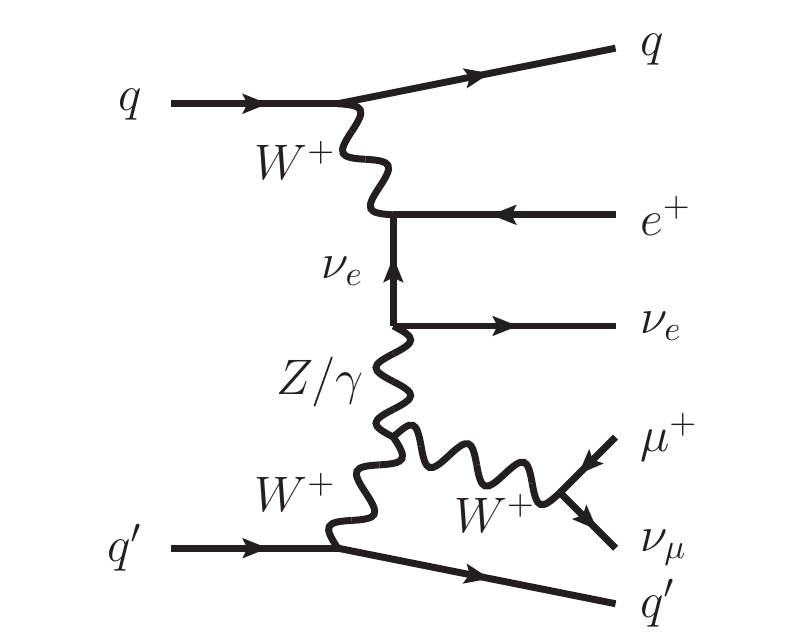} \\[1em]
\end{tabular}
\caption{Typical Feynman diagrams appearing in VBF production. Diagrams
in the left and center columns are for the process $pp \rightarrow e^+ \nu_e
\mu^+ \nu_\mu j j$ (``VBF-$W^+W^+jj$''), while the right column shows
additional contributions appearing for $pp \rightarrow e^+ \nu_e
\mu^- \bar{\nu}_\mu j j$ (``VBF-$W^+W^-jj$''). The upper row contains
the signature diagrams giving rise to the name of the process class,
while in the middle row there are additional diagrams needed to form a
gauge-invariant set. The lower row shows an example for a non-resonant
contribution.
}
\label{fig:fm_wpwplo}
\end{figure}

Typical Feynman diagrams contributing to VBF are depicted in
\fig{fig:fm_wpwplo}. We will mostly use same-sign $W$ pair production
via VBF,
\begin{equation}
pp \rightarrow W^+ W^+ jj \rightarrow e^+ \nu_e \mu^+ \nu_\mu j j + X \quad
\text{(``VBF-$W^+W^+$'')}\,,
\label{proc:w+w+jj}
\end{equation}
as an example in this chapter, pointing out differences or
additional contributions appearing in other processes when necessary.

The Feynman diagrams shown in the upper row of \fig{fig:fm_wpwplo} are
responsible for the name of this process class. Looking only at the
inner part, the two virtual $W$ bosons emitted off the quark or
anti-quark lines and the two final-state $W$ bosons describe 
a $2\rightarrow2$ scattering process. The two incoming bosons
are, however, space-like ($q^2<0$) and not on-shell as in a real scattering
process. The interaction part can either be direct via a quartic gauge
coupling (upper left diagram), or via exchange of a space-like boson
(upper center).  In our example this could be a photon, $Z$ boson or a
Higgs boson. For other final states, also $s$-channel diagrams can appear
(upper right).

To end up with a gauge-invariant set of Feynman diagrams, we need to
add some other contributions as well which are shown in the middle row of
\fig{fig:fm_wpwplo}. The final state $W^+$ bosons can also be emitted
directly off the quark lines, either only one of them (middle left) or
both (middle center). Also diagrams with one boson radiated off a quark
line and then splitting into the two final-state quarks via a triple
gauge coupling (middle right) can appear for other processes.

In general we will consider not only on-shell production of the
final-state bosons, but also include their decays, as already alluded to
by the process definition \eq{proc:w+w+jj}. The vector bosons can either
decay into a pair of leptons or a pair of quarks. The latter mode is
usually called hadronic decays, as the particles observed in the
detector are hadrons. In case of two final-state vector bosons, the
nomenclature is as follows:
\begin{itemize}
\item leptonic: both vector bosons decaying into leptons,
\item hadronic: both vector bosons decaying into quarks,
\item semi-leptonic: one vector boson decaying into leptons, the other into quarks.
\end{itemize}

When considering leptons or quarks as final states, additional
non-resonant diagrams appear. An example is depicted in the lower row of
\fig{fig:fm_wpwplo}. The additional contribution of such diagrams is
typically rather small, of the order of a few percent at maximum, as
these diagrams lack at least one resonant contribution and carry a suppression
factor of order $\Gamma/M$. Exceptions can
however occur where phase-space cuts force one of the bosons off-shell.
The same is true for the contribution of virtual photons, which 
appear in the diagrams instead of a $Z$ boson.

The experimental tell-tale to distinguish this process class from QCD-induced
production mechanisms and others are, however, the two quarks in the final
state, which appear as hadronic jets in the detector, called tagging jets. 
On a historic side note, very early
papers~\cite{Cahn:1983ip,Dawson:1984gx,Duncan:1985vj,Butterworth:2002tt},
targeted for the SSC, a
planned but never finished proton-proton collider in the US with a target
center-of-mass energy of 40 TeV, considered to not resolve these. Instead, one
would define the gauge bosons as additional partons, similar to the recent
inclusion of photons in parton distribution functions (PDFs), and compute a
$2\rightarrow2$ scattering process with on-shell incoming gauge bosons.
Only later was it realized that an explicit inclusion of the tagging
jets is needed~\cite{Cahn:1986zv,Kleiss:1987cj,Barger:1988mr}.

The tagging jets are typically in the forward regions of the detector. One can see
this from the following argument~\cite{Zeppenfeld:1999yd,Plehn:2009nd}.
Let us for simplicity take VBF-Higgs production, and assume a flavor
combination of the external quarks that allows only for $W$ exchange,
$q_1(p_1) q_2(p_2) \rightarrow q_3(p_3) q_4(p_4) H(p_5)$.
Then a short analytic calculation shows that the squared matrix element is proportional
to~\cite{Cahn:1983ip,Dicus:1985zg,Altarelli:1987ue,Kilian:1995tr,Djouadi:2005gi}
\begin{equation}
|\Mfi|^2 \propto \frac{p_1 \cdot p_2 \ p_3 \cdot p_4}{(q_1^2 -
M_W^2)^2 (q_2^2 - M_W^2)^2 } \,,
\end{equation}
where $q_1 = p_1-p_3$ and $q_2 = p_2-p_4$ denote the momenta of the two
virtual $W$ bosons.

In order to maximize this expression for a fixed partonic center-of-mass
energy $\sqrt{\hat{s}}= \sqrt{\mathstrut 2 \; p_1\cdot p_2}$, we have two possibilities. First, we
can increase the numerator, \ie require that $p_3 \cdot p_4$ is large.
This condition means that the invariant mass of the two tagging jets
$m_{jj} = \sqrt{\mathstrut 2\; p_3 \cdot p_4}$ should be fairly high. Second, we should make
the denominator small. As all quarks are taken as massless, we can use 
\begin{equation}
q_1^2 = -2 \; p_1 \cdot p_3 = -2 E_1 E_3 (1-\cos\theta_1) 
= - \frac2{1+\cos\theta_1} \frac{E_1}{E_3} \; p_{T,3}^2 \,, 
\end{equation}
where $\theta_1$ describes the scattering angle between $\vec{p}_1$ and
$\vec{p}_3$, and the absolute value of the transverse component of $\vec{p}_3$
is given by $p_{T,3} = E_3 \sin\theta_1$.
The upper bound of $q_1^2$ is at zero, so the inverse $W$-boson propagator
becomes small for values of $q_1^2$ close to zero, when the scattering
angle becomes small. There, we can rewrite the inverse propagator as 
$(q_1^2 - M_W^2) \simeq - (\frac{E_1}{E_3} p_{T,3}^2 + M_W^2)$.

From this we see that the transverse momentum of the $W$, which is equal to the
transverse momentum of the final-state quark, should not significantly exceed
$M_W$ in order to not incur an additional suppression factor.  At the same
time, the quark should have a large momentum component in the $z$
direction to yield a small scattering angle. The $W$ boson will
typically carry only a fairly small portion of the momentum of the
incoming parton, similar to photon radiation from an energetic electron.
But it also needs to carry enough energy to produce the final-state
Higgs boson, namely around $\frac{M_H}2$. 

Hence, the two tagging jets are expected to have large energies with
only moderate transverse momenta, putting them into detector regions with
fairly small scattering angle or large pseudorapidity. 

\begin{figure}
\begin{center}
\includegraphics[width=0.45\textwidth]{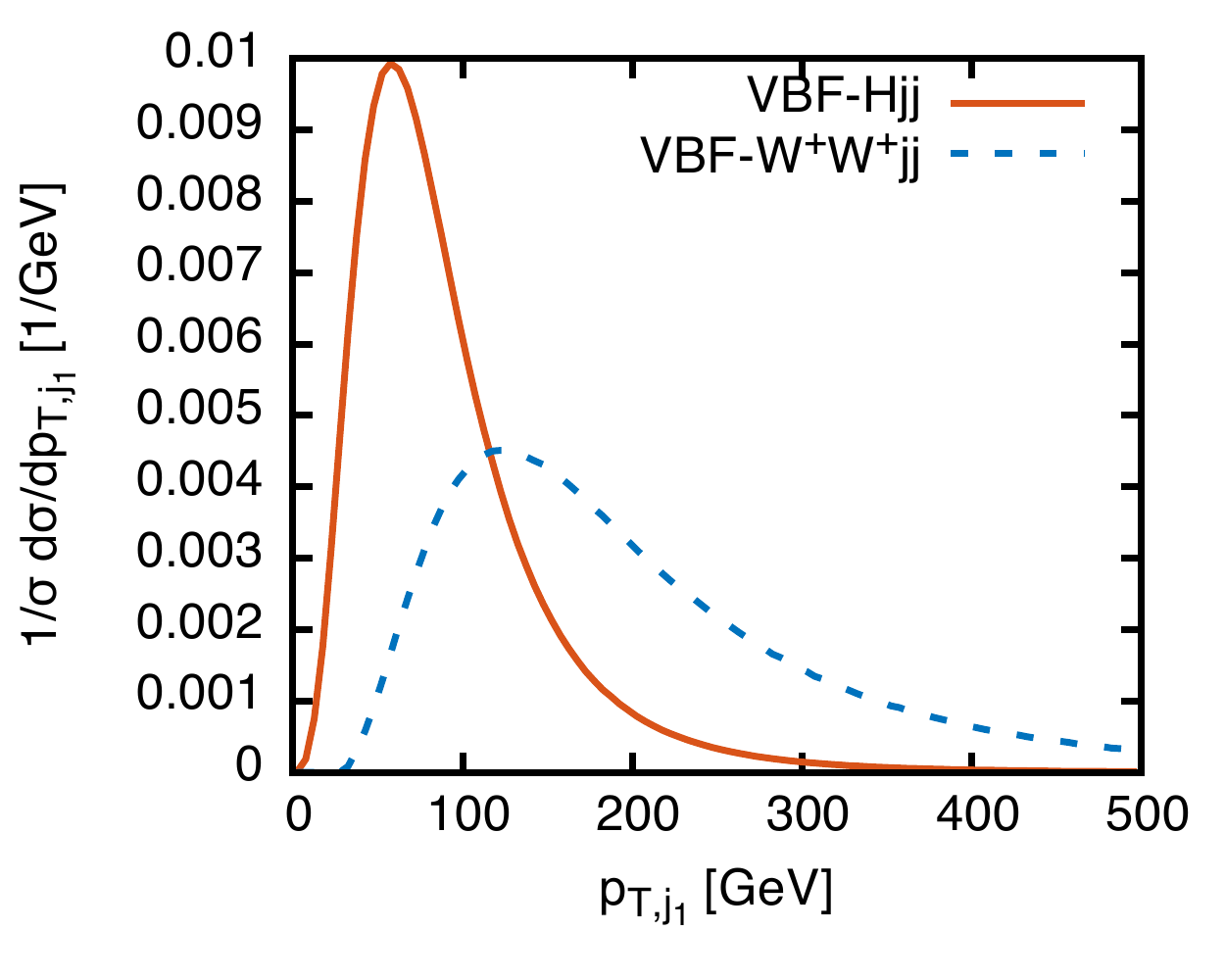} 
\includegraphics[width=0.45\textwidth]{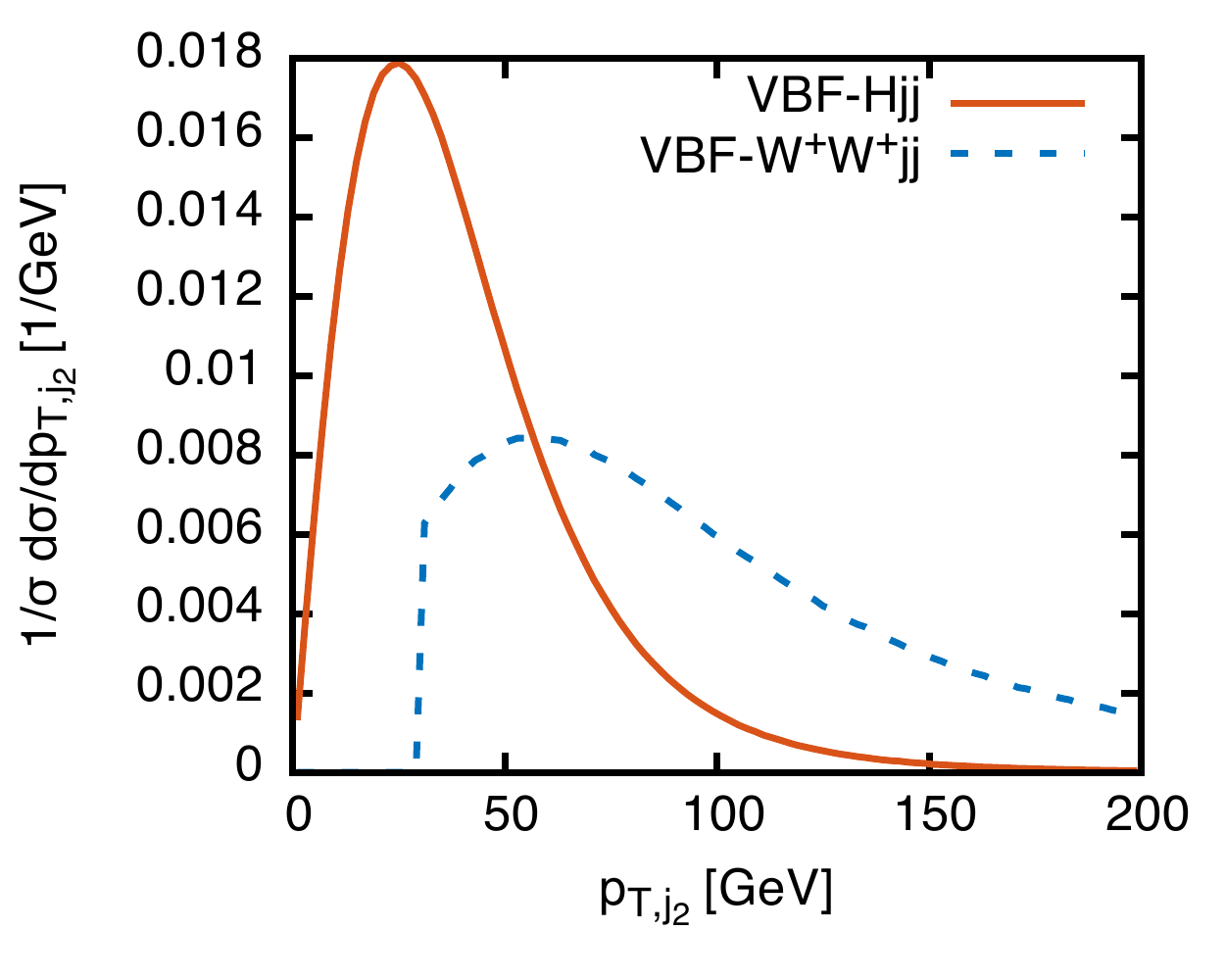} \\[1em]
\includegraphics[width=0.45\textwidth]{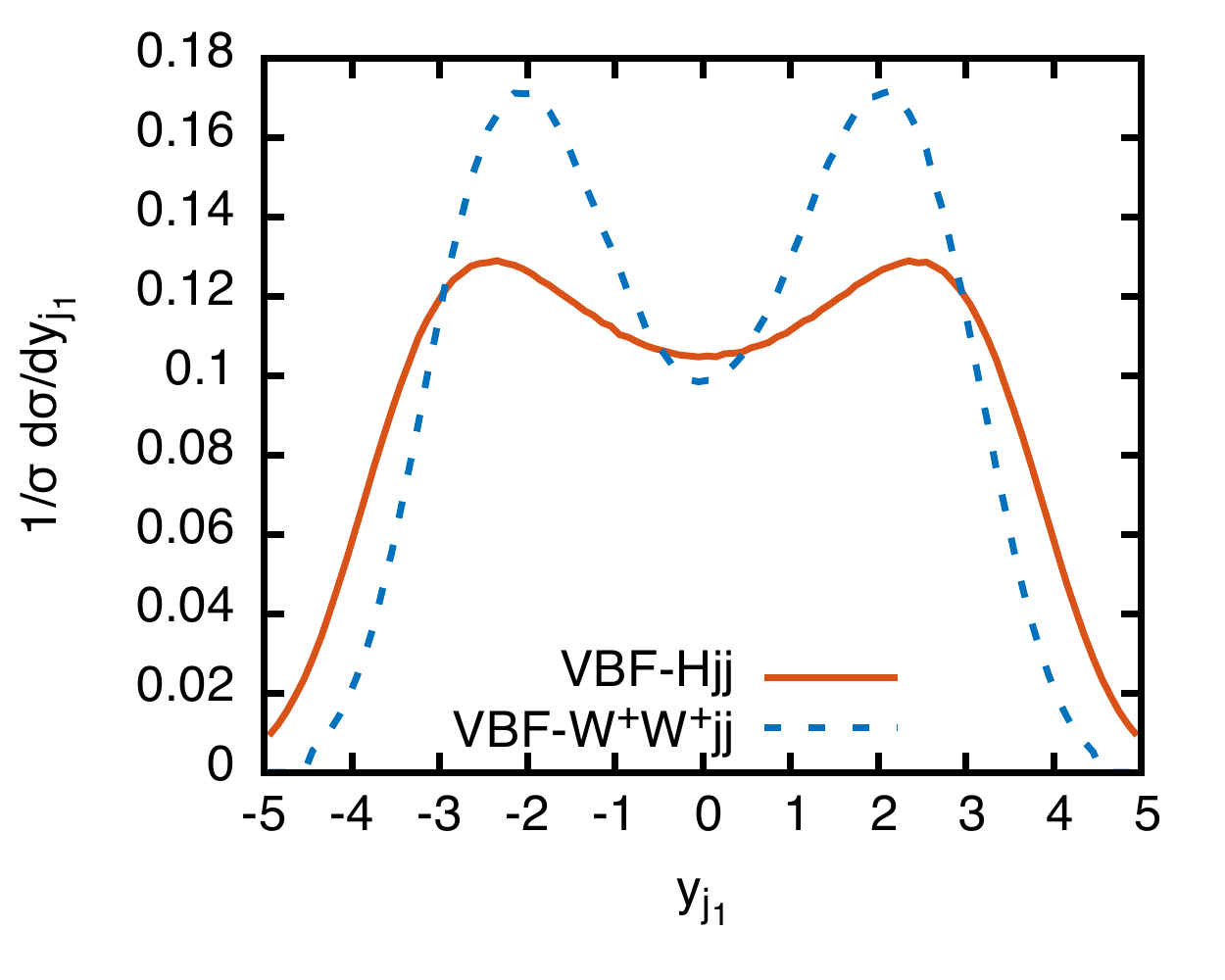} 
\includegraphics[width=0.45\textwidth]{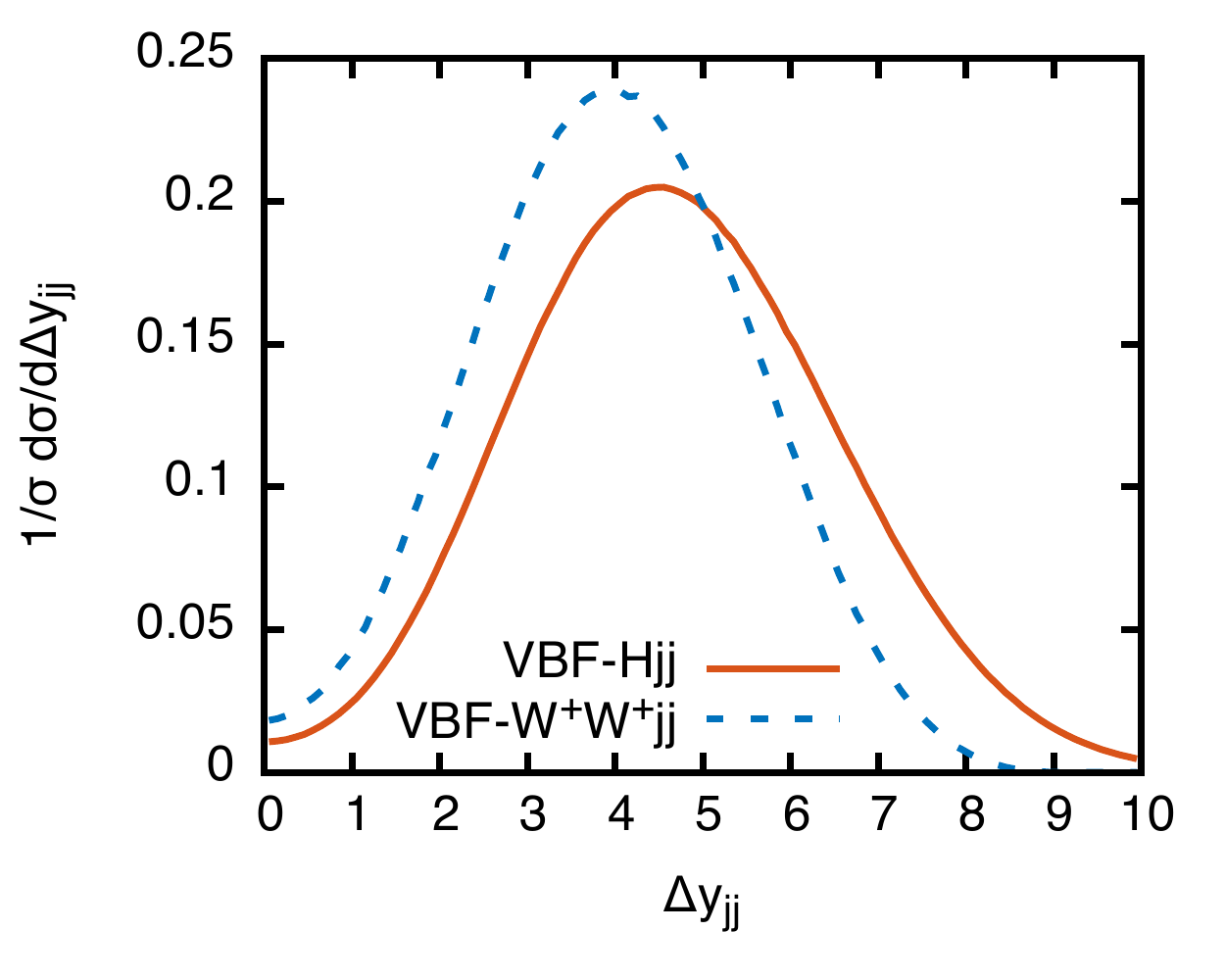} \\[1em]
\includegraphics[width=0.45\textwidth]{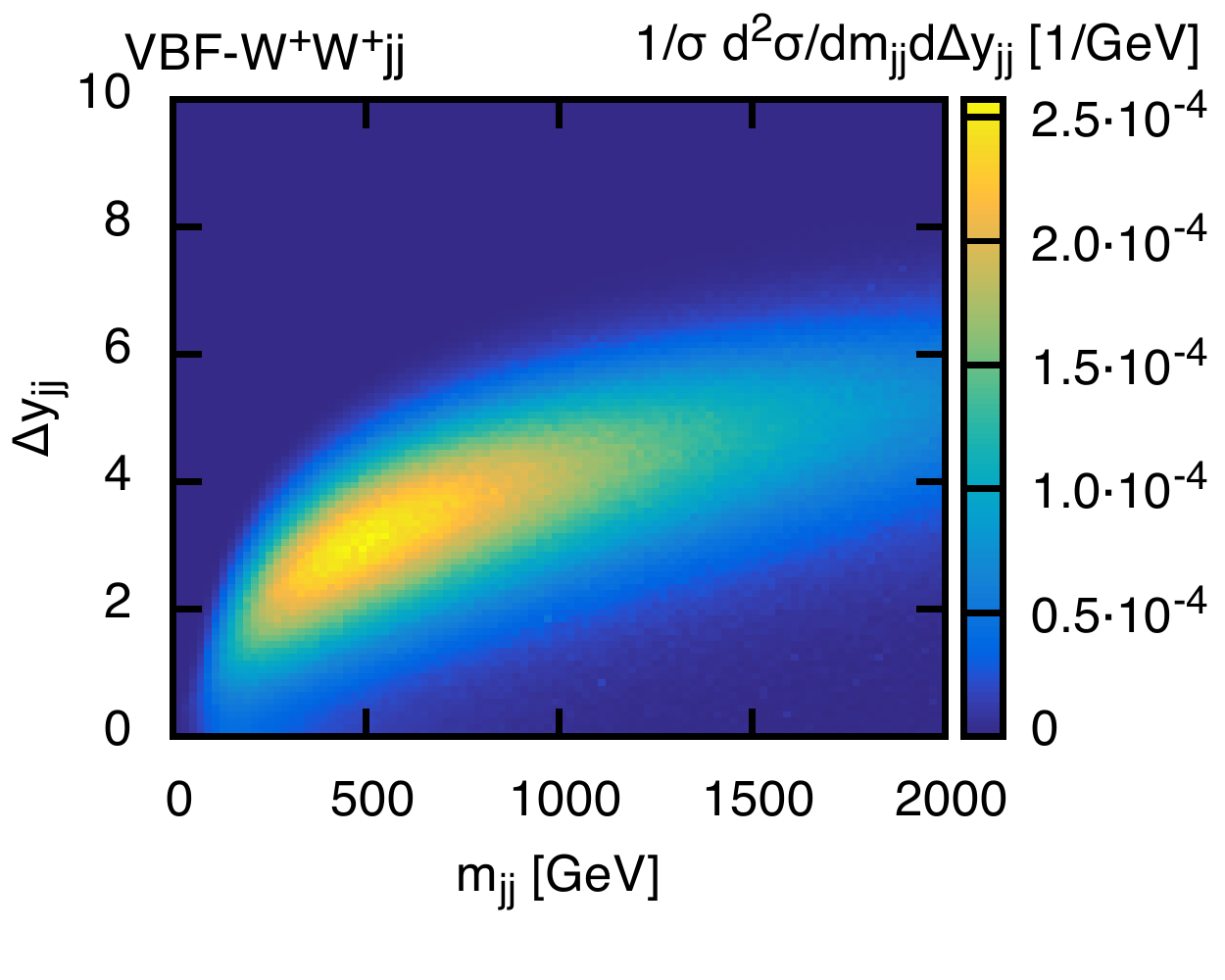} 
\includegraphics[width=0.45\textwidth]{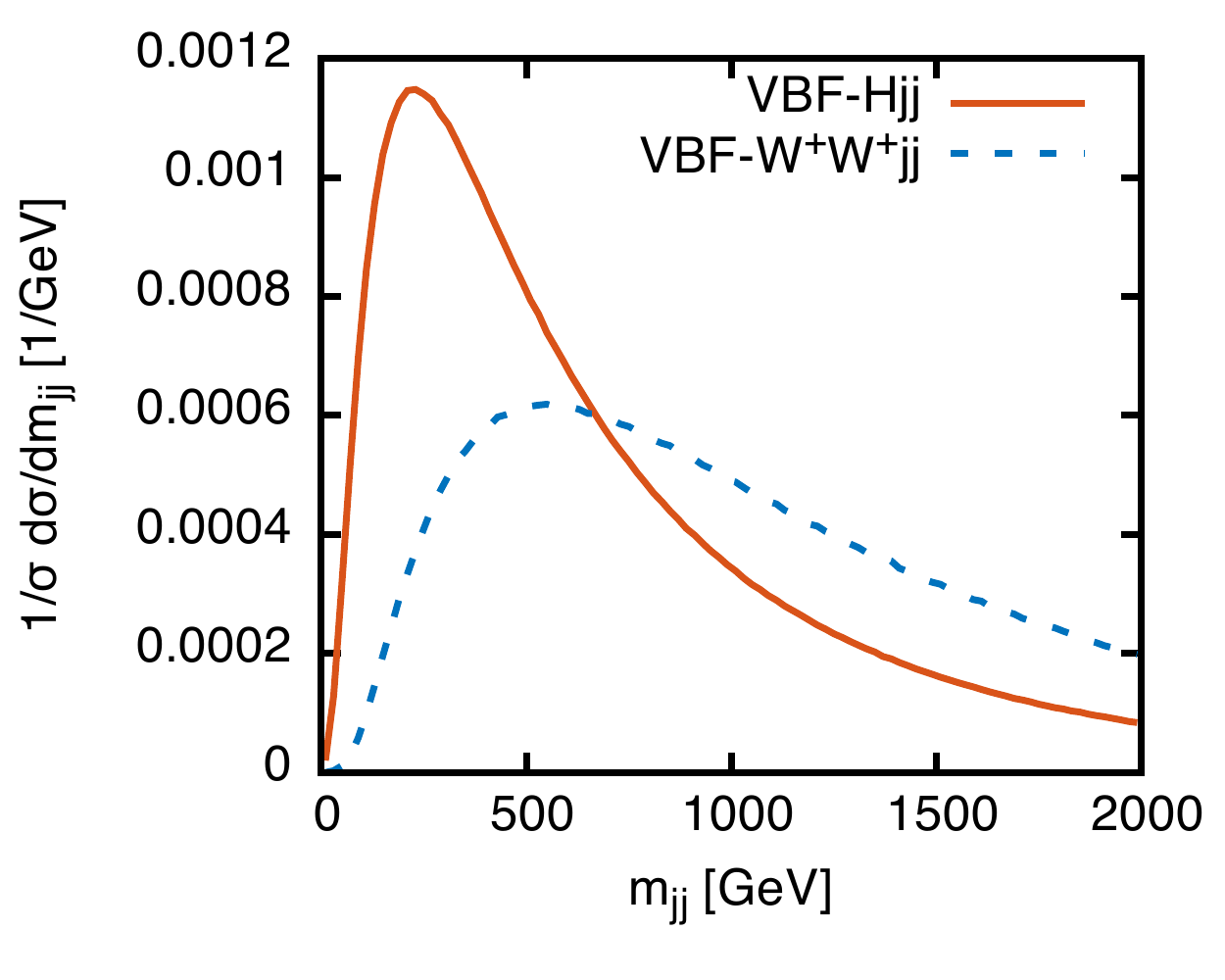} 
\end{center}
\caption{Normalized LO differential distributions for VBF production of
$Hjj$ (\textit{solid red}) and $W^+W^+jj$ (\textit{dashed
blue} and \textit{bottom left}) for the LHC at 13 TeV center-of-mass
energy.
Distributions presented are the transverse momentum of the leading and
the second jet (\textit{top row}), the rapidity of the leading jet
(\textit{middle left}), and the rapidity difference (\textit{middle right})
and invariant mass (\textit{bottom right}) of the two jets. The
bottom left panel shows the correlation plot between invariant mass and
rapidity difference of the two jets for VBF-$W^+W^+jj$. 
}
\label{fig:tagjetprops}
\end{figure}
Several normalized differential distributions illustrating these
features at leading order (LO) are shown in \fig{fig:tagjetprops}. This
and the following figure have been generated with
\VBFNLO~\protect\cite{Arnold:2008rz,Baglio:2014uba,VBFNLO}.  As
processes, we use both VBF-$W^+W^+jj$ and VBF-$Hjj$ production. For the
latter no cuts are applied on the final-state jets. This still gives a
finite and well-defined cross section, as no singular regions appear for
example when the jet transverse momentum approaches zero. To
VBF-$W^+W^+jj$ production we instead apply standard jet cuts, namely
$p_{T,j}>30\GeV$, $|y_j|<4.5$ and $R_{jj} > 0.4$.

In the upper left panel we plot the transverse momentum of the leading
jet, \ie the jet with the largest transverse momentum, and the
transverse momentum of the second jet on the right. Consistent with our
discussion before, both distributions peak at fairly small values of the
transverse momentum. For $Hjj$ the value is smaller and extends less to
larger values, as less energy is required to produce the Higgs boson.
Also in the Higgs case the exchange is dominated by longitudinal gauge
bosons, while they are mostly transverse for same-sign $W$ production.
The different polarizations lead to a relative factor of
$\frac{p_T^2}{M^2}$ for transverse over longitudinal, thus processes
with transverse polarizations prefer larger transverse momenta.
The jet $p_T$ of $W^+W^+jj$ averaged over both jets however also
exhibits a maximum at only 80~GeV, which corresponds to our previous
estimate.
In the middle left panel, where the rapidity of the leading jet is
plotted, we see that central jets around $y=0$ are suppressed compared
to forward jets, with the maximum around $|y|=2$. Consequently, there is
a fairly large rapidity gap between the two jets of about $4$ on
average. The jet cuts impose an upper limit of $\Delta y_{jj}<9$ for
VBF-$W^+W^+jj$ production. The invariant mass of the two jets is plotted
in the lower right panel. Again, this distribution extends to fairly
large values as expected, with $W^+W^+jj$ production generating more
energetic events. Finally, the correlation plot between the invariant
mass and the rapidity difference of the two jets for $W^+W^+jj$ is shown
on the bottom left of \fig{fig:tagjetprops}. These are the two variables
which are typically used for cuts to enhance the VBF contribution over
other processes. The correlation between the two variables can also be
seen from the analytic formula
\begin{equation}
m_{jj}^2 \simeq 2 p_{T,j_1} p_{T,j_2} \left( \cosh(\Delta y_{jj}) -
\cos(\Delta \phi_{jj}) \right) \,,
\end{equation}
where $\Delta \phi_{jj}$ denotes the azimuthal angle difference between
the two jets. The formula becomes exact if the two jets are massless,
which is the case for our LO discussion here.

\begin{figure}
\begin{center}
\includegraphics[width=0.45\textwidth]{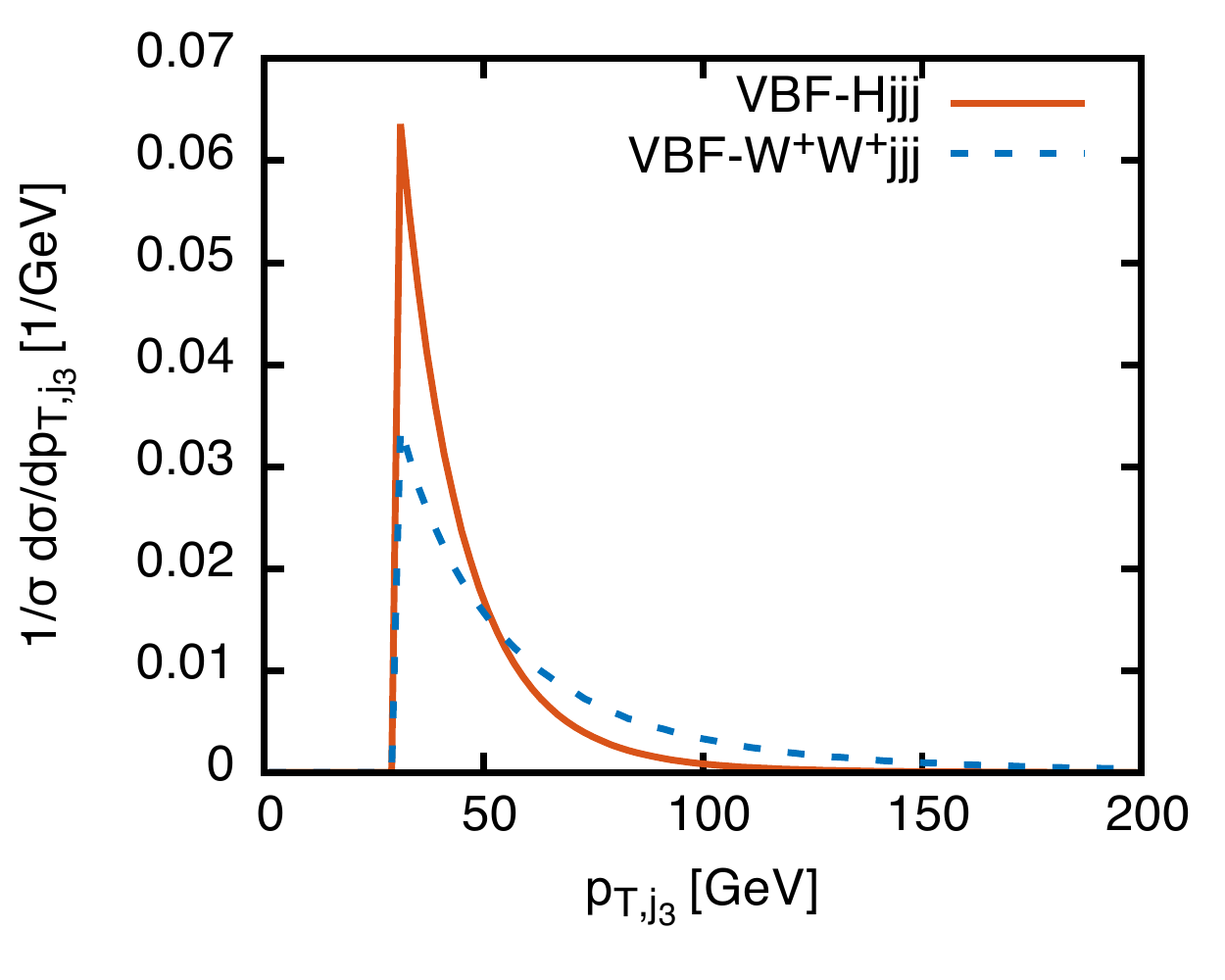} 
\includegraphics[width=0.45\textwidth]{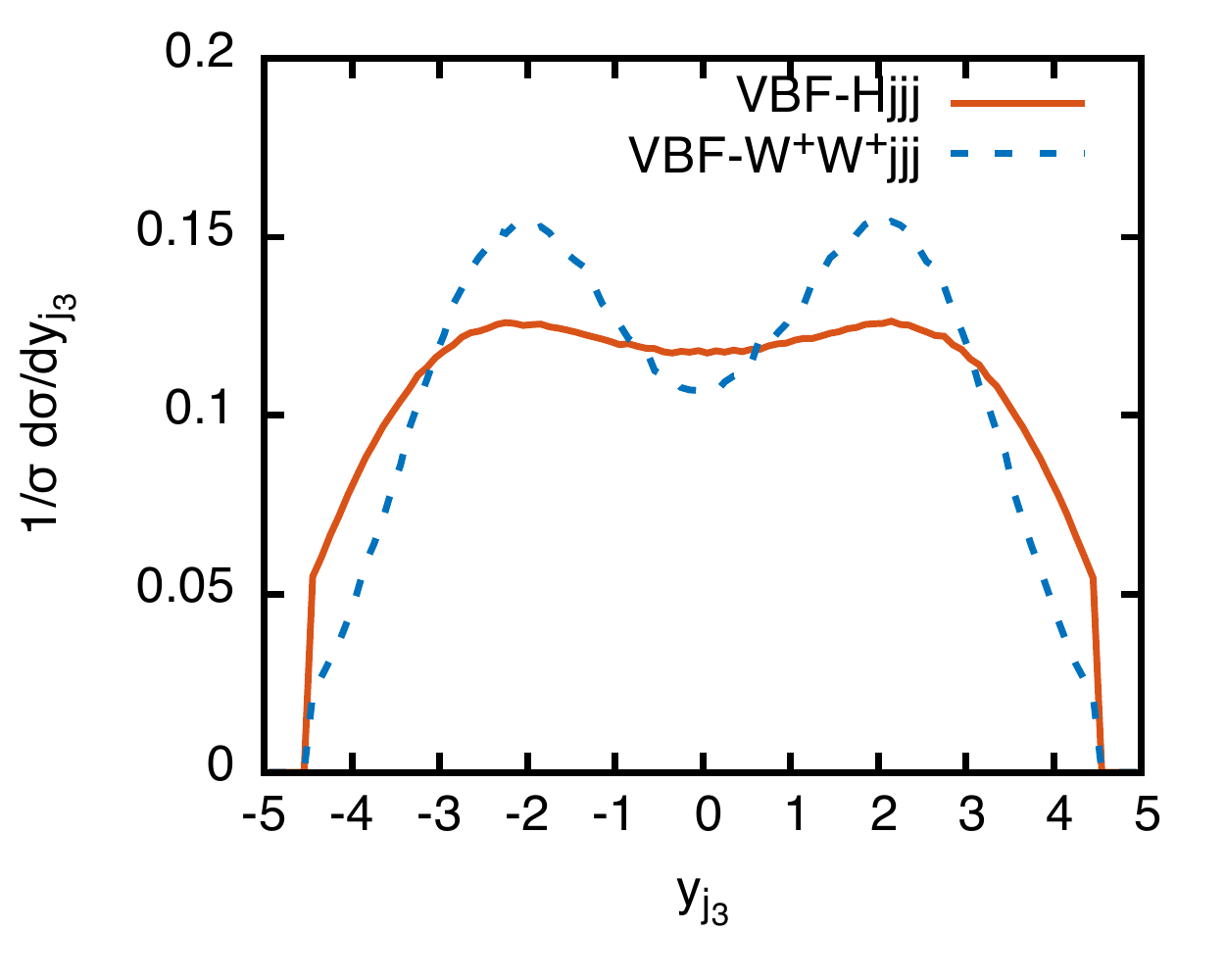} \\[1em]
\includegraphics[width=0.45\textwidth]{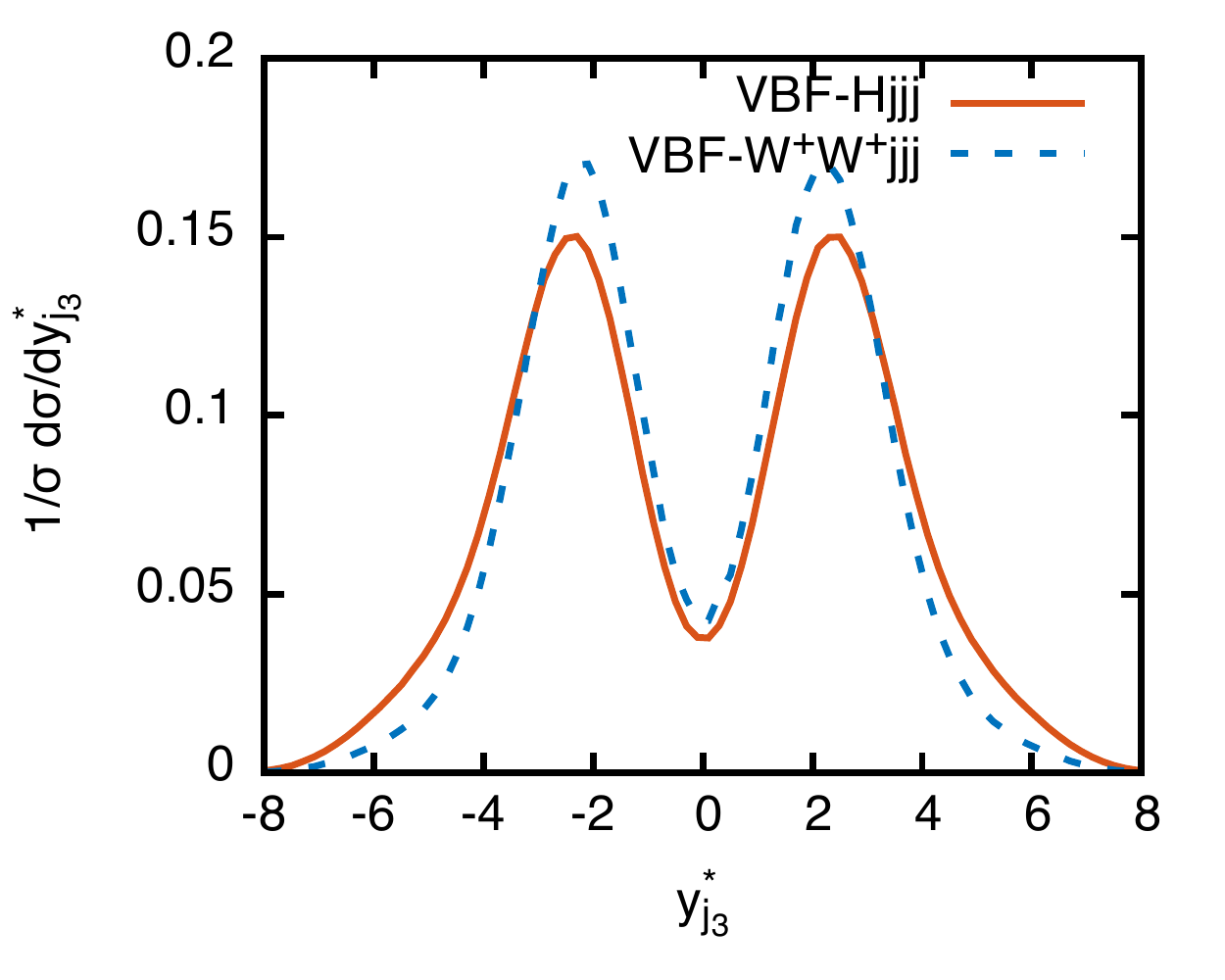} 
\includegraphics[width=0.45\textwidth]{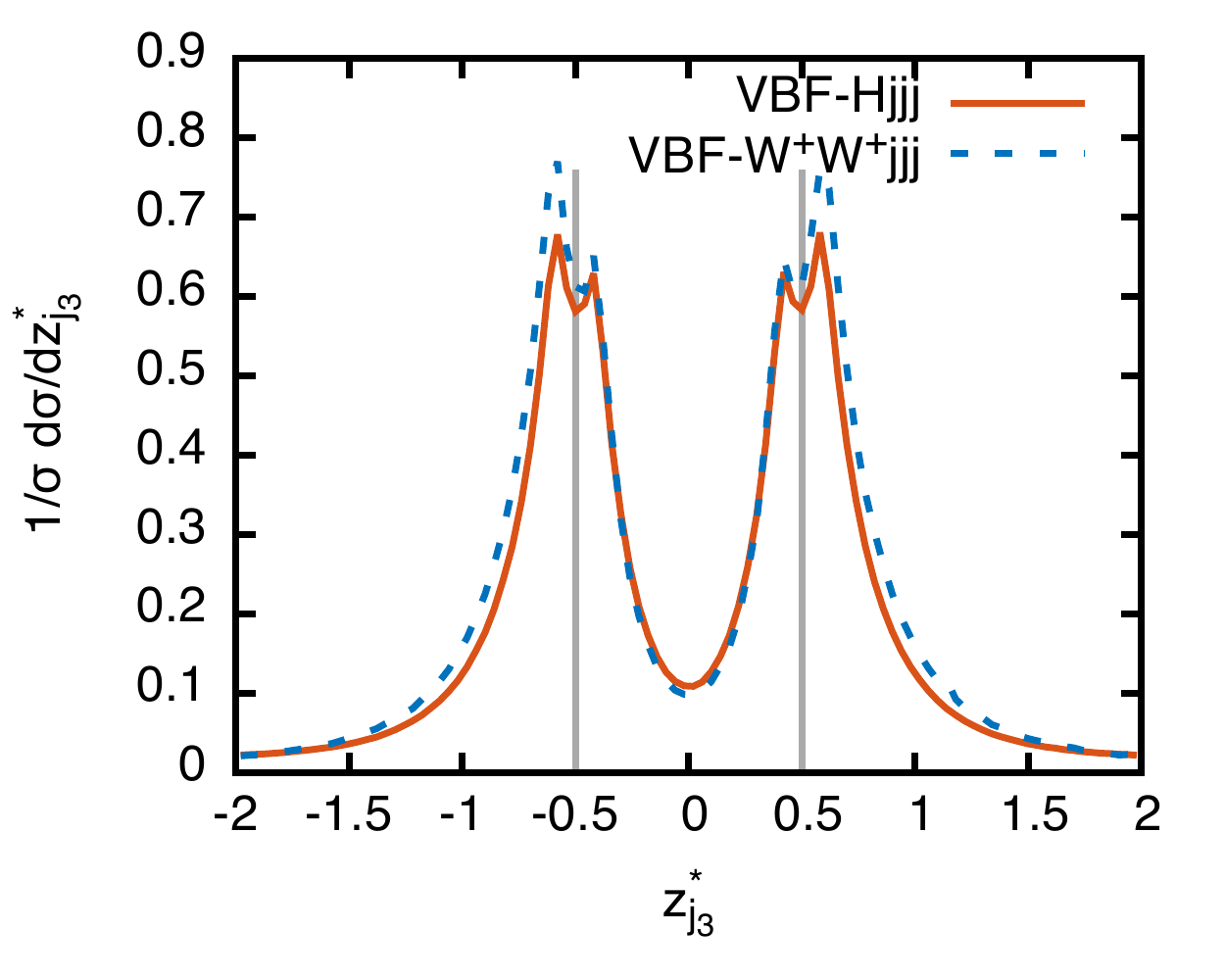} 
\end{center}
\caption{Normalized differential distributions for VBF production plus extra
jet of $Hjjj$ (\textit{solid red}) and $W^+W^+jjj$ (\textit{dashed
blue}) for the LHC at 13 TeV center-of-mass energy.  Distributions show
the transverse momentum (\textit{upper left}) and the rapidity of the
third jet (\textit{upper right}). In the lower panel the third-yet
rapidity relative to the averaged rapidity of the two tagging jets
(\textit{lower left}) and additionally normalized to the position of the
two tagging jets (\textit{lower right}) is plotted.
}
\label{fig:thirdjetprops}
\end{figure}
A crucial question for experimental analyses is whether this central
rapidity gap with reduced jet activity is only an artifact of performing
a LO calculation, or stable also when adding extra jet radiation. For a
first discussion of this issue, we show results for the two processes
with an extra jet emission, \ie electroweak $Hjjj$ and $W^+W^+jjj$
production. In terms of Feynman diagrams, we need to attach a gluon to
the quark lines at all possible positions, keeping in mind that the
interference of two diagrams, where the gluon is attached to different
quark lines, vanishes due to the color structure. Additionally,
processes with the gluon crossed into the initial state need to be taken
into account. As these processes are no longer finite without jet cuts,
we impose the aforementioned cuts now also to $Hjjj$ production.
In the upper left panel of \fig{fig:thirdjetprops}, first the
transverse momentum of the third jet, ordered by decreasing transverse
momentum, is plotted. The extra jet is preferably close to the minimal
cut value of 30~GeV. The upper right panel shows the rapidity of the
third jet, ordered by decreasing transverse momentum. A dip in the
central region is also visible for this distribution, similar in shape
to the one for the two leading jets. There, we have seen that the
difference in rapidity exhibits an even stronger dip at zero than the
rapidities themselves. Similarly, one can define a variable $y^*$,
which gives the rapidity of the $n$-th jet relative to the two tagging
jets~\cite{Rainwater:1996ud},
\begin{equation}
y^*_{j_n} = y_{j_n} - \frac{y_{j_1}+y_{j_2}}2 \,.
\label{eq:ystar}
\end{equation}
This is shown in the lower left panel of \fig{fig:thirdjetprops}, which
indeed exhibits a pronounced dip at zero. As we have seen in
\fig{fig:tagjetprops}, the rapidities of the tagging jets have a maximum
at about $\pm2.5$, which is also the maximum for the $y^*_{j_3}$
distribution. Hence, one can already conclude that the extra jet
emission is preferably close to the position of one of the tagging jets. 
Alternatively, via an additional scaling factor one can make this even
more obvious and fix the position of the two tagging jets by
defining~\cite{Schissler:2014nga}
\begin{equation}
z^*_{j_n} = \frac{y^*_{j_n}}{\left|y_{j_1}-y_{j_2}\right|} \,,
\label{eq:zstar}
\end{equation}
where the two tagging jets are located at $z^*=\pm \frac12$. This
divides the phase space in the central rapidity gap, $|z^*|<\frac12$, and
the region between tagging jets and beam axis, $|z^*|>\frac12$. 
This distribution, plotted in the lower right panel of
\fig{fig:thirdjetprops}, indeed has maxima around $\pm0.5$. The small dip
at exactly $\pm0.5$ is due to the jet separation cut $R_{jj}$, which
removes events where the two jets are very close together.

\subsection{Triboson and QCD-induced Production}
\begin{figure}
\begin{center}
\begin{tabular}{cc}
\includegraphics[width=0.4\textwidth]{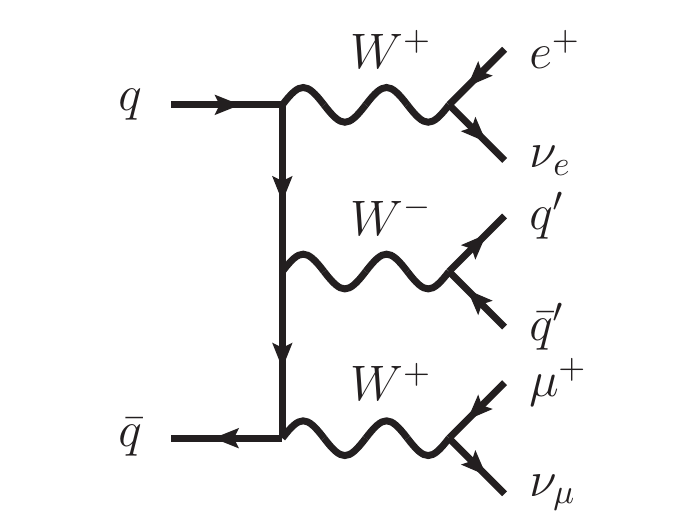} &
\includegraphics[width=0.4\textwidth]{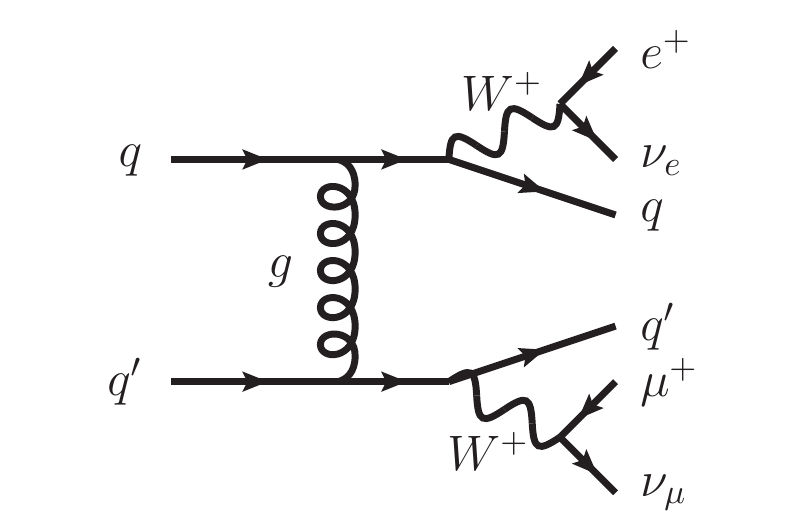} \\[1em]
\includegraphics[width=0.4\textwidth]{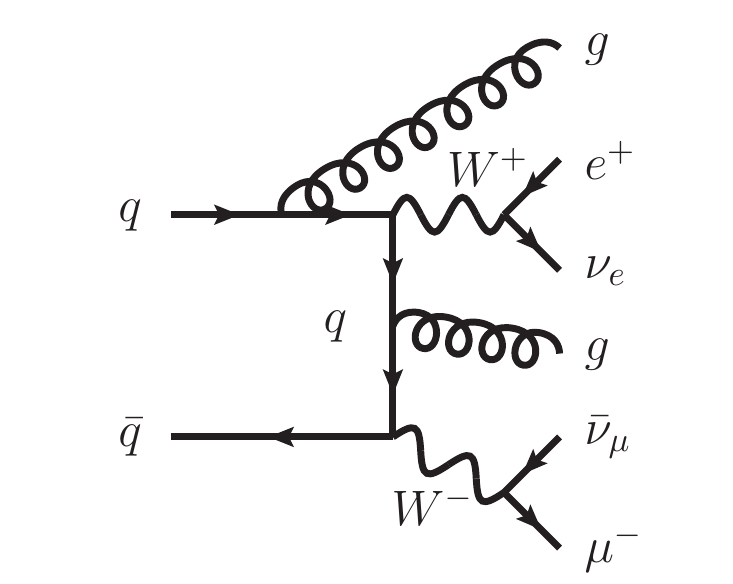} &
\includegraphics[width=0.4\textwidth]{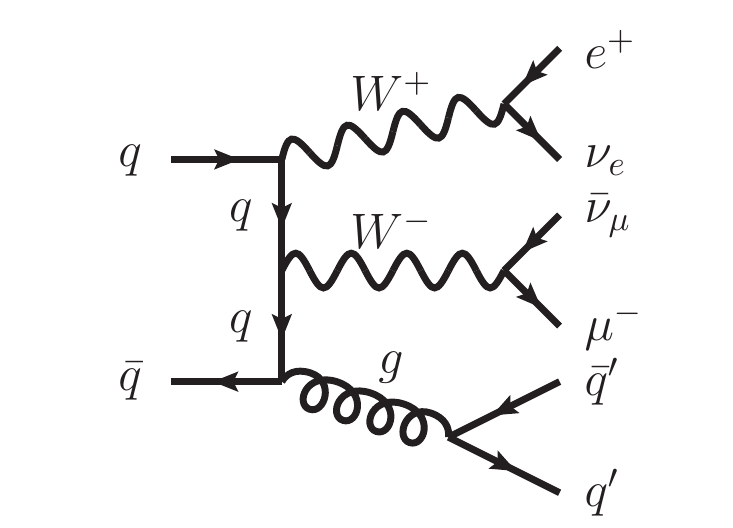} 
\end{tabular}
\end{center}
\caption{Feynman diagrams of other processes contributing to the same
final state as the VBF process. \textit{Top left:} triboson production,
\textit{top right} and \textit{bottom:} QCD-induced $VVjj$ production.
}
\label{fig:fm_nonvbf}
\end{figure}
Looking at the process definition of VBF processes like
\eq{proc:w+w+jj}, one realizes that there are other possibilities to produce
the same final state. One such option is crossing the quark lines
between initial and final state such that the two incoming quarks are
connected, and hence also the two outgoing ones. An example
Feynman diagram is shown in \fig{fig:fm_nonvbf} on the top left. The
final-state quark--anti-quark pair originates from a gauge boson, so
this process class can be seen as triboson production with one hadronic
and two leptonic decays of the three bosons. From the diagram we can
already guess that the characteristics of the jets will be quite
different from those of VBF processes. The invariant mass of the two
jets is expected to follow a Breit-Wigner distribution centered around
the mass of the corresponding gauge boson and with its width, hence at
much smaller values than in VBF processes. Also the rapidity
distribution of the two jets should be much more central. 

Another possibility is QCD-induced $VV$ production
in association with two
jets~\cite{Melia:2010bm,Melia:2011dw,Melia:2011gk,Jager:2011ms,Greiner:2012im,Campanario:2013qba,Gehrmann:2013bga,Campanario:2013gea,Badger:2013ava,Campanario:2014dpa,Bern:2014vza,Alwall:2014hca,Campanario:2014ioa,Campanario:2014wga,Campanario:2015vqa}.
In these processes a gluon
instead of a vector boson gets exchanged between the two quark lines.
Hence, the coupling order of the process has two powers of the strong
coupling constant $\alpha_s$, and two orders of the electromagnetic
coupling $\alpha$ less than the corresponding VBF process, \ie
$\Order{\alpha_s^2 \alpha^4}$ instead of $\Order{\alpha^6}$. The two
diagrams on the right-hand side of \fig{fig:fm_nonvbf} are examples for
this process class.
Except for same-sign $W$ production, partonic subprocesses with two
quarks and two gluons as external colored particles exist as well, shown
on the bottom left of the figure.
The exchange of a colored particle between the two quark lines again
leads to a markedly different phase-space distribution of the jets. The
jets are going to be much more central than for VBF processes. 

\begin{figure}
\begin{center}
\includegraphics[width=0.6\textwidth]{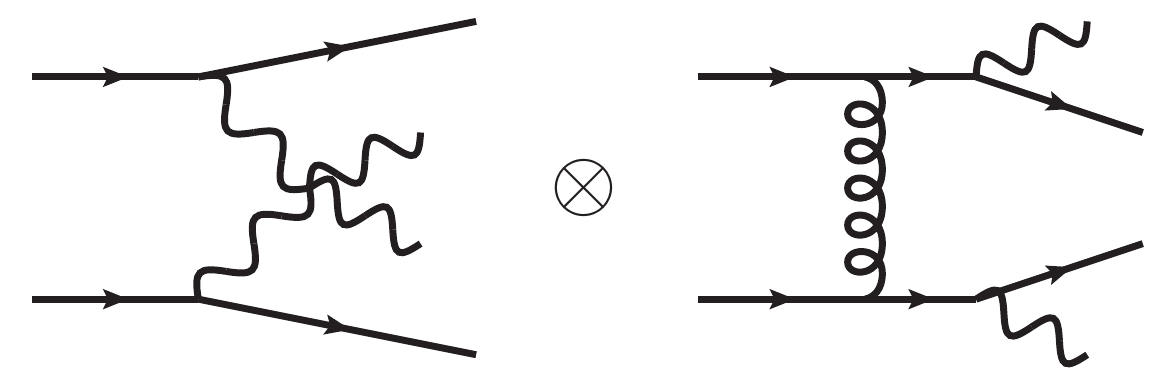} \\[1em]
\includegraphics[width=0.6\textwidth]{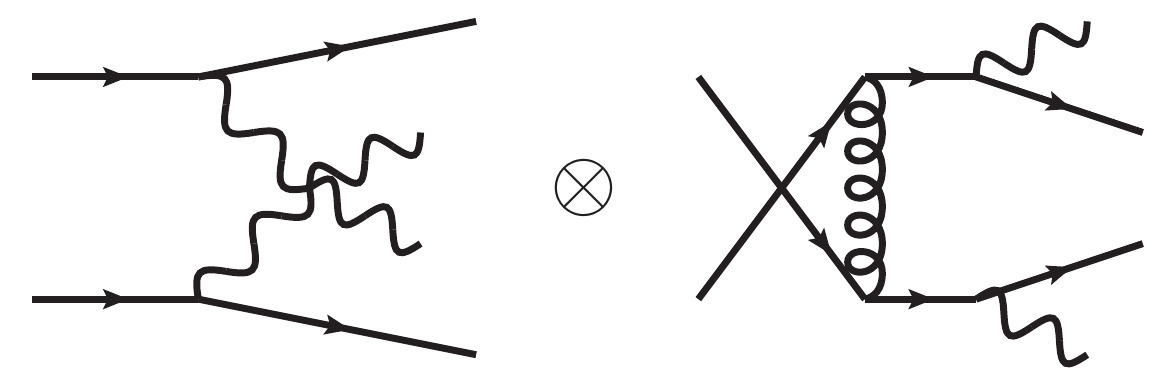} \\[1em]
\includegraphics[width=0.6\textwidth]{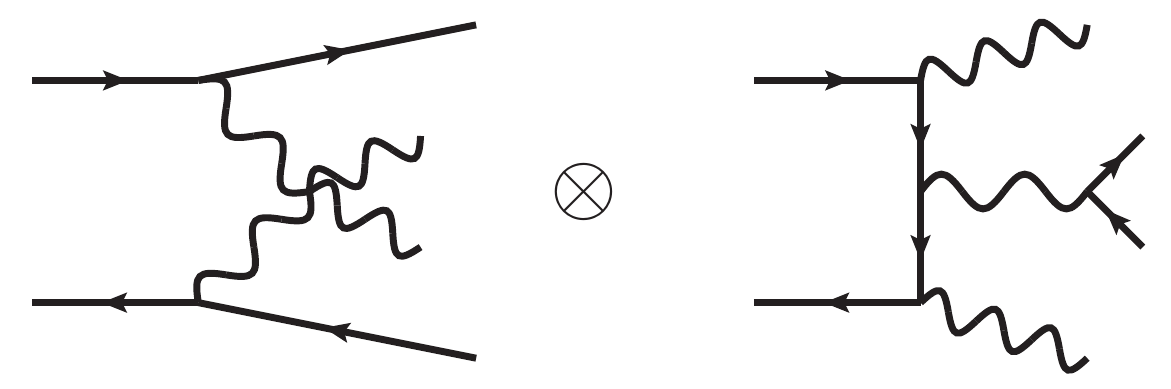} \\[1em]
\end{center}
\caption{Interference contributions between the VBF process and
QCD-induced $VV$ production for different quark flavors (\textit{first
row}), additional contribution for same quark flavors (\textit{middle
row}), and interference with triboson production (\textit{lower
row}).
The decays of the vector bosons have been left out for clarity.
}
\label{fig:fm_interference}
\end{figure}
Interference between these processes and the VBF one needs also to be
taken into account. For partonic subprocesses where the external quark
content is the same, this will in general be non-zero. The
important question is, however, if these effects are large or possibly
small enough to be neglected. 

\subsection{QCD-induced--VBF Interference}
\label{sec:qcdvbfint}

Let us first consider interference between the QCD-induced and VBF
process, when the flavor of the two quark lines is different (upper row
of \fig{fig:fm_interference}). The color part of a quark-gluon vertex is 
$T^a_{ij}$, where $T$ are the generators of $SU(3)$, and propagators are
simply $\delta$-functions of the attached color charges. Hence, at the
upper quark-gluon vertex we find $T^a_{ii} = \Tr[T^a] = 0$.
We can also argue in the following way: The color exchange between the
two quark lines in VBF is always a color singlet. Hence, to have a
non-vanishing interference, the exchange on the right-hand side must be
a color singlet as well. The gluon is however a pure color octet, and so
this contribution actually vanishes exactly. 
In case of identical quark flavors another diagram appears, where the
quark lines are flipped (middle row). In this case we obtain for the
color trace of the color-averaged matrix element squared
$\frac1{N^2} T^a_{ij} T^a_{ji} = \frac{N^2-1}{2N^2} = \frac49$, \ie a
finite contribution. The phase-space characteristics of the final-state
jets is still quite different for the two processes. The preference for
forward jets in VBF would mean for the interference to be maximal that
the jets from the QCD-induced part should mostly go in the backward
region, which is not the case.

Interference between VBF and triboson processes (lower row) can only
happen for quark--anti-quark initial states. The color factor is
$\frac1{N}=\frac13$ for this interference term, compared to $1$ for the
VBF color-averaged squared matrix element. The main reason that this
interference term is also suppressed is the different behavior of the
two-jet invariant mass. The triboson part is strongly peaked at the mass
of the vector boson where the two jets originate from, while the VBF
part again prefers much larger values.

\begin{figure}
 \begin{center}
  \includegraphics[width=0.47\textwidth]{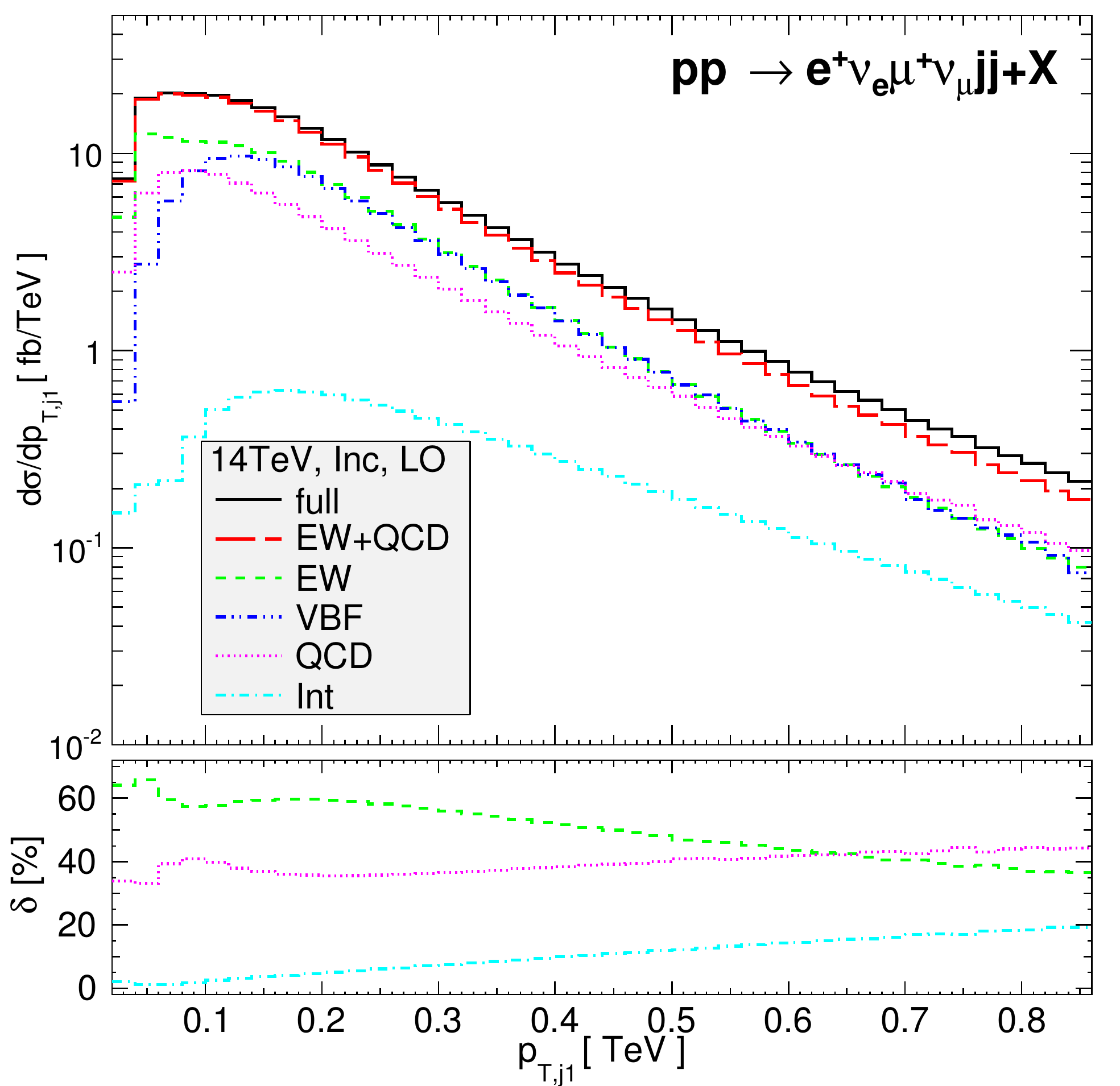}\hfill
  \includegraphics[width=0.47\textwidth]{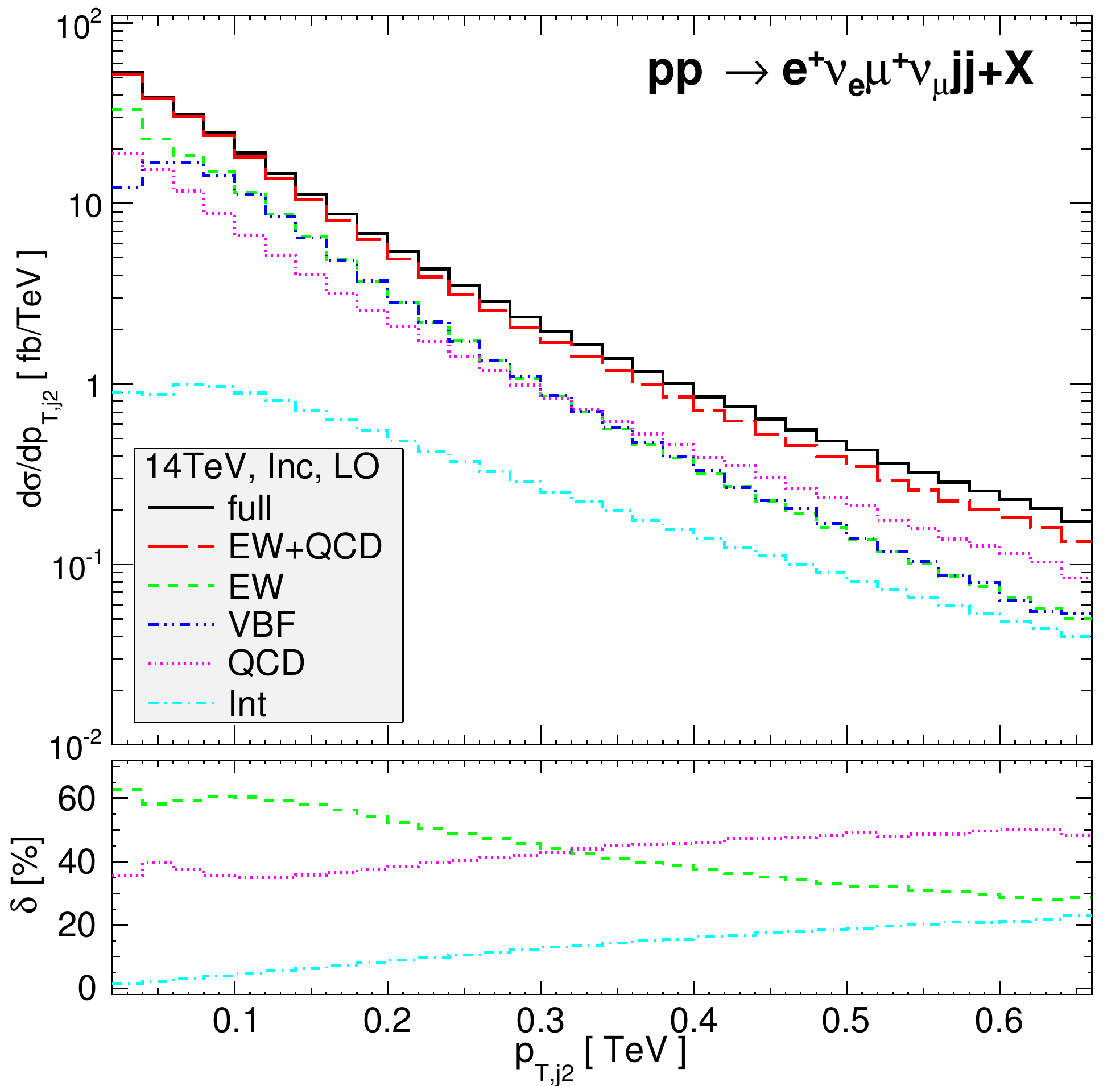}\\
  \includegraphics[width=0.47\textwidth]{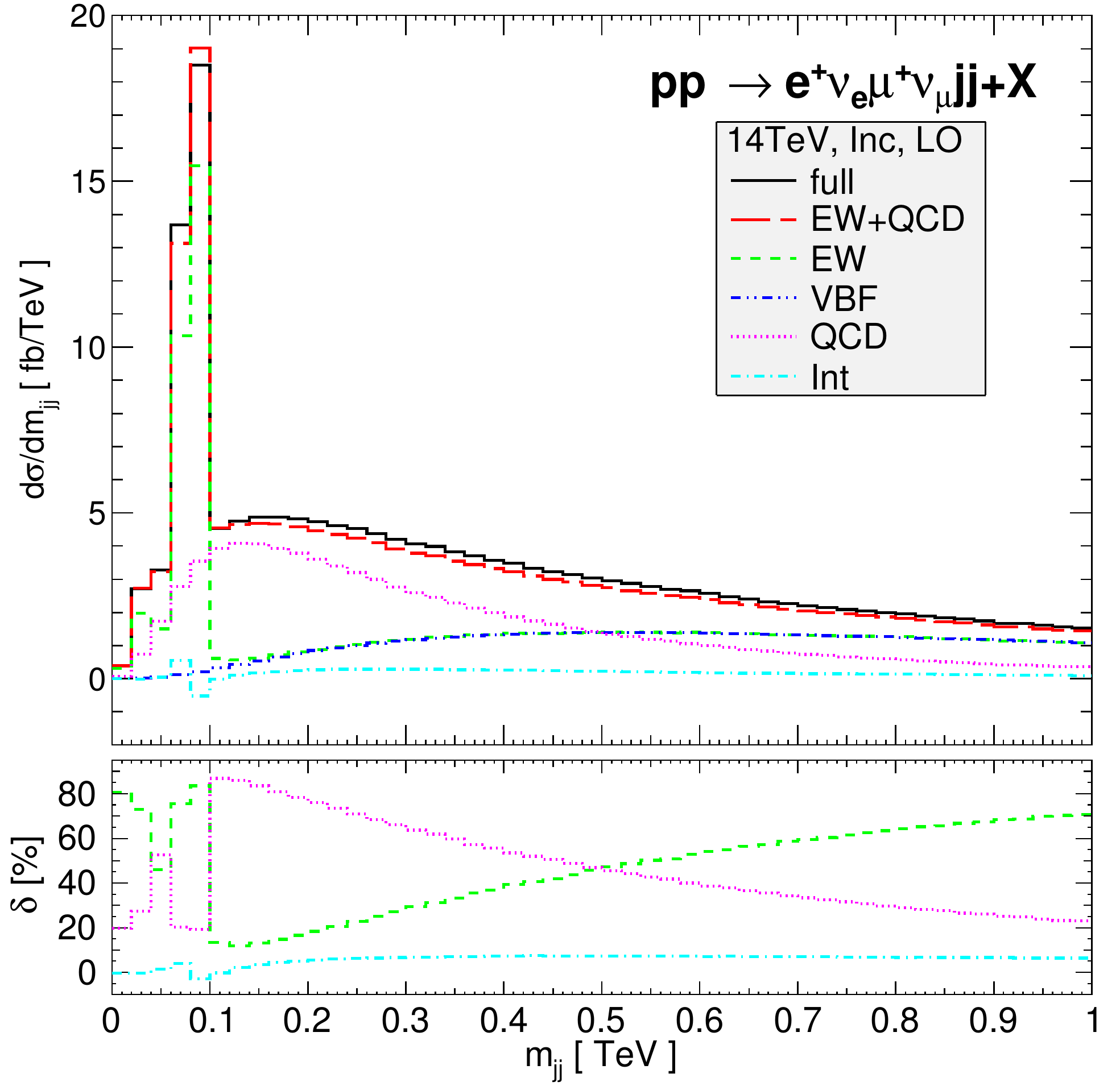}\hfill
  \includegraphics[width=0.47\textwidth]{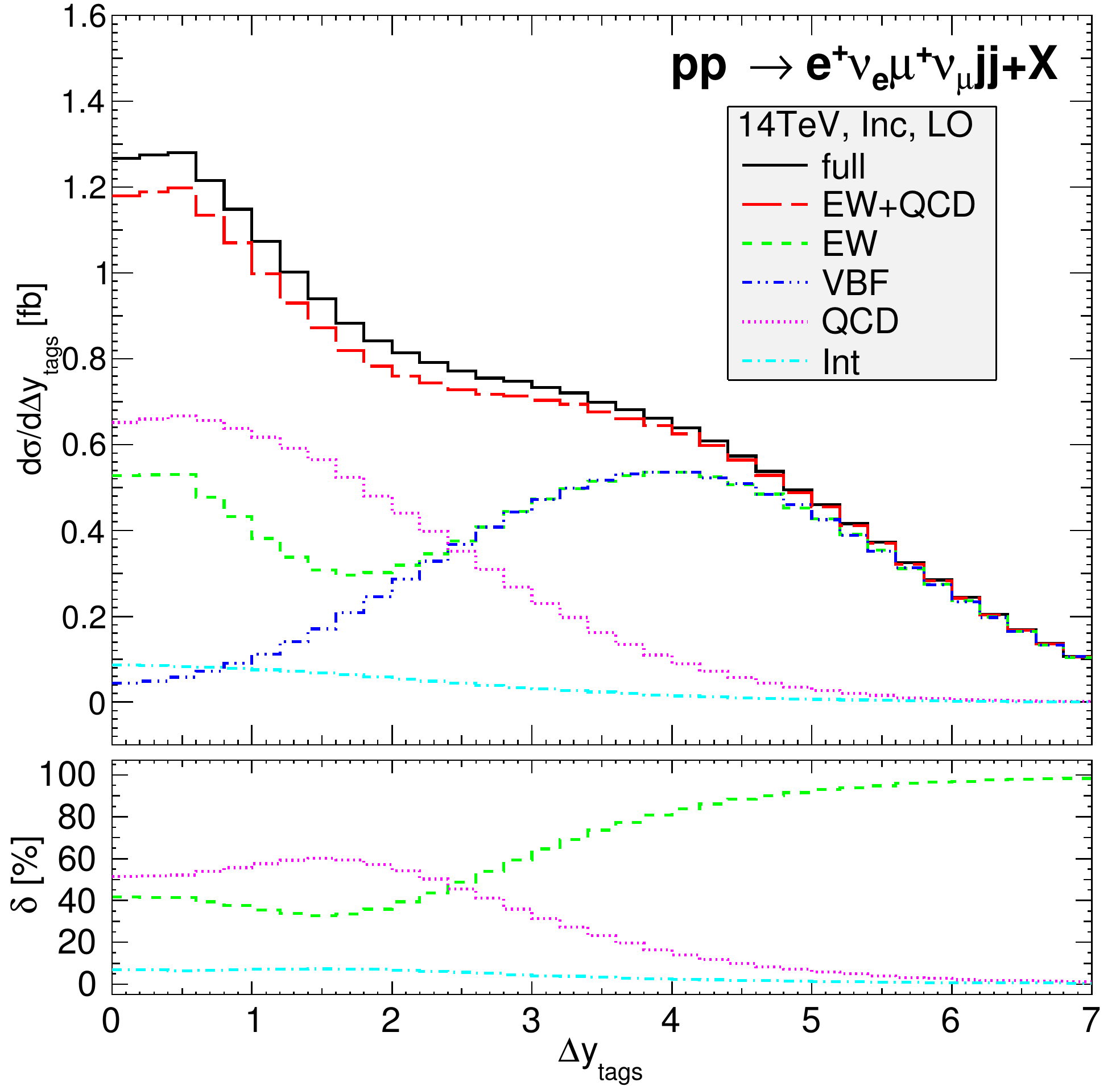}
 \end{center}
  \caption{Differential cross sections with inclusive cuts for the transverse momenta (top row) and 
the invariant mass (bottom left) of the two tagging jets ordered by $p_T$. 
The distributions of the rapidity separation between the two jets are in the bottom right panel. 
The relative EW, QCD and interference contributions compared to the full LO results are also plotted in the small panels.
Figure taken from Ref.~\protect\cite{Campanario:2013gea}.
}
\label{fig:vbfinterference}
\end{figure}
A more quantitative picture is presented in \fig{fig:vbfinterference}, taken
from Ref.~\cite{Campanario:2013gea}. There we plot, at leading order
(LO) for the LHC running at 14~TeV center-of-mass energy, the size of
the different contributions for four important differential
distributions, namely the transverse momenta of the two tagging jets,
the invariant mass of them, and their separation in rapidity. In the
smaller panels below each distribution the relative contribution
compared to the full LO result is shown.
The individual curves in the plots denote the following contributions:
the one labeled ``QCD'' contains the QCD-induced contribution of order
$\Order{\alpha_s^2\alpha^4}$,  ``EW'' contains all contributions of
order $\Order{\alpha^6}$, \ie both the VBF part ($t$- and $u$-channel) and
the triboson part ($s$-channel) and their interferences. ``QCD+EW'' is the
simple sum of these two, while ``full'' also includes their interference
term of order $\Order{\alpha_s\alpha^5}$, also shown separately as
``Int''. Finally, the ``VBF'' curve shows the VBF contribution only.
This part also neglects interference terms between $t$- and $u$-channel. In
Ref.~\cite{Denner:2012dz} it has been shown that these interference
terms are negligible once VBF cuts are applied. This is also expected
from a theory point of view, as in addition to a color factor of
$\frac13$, jets which are forward for one part are backward for the
other one, and hence leads to a strongly suppression.

As already expected from the discussions above, the VBF contribution is
relatively big for large invariant masses of the two tagging jets,
reaching up to 70\% of the full cross section for an invariant mass of
1~TeV. The difference to the full EW contribution is large only at
invariant masses around the $W$ mass, where the Breit-Wigner peak is
clearly visible. For larger invariant masses the effect is negligible. 
From the plots one can also see that QCD-EW interference effects are
biggest when the transverse momenta of the tagging jets are large and
the rapidity separation between them is small. The VBF curve on the
other hand favors a large rapidity separation, where the interference
drops below the 5\% level. In Ref.~\cite{Campanario:2013gea} two
sets of VBF cuts have been defined
\begin{align}
m_{jj} &> 200 \GeV \,, & \Delta y_{jj} &> 2.5 \,, &&&&
\text{(loose)} \\
m_{jj} &> 500 \GeV \,, & \Delta y_{jj} &> 4 \,, & y_{j_1} \cdot y_{j_2}
&< 0 \,. && \text{(tight)}
\end{align}
For the loose cut set, the QCD contribution yields an additional
$20.3\%$ and the QCD-EW interference $3.3\%$ on top of the VBF cross
section of $1.784\fb$. Using the tight VBF cuts, this reduces to $4.1\%$
and $1.3\%$, respectively, with a VBF cross section of $0.971\fb$.
Additional effects due to EW non-VBF diagrams are at or below 1 per mill
in both cases. Therefore, when comparing to experimental measurements,
QCD-induced production is a relevant background process which needs to
be taken into account, but any interference is small once we impose VBF
cuts.

Finally, the question is how we can formally define the VBF
contribution and separate them from the other processes with the same
final state. For the distinction with the QCD process, we can simply
count the order of the coupling constants, which should be
$\Order{\alpha_s^0 \alpha^n}$, where $n$ is the number of
particles in the final state. In our process we have used for most of
the chapter, \eq{proc:w+w+jj}, $n$ would be $6$. In order to separate
triboson production, we assume that two copies of $SU(3)_c$ exist, and
the quark from the first proton carries the usual color charge from one
$SU(3)_c$, but is neutral with respect to the other one, and vice versa
for the quark from the second proton. Then connecting the two incoming
quarks is not possible, as this would violate color charge conservation.
Additionally, $t$-/$u$-channel interference becomes automatically
absent. This scheme is also known as the structure-function approach, as
the $qqV$ vertex can be included in charged-current and neutral-current
hadronic structure functions~\cite{Han:1992hr}.

This picture will be particularly useful in the next chapter,
where we discuss higher-order corrections, as it gives a clear
prescription which terms to include and which are omitted due to the VBF
approximation. Also, the sometimes used statement that VBF can be
seen as a process of two-sided deep inelastic scattering becomes
obvious in this picture.

\section{Higher-order Corrections}
\label{chap:nlo}

In order to get a useful comparison between theoretical predictions and
experimental results, the uncertainty of the former, for example due to
the truncation of the perturbative series, should at least match the
precision of the latter. At the LHC, we can expect a final accuracy of a
few percent for VBF processes (see \chap{chap:experiment} for a more
detailed discussion). Using just LO calculations is not sufficient to
obtain the same level for predictions on cross sections and
distributions.

In this chapter we will discuss the available higher-order corrections
and show their impact on integrated cross sections and differential
distributions.

\subsection{NLO QCD Corrections}

Next-to-leading order corrections in the strong coupling constant
$\alpha_s$, NLO QCD for short, are available for all VBF processes
considered here. They have been first calculated in
Refs.~\cite{Figy:2003nv,Oleari:2003tc,Jager:2006zc,Jager:2006cp,Bozzi:2007ur,Jager:2009xx,Denner:2012dz,Campanario:2013eta}. 

\begin{figure}
\hspace*{-2em}
\includegraphics[width=0.37\textwidth]{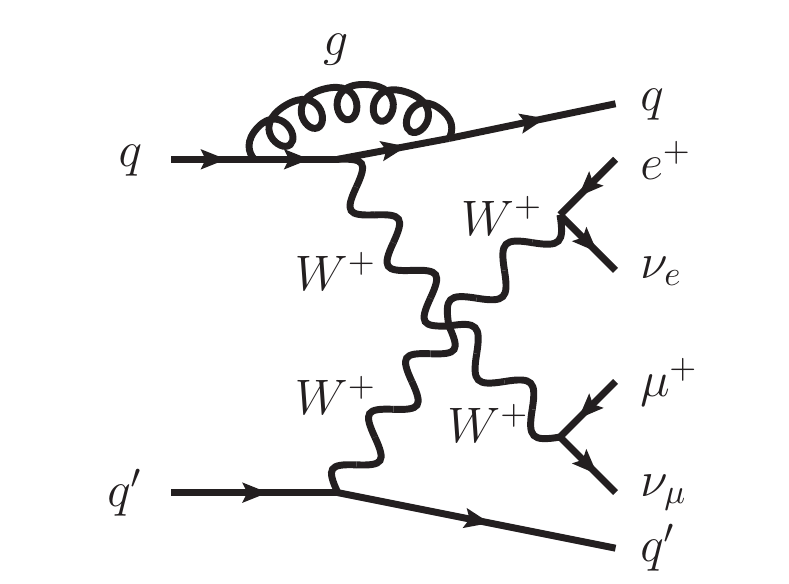} 
\hspace*{-2em}
\includegraphics[width=0.37\textwidth]{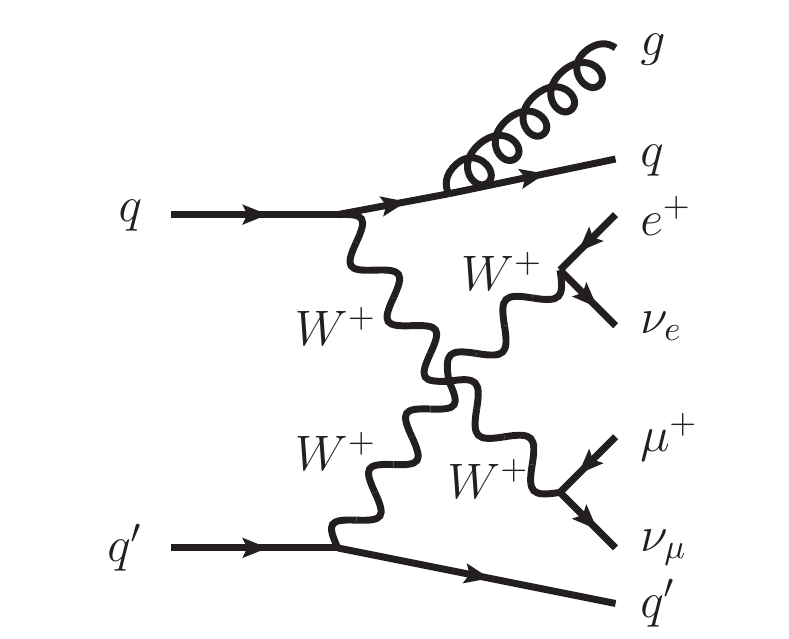} 
\hspace*{-2em}
\includegraphics[width=0.37\textwidth]{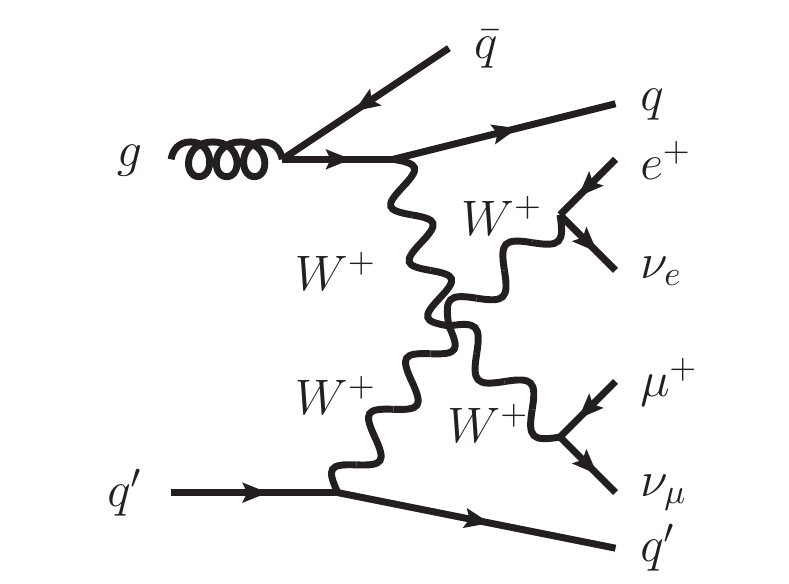} 
\caption{Typical Feynman diagrams appearing in VBF production at NLO QCD. 
The diagram on the left-hand side shows a virtual correction to the
upper quark line. The other two diagrams are examples for real emission
processes, either with final-state gluon radiation (\textit{center}) or
gluon-initiated (\textit{right}).
}
\label{fig:fm_wpwpnlo}
\end{figure}

\fig{fig:fm_wpwpnlo} shows examples of Feynman diagrams contributing.
These can either be virtual corrections (left diagram), which contribute
via an interference with the Born matrix element, or real-emission
diagrams, where an additional gluon is radiated off the quark line
(center diagram). Crossing of initial and final state then also leads to
contributions where the gluon is in the initial state (right diagram).
Diagrams for the lower quark line and where initial and final state of a
quark line are exchanged are analogous. For the moment, we will restrict
our discussion to the diagrams where no final-state $W$ is radiated off
the quark lines, and come back to these separately later. This
corresponds for example to the case of VBF-$H$ production, where only
such diagrams are present.

In this case the only virtual contributions are corrections to the $qqV$
vertex as shown in \fig{fig:fm_wpwpnlo}. For definiteness, let us
introduce momenta in the following way: $q(p_1) q'(p_2) \rightarrow
q(p_3) q'(p_4) e^+(k_1) \nu_e(k_2) \mu^+(k_3) \nu_\mu(k_4)$. 
Additionally, we define $q_1 = p_1-p_3$ and $q_2=p_2-p_4$ as well as
$Q_{1,2}^2 = |-q_{1,2}^2|$.  All quarks are taken as massless. 
Then the upper fermion line can be written as 
\begin{equation}
\mathcal{M}^\mu = \mu^{4-D} \int \frac{\di^D\ell}{(2\pi)^D} \bar{u}(p_2)
i g_s \gamma^\nu \frac{i}{\slashed{\ell}+\slashed{p}_2} i g \gamma^\mu
P_L \frac{i}{\slashed{\ell}+\slashed{p}_1} i g_s \gamma^\rho u(p_1) \frac{-i
g_{\nu\rho}}{\ell^2} T^a_{jk}T^a_{ki} \,.
\end{equation}
The last two factors are the SU(3) generators $T$. They yield 
$T^a_{jk}T^a_{ki} = C_F \delta_{ij} = \frac43 \delta_{ij}$, where $i$
and $j$ are the color indices of the incoming and outgoing quark,
respectively.
An explicit calculation, either by hand or using modern
tools~\cite{Hahn:2000kx,Hahn:1998yk,Hahn:2002vc,Hahn:2004rf,Hahn:2005vh,Hahn:2006qw,Hahn:2006ig,Hahn:2006zy,vanOldenborgh:1989wn}
shows that these corrections factorize
against the Born matrix element $\mathcal{M}_B$ and the whole amplitude
in dimensional regularization is given by
\begin{align}
\mathcal{M}_V 
&= \mathcal{M}_B \frac{\alpha_s}{4\pi} C_F \nonumber\\
&\qquad \cdot
 \Bigl[ B_0(0) - 2 Q_1^2 \bigl( C_1(-Q_1^2,0,0)
+ C_2(-Q_1^2,0,0) \bigr) - 4 C_{00}(-Q_1^2,0,0) \Bigr] \nonumber\\
&\stackrel{PV}{=} \mathcal{M}_B \frac{\alpha_s}{4\pi} C_F 
  \left( 4 B_0(0) -3 B_0(-Q_1^2) + 2 Q_1^2 C_0(-Q_1^2,0,0) -2 \right) \,.
\label{eq:matv}
\end{align}
All mass arguments of the loop functions are zero and have been left
out. In the last step, we have converted all loop functions to the basic
scalar integrals using the Passarino-Veltman (PV) reduction
method~\cite{Passarino:1978jh}. Their explicit form is
\begin{align}
B_0(0) &= \frac{\left(4\pi\right)^{\epsilon_\text{UV}}}{\Gamma(1-\epsilon_\text{UV})} \frac1{\epsilon_\text{UV}} - 
\frac{\left(4\pi\right)^{\epsilon_\text{IR}}}{\Gamma(1-\epsilon_\text{IR})} \frac1{\epsilon_\text{IR}} \\
B_0(p^2) &= \left(\frac{4\pi\mu^2}{-p^2}\right)^{\epsilon_\text{UV}} \frac1{\Gamma(1-\epsilon_\text{UV})} \left[ \frac1{\epsilon_\text{UV}} +2 \right] \\
C_0(p^2,0,0) &= \left(\frac{4\pi\mu^2}{-p^2}\right)^{\epsilon_\text{IR}} \frac1{\Gamma(1-\epsilon_\text{IR})} \frac1{p^2} \frac1{\epsilon^2_\text{IR}} \,.
\end{align}
For the poles in $\epsilon$ we have indicated in each loop function whether this originates
from the UV region, where the loop momentum $\ell$ goes to infinity and we need
to approach the limit from the positive side, \ie $\lim_{\epsilon_\text{UV}\rightarrow
0^+}$, or they are from the IR region, $|\ell| \rightarrow 0$, with the
limit $\lim_{\epsilon_\text{IR}\rightarrow 0^-}$.

The total expression is both UV- and IR-divergent. The UV-divergent part
is given by
\begin{equation}
\mathcal{M}_V\Bigr|_{\text{UV}} = \mathcal{M}_B
\frac{\alpha_s}{4\pi} C_F \Delta_{\text{UV}}
\end{equation}
with
\begin{equation}
\Delta_{\text{UV}} = \frac{\left(4\pi\right)^{\epsilon_\text{UV}}}{\Gamma(1-\epsilon_\text{UV})}
\frac1{\epsilon_{\text{UV}}} = \frac1{\epsilon_{\text{UV}}} - \gamma_E +
\ln(4\pi) \,,
\end{equation}
which encompasses also some finite terms which are removed together with the
pole in the $\MSbar$ scheme. 
To get rid of this divergence, we need to perform renormalization on the
$qqW$-vertex and add the corresponding counter term
\begin{equation} 
\mathcal{M}_{CT} = - \left(\frac{Z_g}{\sqrt{Z_q^2 Z_W}}-1\right) \mathcal{M}_B
\Biggr|_{\text{NLO}} = - \left( \delta Z_g - \delta Z_q - \frac{\delta
Z_W}2\right) \mathcal{M}_B \,.
\end{equation} 
The renormalization constants for the weak coupling $g$ and the $W$
field do not receive any QCD contributions at one-loop
order, 
\begin{equation}
\delta Z_g = \delta Z_W = 0 \,,
\end{equation}
while for $\delta Z_q$ we obtain\footnote{For an overview on
renormalization see \eg Ref.~\cite{Denner:1991kt}.}
\begin{align}
\delta Z_q &= - \frac{\alpha_s}{4\pi} C_F B_0(0) \,.
\end{align}
Note that this wave function renormalization acts on external particles,
so we have to include the full expression despite working in the
$\MSbar$ scheme to guarantee that the residue of the propagator pole is
correctly normalized to unity. So the renormalized expression becomes
\begin{equation}
\mathcal{M}_V\Bigr|_{\text{UV}} + \mathcal{M}_{CT} =
\mathcal{M}_B \frac{\alpha_s}{4\pi} C_F
 \frac{\left(4\pi\right)^{\epsilon_\text{IR}}}{\Gamma(1-\epsilon_\text{IR})}
\frac1{\epsilon_\text{IR}} \,,
\end{equation}
\ie effectively in \eq{eq:matv}, all $\epsilon_\text{UV}$ have been
replaced by $\epsilon_\text{IR}$, and we will drop the subscript from
now on.

In total we obtain for the Born-Virtual interference~\cite{Figy:2003nv}
\begin{align}
&2\Re\left[\mathcal{M}_B^* \mathcal{M}_{V+CT}\right]
 =
\left|\mathcal{M}_B\right|^2
\frac{\alpha_s(\mu)}{2\pi} C_F
\left(\frac{4\pi\mu^2}{Q_1^2}\right)^\epsilon \frac1{\Gamma(1-\epsilon)} \left[
-\frac2{\epsilon^2}-\frac3{\epsilon}-8 \right] \nonumber\\
&\qquad = C_\epsilon \frac{\alpha_s}{2\pi} C_F 
\left|\mathcal{M}_B\right|^2 \nonumber\\
&\qquad\qquad \cdot
\left( -\frac2{\epsilon^2} \
       -\frac{3+2 \ln \frac{\mu_R^2}{2p_1\cdot p_3}}{\epsilon} 
       -8 - 3\ln \frac{\mu_R^2}{2p_1\cdot p_3} 
         - \ln^2 \frac{\mu_R^2}{2p_1\cdot p_3} 
\right)
\,,
\end{align}
with
\begin{equation}
C_\epsilon=\frac1{\Gamma(1-\epsilon)}\left( \frac{4\pi\mu^2}{\mu_R^2}
\right)^\epsilon
\,.
\end{equation}
In the last step the constant $C_\epsilon$ has been pulled out, as is
commonly done in loop calculations~\cite{Binoth:2010xt,Alioli:2013nda},
and the rest fully expanded in $\epsilon$.  In dimensional reduction the
number in the square bracket would be $-7$ instead of $-8$.

For sufficiently inclusive quantities a theorem by Kinoshita, Lee
and Nauenberg~\cite{Kinoshita:1962ur,Lee:1964is} states that infrared
divergences must cancel. The necessary additional contribution is
exactly the real-emission diagrams of \fig{fig:fm_wpwpnlo}. If we look
at the center diagram with final-state radiation and denote the momentum
of the gluon by $p_5$, the propagator of the quark splitting into the
final-state quark and the gluon is given by
\begin{equation}
\frac{i(\slashed{p}_3+\slashed{p}_5)}{(p_3+p_5)^2} =
\frac{i(\slashed{p}_3+\slashed{p}_5)}{ 2 p_3 \cdot p_5} \,,
\label{eq:redivprop}
\end{equation}
where we have used that both particles are massless. 
The denominator can be written in any reference frame as
\begin{equation}
2 p_3 \cdot p_5 = 2 E_3 E_5 (1-\cos\theta) \,.
\end{equation}
In the full matrix element, this propagator is attached to the
expression $\bar{u}(p_3)\slashed{\epsilon}(p_5)$ from the right. So we
get effectively another factor of $\sqrt{E_3}$ from the fermion wave
function, while no such term appears for the gluon. This additional
factor breaks the apparent symmetry between the two in
\eq{eq:redivprop}.

There are two important limits here. First, $p_5$ can go to zero such
that the direction stays constant $p_5 \rightarrow \lambda p_5$,
$\lambda\rightarrow0$. Then $E_5 \rightarrow \lambda E_5$, while the
angle $\theta$ stays constant. This denotes the soft limit. In the
collinear limit, $E_5$ is unchanged, but the two particles become
collinear, \ie $\theta\rightarrow0$, which also leads to a divergence of
the propagator. When integrated over the phase space, these lead to
corresponding divergences of the cross section, which exactly cancel the
corresponding divergences in the virtual amplitudes. For the quark, the
additional factor from the external wave function leads to finite
expressions after phase-space integration.

If we want to do a numerical Monte-Carlo implementation of the cross
section, we first need to cancel these divergences explicitly so that
only finite results appear. This procedure is implemented via a
subtraction scheme. Popular choices for NLO QCD are
the Frixione-Kunszt-Signer (FKS)~\cite{Frixione:1995ms} or
Catani-Seymour (CS)~\cite{Catani:1996vz} scheme. We will follow the
conventions of the latter one here. The idea in all schemes is to
subtract a function $A$ from the real-emission part which shows the same
divergent behavior in the singular regions and is small in other
regions. To not change the overall result, the same function needs to be
subtracted again from the virtual part. If the function can be
integrated analytically over the phase space of the extra emission, then
this will produce poles in $\epsilon$ which exactly cancel the ones from
the loop integrals. Schematically, this can be written in the following
way~\cite{Catani:1996vz}:
\begin{align}
\sigma^\text{NLO} &= \int_m \di \sigma_B 
  + \int_m \di \sigma_V + \int_{m+1} \di \sigma_R \nonumber\\
&= \int_m \di \sigma_B 
  + \int_m \left( \di \sigma_V 
    + \int_1 \sum_\text{dipoles} (\di \sigma_B \otimes \di V_\text{dipole})
     \right)_{\epsilon=0} \nonumber\\
&\quad
  + \int_{m+1} \left( \di \sigma_R - \sum_\text{dipoles} \di \sigma_B \otimes \di
V_\text{dipole} \right)_{\epsilon=0}
\,.
\end{align}
Thereby, $\int_m$ denotes the integration over the $m$-particle
phase space, $\di \sigma_B$, $\di \sigma_V$ and $\di \sigma_R$ are the
differential Born, Born-virtual interference and real-emission cross
sections, respectively. The added function is chosen to factorize
into the corresponding Born cross section and the so-called dipoles.
These are independent of the details of the process. The name dipole
originates from the fact that starting from the Born process we have one
particle where the emission of the extra parton happens. To ensure
overall momentum conservation for an emission with finite energy and
angle, a second parton, the so-called spectator, is needed. 

For our process of VBF production, only the other parton on the same
quark line needs to be considered as a spectator, as color correlations
present in the dipoles will make the contributions where the spectator
is part of the other quark line vanish. Therefore, for a final-state
emission of the gluon on the upper line, there are two dipoles,
${\mathcal{D}_{35}}^1$ and ${\mathcal{D}_{3}}^{15}$ in the notation of
Ref.~\cite{Catani:1996vz}: 
\begin{align}
{\mathcal{D}_{35}}^1 &= \frac1{2 p_3 \cdot p_5} \frac1{x_{35,1}}
8\pi \alpha_s C_F \left( 
\frac2{1-\tilde{z}_3 + (1-x_{35,1})} - (1+\tilde{z}_3) 
\right) \left|\mathcal{M}_B(\tilde{p})\right|^2 \nonumber\\
{\mathcal{D}_{3}}^{15} &= \frac1{2 p_1 \cdot p_5} \frac1{x_{35,1}}
8\pi \alpha_s C_F \left(
\frac2{1-x_{15,3}+u_5)} - (1+x_{15,3}) 
\right) \left|\mathcal{M}_B(\tilde{p})\right|^2 \,,
\end{align}
with
\begin{align}
x &\equiv x_{35,1} = \frac{p_1 \cdot p_3 + p_1 \cdot p_5 - p_3 \cdot p_5}{p_1
\cdot p_3 + p_1 \cdot p_5} \,, \nonumber\\
z &\equiv \tilde{z}_3 = 1-u_5 = \frac{p_1 \cdot p_3}{p_1 \cdot p_3 + p_1 \cdot p_5}  
\,,
\end{align}
and the Born momenta are 
\begin{align}
\tilde{p}_1 &= x p_1 \,, &  
\tilde{p}_3 &= p_3 + p_5 - (1-x) p_1 
\end{align}
with all other momenta unchanged.
Combining everything yields our subtraction matrix element
\begin{equation}
\left|\mathcal{M}_{\mathcal{D},f}\right|^2 = {\mathcal{D}_{35}}^1 +
{\mathcal{D}_{3}}^{15} 
=8\pi \alpha_s C_F \frac1{Q^2} 
\frac{x^2+z^2}{(1-x)(1-z)}
\left|\mathcal{M}_B(\tilde{p})\right|^2
\,.
\end{equation}
Combining everything and also adding the convolution with the parton
densities gives for the subtracted real-emission process with
final-state gluon emission
\begin{align}
\sigma_{R,\text{subtr}} &= \int_0^1 \di x_a \int_0^1 \di x_b \ 
f_{q/p}(x_a,\mu_F) f_{q'/p}(x_b,\mu_F) \nonumber\\
&\quad \cdot \frac1{4 p_1 \cdot p_2}\int \di\text{PS}_{2\rightarrow7}
\left(
\left|\mathcal{M}_R\right|^2 F_J^{(3)} -
\left|\mathcal{M}_{\mathcal{D},f}\right|^2 F_J^{(2)}(\tilde{p})
\right) \,,
\end{align}
where $\mathcal{M}_R$ denotes the real-emission matrix element and
$F_J^{(3)}$ and $F_J^{(2)}$ is a infrared soft and collinear safe jet
algorithm for the 3 and 2-parton final state, where the latter uses the
momenta $\tilde{p}$ as input.

Similarly, for the real emission process where the gluon is in the
initial state,
$g(p_1) q'(p_2) \rightarrow
q(p_3) \bar{q}(p_5) q'(p_4) e^+(k_1) \nu_e(k_2) \mu^+(k_3)
\nu_\mu(k_4)$, we find~\cite{Catani:1996vz,Figy:2003nv}:
\begin{align}
\left|\mathcal{M}_{\mathcal{D},i}\right|^2
&= {\mathcal{D}_{3}}^{15} + {\mathcal{D}_{5}}^{13} \nonumber\\
&= 8\pi \alpha_s T_R 
\frac{1-2x(1-x)}{x} \left(
\frac1{2 p_1 \cdot p_5} 
\left|\mathcal{M}_B^{q}(\tilde{p})\right|^2 
+ \frac1{2 p_1 \cdot p_3} 
\left|\mathcal{M}_B^{\bar{q}}(\tilde{p})\right|^2 \right) \nonumber\\
&= 
8\pi \alpha_s T_R 
\frac{x^2+(1-x)^2}{Q^2} \left(
\frac1{1-z} 
\left|\mathcal{M}_B^{q}(\tilde{p})\right|^2 
+ \frac1{z} 
\left|\mathcal{M}_B^{\bar{q}}(\tilde{p})\right|^2 \right) 
\,.
\end{align}
The color factor $T_R=\frac12$ and $\mathcal{M}_B^{\bar{q}}$ is the Born
matrix element with a $\bar{q}$ as incoming parton on the upper line.

As said before, we need to integrate these dipoles and add them back to
the born-virtual interference. This contribution is called the
$I$-operator. The term operator is chosen because it contains color
matrices which act on the Born matrix element. As their effect is
trivial for our VBF process, we have already inserted these into the
expression and get for the squared matrix element
\begin{align}
\left|\mathcal{M}_{I}\right|^2 &=
\frac{\alpha_s}{2\pi} C_F \frac1{\Gamma(1-\epsilon)} 
\left( \frac{4\pi\mu^2}{2 p_1 \cdot p_3} \right)^\epsilon
\left(\frac2{\epsilon^2} + \frac3{\epsilon} + 10 - \pi^2 \right)
\left|\mathcal{M}_B\right|^2 
\nonumber\\
&= C_\epsilon \frac{\alpha_s}{2\pi} C_F 
\left|\mathcal{M}_B\right|^2 \nonumber\\
&\qquad\cdot
\left( \frac2{\epsilon^2} + \
       \frac{3+2 \ln \frac{\mu_R^2}{2p_1\cdot p_3}}{\epsilon} +
       10 - \pi^2 + 3\ln \frac{\mu_R^2}{2p_1\cdot p_3} 
         + \ln^2 \frac{\mu_R^2}{2p_1\cdot p_3} 
\right)
\,,
\end{align}
where the momenta are those of the Born process. In the last step the
constant $C_\epsilon$ has been pulled out again and the rest has been
fully expanded in $\epsilon$.

The poles in $\frac1{\epsilon^2}$ and $\frac1{\epsilon}$ now cancel
exactly with the corresponding parts from the virtual amplitude and the
Born and renormalized virtual part of the cross section is given by
\begin{align}
\sigma_{B+V} &= \int_0^1 \di x_a \int_0^1 \di x_b \ 
f_{q/p}(x_a,\mu_F) f_{q'/p}(x_b,\mu_F) \nonumber\\
&\quad \cdot \frac1{4 p_1 \cdot p_2}\int \di\text{PS}_{2\rightarrow6}
\left|\mathcal{M}_B\right|^2 F_J^{(2)}
\left(
1 + \frac{\alpha_s(\mu_{R1})+\alpha_s(\mu_{R2})}{2\pi} C_F \left( 2 - \pi^2 \right)
\right)
\,.
\end{align}
In this expression the virtual effects on the lower line have been
included as well and we have introduced the option to choose different
renormalization scales for the two corrections, which we are going to
exploit later.

\subsubsection{Initial-state Collinear Divergences}

A slight complication actually occurs due to the presence of partons in
the initial state, and is related to the fact that their momentum is
fixed. The corresponding Born process with momentum $\tilde{p}$ has a
center-of-mass energy of $\sqrt{x 2 p_1 \cdot p_2}$ instead of $\sqrt{2
p_1 \cdot p_2}$ and hence the divergences of virtual correction and real
emission do not match. In the soft limit $x\rightarrow1$ and everything
is fine. In the collinear limit the solution is to include the remaining
divergence in a redefinition of the PDFs. These then also become
scheme-dependent, as there is an ambiguity which finite parts are absorbed
into the PDFs as well. The most common choice for hadron-hadron
collisions is the $\MSbar$ scheme. This induces an additional
collinear-subtraction counterterm in our expression
\begin{align}
\left|\mathcal{M}_C\right|^2 &= - \frac{\alpha_s}{2\pi}
\frac1{\Gamma(1-\epsilon)} \int_0^1 \di x \left[ - \frac1\epsilon
\left(\frac{4\pi\mu^2}{\mu_F^2}\right)^\epsilon P^{qg}(x) + K^{qg}(x)\right] 
\left|\mathcal{M}_B^{q}(\tilde{p})\right|^2 \,.
\end{align}
Here, $\mu_F$ denotes the factorization scale, $P^{qg}(x) = C_F
\frac{1+(1-x)^2}{x}$ is the
DGLAP~\cite{Dokshitzer:1977sg,Gribov:1972ri,Altarelli:1977zs} splitting kernel,
and
$K^{qg}(x)\stackrel{\MSbar}{=}0$ denotes a possible additional finite
part. Because of these two symbols appearing, this contribution is
sometimes also referred to as the $PK$-operator.

The corresponding cross section contribution then yields~\cite{Figy:2003nv}
\begin{align}
\sigma_{C} &= \int_0^1 \di x_a \int_0^1 \di x_b \ 
\frac{\alpha_s}{2\pi} \int_{x_a}^1 \frac{\di x}{x} \tilde{f}_{q/p}(x_a,x,\mu_F) f_{q'/p}(x_b,\mu_F) \nonumber\\
&\quad \cdot \frac1{4 p_1 \cdot p_2}\int \di\text{PS}_{2\rightarrow6}
\left|\mathcal{M}_B\right|^2 F_J^{(2)}
\,,
\end{align}
where the additional contribution has been put into
\begin{align} 
\tilde{f}_{q/p}(x_a,x,\mu_F) &= 
f_{g/p}\left(\frac{x_a}x,\mu_F\right) A(x) \nonumber\\
&\quad + \left[ f_{q/p}\left(\frac{x_a}x,\mu_F\right) - x f_{q/p}\left(x_a,\mu_F\right)
  \right] B(x)
\nonumber\\
&\quad+ f_{q/p}\left(\frac{x_a}x,\mu_F\right) C(x)
+ f_{q/p}\left(x_a,\mu_F\right) \frac{D(x_a)}{1-x_a} 
\label{eq:fcoll}
\intertext{with}
A(x) &= T_R \left[ x^2 + (1-x)^2 \right] \ln \frac{(1-x)Q^2}{x \mu_F^2}
+ 2 T_R x (1-x) \\
B(x) &= C_F \left[ \frac1{1-x} \ln \frac{(1-x)Q^2}{\mu_F^2} - \frac32
\frac1{1-x} \right] \\
C(x) &= C_F \left[ 1-x-\frac2{1-x}\ln x - (1+x) \ln \frac{(1-x)Q^2}{x
\mu_F^2}\right] \\
D(x_a) &= C_F \Bigl[ \frac32 \ln\frac{Q^2}{(1-x_a)\mu_F^2} + 2 \ln(1-x_a)
\ln \frac{Q^2}{\mu_F^2} + \ln^2(1-x_a) \nonumber\\
&\quad\phantom{C_F \Bigl[\ } - \frac{11}2 + \frac{2\pi^2}3
\Bigr] \,.
\end{align} 

Since this expression originates from a limit of the real-emission phase
space, it is possible to rewrite the extra $\int \di x$ integration into
an integral $\int \di^3 p_5$, such that this contribution can be
combined together with the real-emission processes into a single
integration of the real-emission phase space.

Then we obtain~\cite{Figy:2003nv}
\begin{align}
\sigma_{C} &= \int_0^1 \di x_a \int_0^1 \di x_b \ 
\frac1{4 p_1 \cdot p_2} \nonumber\\
&\quad \cdot \int \di\text{PS}_{2\rightarrow7}
\tilde{f}_{q/p}(x x_a,x,\mu_F) f_{q'/p}(x_b,\mu_F) 
\frac{8\pi \alpha_s}{Q^2} \left|\mathcal{M}_B\right|^2 F_J^{(2)}
\,,
\end{align}
with $\tilde{f}_{q/p}$ given in \eq{eq:fcoll}. This procedure is used
for example in the \VBFNLO
implementation~\cite{Arnold:2008rz,Arnold:2011wj,Arnold:2012xn,Baglio:2014uba}
of the VBF processes.

The total NLO cross section is then finally given by the sum of all
contributions
\begin{equation}
\sigma_{\text{NLO}} = \sigma_{B+V} + \sigma_{R,\text{subtr}} +
\sigma_{C} \,.
\label{eq:nlocs}
\end{equation}

\begin{figure}
\begin{center}
\includegraphics[width=0.9\textwidth]{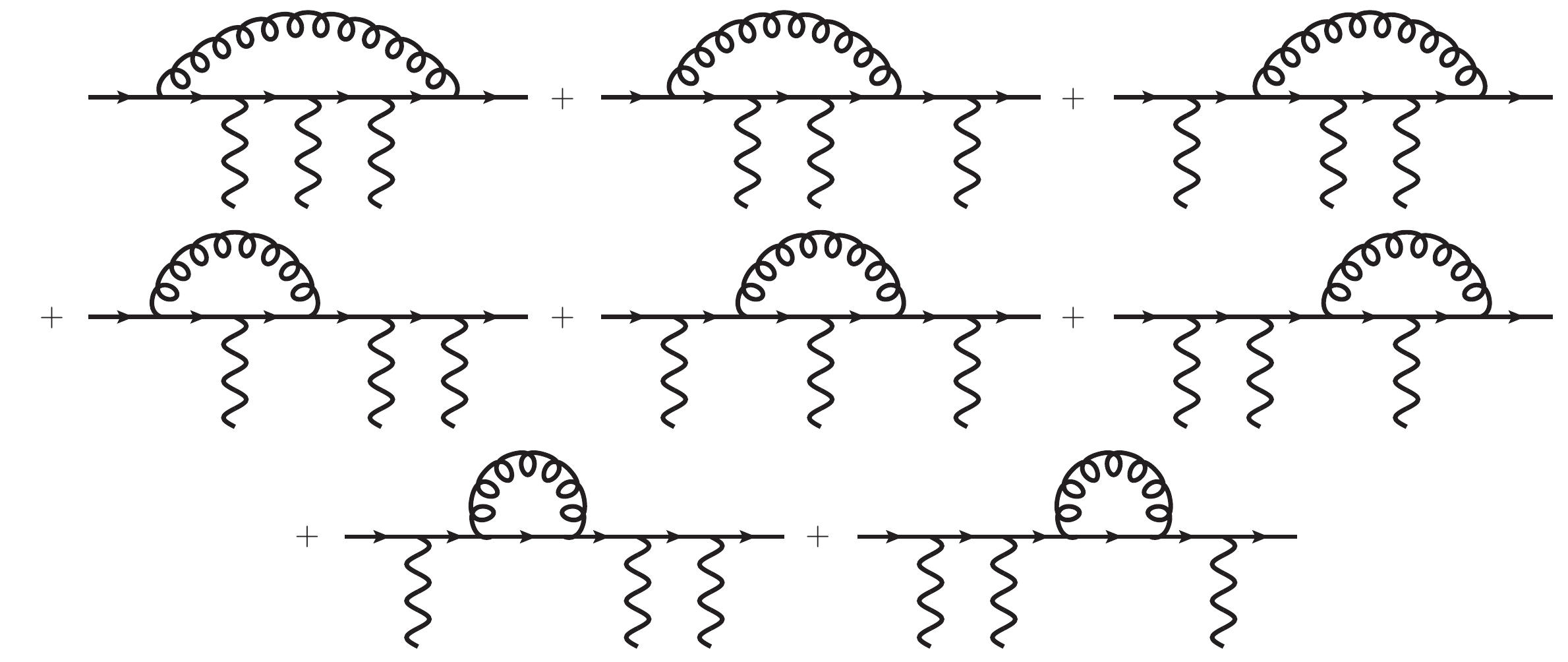} 
\end{center}
\caption{NLO QCD loop corrections to the quark line with three attached
gauge bosons. The order of the gauge bosons is the same between all
diagrams.
}
\label{fig:fm_penline}
\end{figure}
As mentioned before, for general VBF processes the final-state gauge
bosons can also be radiated off the quark lines. In this case loop
corrections with up to five external particles, \ie pentagons can
appear. It is convenient to group all loop corrections to the quark line
together~\cite{Jager:2006zc}, as shown in \fig{fig:fm_penline}. There
the three vector bosons can be either the $t$-channel or final-state
bosons, with the other quark line or the final-state leptons removed for
clarity. The order of the vector bosons is fixed, a permutation of them
is another fermion line. The divergent pieces of this diagram sum
factorize against the Born again, while for the finite terms an
additional contribution appears, $\widetilde{\mathcal{M}}_5$,
\begin{align}
\mathcal{M}_{V,5}\Bigr|_{\text{renorm}} 
 &=
\frac{\alpha_s(\mu)}{4\pi} C_F \ \Biggl(
\mathcal{M}_B
\left(\frac{4\pi\mu^2}{Q_1^2}\right)^\epsilon \frac1{\Gamma(1-\epsilon)} \left[
-\frac2{\epsilon^2}-\frac3{\epsilon}-8 \right] 
+ \widetilde{\mathcal{M}}_5 \Biggr)
\,.
\end{align}
The exact choice of splitting the finite part is arbitrary, but taking
the part not included in $\widetilde{\mathcal{M}}_5$ to be the same as for
the pure vertex correction is convenient for numerical implementations.
Then the part proportional to the Born matrix element is the same for
all virtual corrections. Also it turns out that with this choice the
contribution of $\widetilde{\mathcal{M}}_5$ is fairly small. As its
evaluation is rather time-consuming due to the presence of pentagon loop
functions, this allows us to integrate this part separately with less
statistics without affecting the overall accuracy of the result. 

These divergences are then canceled by the corresponding real-emission
diagrams, where a gluon is emitted from the initial- or
final-state quark. Gluon emission from internal quark lines yields only
finite contributions.

\subsubsection{Phenomenological Impact}
\label{sec:nloqcdpheno}

Having everything in place, we can now evaluate NLO cross sections. A
summary of reference values is given in Tables~\ref{tab:nlocs8}
and~\ref{tab:nlocs13}, providing an
updated version of Ref.~\cite{Campanario:2015vqa}.
\begin{table}
	\centering
{
\renewcommand{\arraystretch}{1.25}
\small{
\begin{tabular}{l|lll}
LHC & \multicolumn{3}{c}{$\sqrt{s} = 8 \TeV$}  \\
Process & $\sigma_{\text{LO}}$ & $\sigma_{\text{NLO}}$ & K \\\hline
$pp \rightarrow H jj $ (``VBF-$H$'') &
  335.46(4) fb   & 316.23(2) fb & 0.94 \\ 
$pp \rightarrow H jjj $ (``VBF-$H$+jet'') &
  46.516(14) fb   & 37.56(18) fb   & 0.81 \\ 
$pp \rightarrow HH jj $ (``VBF-$HH$'') &
  0.14812(3) fb  & 0.13821(10) fb & 0.93 \\ 
$pp \rightarrow H\gamma jj $ (``VBF-$H\gamma$'') &
  4.9768(7) fb   & 4.696(4) fb   & 0.94 \\ 
\hline
$pp \rightarrow \ell^+ \ell^- jj $ (``VBF-$Z_\ell$'') &
  91.98(5) fb    & 93.77(17) fb  & 1.02 \\
$pp \rightarrow \nu \bar{\nu} jj $ (``VBF-$Z_\nu$'') &
  230.11(12) fb  & 239.7(4) fb   & 1.04 \\
$pp \rightarrow \ell^+ \nu jj $ (``VBF-$W^+$'') &
  882.77(16) fb  & 867.2(6) fb   & 0.98 \\ 
$pp \rightarrow \ell^- \bar{\nu} jj $ (``VBF-$W^-$'') &
  471.24(9) fb   & 483.8(4) fb   & 1.03 \\ 
$pp \rightarrow \gamma jj $ (``VBF-$\gamma$'') &
  2090.3(6) fb   & 2139(2) fb    & 1.02 \\ 
$pp \rightarrow \ell_1^+ \nu_{\ell_1} \ell_2^- \bar{\nu}_{\ell_2} jj $ (``VBF-$W^+ W^-$'') &
  6.860(4) fb    & 6.704(13) fb  & 0.98 \\ 
$pp \rightarrow \ell_1^+ \ell_1^- \ell_2^+ \ell_2^- jj $ (``VBF-$Z_\ell Z_\ell$'') &
  59.79(5) ab    & 61.11(15) ab  & 1.02 \\ 
$pp \rightarrow \ell_1^+ \ell_1^- \nu_2 \bar{\nu}_2 jj $ (``VBF-$Z_\ell Z_\nu$'') &
  0.4279(3) fb   & 0.4313(13) fb & 1.01 \\ 
$pp \rightarrow \ell_1^+ \nu_{\ell_1} \ell_2^+ \ell_2^- jj $ (``VBF-$W^+ Z_\ell$'') &
  0.48291(18) fb & 0.4718(8) fb  & 0.98 \\ 
$pp \rightarrow \ell_1^- \bar{\nu}_{\ell_1} \ell_2^+ \ell_2^- jj $ (``VBF-$W^- Z_\ell$'') &
  0.22909(5) fb  & 0.2373(3) fb  & 1.04 \\ 
$pp \rightarrow \ell_1^+ \nu_{\ell_1} \ell_2^+ \nu_{\ell_2} jj $ (``VBF-$W^+ W^+$'') &
  1.6778(5) fb   & 1.6230(14) fb & 0.97 \\ 
$pp \rightarrow \ell_1^- \bar{\nu}_{\ell_1} \ell_2^- \bar{\nu}_{\ell_2} jj $ (``VBF-$W^- W^-$'') &
  0.39862(11) fb & 0.4411(12) fb & 1.11 \\ 
$pp \rightarrow \ell^+ \nu \gamma jj $ (``VBF-$W^+\gamma $'') &
  11.004(3) fb   & 10.694(15) fb & 0.97 \\ 
$pp \rightarrow \ell^- \bar{\nu} \gamma jj $ (``VBF-$W^-\gamma $'') &
  5.5906(15) fb  & 5.695(7) fb   & 1.02 \\ 
$pp \rightarrow \ell^+ \ell^- \gamma jj $ (``VBF-$Z_\ell \gamma $'') &
  2.2749(9) fb   & 2.310(4) fb   & 1.02 \\ 
$pp \rightarrow \nu \bar{\nu} \gamma jj $ (``VBF-$Z_\nu \gamma $'') &
  5.2025(13) fb  & 5.335(6) fb   & 1.03 \\ 
\hline
$pp \rightarrow q \bar{q} \ell^- \bar{\nu}_{\ell} jj $ (``VBF-$W_\text{had}^+ W^-$'') &
  6.138(5) fb    & 6.68(2) fb    & 1.09  \\ 
$pp \rightarrow \ell^+ \nu_{\ell} q \bar{q} jj $ (``VBF-$W^+ W_\text{had}^-$'') &
  5.782(5) fb    & 6.31(3) fb    & 1.09  \\ 
$pp \rightarrow \ell^+ \ell^- q \bar{q} jj $ (``VBF-$Z_\ell Z_\text{had}$'') &
  0.5284(6) fb   & 0.588(3) fb   & 1.11  \\ 
$pp \rightarrow q \bar{q} \ell^+ \ell^- jj $ (``VBF-$W_\text{had}^+ Z_\ell$'') &
  0.5592(3) fb   & 0.5992(10) fb & 1.07  \\ 
$pp \rightarrow \ell^+ \nu_{\ell} q \bar{q} jj $ (``VBF-$W^+ Z_\text{had}$'') &
  2.3829(12) fb  & 2.500(6) fb   & 1.05  \\ 
$pp \rightarrow q \bar{q} \ell^+ \ell^- jj $ (``VBF-$W_\text{had}^- Z_\ell$'') &
  0.24536(12) fb & 0.2763(5) fb  & 1.13  \\ 
$pp \rightarrow \ell^- \bar{\nu}_{\ell} q \bar{q} jj $ (``VBF-$W^- Z_\text{had}$'') &
  1.0994(6) fb   & 1.238(2) fb   & 1.13  \\ 
$pp \rightarrow q \bar{q} \ell^+ \nu_{\ell} jj $ (``VBF-$W_\text{had}^+ W^+$'') &
  3.8878(15) fb  & 4.042(6) fb   & 1.04  \\ 
$pp \rightarrow q \bar{q} \ell^- \bar{\nu}_{\ell} jj $ (``VBF-$W_\text{had}^- W^-$'') &
  0.8494(3) fb   & 1.0159(13) fb & 1.20  \\
\end{tabular} }
}
\caption{Integrated cross sections for VBF production processes for the
LHC running at a center-of-mass energy of 8 TeV using the cuts given in
\eq{eq:generalcuts}. Results are given summed over all three lepton
generations and, in case of quarks, all combinations which do not
involve a top quark. The error in brackets is the statistical error from
Monte Carlo integration.}
\label{tab:nlocs8}
\end{table}
\begin{table}
	\centering
{
\renewcommand{\arraystretch}{1.25}
\small{
\begin{tabular}{l|lll}
LHC & \multicolumn{3}{c}{$\sqrt{s} = 13 \TeV$} \\
Process & $\sigma_{\text{LO}}$ & $\sigma_{\text{NLO}}$ & K \\\hline
$pp \rightarrow H jj $ (``VBF-$H$'') &
  960.61(10) fb &  906.2(5) fb  & 0.94 \\ 
$pp \rightarrow H jjj $ (``VBF-$H$+jet'') &
  163.73(5) fb   & 131.9(8) fb   & 0.81 \\ 
$pp \rightarrow HH jj $ (``VBF-$HH$'') &
  0.60364(12) fb & 0.5613(4) fb  & 0.93 \\ 
$pp \rightarrow H\gamma jj $ (``VBF-$H\gamma$'') &
  15.596(2) fb   & 14.666(11) fb & 0.94 \\ 
\hline
$pp \rightarrow \ell^+ \ell^- jj $ (``VBF-$Z_\ell$'') &
  265.69(15) fb  & 274.4(5) fb   & 1.03 \\
$pp \rightarrow \nu \bar{\nu} jj $ (``VBF-$Z_\nu$'') &
  714.3(4) fb    & 750.8(13) fb  & 1.05 \\
$pp \rightarrow \ell^+ \nu jj $ (``VBF-$W^+$'') &
  2319.4(4) fb   & 2317.4(15) fb & 1.00 \\ 
$pp \rightarrow \ell^- \bar{\nu} jj $ (``VBF-$W^-$'') &
  1387.3(3) fb   & 1432.1(9) fb  & 1.03 \\ 
$pp \rightarrow \gamma jj $ (``VBF-$\gamma$'') &
  5327.1(15) fb  & 5528(9) fb    & 1.04 \\ 
$pp \rightarrow \ell_1^+ \nu_{\ell_1} \ell_2^- \bar{\nu}_{\ell_2} jj $ (``VBF-$W^+ W^-$'') &
  22.691(13) fb  & 22.30(4) fb   & 0.98 \\ 
$pp \rightarrow \ell_1^+ \ell_1^- \ell_2^+ \ell_2^- jj $ (``VBF-$Z_\ell Z_\ell$'') &
  228.90(16) ab  & 234.0(7) ab   & 1.02 \\ 
$pp \rightarrow \ell_1^+ \ell_1^- \nu_2 \bar{\nu}_2 jj $ (``VBF-$Z_\ell Z_\nu$'') &
  1.5654(9) fb   & 1.581(4) fb   & 1.01 \\ 
$pp \rightarrow \ell_1^+ \nu_{\ell_1} \ell_2^+ \ell_2^- jj $ (``VBF-$W^+ Z_\ell$'') &
  1.7056(7) fb   & 1.679(2) fb   & 0.98 \\ 
$pp \rightarrow \ell_1^- \bar{\nu}_{\ell_1} \ell_2^+ \ell_2^- jj $ (``VBF-$W^- Z_\ell$'') &
  0.9164(3) fb   & 0.9424(9) fb  & 1.03 \\ 
$pp \rightarrow \ell_1^+ \nu_{\ell_1} \ell_2^+ \nu_{\ell_2} jj $ (``VBF-$W^+ W^+$'') &
  5.7321(17) fb  & 5.616(4) fb   & 0.98 \\ 
$pp \rightarrow \ell_1^- \bar{\nu}_{\ell_1} \ell_2^- \bar{\nu}_{\ell_2} jj $ (``VBF-$W^- W^-$'') &
  1.7388(5) fb   & 1.8785(18) fb & 1.08 \\ 
$pp \rightarrow \ell^+ \nu \gamma jj $ (``VBF-$W^+\gamma $'') &
  32.412(10) fb  & 31.98(3) fb   & 0.99 \\ 
$pp \rightarrow \ell^- \bar{\nu} \gamma jj $ (``VBF-$W^-\gamma $'') &
  18.496(5) fb   & 18.88(2) fb   & 1.02 \\ 
$pp \rightarrow \ell^+ \ell^- \gamma jj $ (``VBF-$Z_\ell \gamma $'') &
  7.406(3) fb    & 7.560(14) fb  & 1.02 \\ 
$pp \rightarrow \nu \bar{\nu} \gamma jj $ (``VBF-$Z_\nu \gamma $'') &
  18.123(5) fb   & 18.62(3) fb   & 1.03 \\ 
\hline
$pp \rightarrow q \bar{q} \ell^- \bar{\nu}_{\ell} jj $ (``VBF-$W_\text{had}^+ W^-$'') &
  22.384(18) fb   & 24.97(7) fb   & 1.12 \\ 
$pp \rightarrow \ell^+ \nu_{\ell} q \bar{q} jj $ (``VBF-$W^+ W_\text{had}^-$'') &
  21.075(17) fb   & 23.44(9) fb   & 1.11 \\ 
$pp \rightarrow \ell^+ \ell^- q \bar{q} jj $ (``VBF-$Z_\ell Z_\text{had}$'') &
  2.186(3) fb     & 2.455(19) fb  & 1.12 \\ 
$pp \rightarrow q \bar{q} \ell^+ \ell^- jj $ (``VBF-$W_\text{had}^+ Z_\ell$'') &
  2.0880(10) fb   & 2.304(4) fb   & 1.10 \\ 
$pp \rightarrow \ell^+ \nu_{\ell} q \bar{q} jj $ (``VBF-$W^+ Z_\text{had}$'') &
  9.046(5) fb     & 9.71(2) fb    & 1.07 \\ 
$pp \rightarrow q \bar{q} \ell^+ \ell^- jj $ (``VBF-$W_\text{had}^- Z_\ell$'') &
  1.0516(5) fb    & 1.1991(19) fb & 1.14 \\ 
$pp \rightarrow \ell^- \bar{\nu}_{\ell} q \bar{q} jj $ (``VBF-$W^- Z_\text{had}$'') &
  4.670(2) fb     & 5.347(11) fb  & 1.14 \\ 
$pp \rightarrow q \bar{q} \ell^+ \nu_{\ell} jj $ (``VBF-$W_\text{had}^+ W^+$'') &
  13.942(5) fb    & 14.960(15) fb & 1.07 \\ 
$pp \rightarrow q \bar{q} \ell^- \bar{\nu}_{\ell} jj $ (``VBF-$W_\text{had}^- W^-$'') &
  3.9314(15) fb   & 4.685(7) fb   & 1.19 \\
\end{tabular} }
}
\caption{Integrated cross sections for VBF production processes for the
LHC running at a center-of-mass energy of 13 TeV using the cuts given in
\eq{eq:generalcuts}. Results are given summed over all three lepton
generations and, in case of quarks, all combinations which do not
involve a top quark. The error in brackets is the statistical error from
Monte Carlo integration.}
\label{tab:nlocs13}
\end{table}

As input parameters in the electroweak sector we choose the Fermi
constant and $W$, $Z$ and the Higgs boson mass. The other electroweak
parameters, namely the electromagnetic coupling and the weak mixing
angle, are then fixed via tree-level relations. As numerical values we
use~\cite{Agashe:2014kda}
\begin{align}
M_W &= 80.385  \GeV \,, & \Gamma_W &= 2.097 \GeV \,, \nonumber\\
M_Z &= 91.1876 \GeV \,, & \Gamma_Z &= 2.508 \GeV \,, \nonumber\\
M_H &= 125.0   \GeV \,, & \Gamma_H &= 4.070 \MeV \,, \nonumber\\
G_F &= 1.16638\cdot 10^{-5}\GeV^{-2} \,, \nonumber\\
\alpha^{-1} &\equiv \alpha_{G_F}^{-1} = 132.23308 \,, &
\sin^2(\theta_W) &= 0.22290 \,.
\label{eq:EWpara}
\end{align}
A set of minimal cuts is imposed on the transverse momenta and
rapidities of the final-state charged leptons and photons as well as on
their separation to simulate the capabilities of the experimental
detectors. 
\begin{align}
p_{T,\ell(\gamma)} &> 20 \GeV , &
|y_{\ell(\gamma)}| &< 2.5 , \nonumber\\
R_{j\ell} &> 0.4 , &
R_{\ell\gamma} &> 0.4 , \nonumber\\
R_{j\gamma} &> 0.7 , &
R_{\gamma\gamma} &> 0.4 , \nonumber\\
M_{\ell^+\ell^-} &> 15 \GeV .
\label{eq:generalcuts}
\end{align}
The last cut is chosen to remove singularities from a virtual photon
splitting into a pair of charged leptons, $\gamma^* \rightarrow \ell^+
\ell^-$.

All final-state partons with pseudorapidity $\eta<5.0$ are clustered
into jets with the anti-$k_T$
algorithm~\cite{Catani:1993hr,Ellis:1993tq} using an $R$ separation
parameter of $0.4$. Jets are required to have a transverse momentum
$p_{T,j} > 30 \GeV$ and rapidity $|y_j|<4.5$. 
A fixed $R$ separation cut between partons and jets would spoil the
cancellation of infrared singularities. Therefore, to separate
final-state jets and photons, we employ the procedure suggested by
Frixione in Ref.~\cite{Frixione:1998jh}. 
An event is only accepted, if the condition
\begin{align}
\sum_i E_{T,i} \Theta(\delta-R_{i\gamma}) &\le p_{T,\gamma}
\frac{1-\cos\delta}{1-\cos\delta_0} & \forall \delta & \le \delta_0
\label{eq:Frixione}
\end{align}
is fulfilled. Thereby, $E_{T,i}$ denotes the transverse energy of parton
$i$, $p_{T,\gamma}$ the transverse momentum of the photon and
$R_{i\gamma}$ their separation. In our setup we choose the separation
parameter $\delta_0=0.7$. The formula given above avoids the QED IR
divergence from collinear emission of a photon from a quark, while at
the same time allowing final-state gluons arbitrarily close to the
direction of the photon as long as these are soft enough, thus retaining
the full QCD pole.

\begin{figure}
\begin{center}
\begin{tabular}{cc}
\includegraphics[width=0.4\textwidth]{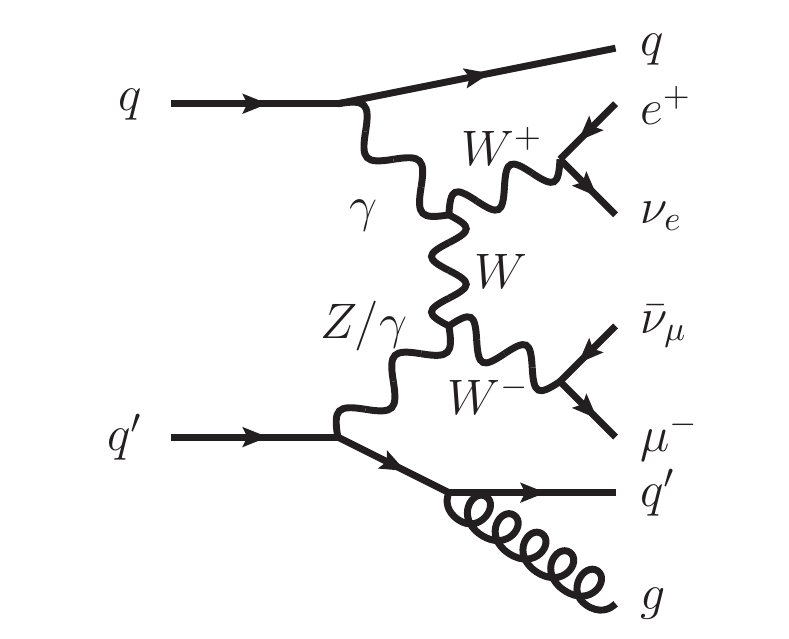} &
\includegraphics[width=0.4\textwidth]{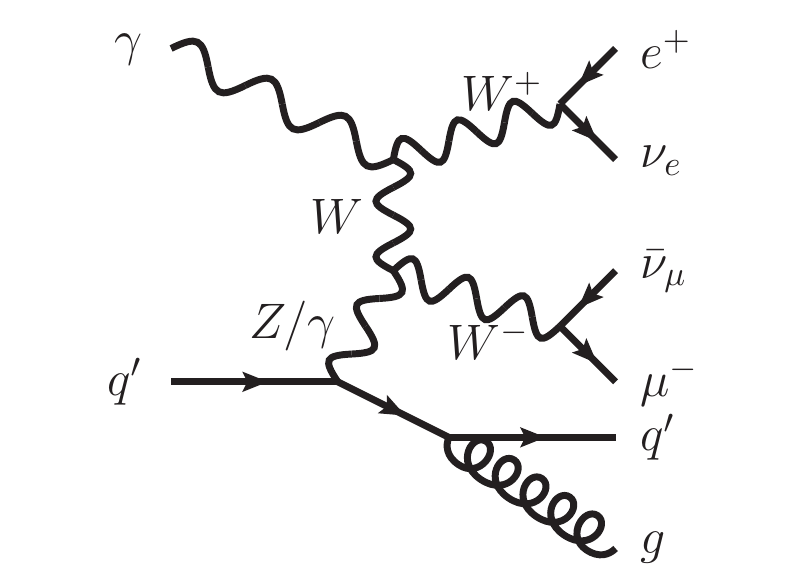} \\[1em]
\end{tabular}
\end{center}
\caption{Example Feynman diagrams of a VBF process (VBF-$W^+W^-$
production)where a $t$-channel photon can develop a QED divergence.
\textit{Left:} real-emission diagram with divergence, \textit{right:}
corresponding Born diagram which would provide the contribution to
cancel the divergence.
}
\label{fig:fm_gammaproblem}
\end{figure}
Another issue arises in VBF processes where an exchanged virtual photon
can become on-shell. An example Feynman diagram where this can occur is
given on the left-hand side of \fig{fig:fm_gammaproblem}, taking
VBF-$W^+W^-$ production as an example. If both the final-state quark on
the lower line and the gluon are hard enough and well separated to be
identified as the two tagging jets of the process, the transverse
momentum of the upper quark is no longer restricted. In particular it
can become collinear to the beam axis, \ie $p_1 \parallel p_3$. In this case, the
momentum transfer becomes zero and the photon propagator develops a
divergence. Note that this issue can only appear in the real-emission
part, as for Born kinematics both quarks need to have a finite
transverse momentum to be identified as the tagging jets. Also, if only
massive bosons can appear in the $t$-channel, like in VBF-$H$ or
VBF-$W^+W^+$ production, the divergence is regulated by the mass
of the exchanged boson. The origin of this QED divergence is the fact
that this diagram is also the real-emission correction to another
process, namely $\gamma q' \rightarrow q' g e^+ \nu_e \bar{\nu}_\mu
\mu^-$. The corresponding Feynman diagram is depicted on the right-hand
side of \fig{fig:fm_gammaproblem}. The virtual corrections to this
process and the $\Order{\alpha}$ corrections to the photon PDF will then
generate the same terms, but with a relative minus
sign, so that the divergences cancel. To calculate this, we would also
need the parton density of photons in the proton. Such PDF sets are in
principle available nowadays, but have quite large errors on the photon
densities~\cite{Martin:2004dh,Ball:2013hta,Schmidt:2015zda}. To avoid
adding this contribution altogether, we can alternatively introduce a
technical cut on the photon virtuality. In our case we choose
$Q^2>4\GeV^2$. Varying this cut then gives an estimate on the error
introduced by this procedure, which turns out to be small and overall
negligible~\cite{Oleari:2003tc}. One can understand the smallness also
from the kinematic structure of these events. As $Q^2$ approaches zero,
the final-state quark has no transverse momentum and is lost in the
beam pipe. Therefore, the extra gluon emission must form the second jet
of the VBF signature. For both the quark and the gluon radiated from it
to be identified as tagging jets, the latter needs to be a very hard and
wide-angle emission, which induces a large suppression factor from the
matrix element.

In order to exploit the particular properties of our VBF signature, we
require the presence of at least two jets in the final state. The two
jets with the largest transverse momentum are labeled tagging jets and
must additionally fulfill 
\begin{align}
|\Delta y_{jj}| &> 3.6 \,, & m_{jj} &> 600 \GeV \,.
\label{eq:vbfcuts}
\end{align}
For processes with semi-leptonic decays of the vector bosons, where
additional jets due to these decays appear in the final-state, the jet
pair with an invariant mass closest to the vector boson mass is removed
first before applying the tagging jet criterion above.

As PDFs we choose the CT14llo set~\cite{Dulat:2015mca} with $\alpha_s(m_Z) =
0.130$ at LO and the central set of PDF4LHC15\_nlo at NLO having
$\alpha_s(m_Z) = 0.118$, using the implementation provided by
LHAPDF~\cite{Buckley:2014ana}. 

Looking at the results for the LHC running 8 and 13 TeV center-of-mass
energy given in Tables~\ref{tab:nlocs8} and~\ref{tab:nlocs13},
respectively, we see that the NLO QCD corrections are modest, typically
ranging in the $\pm10\%$ range.
Comparing with the results in Ref.~\cite{Campanario:2015vqa}, the more relaxed jet cuts
give in general larger cross sections. Also the CT14llo set produces
larger PDF values in the relevant $x$ regions than the older CTEQ6L1
sets used in Ref.~\cite{Campanario:2015vqa}, while the difference in the used NLO sets
is smaller. Hence, the $K$ factors are consistently smaller here.

Factorization and renormalization scale have been taken as the
virtuality of the $t$-channel gauge bosons,
$\mu_{F,i}=\mu_{R,i}=\sqrt{Q_i^2}$. Thereby, we use independent values
for the PDFs, loop corrections or gluon emissions on the upper line and
on the lower line. As we have already seen in the analytic calculations
above, \eg \eq{eq:matv}, this is the relevant scale appearing in the
process, so should be a good choice. 

\begin{figure}
\includegraphics[width=0.47\textwidth]{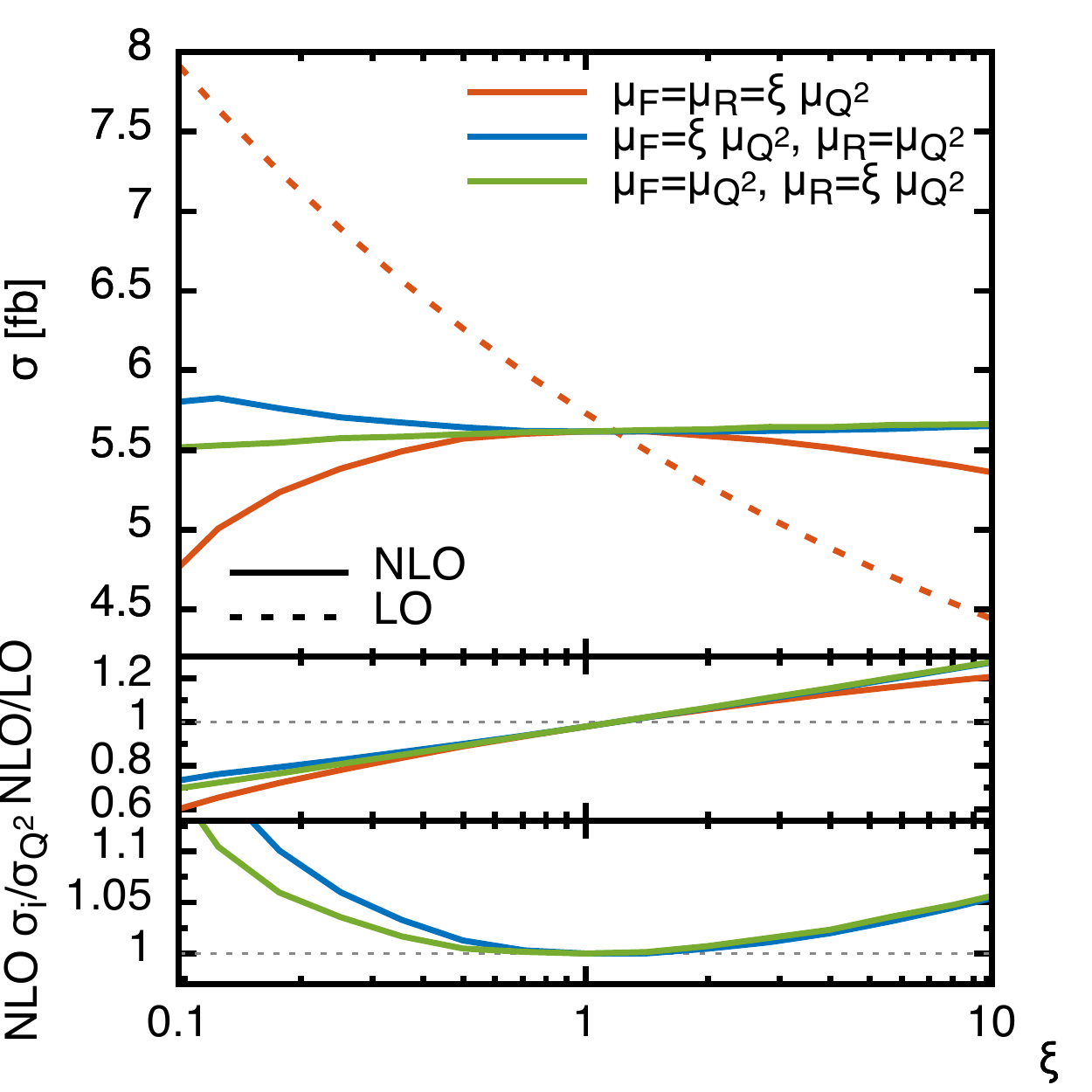} \quad
\includegraphics[width=0.47\textwidth]{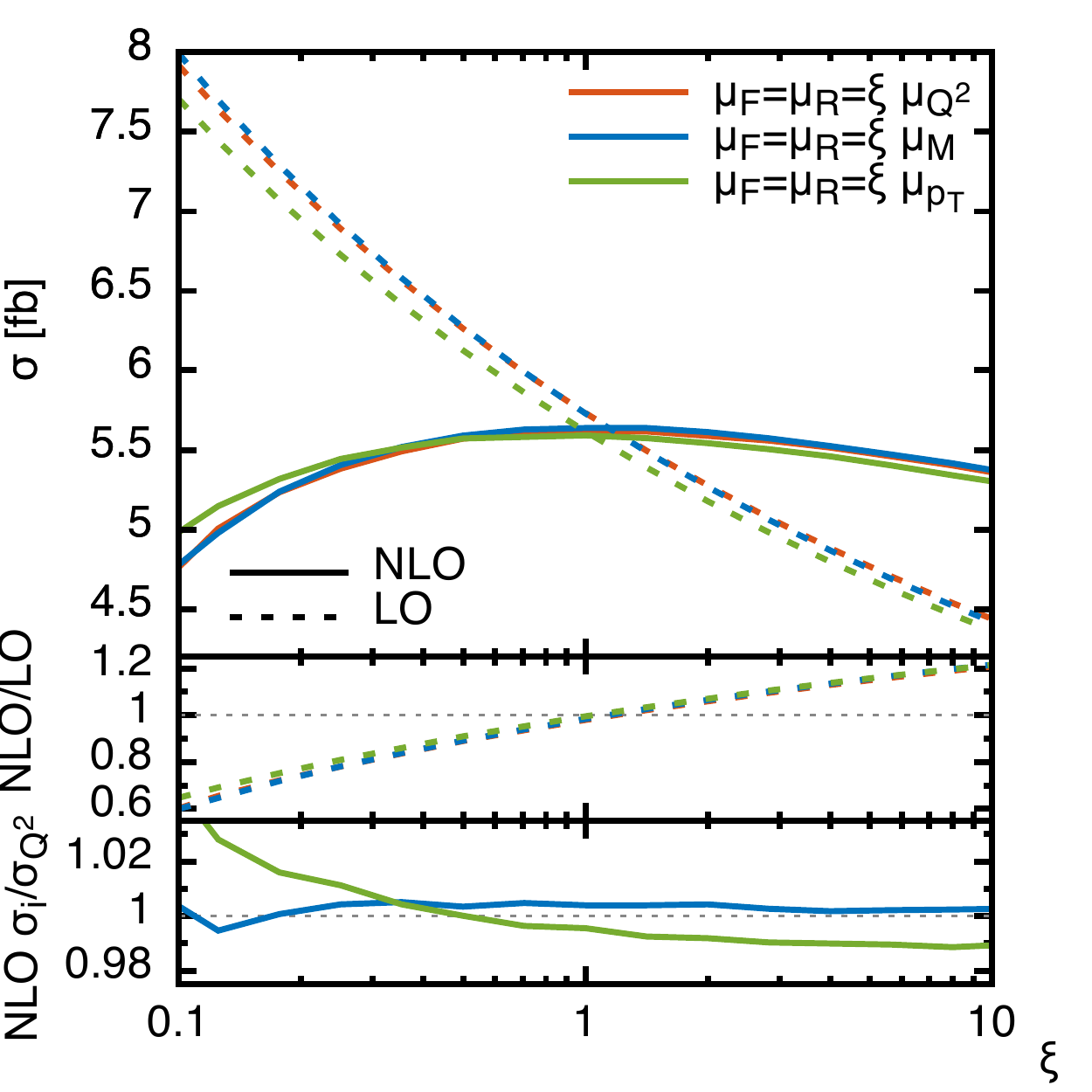} \\
\caption{Scale variation of the LO and NLO cross sections for the
process $pp \rightarrow e^+ \nu_e \mu^+ \nu_\mu j j$
(``VBF-$W^+W^+jj$'') for the LHC at 13 TeV center-of-mass energy.
\textit{Left:} Joint and independent variation of the factorization
scale $\mu_F$ and renormalization scale $\mu_R$ using the virtuality
$Q^2$ of the exchanged bosons as central value.
\textit{Right:} Comparison of three different choices for the central
scale: virtuality $Q^2$, fixed value $\mu_M=2 M_W$ and transverse
momentum $p_T$ of the leading jet.
The smaller panels in both plots show the ration of NLO over LO cross
sections in the upper one and in the lower one the cross section ratio
over a joint scale variation using $\mu_{Q^2}$ as the central scale.
}
\label{fig:scalevar}
\end{figure}
An estimate of missing higher-order corrections is given by varying
factorization and renormalization scale or taking other sensible
choices, as the scale dependence of cross sections is an artifact of the
truncation of the perturbative series and vanishes for calculations to
all orders. Besides the virtuality of the exchanged bosons, we consider
also two other choices. One is simply a fixed scale, where as central
value we use $2 M_W$, taking again our example process of VBF-$W^+W^+jj$
production. The second option we are going to use is the transverse
momentum of the leading jet. This observable has the advantage that it
is directly observable and also well-defined once we include additional
parton-shower and hadronization effects. 

Numerical results are shown in \fig{fig:scalevar}. On the left we use
the virtuality of the exchanged bosons as central value $\mu_{Q^2}$ and
vary this with a factor $\xi$ from 0.1 to 10. At LO, as no factors of
$\alpha_s$ are present in the cross section, the dependence is purely on
the factorization scale. A conventional estimate for the associated
errors is given by a scale variation with a factor between $\frac12$ and
2, which here yields a cross section variation of $+9.3\%$ and $-8.0\%$,
respectively. At NLO, the scale dependence is very flat in the central
region, with a remaining scale variation uncertainty of around $-0.8\%$
when varying both scales either jointly or only one of them. 
In the two smaller panels below we plot ratios of cross sections. The
upper of those shows the $K$ factor, the ratio of NLO over the LO cross
section. While in the central range around $\xi=1$ the two agree
reasonably well, for smaller or larger values the strong scale
dependence of the LO cross section induces $K$ values significantly
different from 1. The lowest panel contains the ratio of individual
scale variations over the joint one, again showing a very small spread
in the different predictions except for very small or large values of
$\xi$.

The right plot in \fig{fig:scalevar} compares the three different scale
choices, virtuality $Q^2$, fixed value $M$ and leading-jet transverse
momentum $p_T$ for a joint factorization and renormalization scale
variation. Similar to the previous case, the quite significant scale
dependence at LO is vastly reduced for the NLO results, namely around
one percent for both fixed value and momentum transfer. The middle panel
shows again the $K$ factor, which is basically independent of the scale
type choice. In the lowest panel we finally compare cross section ratios
of the three scale types. Choosing different scale types can serve as
another method of obtaining a scale variation error. However, also there
we only get a difference of $0.4\%$ when comparing the central values
and $1.3\%$ when additionally taking a variation within $\xi \in[0.5;2]$
into account.

Such tiny numbers should however be taken with a grain of salt, as the
variation by a factor 2 is purely conventional and there are numerous
examples where this underestimates known higher-order corrections. Also,
the effect of new topologies opening up, like double-gluon exchange
between the quark lines, is not covered by this scale variation
procedure at all. We will see in the last section of this chapter how
the known 2-loop corrections in VBF-$H$ production modify this picture.

Finally, the central values of the PDFs and $\alpha_s$ are only
determined with a finite accuracy, as they are extracted from a large
number of different experimental measurements. Modern PDF sets contain
different member sets besides the best-fit one, where the internal
parameters are varied along the eigenvectors of their correlation
matrix within the uncertainties given by the experimental input. Taking
the minimum and maximum value of the different results then yields the
PDF variation uncertainty. Using the PDF4LHC15\_nlo\_30\_pdfas
set~\cite{Butterworth:2015oua}, which is a
compressed~\cite{Watt:2012tq,Gao:2013bia,Carrazza:2015aoa} set combining the
latest results of the three main PDF fitting groups CTEQ~\cite{Dulat:2015mca},
MMHT~\cite{Harland-Lang:2014zoa} and NNPDF~\cite{Ball:2014uwa}, we obtain
variations of $-0.7\%$ and $+1.2\%$ compared to the central set. Consistently
varying $\alpha_s(M_Z)$ to $0.0165$ and $0.0195$ both in the PDF and the matrix
element changes the cross section by $\pm 0.2\%$, respectively.

\subsection{NLO Electroweak Corrections}
\label{sec:nloew}

Corrections beyond the NLO QCD level are known so far only for VBF-$H$
production. While a definitive statement about these effects for other
VBF processes can only be made once they have been explicitly
calculated, the expectation is that the general features will be similar
in all VBF processes.

\begin{figure}
\begin{center}
\begin{tabular}{cc}
\includegraphics[width=0.45\textwidth]{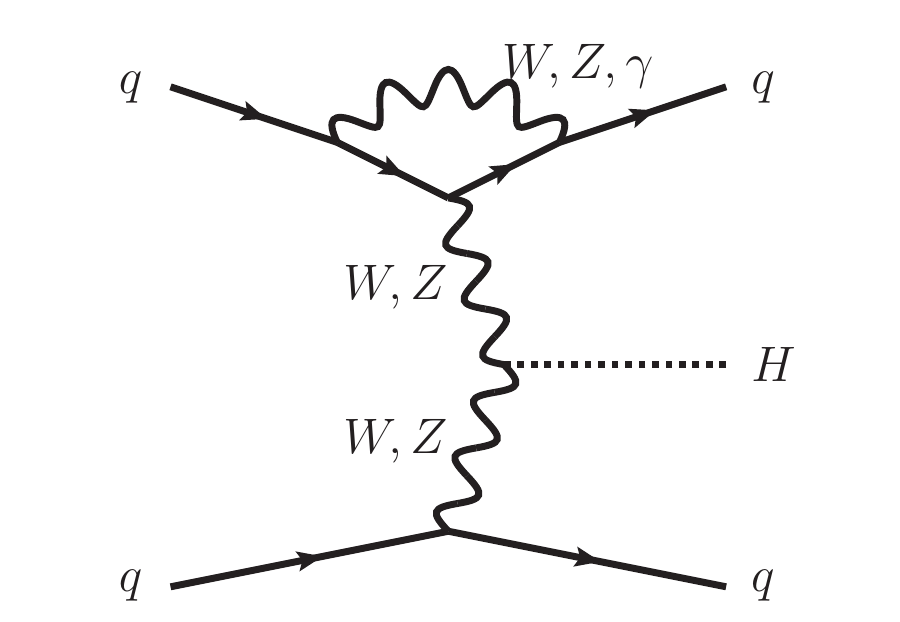} &
\includegraphics[width=0.45\textwidth]{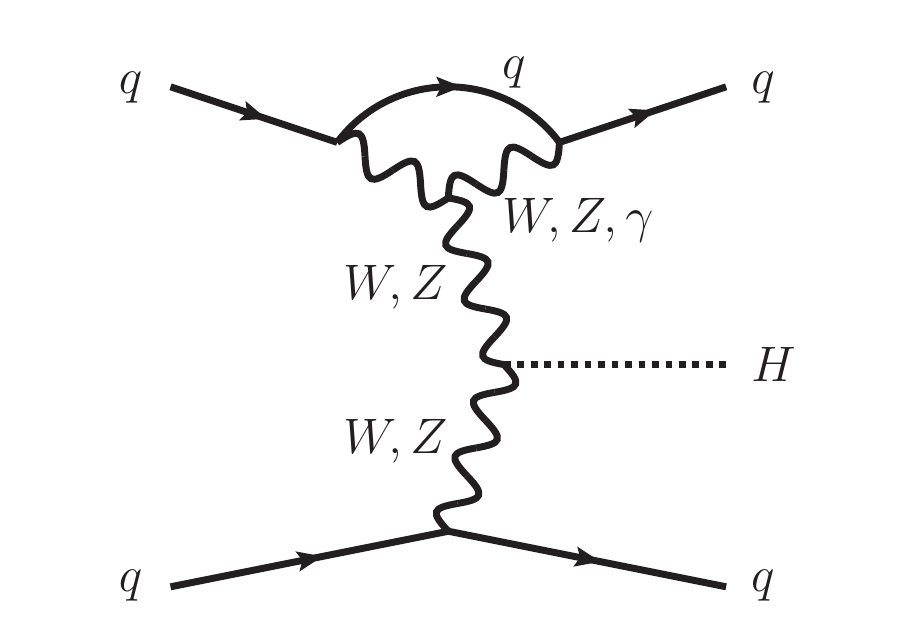} \\[1em]
\includegraphics[width=0.45\textwidth]{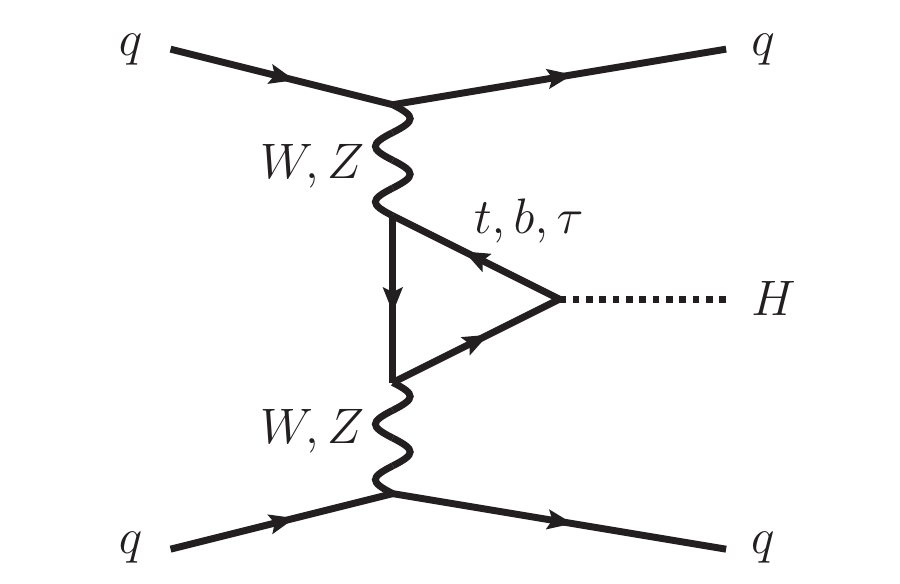} &
\includegraphics[width=0.45\textwidth]{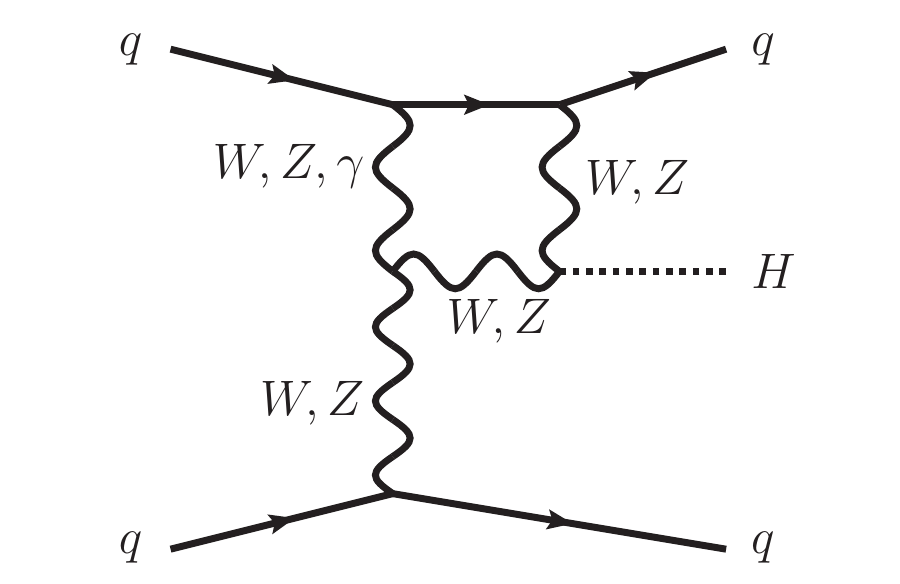} \\[1em]
\includegraphics[width=0.45\textwidth]{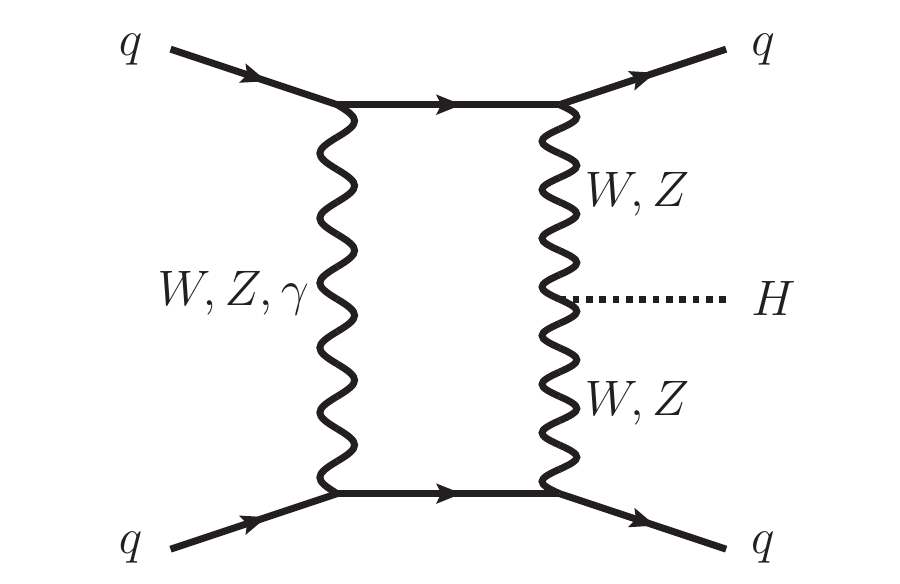} &
\includegraphics[width=0.45\textwidth]{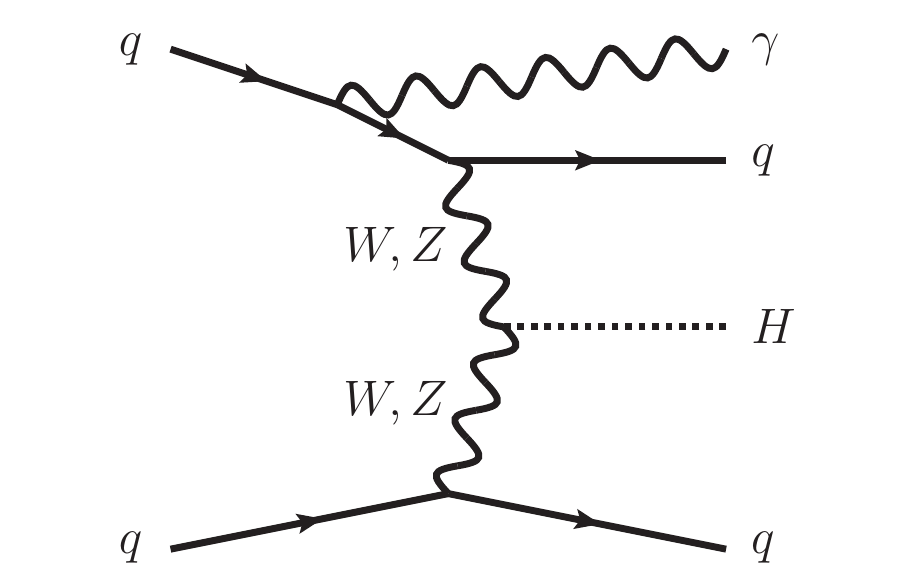} \\[1em]
\end{tabular}
\end{center}
\caption{Example Feynman diagrams of electroweak corrections in VBF-$H$
production. 
When selecting specific values for the quark flavors or
gauge bosons, electromagnetic charge conservation needs to be taken into
account.
}
\label{fig:fm_vbfh_ew}
\end{figure}
The first type we are going to discuss are electroweak corrections to
VBF-$H$ production~\cite{Ciccolini:2007jr,Ciccolini:2007ec,Figy:2010ct}. The
virtual corrections are much richer than for the QCD case. Some example Feynman
diagrams are shown in
\fig{fig:fm_vbfh_ew}.
Besides loop contributions to the $qqV$ vertices, now
also the $VVH$ vertex receives corrections, either by a loop of
electroweak gauge bosons or a closed loop of heavy, third-generation
fermions. Gauge bosons can also connect the two quark lines, giving rise
to loop diagrams up to the pentagon level. Whenever a photon is attached
to an external quark, the corresponding diagram shows an infrared
divergence. These need to be canceled by corresponding real-emission
diagrams with an additional external photon. These can be either in the
final-state or appearing in the initial-state as partons coming from
the proton. For the latter case, corresponding QED PDFs are necessary for
the calculation.
Emission of heavy vector bosons does not need to be taken into account.
Due to their finite mass, no divergences appear in the soft or collinear
limit, and their decay products leave extra signatures in the detector,
so these processes are clearly distinguishable.
The diagrams also point to the main difficulty in calculating the
electroweak corrections for other VBF processes. There will always be 
diagrams where all external particles are connected by one big loop like
the pentagon diagram for VBF-$H$. So for single $V$ production these are
hexagons and for vector-boson scattering with decays one needs to
compute octagons, loops with eight external particles. Both of them are
challenging in terms of numerical stability as well as run time.

For a Higgs boson of 125~GeV, the size of the NLO electroweak correction
on the total VBF-$H$ cross section is around $-5\%$~\cite{Ciccolini:2007jr,Ciccolini:2007ec,Figy:2010ct}, and
therefore of the same size as the NLO QCD corrections. Applying typical
VBF cuts changes this value only mildly. The photon-induced
real-emission processes thereby lead to a small reduction of the
corrections by about $1\%$ of the Born cross section.

\begin{figure}
\begin{center}
\begin{tabular}{cc}
\includegraphics[width=0.45\textwidth]{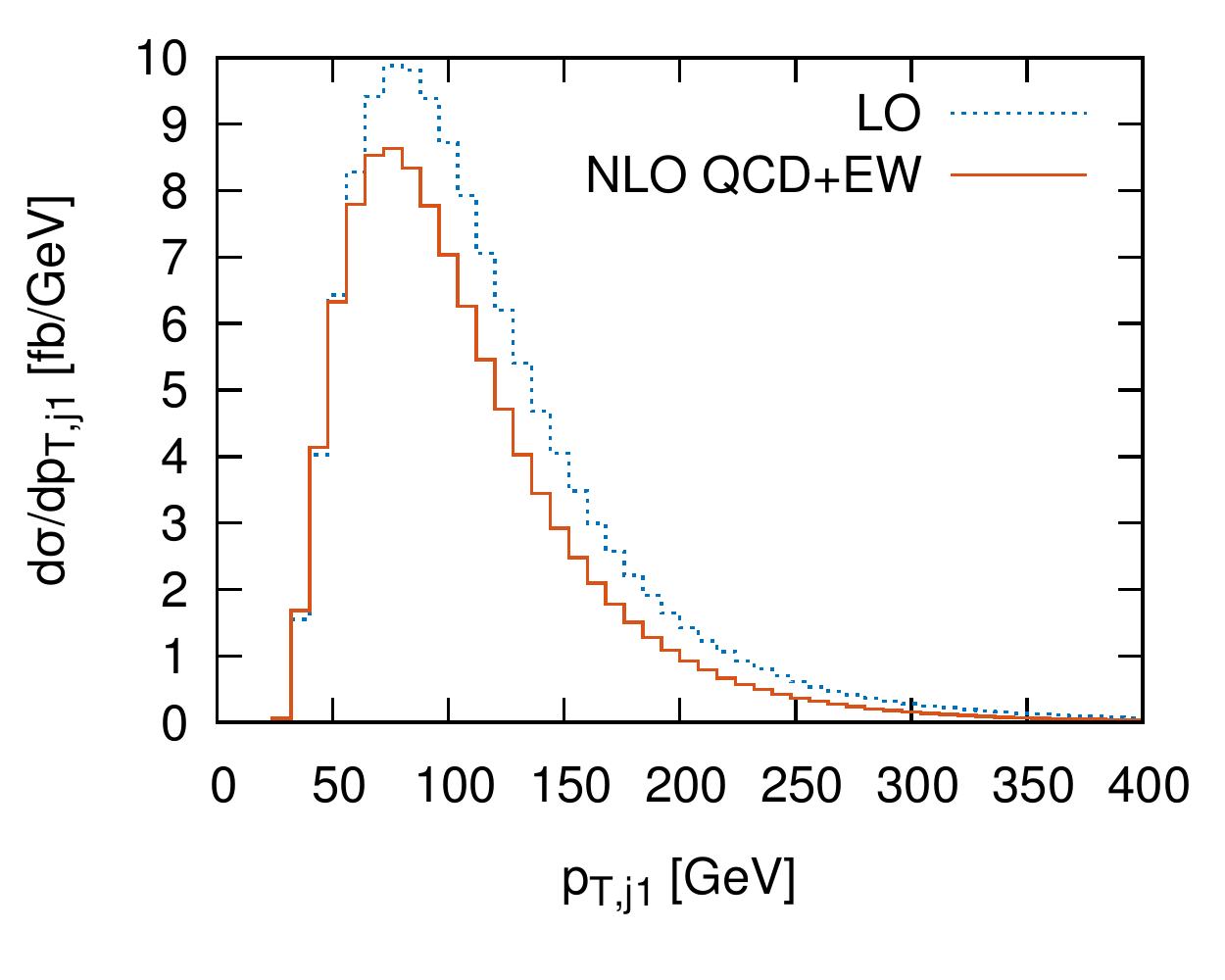} &
\includegraphics[width=0.45\textwidth]{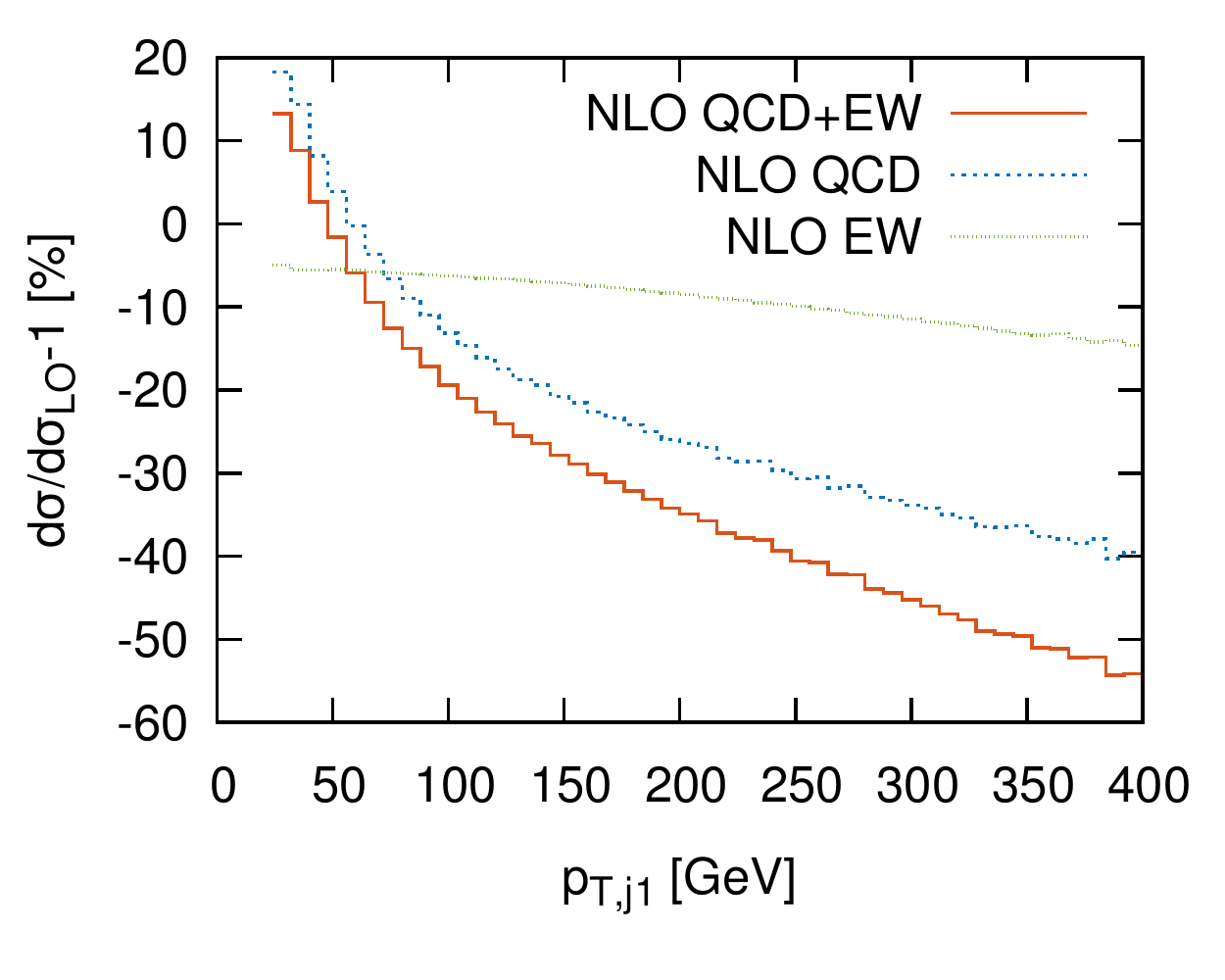} \\[2em]
\includegraphics[width=0.45\textwidth]{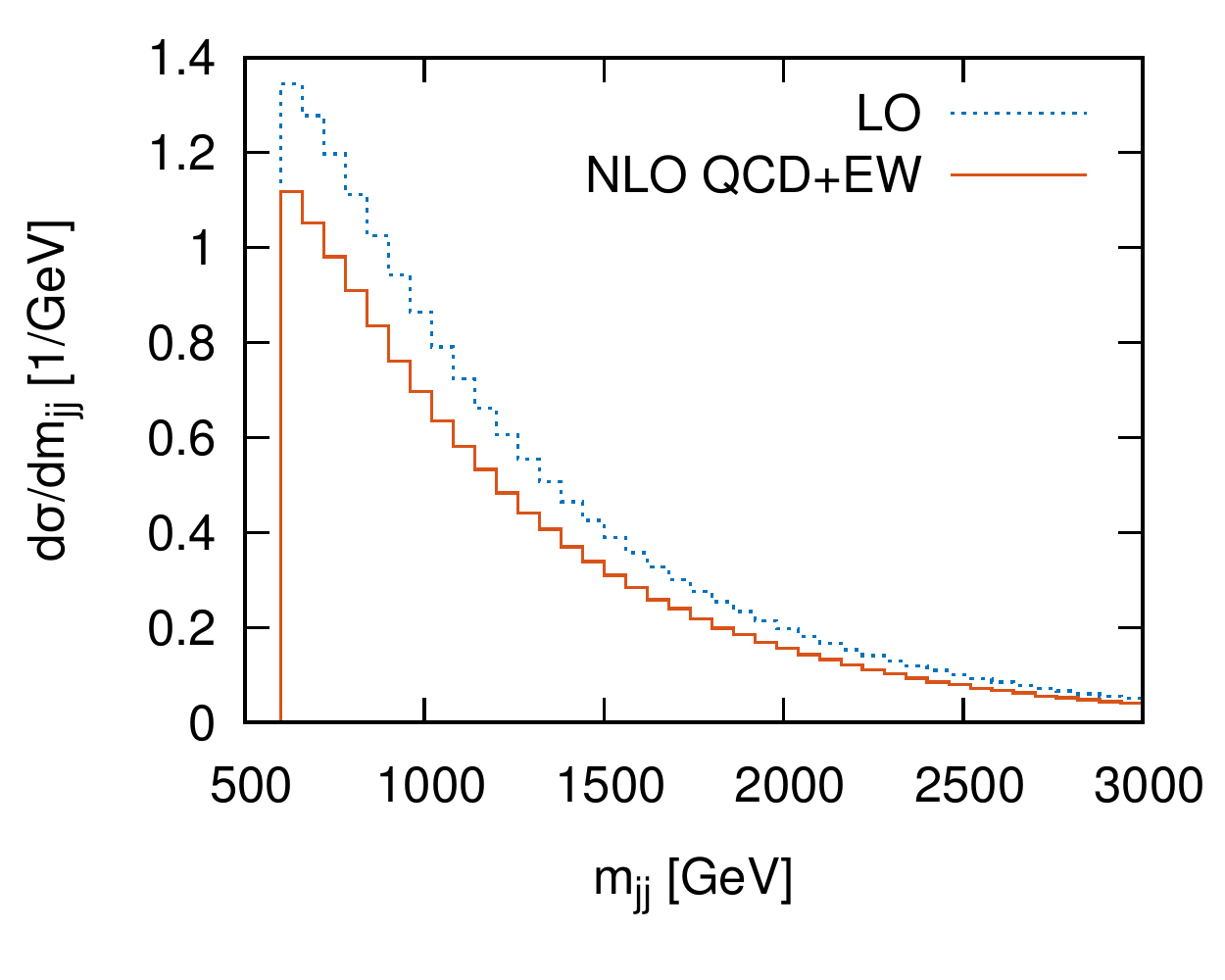} &
\includegraphics[width=0.45\textwidth]{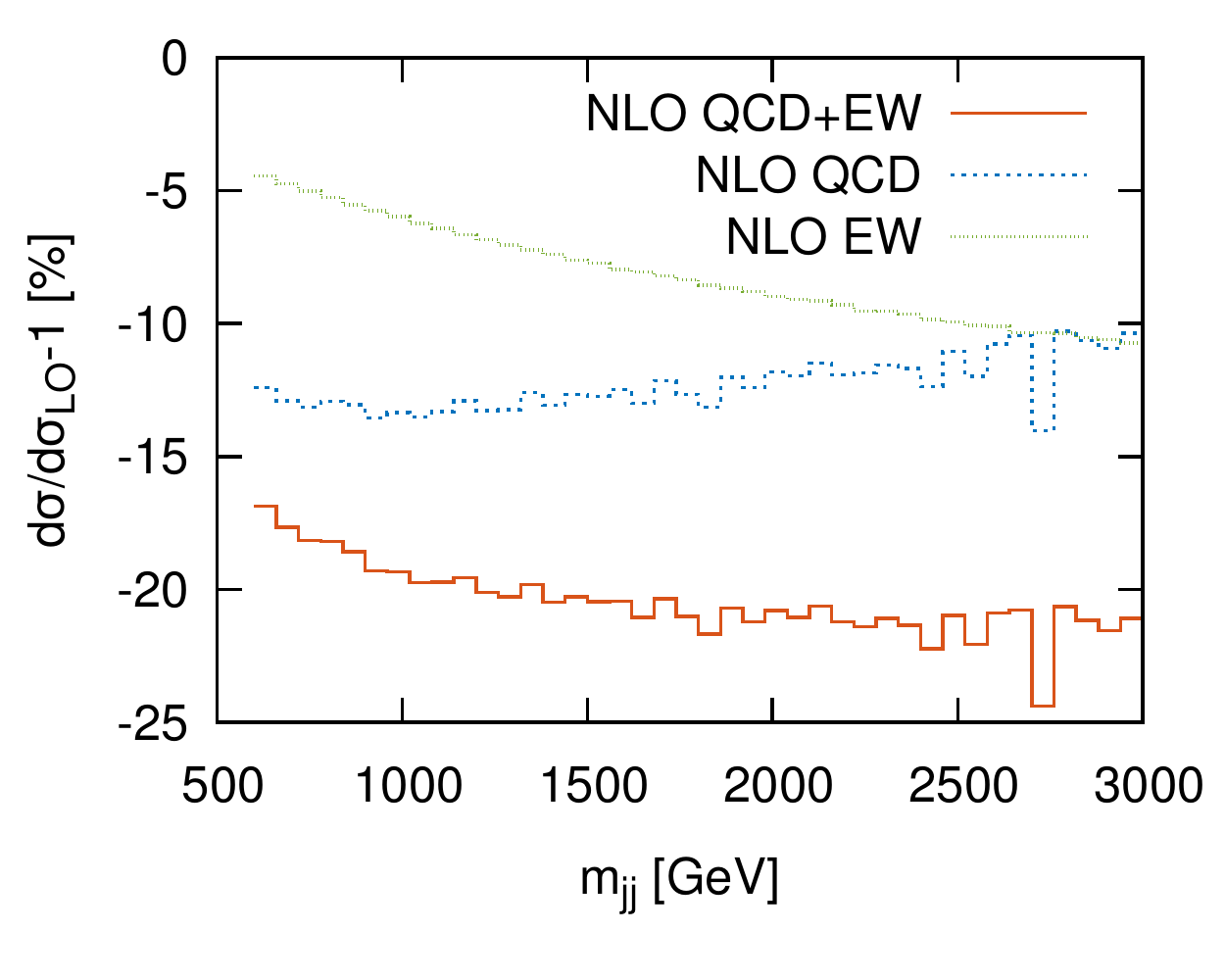} \\[1em]
\end{tabular}
\end{center}
\caption{NLO QCD and EW effects in VBF-$H$ production with VBF cuts
for the LHC with 13 TeV center-of-mass energy.
\textit{Top row:} Transverse momentum distribution of the leading jet.
\textit{Bottom row:} Invariant mass distribution of the two tagging
jets. 
}
\label{fig:nloew}
\end{figure}
In \fig{fig:nloew}, generated with \HAWK~\cite{HAWK}, we show the effects
on two important distributions, namely the transverse momentum of the
leading jet and the invariant mass of the two tagging jets. The left
panels in the figure show the absolute distributions both for LO and the
combined NLO QCD and EW contribution and the right panels their relative
effects both separately and jointly. Looking at the various curves, we
see that for the transverse-momentum distribution the two types of
corrections show a similar behavior. They are fairly modest for small
values and rise significantly when going to larger transverse momenta,
reaching almost 60\% for a value of 400~GeV. The same effect also
happens for the second tagging jet, inducing corrections of more than
$-40\%$ for transverse momenta larger than 150~GeV. EW corrections hence
lead to a significantly larger suppression of tagging jets with large
transverse momenta compared than would be expected from QCD effects
alone. The situation is different for the invariant mass of the two
tagging jets, shown in the bottom row of \fig{fig:nloew}. Here the mass
dependence of QCD and EW corrections approximately cancels between the
two for large invariant-mass values, leading to a flattening of the
correction with a numerical value of about $-20\%$.

\subsection{QCD Corrections beyond NLO}
\label{sec:nnloqcd}

Another class of contributions beyond NLO QCD are the
NNLO QCD corrections. As with the NLO
EW corrections, these have been considered only for VBF-$H$ production
so far. Also, the results are known only in the VBF approximation
discussed at the end of \chap{chap:processclass}, where
the two quarks coming from the proton are assumed to be in two separate
copies of the color $SU(3)$ group. Hence, the only contributions are
loop corrections to the $qqV$ vertices and real-emission diagrams
without interference of emissions from the upper and lower line. The
two-loop virtual diagram with two gluons exchanged between the quark
lines, which is color-suppressed compared to other two-loop matrix
elements, is neglected in this structure-function approach. 
The NNLO QCD corrections to the inclusive cross section have been
computed in Ref.~\cite{Bolzoni:2010xr,Bolzoni:2011cu} and are found to
be very small, on the order of $0.4\%$. The uncertainty determined by
varying factorization and renormalization scale around the central
choice of momentum transfer $Q$ is further reduced compared to the NLO
QCD value. Recently, also the N3LO QCD corrections for the inclusive cross
section in the structure-function approach have been
calculated~\cite{Dreyer:2016oyx,Karlberg:2016zik}. Their impact is tiny,
at the level of $0.1-0.2\%$, which is well covered by the scale
variation band of the NNLO QCD calculation. The remaining scale
uncertainty is also at the per mill level.

The NNLO QCD calculation has been recently extended to differential cross
sections in Ref.~\cite{Cacciari:2015jma}. The calculation uses a trick,
which is possible due to the rather simple QCD structure of the diagram.
It is based on the fact that the knowledge of the momentum-transfer
vectors $q_i$, when combined with both quarks being on-shell and the
incoming ones along the $z$-axis, allows to fully reconstruct the
four-vectors of the external quarks in the Born diagram.
The inclusive corrections given by the structure function approach yield
the loop contributions and the corresponding single and double
real-emission contributions, with extra radiation integrated out,
combined in one single number per Born event. 
This can then be augmented by the contributions with a single real
emission and a one-loop vertex correction and the ones with two real
emissions. These contain the extra emissions explicitly and therefore
have the correct momentum structure of the real-emission events instead
of the integrated out net effect. The necessary ingredients are given
exactly by the NLO QCD calculation of electroweak $H+3\text{ jets}$
production~\cite{Figy:2007kv,Jager:2014vna,Campanario:2013fsa}. To
remove the double-counting just introduced, we finally need to project
these extra-emission events back to their corresponding Born structure.
To this end we require that the momentum transfer $q_i$ stays unchanged
and then recompute the external quark momenta. These projected events
then enter with the same absolute weight as the real-emission one but
opposite sign and thus cancel their corresponding contribution in the
structure-function part.

\begin{figure}
\begin{center}
\begin{tabular}{c@{\hspace*{3em}}c}
\includegraphics[height=0.5\textheight,page=1]{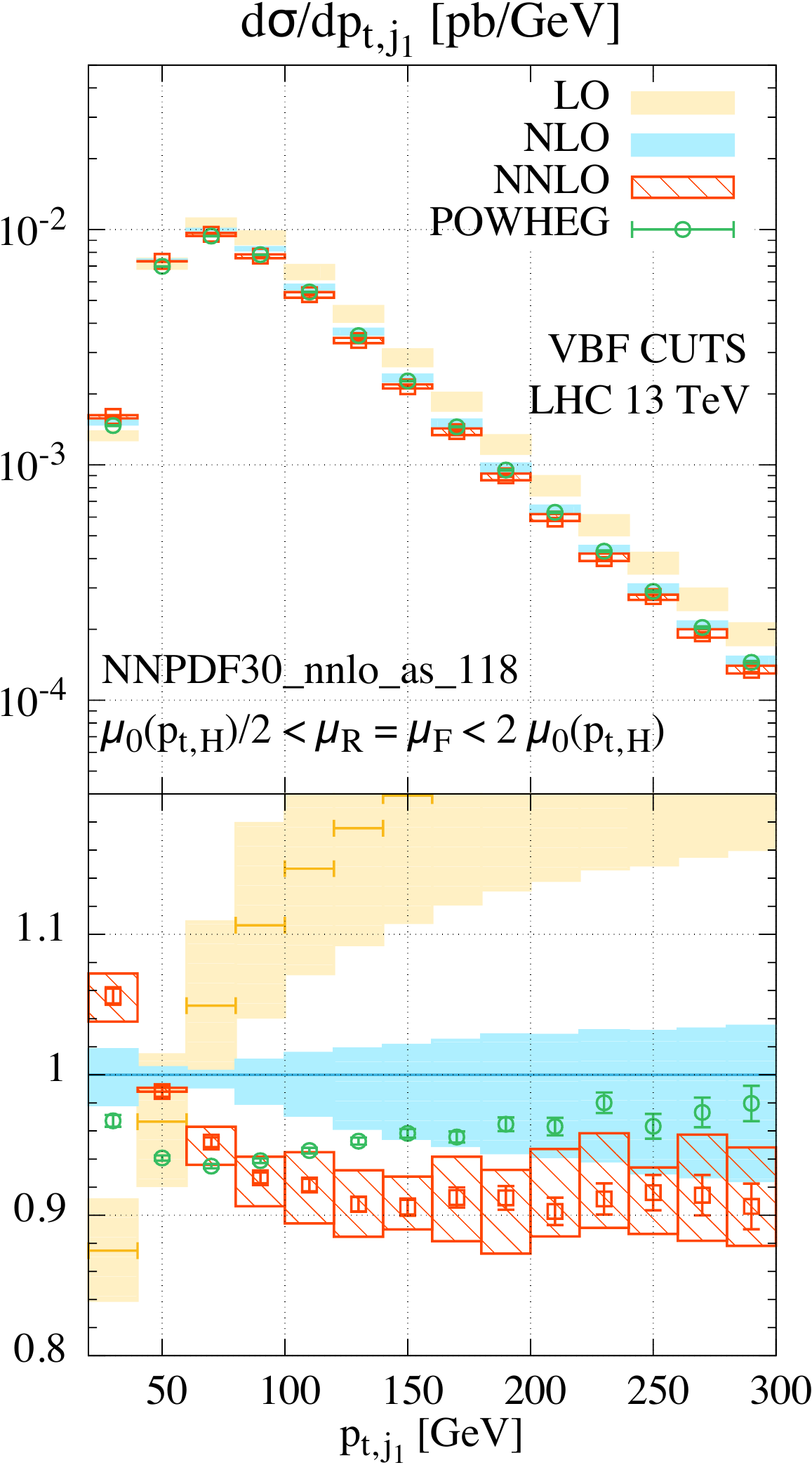} &
\includegraphics[height=0.5\textheight,page=4]{1506.02660/NLO-NNLO-crop.pdf} \\[1em]
\end{tabular}
\end{center}
\caption{
Differential cross section for VBF-$H$ production at the LHC up to NNLO
QCD and NLO QCD combined with parton shower (POWHEG-BOX and Pythia~6).
Shown distributions are the transverse momentum of the leading jet
(\textit{left}) and the rapidity difference of the two tagging jets
(\textit{right}). The vertical bars denote the statistical error, and
the boxes give the uncertainty from varying factorization and
renormalization scales jointly by factors $\frac12$ and $2$ around the
central scale $\mu_0=\frac{M_H}2 \sqrt{\frac{M_H^2}4 + p_{T,H}^2}$. 
Figure taken from Ref.~\protect\cite{Cacciari:2015jma}.
}
\label{fig:nnloqcd}
\end{figure}
In \fig{fig:nnloqcd} we present differential distributions, taken from
Ref.~\cite{Cacciari:2015jma}. Plotted results are the LO, NLO and NNLO QCD
differential cross sections, all evaluated in the VBF approximation and
with VBF cuts applied. The fourth set of bins is the NLO QCD calculation
combined with parton shower effects (see \chap{chap:partonshower} for
more details), using the POWHEG-Box implementation~\cite{Nason:2009ai}
combined with Pythia~6.428 with the Perugia~2012
tune~\cite{Skands:2010ak}. 
The left panel shows again the transverse momentum of the leading jet.
From the lower part, where the relative difference to the NLO QCD curve
is plotted, we see that the general feature of the NNLO QCD corrections
is a reduction in the cross section of up to 10\%, except for small
transverse momenta where instead an enhancement occurs. The underlying
cause for this is assumed to come from a redistribution of the jets from
higher to lower transverse momenta, and this makes it more difficult to
fulfill the VBF cuts, thus reducing the cross
section~\cite{Cacciari:2015jma}.
These corrections are generally outside the scale variation bands of the
NLO cross section and only at larger transverse-momentum values do they
start to overlap. Comparing to the NLO plus parton-shower curve instead
yields smaller deviations, but still outside the scale variation bands.

In the right panel the rapidity difference between the two tagging jets
is plotted. The NNLO compared to the NLO curve yields corrections of up
to 10\% again, with negative values for small rapidity differences and
rising to positive values for large ones. As before the corrections are
not covered by the scale variation estimates. For this distribution
adding parton-shower effects does not help at all, but instead increases
the deviations when compared to the NNLO curve.

Overall, additional effects beyond the NLO QCD accuracy are expected
to yield corrections to inclusive cross sections and in particular to
distributions of the order of several percent. In the tails of
distributions effects of up to 10\% have been observed. The uncertainty
estimate from performing scale variation clearly underestimates these
corrections. As mentioned before, the small values obtained there should
be taken with the necessary caution.

\section{Parton Shower Effects}
\label{chap:partonshower}

For a useful comparison between theory and experiment, it is necessary to
bring the fixed-order parton-level calculations, which we have
considered so far, closer to quantities actually measured by
experiments. In the language of Monte-Carlo event generators, the
fixed-order parton-level process, calculated via perturbation theory is
often called the hard process. At LO, the final-state partons are
directly identified with the jets, streams of clustered hadrons,
observed in the detector. At NLO, as we have already shortly mentioned
in the previous chapter, we need to apply a so-called jet algorithm.
This ensures that when radiation becomes soft or collinear, the event
asymptotically approaches its corresponding Born configuration, thereby
ensuring the cancellation of the poles with their corresponding
counterparts from the virtual expression. An extra jet can only appear
for hard, wide-angle emissions.  This behavior is known as the
infrared-safeness of observables. The typical scale of the hard process
is given by the factorization scale. They give us a good description of
inclusive observables with hard and well-separated jets. 

What the detector ultimately observes is inclusive only within a size
the order of a single detector cell, so actually a very fine-grained
picture. When we reduce the energy and angular-separation requirements,
we will gradually start to see more and more jets emerging. To model
this from the theory side with corresponding hard processes quickly
becomes cumbersome. Therefore, a different approach is used, which
starts from the hard process and generates additional radiation by a
unitary procedure, the so-called parton shower. We will describe the
details necessary for our VBF discussion in the following section.

After several steps, eventually the energy of the partons has dropped to
a level where the strong coupling constant $\alpha_s$ becomes very large
and any perturbative description breaks down. The usual scale for this
to happen is around 1~GeV. There, the transition from colored partons
into color-neutral hadrons takes place. Combining the partons is
modeled non-perturbatively, taking into account information from
momenta, quark flavor and color assignment of the partons. Finally,
these hadrons, which at this stage also contain excited and short-lived
states, decay into the long-lived objects that are able to hit the
detector elements, namely protons, neutrons, pions and Kaons, together
with electrons, muons and photons.

These two steps, parton shower and hadronization, are the task of Monte
Carlo event generators. The most common choices nowadays are
Herwig~7~\cite{Bahr:2008pv,Bellm:2015jjp}, with its predecessors
HERWIG~\cite{Corcella:2000bw,Corcella:2002jc} and
Herwig++~\cite{Bellm:2013hwb}, Pythia~8~\cite{Sjostrand:2007gs} and the
previous Pythia~6~\cite{Sjostrand:2006za}, and
Sherpa~\cite{Gleisberg:2008ta}.

In the following, we will first introduce the main concepts and formulae
of parton showers, and then discuss the consequences for VBF production
processes. Hadronization effects do not play any particular role, so we
are not going to cover this topic further. This, and a more in-depth
discussion of general features of Monte Carlo event generators, can be
found in
Refs.~\cite{Ellis:1991qj,Salam:2009jx,Buckley:2011ms,Gieseke:2013eva}.
The discussion in the next section follows the presentation in
Ref.~\cite{Gieseke:2013eva}.

\subsection{Parton Shower Overview}

\subsubsection{Parton Shower with Born Matrix Elements}

The main properties on which parton showers are based are unitarity
and collinear factorization. The latter states that in the collinear
limit, the emission of an extra parton $j$ from an emitter $i$ can be
written as
\begin{equation}
\di\sigma = \sigma_B \sum_{i,j} \frac{\alpha_s}{2\pi}
\frac{\di\theta^2}{\theta^2} P_{ij}(z) \di z \di\phi \,.
\label{eq:collfac}
\end{equation}
Here, $z$ is the energy fraction of the emission $j$, and $\theta$ the
emission angle, which can be replaced by any other variable linearly
dependent on the angle, like the transverse momentum $p_T$ or the
virtuality $Q$, without changing the expression. Eq.~\ref{eq:collfac}
neglects interference effects between different emissions, which do not
play a role, and so we can build up the whole evolution as a sequence of
single emissions. 
$P_{ij}$ are the DGLAP splitting
kernels~\cite{Dokshitzer:1977sg,Gribov:1972ri,Altarelli:1977zs}, which we
have already seen in the previous chapter. Their explicit form, averaged
over azimuthal angles and spins, is
\begin{align}
P_{qq}(z) &= C_F \frac{1+z^2}{1-z} \,, \nonumber\\
P_{gq}(z) &= C_F \frac{1+(1-z)^2}{z} \,, \nonumber\\
P_{qq}(z) &= C_A \frac{(1-z(1-z))^2}{z(1-z)} \,, \nonumber\\
P_{qq}(z) &= T_R (1-2z(1-z)) \,,
\end{align}
with the color factors $C_F=\frac43$, $C_A=3$ and $T_R=\frac12$. 
The poles when integrating \eq{eq:collfac} would again
be canceled by corresponding virtual loop diagrams. To avoid their
explicit calculation, we can employ two arguments. First, in the limit
where the energy fraction $z$ approaches $0$ or $1$ or the angle
$\theta$ goes to $0$, whether an emission happens or not becomes
indistinguishable. So we can introduce cutoffs $z_{\pm}(\theta)$ that
eliminate these regions. Second, unitarity tells us that the combined
probability of having an emission or not having one must sum to $1$.
This also ensures that the leading logarithmic behavior of the virtual
corrections is correctly modeled by the parton shower. 
Finally, the hard process gives the starting value for the parton
shower, \eg for the opening angle $\theta$, $\theta_{\max}$.
The probability density for the first branching happening at an angle
$\theta$, $\di \Delta_i(\theta_{\max},\theta)$, is given by the
probability of having no emission up to $\theta$ times the emission
probability within an infinitesimal interval $\di\theta$ around
$\theta$, which is given by $\frac{\alpha_s}{2\pi}
\frac{\di\theta^2}{\theta^2} \int_{z_{-}(\theta)}^{z_{+}(\theta)}
P_{ij}(z) \di z$. This differential expression can be integrated and
yields the so-called Sudakov form factor
\begin{equation}
\Delta_i(\theta_{\max},\theta) = \exp \left( - \sum_j
\int_{\theta^2}^{\theta_{\max}^2}
\frac{\di\tilde{\theta}^2}{\tilde{\theta}^2}
\int_{z_{-}(\tilde{\theta})}^{z_{+}(\tilde{\theta})}
\di z \ \frac{\alpha_s}{2\pi} P_{ij}(z) \right) \,,
\label{eq:sudakov}
\end{equation}
the probability of not having any resolvable emission between
$\theta_{\max}$ and $\theta$. The $\alpha_s$ factor has been moved in
the innermost integral, as it is a function of the renormalization
scale, whose choice may depend on both $z$ and $\theta$.
The probability for an emission at an angle $\theta_1$ is then given by
$\frac{\di \Delta_i(\theta_{\max},\theta_1)}{\di \theta_1^2}$. The individual
branchings should be ordered. Hence, the probability for the second
emission $\theta_2$ has now $\theta_1$ as its starting value. With this,
we can build up the full tower of emissions until we reach the lower
cutoff, where we can no longer resolve any emissions.

The expressions we have discussed so far are for radiation off
final-state particles. But extra parton emissions can also happen from
the initial-state partons, occuring before the hard process.
Technically, the parton-shower evolution is performed starting with the
hard process and then gradually going backwards in time and adding
further emissions to the current initial-state particle, again going
from large angles to small angles. As the extra emission requires more
energy to be extracted from the parton, an extra factor, the
corresponding ratio of the PDFs, appears in the Sudakov form factor,
such that it reads 
\begin{equation}
\Delta_i(\theta_1,\theta_2,x) = \exp \left( - \sum_j
\int_{\theta_2^2}^{\theta_1^2}
\frac{\di\tilde{\theta}^2}{\tilde{\theta}^2}
\int_{z_{-}(\tilde{\theta})}^{z_{+}(\tilde{\theta})}
\di z \frac{\frac{x}{z} f_j(\frac{x}{z},q^2)}{x f_i(x,q^2)} \frac{\alpha_s}{2\pi} P_{ij}(z) \right) \,,
\end{equation}
with proton momentum fraction $x$ as an extra argument and $q^2$ is the
extraction scale corresponding to the angle $\theta_2$. Any radiation
generated as such will then of course also undergo the normal final-state
shower discussed before.

Soft emission already factorizes on the amplitude level. This also means
that these emissions appear equally well for interference terms and one
cannot interpret the results in terms of individual Feynman diagrams.
The corresponding emission cross section of a soft gluon with
four-momentum $q$ in solid angle $\Omega$ can be written as
\begin{equation}
\di\sigma = \di\sigma_B \frac{\di q_0}{q_0}{\di\Omega}\ {2\pi}
\frac{\alpha_s}{2\pi} \sum_{i,j} C_{ij} q_0^2 \frac{p_i \cdot
p_j}{(p_i\cdot q)(p_j \cdot q)} \,,
\end{equation}
where the sum is over all pairs $(i,j)$ of colored external particles
and $C_{ij}$ is their color factor,  $C_F$ for a quark or anti-quark
emitter, $C_A$ for a gluon emitter splitting into a quark pair, and
$C_A/2$ for a gluon splitting into gluons, where the additional factor
accounts for the symmetry that either emitted gluon can be soft.
From this expression one can deduce that after azimuthal averaging the
emission of wide-angle soft gluons predominantly takes place within a
cone spanned by the two emitting particles. 

\subsubsection{Combining Parton Shower and NLO Calculations}

So far, we have discussed adding parton showers on top of Born-level
cross section calculations. For fixed-order, NLO accuracy can be seen as
more or less standard nowadays, so the next step is to discuss the
necessary modifications when adding parton showers to NLO cross sections
instead. 

As we have seen in the previous section, the parton shower generates
extra emissions of final-state partons, which are subsequently smaller
in the ordering parameter, which we have taken as $\theta$. The
real-emission part of our NLO calculation, however, also contains
already one extra emission compared to the Born process. In order to
obtain a reasonable prediction, this has to be taken into account to
avoid double-counting. This procedure is also known under the term of
matching NLO calculations with parton showers. 

In the previous chapter, we have discussed the individual pieces
entering an NLO calculation, with the final expression given by
\eq{eq:nlocs}. We now come back to this, but rewrite it slightly for the
following discussion, considering its effect on an observable
$\obs$, such that $\langle\obs\rangle_{\text{LO}} = B \obs(0)$. Then we
obtain
\begin{equation}
\langle\obs\rangle_{\text{NLO}} = B\obs(0) + \left( V + \int \di x D(x)
\right) \obs(0) + \int_0^1 \di x \bigl( R(x)\obs(x) - D(x)\obs(0) \bigr)
\,.
\label{eq:nlocs2}
\end{equation}
The first term is the Born contribution, the second one the virtual part
with the integrated dipoles added to yield an infrared-finite result,
and the last term is the real-emission contribution from which the
dipole terms are subtracted. The variable $x\in[0;1]$, which also is the
argument of $\obs$, denotes the phase space of the extra emission in a
symbolic notation. In the limit $x\rightarrow0$, it approaches the Born
phase space. As the dipole subtraction terms are evaluated with the
corresponding tilde kinematics, the observable is hence evaluated at
$x=0$.

The Sudakov form factor, \eq{eq:sudakov}, can be expressed in the
following schematic form with the splitting function $P(x)$
\begin{equation}
\Delta(x) = \exp\left( -\int_\mu^x \di x' P(x') \right) \simeq 1 -\int_\mu^x \di
x' P(x') \,.
\label{eq:sudakovexpanded}
\end{equation}
The parameter $\mu$ is the lower cutoff, where extra emissions are not
resolved explicitly anymore, and on the right-hand the expression has
been expanded up to the NLO order, as $P(x)$ contains an implicit factor
of $\alpha_s$. As the next step we now apply the parton shower to the Born
and consider up to one extra emission only. Then for the observable it
follows that
\begin{equation}
\langle\obs\rangle_\text{PS,$\leq$ 1 emission} = B\obs(0)\Delta(1) + 
\int_\mu^1 \di x B\obs(x) P(x) \frac{\Delta(1)}{\Delta(x)} \,.
\label{eq:obsps1}
\end{equation}
The first term describes the no-emission contribution and gets weighted
by the Sudakov factor, while the second term has one emission at $x$,
giving the factor $P(x)$, and the lower bound of the integral in the
Sudakov factor becomes $x$, which can be formally also written as the
ratio of Sudakov factors in the expression above.
The expansion of the Sudakov factors in \eq{eq:obsps1}, dropping all
terms higher than NLO in $\alpha_s$, yields
\begin{equation}
\langle\obs\rangle_\text{PS,$\leq$ 1 emission} = 
  B\obs(0) \left( 1 - \int_\mu^1 \di x P(x) \right)
+ \int_\mu^1 \di x B\obs(x) P(x) \,.
\end{equation}
This expression can now be inserted in our NLO cross section,
\eq{eq:nlocs2}, again dropping terms which are beyond the NLO level,
\begin{align}
\langle\obs\rangle_{\text{NLO+1 emission}} &= B\obs(0) + \left( V + \int
\di x D(x) \right) \obs(0) \nonumber\\
&\quad + \int_0^1 \di x \bigl( R(x)\obs(x) -
D(x)\obs(0) \bigr) \nonumber\\
&\quad - B\obs(0) \int_\mu^1 \di x P(x) + \int_\mu^1 \di x
B\obs(x) P(x) \,.
\end{align}
The first emission of the parton shower has generated the last two extra
terms. However, one condition of our parton shower was unitarity, \ie the
total effect, integrating out any extra emissions generated, should
vanish. Therefore, these two terms are exactly the double-counting
effect which has been mentioned in the introduction, and these must be
subtracted from the NLO cross section before applying the parton shower.
Hence, the expression reads
\begin{align}
\langle\obs\rangle_{\text{NLO,PS}} &= 
\left( 
 B + \left( V + \int \di x D(x) \right) 
+ \int_0^1 \di x \bigl( B\cdot P(x) - D(x) \bigr)
\right) \obs(0) \nonumber\\
&\quad + \int_0^1 \di x \bigl( R(x) - B\cdot P(x) \bigr) \obs(x) \,.
\label{eq:nlops}
\end{align}
Here we have used the fact that the splitting kernel times the Born also
approaches the real-emission matrix element in the singular limits, and
therefore both parts are regularized. 

In \eq{eq:nlops} we still have the freedom of how to exactly choose the
splitting kernels. The only relevant property is that in the soft or
collinear limit, the correct form of the divergence is reproduced, but
the finite terms away from this limit can in principle be chosen
arbitrary. The most obvious possibility is to take the DGLAP splitting
kernels, which are also used in the parton shower. This scheme is known
as the MC@NLO method~\cite{Frixione:2002ik}, or subtractive matching.
The second term of \eq{eq:nlops} contains the difference $R(x) -
B\cdot P(x)$. This expression is only guaranteed to be positive when
evaluated as part of a physical observable, as it appears as result on
the left-hand side of the equation. Individual events can however have
negative weights, and these are indeed observed in physics calculations.
Their effect is balanced by corresponding events with positive weight
which enter the same histogram bin, thus giving an overall positive
result.

An alternative approach, which has been specifically designed to
circumvent the negative-weights issue, is the POWHEG (POsitive Weight
Hard Emission Generator)~\cite{Nason:2004rx,Frixione:2007vw} method.
Here, the splitting kernel is chosen as 
\begin{equation}
P(x)_\text{POWHEG} = \frac{R(x)}{B} \,,
\end{equation}
the ratio of real-emission over Born matrix element. This gives a
very simple expression for \eq{eq:nlops}, as the last term just drops
out. The parton shower input now only consists of events with Born
kinematics. But this simplicity at the level of the matrix elements
comes with a price for the parton shower. The splitting kernel in the
Sudakov form factor, which governs the appearance of extra emissions,
now contains the full real-emission matrix element. This is except for
very simple cases known numerically only and takes a noticeable amount
of time to evaluate. Hence, the actual parton shower generation is much
more complicated. A way out of this is to consider the first emission
separately from subsequent ones. For the first one, $P(x)_\text{POWHEG}$
is used, so that subtracting the double-counting does not introduce
negative weights. For the subsequent emissions then only the simpler
standard splitting kernels of the parton shower are used. The difference
between the two is of higher order and therefore the two approaches are
equivalent up to the next-to-leading order considered here.

Also, in the POWHEG approach the Sudakov form factor takes a special
ordering variable, namely the transverse momentum, so that it reads in
the symbolic form
\begin{equation}
\Delta_{\text{POWHEG}} = \exp\left( -\int_\mu^x \di x' \frac{R(x')}{B} 
\Theta(p_T(x)-p_T) \right) \,.
\label{eq:sudakovpowheg}
\end{equation}
This makes the POWHEG-generated emission the hardest emission in the
process.  In the subsequent parton shower this ordering has to be
respected as well. For parton showers which are ordered in transverse
momentum, this poses no problem, as one can simply choose the
appropriate $p_T$ as the starting scale. For angular-ordered showers,
like the default shower of Herwig, this is however not the case. The
hardest emission generated by the parton shower will occur somewhere in
the middle of the sequence. To preserve the approximate color structure
of the emissions, we need to keep the ordering of the emissions.
Therefore, the standard technique in these cases is to split the parton
shower into two separate parts. 
First one generates the soft, wide-angle emissions before the hardest
one and stops the parton shower at that value of the evolution variable
where the emission of the POWHEG-generated one takes place in the
ordered sequence. This restriction of the parton shower is known under
the name truncated shower. Then the parton shower is run again with the
POWHEG-generated emission as starting one, and any harder emissions
which might appear afterwards are vetoed. Hence this restriction is also
called a vetoed shower. The application of both is necessary to obtain
correct results~\cite{Frederix:2011ss}.

A third possibility, which has appeared in recent
years~\cite{Giele:2007di,Hoche:2010pf,Hoche:2010kg,Platzer:2011bc,Hoeche:2012ft,Ritzmann:2012ca},
uses the subtraction dipoles of the NLO calculation as splitting kernel,
\ie $P_{\text{subtr}}(x) = \frac{D(x)}{B}$.  Looking at \eq{eq:nlops},
we see that the matching becomes particularly easy, as the only
necessary ingredients are known from the fixed-order NLO calculation
anyway. Plugging $P_{\text{subtr}}(x)$ in, we see that the only
difference is the evaluation of $\obs$ for the real-emission dipole.
Instead of the Born-type tilde kinematics, it is now calculated at the
corresponding real-emission phase-space point. Since the subtraction
dipoles are known analytically, building a corresponding parton shower
out of them becomes a feasible task.

Parton showers in general resum the leading logarithmic terms and the
leading-color part of the next-to-leading logarithmic terms. This level
of accuracy is sufficient to combine them with LO and NLO calculations
of the hard matrix element.

\subsection{Applying Parton Showers to LO VBF Processes}
\label{sec:partonshowerLO}

Having discussed the general basics of parton shower algorithms, we can
now apply them to the VBF processes. In this section we will shortly
discuss the combination of LO calculations and parton shower, before
turning to the NLO case in the next section. As the aim of this part is
mainly to motivate some discussions of special features in the next one,
we will restrict ourselves to considering VBF-$H$ production as a simple
example. 

\begin{figure}
\begin{center}
\begin{tabular}{cc}
\includegraphics[width=0.45\textwidth]{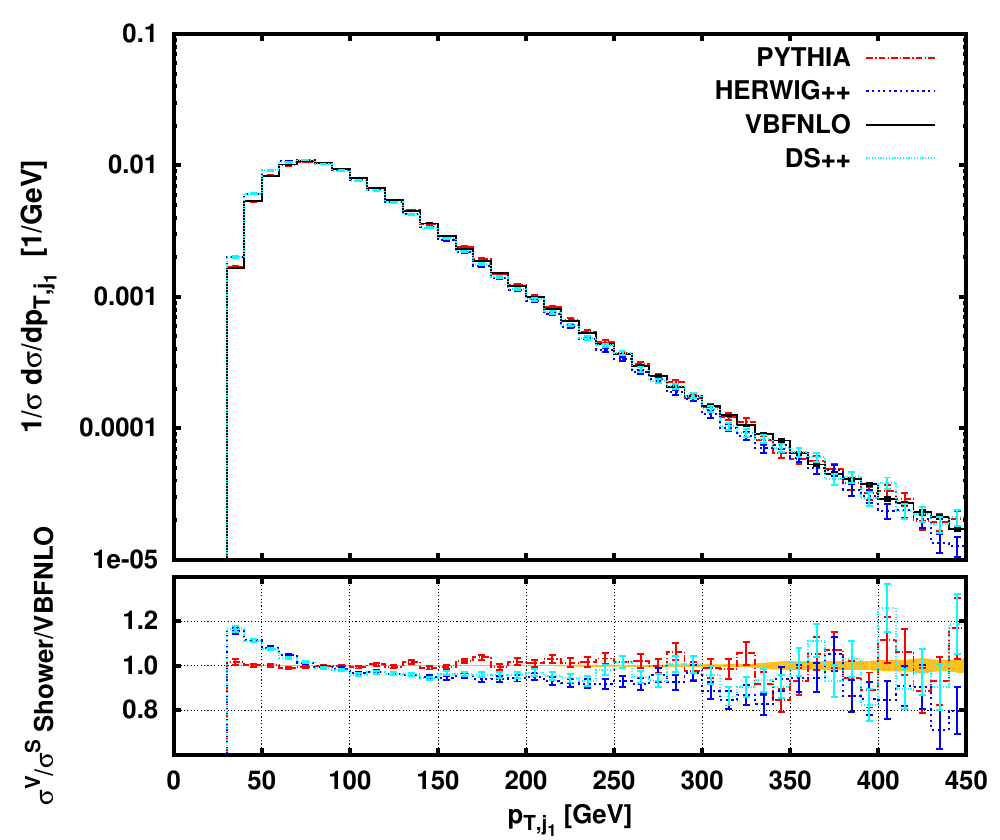} &
\includegraphics[width=0.45\textwidth]{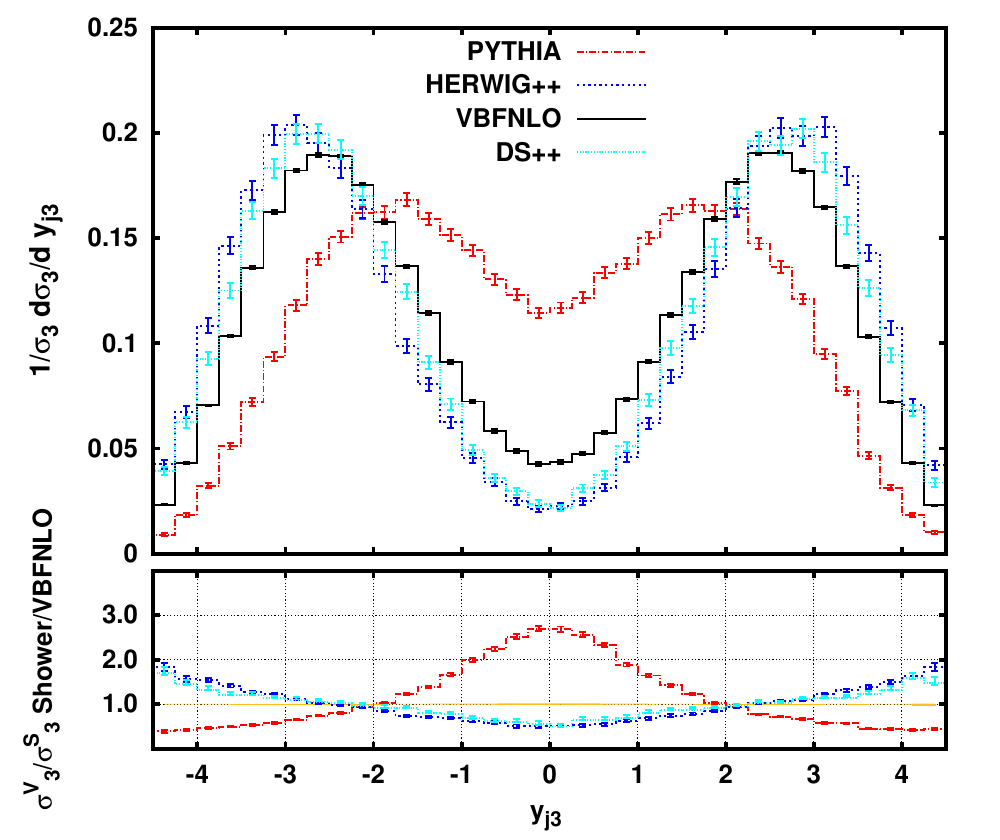} 
\end{tabular}
\end{center}
\caption{
Normalized differential distributions of the transverse momentum of the
leading jet (\textit{left}) and the rapidity of the third jet
(\textit{right}) in VBF-$H$ production. Shown are distributions for
LO calculated with \VBFNLO (using electroweak $H+3$ jets
production for the right plot) and LO plus three different parton
showers. The lower ratio plot is normalized to the respective LO cross
section. Error bands and bars are statistical only.
Figure taken from Ref.~\protect\cite{Schissler:2014nga}.
}
\label{fig:vbfh_lops}
\end{figure}

In \fig{fig:vbfh_lops}, taken from Ref.\cite{Schissler:2014nga}, we show
in the left panel the transverse momentum distribution of the leading
jet. The black LO curve has been generated with
VBFNLO~\cite{Arnold:2008rz,Baglio:2014uba,VBFNLO} and is labeled as
such.
Additionally, three different parton showers have been applied on top of
this. They are Pythia 6.4.25~\cite{Sjostrand:2006za} with the Perugia
0-tune~\cite{Skands:2010ak} in red, which is a $p_T$-ordered shower, the
angular-ordered default shower of Herwig++~2.7.0~\cite{Bellm:2013hwb} in
blue, and the $p_T$-ordered dipole shower~\cite{Platzer:2011bc} of
Herwig++ in cyan, labeled as DS++ here. 
We note that the angular-ordered shower of Herwig++ does not implement
truncation, so some soft, wide-angle radiation is not present there. The
comparison with the dipole shower allows for an estimate of the
importance of this contributions which, as we will see, turns out to be
negligible.
Each curve in the distribution is normalized to its respective
integrated cross section. The lower panel contains the ratio of the
showered samples to the LO one.  The bands and error bars are
statistical.
Standard VBF cuts have been used for the generation, namely
\begin{align}
p_{T,j,\text{tag}} &> 30\text{ GeV} \,, & 
p_{T,j,\text{other}} &> 20\text{ GeV} \,, &
|y_j| &< 4.5 \,, \nonumber\\
m_{jj,\text{tag}} &> 600 \text{ GeV} \,, &
\Delta y_{jj,\text{tag}} &> 4 \,, &
y_{j_1} \cdot y_{j_2} &< 0  \,.
\label{eq:schisslercuts}
\end{align}
For the showered samples, looser generation cuts have been used to allow
for migration effects. We observe that the Pythia shower induces almost
no shape changes to this distribution. For the two Herwig++ showers,
the distribution is shifted towards smaller values of the leading-jet
transverse momentum. This is in accordance with the corresponding NLO
distribution (not shown), which generates a similar shift. 

In the right panel of \fig{fig:vbfh_lops}, we show the rapidity
distribution of the third jet. This one is generated purely by the
respective parton shower. For the LO curve, we therefore use the
prediction for electroweak $H+3$ jets production. This rapidity
distribution has important applications for the experimental detection
of VBF processes as already discussed in
\chap{chap:processclass}. 
The NLO cross section, which is equivalent to the LO electroweak $Hjjj$,
predicts that also the third jet is generated predominantly in the
forward direction, and radiation in the central region is strongly
reduced. This is in contrast to QCD-induced production mechanisms, where
more jet activity happens in the central region. 
This can be exploited by applying a central jet veto
for VBF
processes~\cite{Dokshitzer:1986ec,Bjorken:1992er,Barger:1994zq,Rainwater:1996ud,Khoze:2002fa}.
The additional radiation from
the parton shower could however also fill up the central region, thereby
invalidating this criterion. The result of \fig{fig:vbfh_lops} shows a
quite different answer to this question depending on which parton shower
is used. The Pythia shower predicts that the gap is filled with a
significant amount of radiation, more than $2.5$ times the value of the
LO $H+3$ jets cross section at $y_{j3}=0$. The two Herwig++ showers in
contrast exhibit only a very mild jet activity, even slightly below the
fixed-order curve. From the parton-shower side, there is no reason to
prefer one implementation over the other, as all should be equivalent up
to corrections higher than the considered order. Hence, in first
approximation one would take the envelope of the predictions as
uncertainty on the central jet veto, thus removing any possible
discrimination power. 

To investigate this issue further, we should combine the NLO prediction
with a parton shower. In that way, one gets the correct description of
the large transverse-momentum behavior and the correct normalization of
the cross section from the fixed-order calculation. The parton shower
augments this with the correct description of soft and collinear
radiation and the resummation of the leading logarithms.

\subsection{Parton Showers Matched to NLO VBF Processes}

NLO calculations of VBF processes matched to parton showers have been
first presented using the POWHEG-BOX
framework~\cite{Nason:2004rx,Frixione:2007vw,Alioli:2010xd}. Results can
also be obtained nowadays using the available automated frameworks,
\eg via Sherpa~\cite{Gleisberg:2008ta}, with generating events by
MadGraph5\_aMC@NLO~\cite{Alwall:2014hca} and feeding them through a
parton-shower and hadronization program, or from 
Herwig~7~\cite{Bahr:2008pv,Bellm:2015jjp}, obtaining the required
amplitudes either from automatic generators like
MadGraph5\_\-aMC@NLO~\cite{Alwall:2014hca} or
GoSam~\cite{Cullen:2011ac,Cullen:2014yla}, or dedicated tools like
VBFNLO~\cite{Arnold:2008rz,Baglio:2014uba,VBFNLO} or
HJets++~\cite{Campanario:2013fsa} for electroweak Higgs-boson production
in association with two or three jets.

\subsubsection{The POWHEG-BOX Approach}
\label{sec:partonshowerNLOPOWHEG}

The POWHEG-BOX framework is a semi-automated approach implementing the
POWHEG matching scheme. For each process one needs to implement the
following ingredients:
\begin{itemize}
\item the squared matrix elements for all Born subprocesses,
\item the interference between the renormalized virtual amplitude and
the Born amplitude with integrated subtraction dipoles in the FKS
scheme~\cite{Frixione:1995ms} added,
\item the squared matrix elements for all real-emission subprocesses,
\item the phase space for the Born kinematics,
\item the flavor structure of all Born and real-emission subprocesses,
\item the color structure of the Born processes in the limit of a large
number of colors.
\end{itemize}

The framework itself takes care of all other parts, like the steering of
the phase-space integration, generating the real-emission phase space
with one extra emission, and writing out Les Houches event
files~\cite{Boos:2001cv,Alwall:2006yp}. These can then be used as input for a
transverse-momentum-ordered or a truncated and vetoed angular-ordered
parton shower. 

Implementations are available for the VBF production of
$Hjj$~\cite{Nason:2009ai}, $W^\pm jj$\cite{Schissler:2013nga},
$Zjj$~\cite{Jager:2012xk,Schissler:2013nga}, $W^+W^-jj$~\cite{Jager:2013mu}, $W^\pm
W^\pm jj$~\cite{Jager:2011ms}, $ZZjj$~\cite{Jager:2013iza} and electroweak
production of $Hjjj$~\cite{Jager:2014vna}. 

Coming back to the discussion about additional jet activity in the
central region between the two tagging jets, we now look at the
corresponding distributions at NLO combined with parton shower accuracy,
taking VBF-$W^+jj$, $W^+ \rightarrow \ell^+ \nu_\ell$, $\ell=e,\mu$ as
example.
\begin{figure}
\begin{center}
\begin{tabular}{cc}
\includegraphics[angle=270,width=0.45\textwidth]{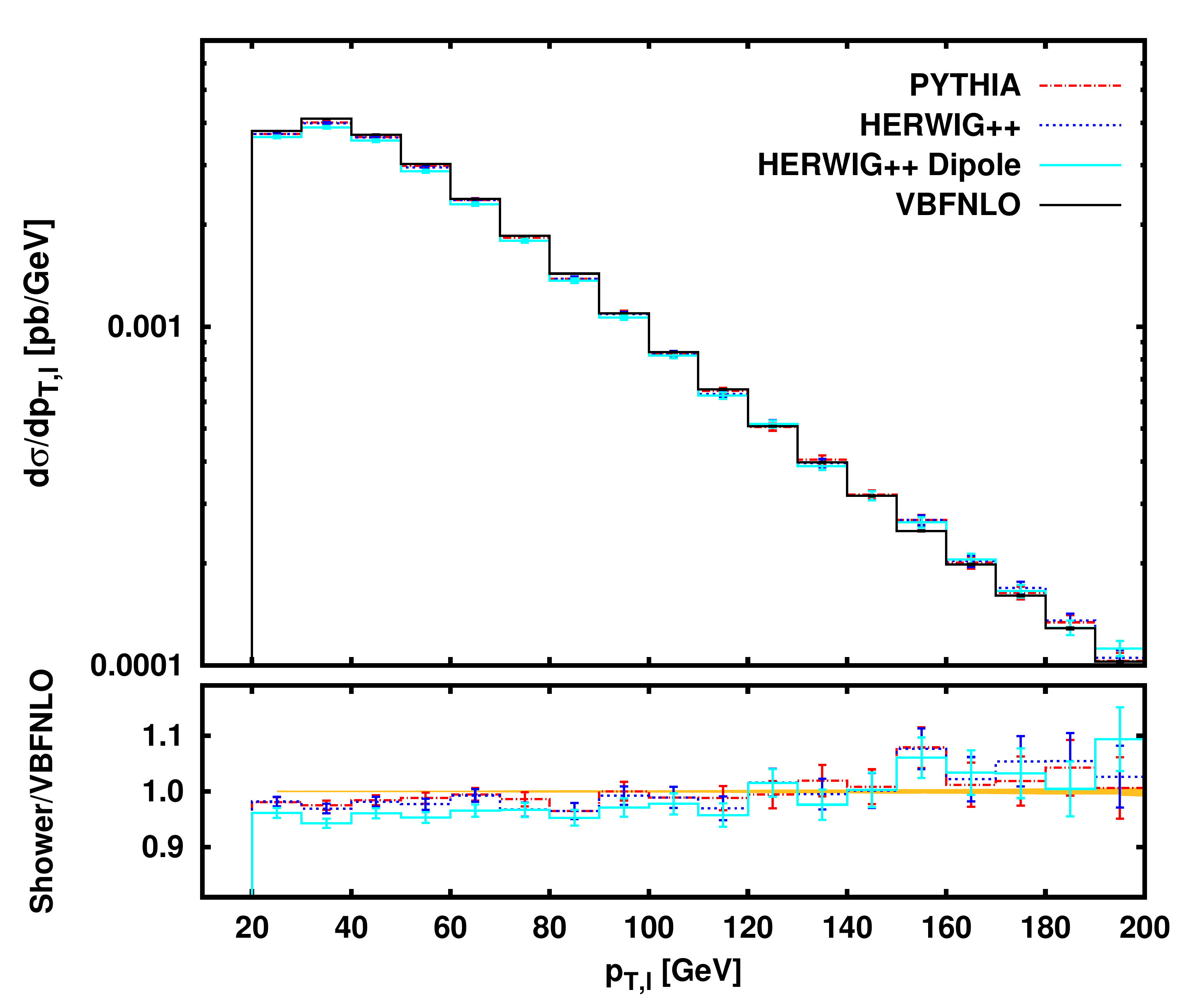} &
\includegraphics[angle=270,width=0.45\textwidth]{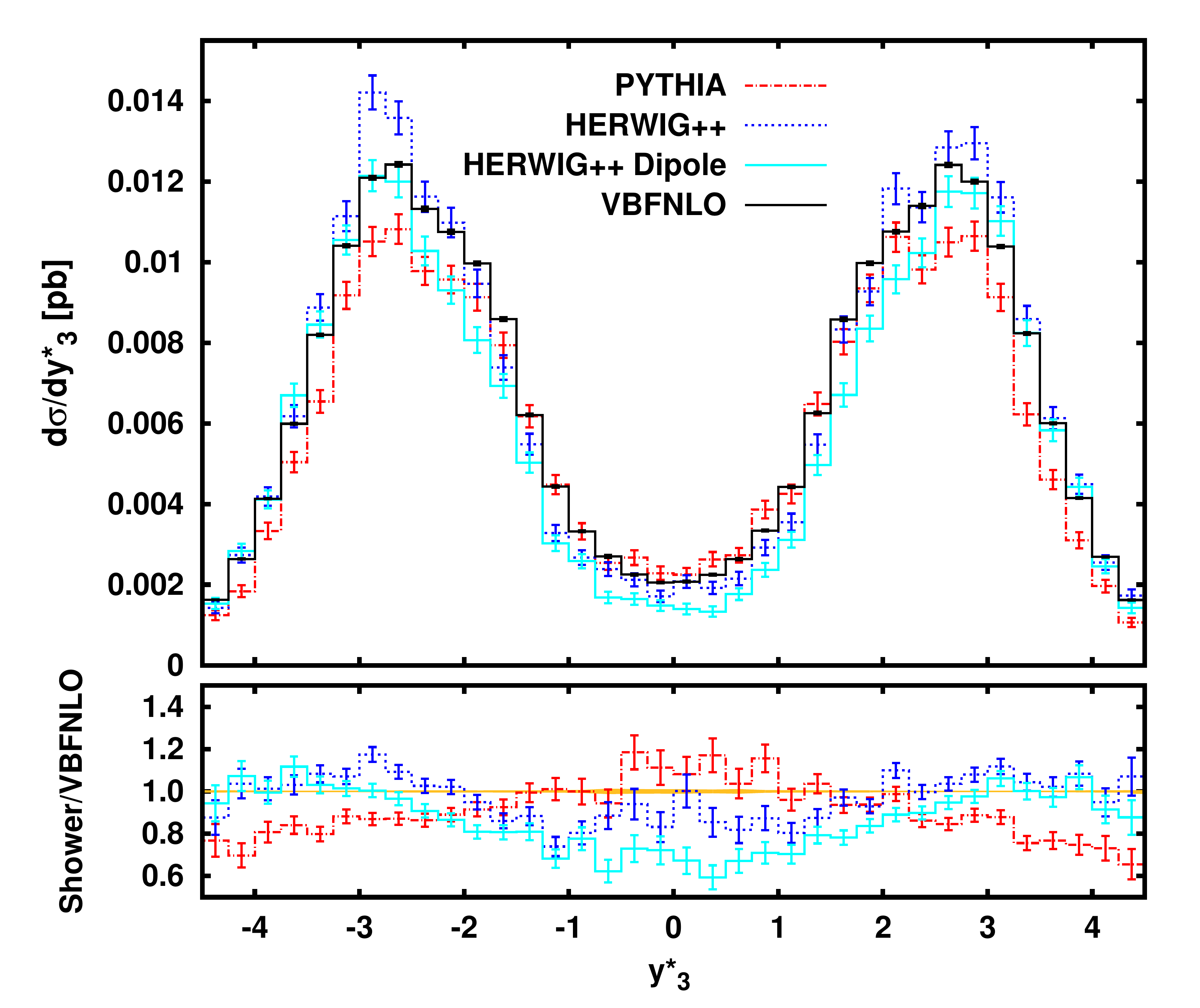} \\[1em]
\includegraphics[angle=270,width=0.45\textwidth]{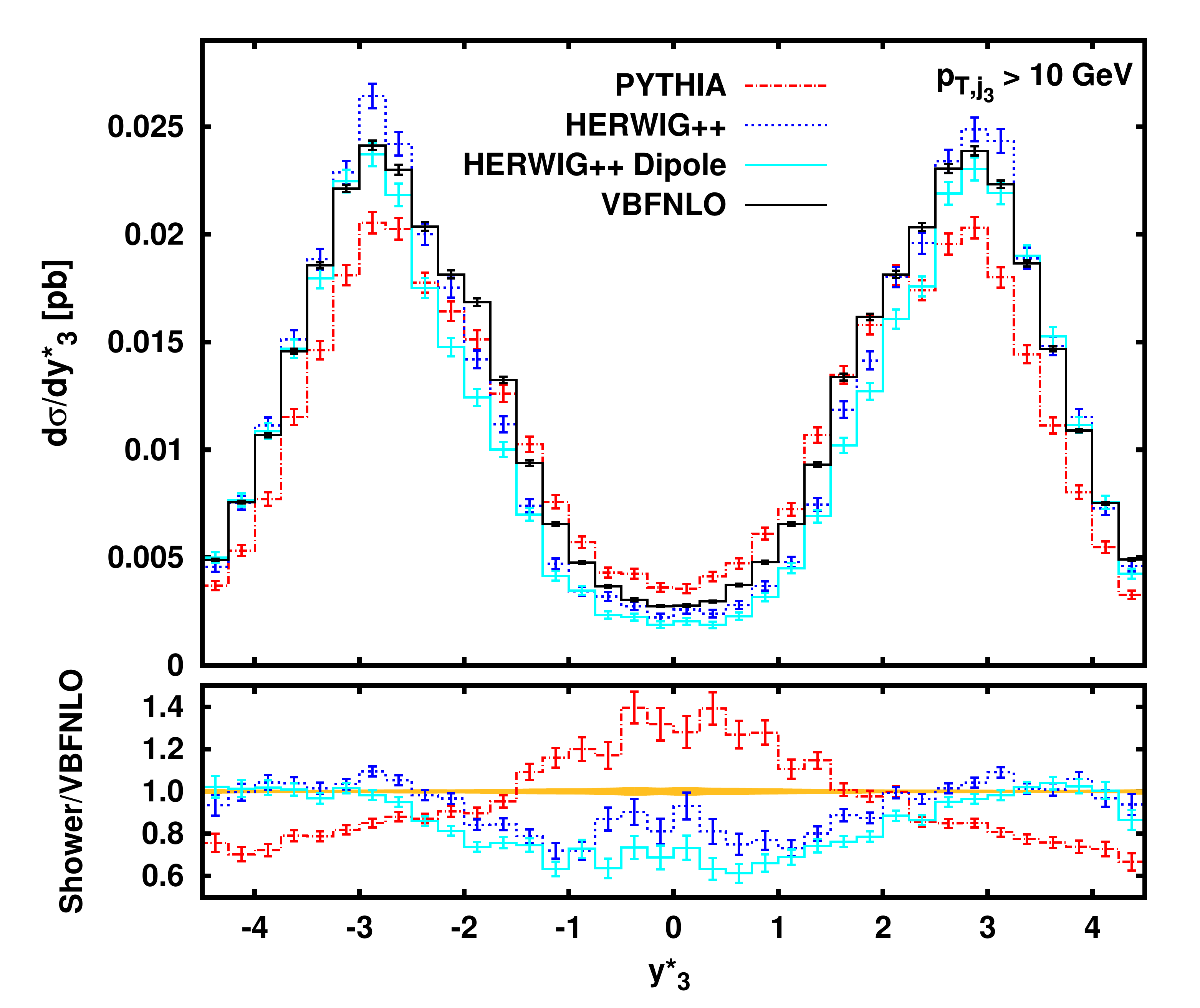} &
\includegraphics[angle=270,width=0.45\textwidth]{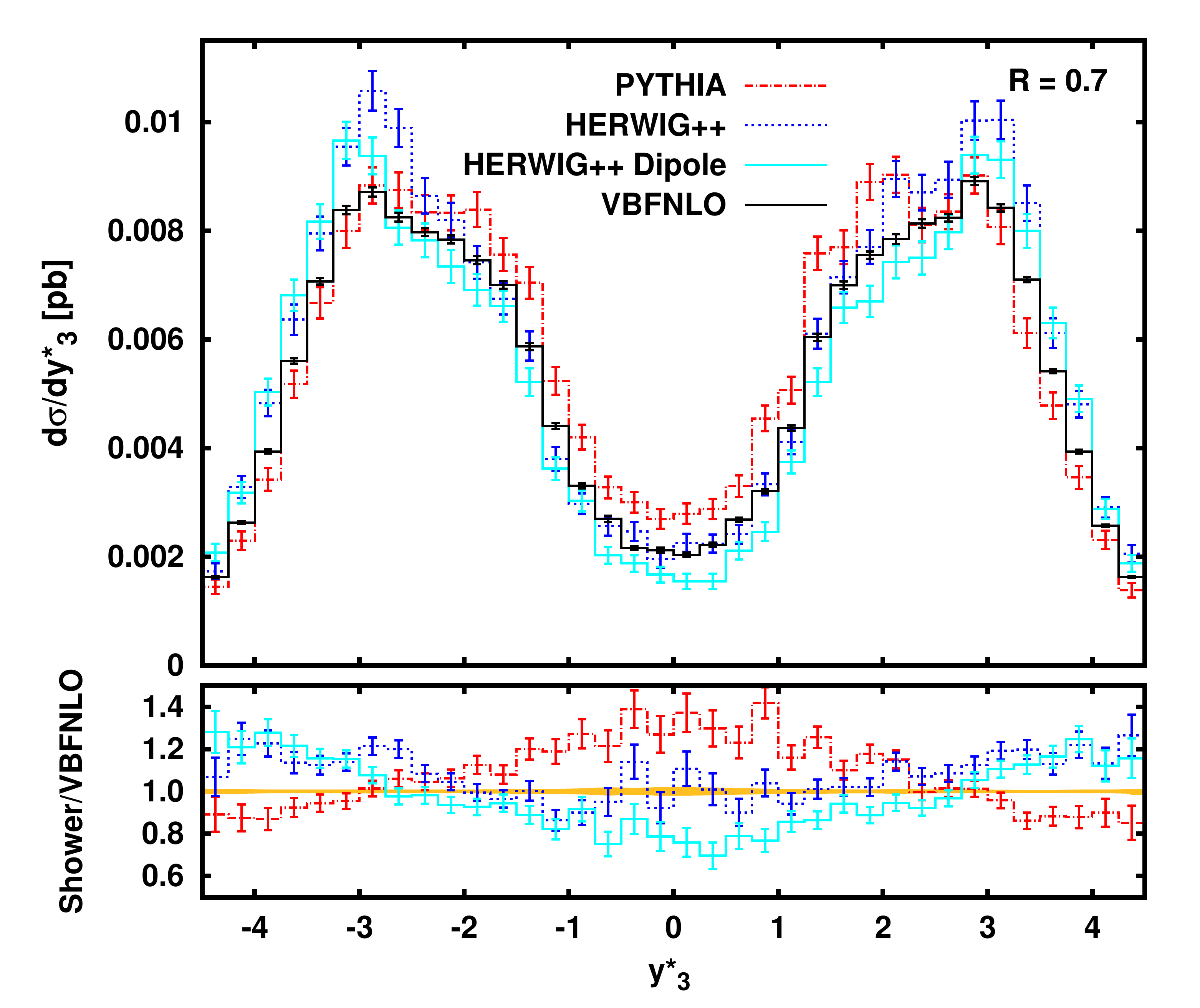} 
\end{tabular}
\end{center}
\caption{
Differential distributions of the transverse momentum of the
charged lepton (\textit{upper left}) and the rapidity of the third jet
relative to the average tagging jet position (\textit{upper right and
both lower panels}) in VBF-$W^+$ production with leptonic decay. 
In the lower panels, the $p_{T,j3}$ cut has been lowered from 20 to 10
GeV (\textit{lower left}), or the jet clustering parameter $R_{jj}$
increased from $0.5$ to $0.7$ (\textit{lower right}). 
Shown are distributions for fixed-order NLO calculated with \VBFNLO and
NLO plus three different parton showers using the POWHEG matching
scheme. 
The lower ratio plot is normalized to the fixed-order NLO cross
section. Error bands and bars are statistical only.
Figure taken from Ref.~\protect\cite{Schissler:2013nga}.
}
\label{fig:vbfwp_nlops}
\end{figure}
In the upper left panel of \fig{fig:vbfwp_nlops}, taken
from Ref.~\cite{Schissler:2013nga}, we first show the transverse momentum of the
charged lepton as example of a distribution which should receive only
mild corrections from the parton shower. From the plot we see that this
is indeed the case. 
The labeling of the curves is similar to \fig{fig:vbfh_lops}.
The black NLO curve has been generated with
VBFNLO~\cite{Arnold:2008rz,Baglio:2014uba,VBFNLO} and is
labeled such. Additionally, three different parton showers have been
applied on top of this. They are Pythia
6.4.25~\cite{Sjostrand:2006za} with the Perugia
0-tune~\cite{Skands:2010ak} in red, which is a $p_T$-ordered shower, the
angular-ordered default shower of Herwig++~2.7.0~\cite{Bellm:2013hwb} in
blue, and the $p_T$-ordered dipole shower~\cite{Platzer:2011bc} of
Herwig++ in cyan, labeled as Herwig++ Dipole here. 
Each curve in the distribution is normalized to its respective
integrated cross section. The lower panel contains the ratio of the
showered samples to the NLO one. The bands and error bars are
statistical.
Cuts are as in \eq{eq:schisslercuts}, with additionally requiring for
the lepton
\begin{align}
p_{T,\ell} &> 20\text{ GeV} \,, & 
|y_\ell| &< 2.5 \,, \nonumber\\
\Delta R_{j\ell} &> 0.4 \,, &
y^{\min}_{j,\text{tag}} &< y_\ell < y^{\max}_{j,\text{tag}} \,. 
\label{eq:schisslercuts2}
\end{align}
From the ratio plot in the lower part we see that the parton-shower
results exhibit a slightly larger cross section for larger transverse
momentum. This can be easily understood from additional initial-state
radiation, which then gives a transverse momentum boost to the hard
process. As the cross section is falling with larger transverse momenta,
migration effects towards larger values dominate over those to smaller
values and thus increase the differential cross section there. Unitarity
of the parton shower then leads to the observed decrease for small
transverse momenta.

The upper right plot of \fig{fig:vbfwp_nlops} shows the rapidity of the
third jet relative to the two tagging jets, using the variable
$y^*_{j,3}$ defined in \eq{eq:ystar}. Overall, the agreement between the
fixed-order result and the parton-shower ones is much better than for
LO, as the parton-shower results get corrected by the $W+3$ jet matrix
element, which enters through the real-emission part of the NLO process.
Significant differences between the parton-shower results are
nevertheless still present. In the central region, the Pythia shower
predicts a differential cross section enhanced by about 10\% compared to
the NLO one. The two Herwig++ showers in contrast show a reduction, in
the case of the dipole shower up to 30\%. The behavior gets reversed
when looking at the regions further forward than the tagging jets at
around $\pm2.7$. There the two Herwig++ showers hardly modify the NLO
result, while Pythia exhibits significantly lower values. The difference
becomes even more pronounced when lowering the minimum transverse
momentum cut of the third jet to 10 GeV (lower left panel of
\fig{fig:vbfwp_nlops}) or increasing the jet clustering parameter
$R_{jj}$ to 0.7 (lower right panel). 

\begin{figure}
\begin{center}
\includegraphics[width=0.7\textwidth]{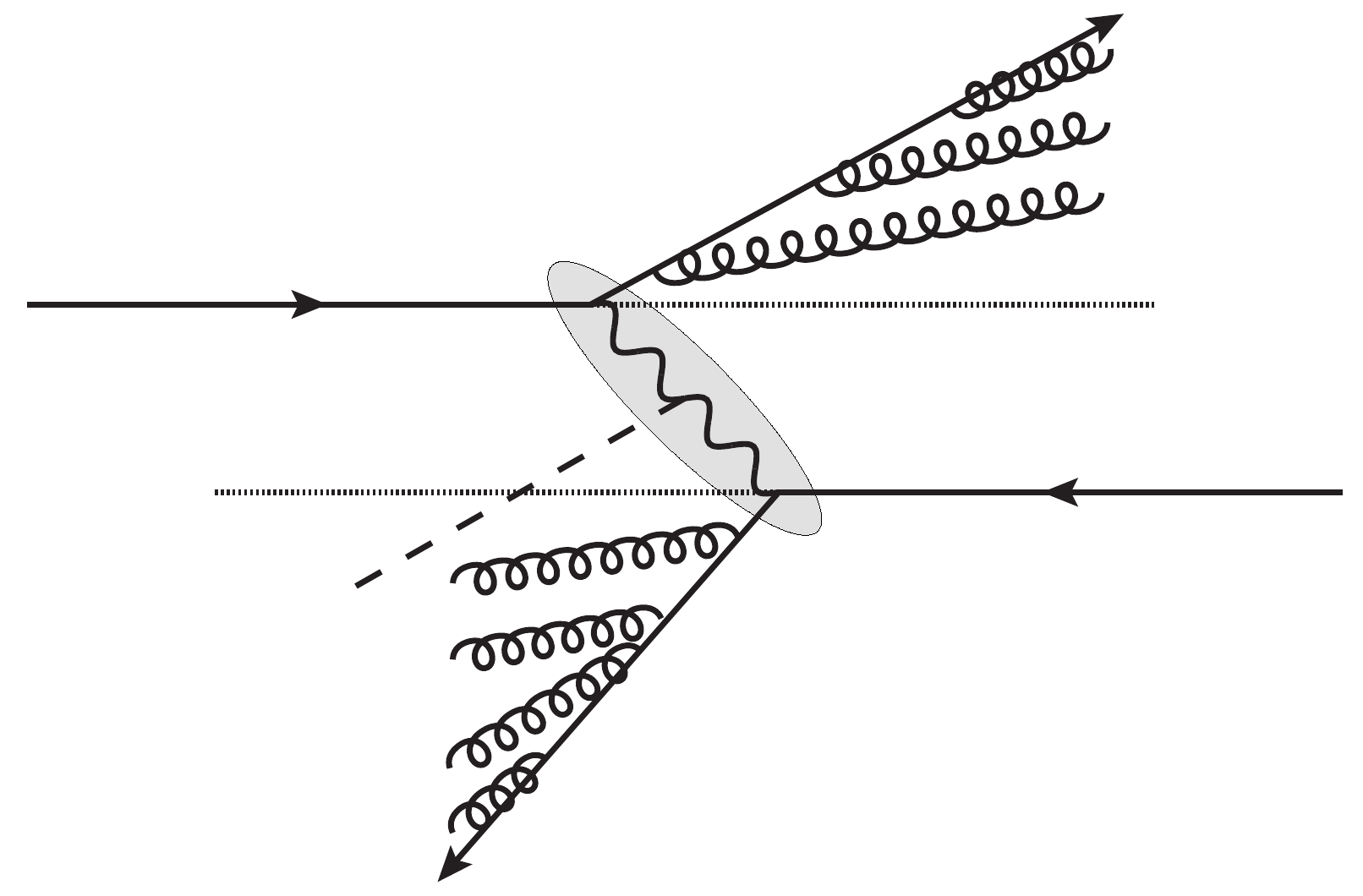}
\end{center}
\caption{
Schematic picture of additional radiation generated by parton showers
for VBF processes.
}
\label{fig:VBFradiation}
\end{figure}
Let us consider again the color structure of VBF processes. 
The exchange of a color-singlet electroweak boson between the two quark
lines means that in the hard process color connections only exist
between the parton forming a tagging jet and its corresponding
initial-state parton connected by a fermion line. Going to a pure
final-state picture, the conjugate color charge of the initial-state
parton is carried by the proton remnant moving along the beam line.
Hence, we expect that color correlations should enhance additional
radiation between the tagging jet and the corresponding positive or
negative $z$-axis, but not in the central region.
A schematic picture is shown in \fig{fig:VBFradiation}.

Pythia tends to generate more additional soft partons in the central
region than the two Herwig++ showers, which emit more into the forward
regions. Therefore, for the Herwig++ showers the jet clustering
algorithm mostly picks up additional radiation from the forward region
and hence the jet moves in that direction, while for Pythia the
enhanced central-region radiation pulls the jet there. 

One can also look at the differential shape of the third jet, defined
as~\cite{Ellis:1992qq,Aad:2011kq,Schissler:2013nga}
\begin{equation} 
\rho(r) = \frac1{\Delta r} 
\sum_{\substack{\text{parton}\in j_3 \\ 
  r_{\text{parton}} \in [r-\frac{\Delta r}2,r+\frac{\Delta r}2]}}
\frac{p_{T,\text{parton}}}{p_{T,j_3}} \,,
\end{equation} 
where $r$ denotes the $R$ separation from the center of the third jet.
Its value can range from $\frac{\Delta r}2$ to $R_{jj}-\frac{\Delta
r}2$. For the calculation of $\rho(r)$, all partons belonging to the
third jet within a cone from $r-\frac{\Delta r}2$ to $r+\frac{\Delta
r}2$ are used. The formula above automatically yields the
normalization $\int_0^{R_{jj}} \di r \rho(r) = 1$.

\begin{figure}
\begin{center}
\includegraphics[angle=270,width=\textwidth]{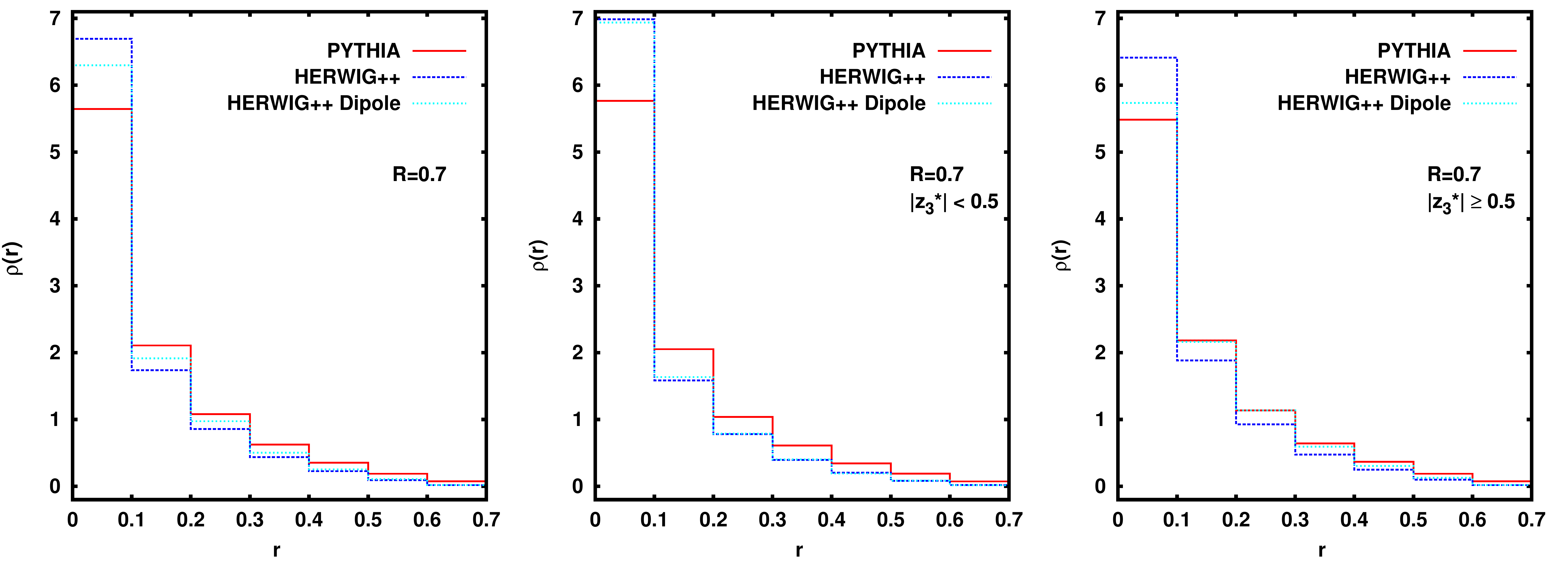} 
\end{center}
\caption{
Differential jet shape distribution $\rho(r)$ for $R_{jj}=0.7$, for the
full phase-space region (\textit{left}), and separated into the central
region between the two tagging jets (\textit{center}) and the forward
regions outside the tagging jets (\textit{right}).
Figure taken from Ref.~\protect\cite{Schissler:2013nga}.
}
\label{fig:vbfwp_nlops_rhor}
\end{figure}
In \fig{fig:vbfwp_nlops_rhor} we show the distribution of this variable,
using $\Delta r = 0.1$. The left panel shows this variable plotted over
the whole phase space after cuts. We observe that Pythia generates much
wider jets than the two Herwig++ showers. The density $\rho(r)$ for
$r>0.1$ is considerably enhanced. This behavior is even more pronounced
when considering only jets which are located between the two tagging
jets in rapidity, shown in the center panel of
\fig{fig:vbfwp_nlops_rhor}. The selection is done by requiring
$|z^*_{j_3}|<0.5$, with $|z^*_{j_3}|$ defined in \eq{eq:zstar}. For
third jets in the forward region on the other hand (right panel), the
distributions differ less, with the Herwig++ dipole shower almost equaling
the Pythia results. 
The reason for the observed differences is the already mentioned
generation of more soft, wide-angle radiation in Pythia. Collinear
radiation happens predominantly close to the emitting parton, so for
small values of $r$, while soft emission is less correlated to it and
leads equally well to jet contributions at larger $r$ values.
This can also explain the broader jets of the Herwig++ dipole shower
compared to the default shower in the forward region, as a lower IR
cutoff of the Sudakov form factor generates more soft radiation.

When considering distributions of the third jet in VBF processes, the
corresponding matrix elements are those of the real-emission process and
hence of LO accuracy only. To study these distributions at NLO accuracy,
one needs to consider electroweak production in association with three
jets. In Ref.~\cite{Jager:2014vna}, this task has been performed for
electroweak Higgs plus three jets production using again the POWHEG
method, again comparing Pythia and the two showers available in
Herwig++~2.7.0. The rapidity distribution of the third jet shows good
agreement between the showered and the fixed-order NLO results. As
before, Pythia predicts slightly more radiation in the central region,
while the two Herwig++ showers show a mild increase in the forward region.
Larger differences then again occur for rapidity distributions of
further jets which are corrected by LO matrix elements only (fourth
jet) or originate purely from the parton shower (fifth and higher jets).

\subsubsection{Shower Uncertainties}
\label{sec:partonshowerNLOunc}

The existence of different parton-shower approaches and different
matching methods allows for comparing the results between them.
Possible deviations are formally effects of higher order, and so the
envelope of the predictions can also serve as an estimate for the
impact of these corrections. A study for VBF-$H$ production, comparing
the MC@NLO and POWHEG matching schemes as well as changes in the
factorization and renormalization scheme, has been performed in
Ref.~\cite{Frixione:2013mta}.

\begin{figure}
\begin{center}
\includegraphics[width=0.85\textwidth]{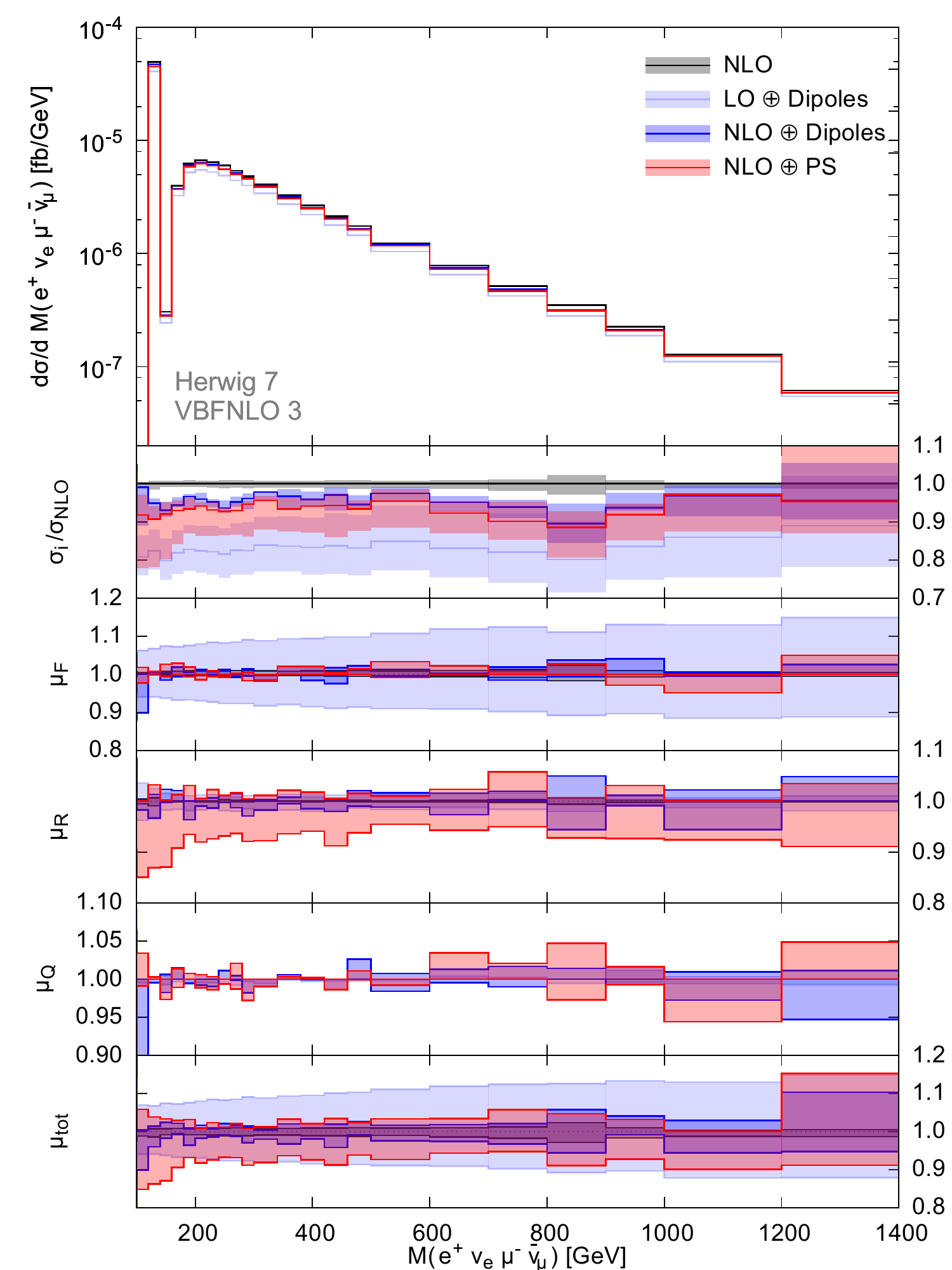} 
\end{center}
\caption{
Invariant mass distribution of the four leptons in
$e^+\nu_e\mu^-\bar{\nu}_\mu jj$ production via VBF. 
The upper panel shows the differential cross section for various
combinations of fixed-order result and parton shower. The four lowest
panels show the uncertainty band when varying the factorization scale
$\mu_F$, renormalization scale $\mu_R$, shower scale $\mu_Q$, and all
three $\mu_{\text{tot}}$ by a factor $2^{\pm1}$ around the central scale
$\mu_0=p_{T,j1}$. The second uppermost panel shows the ratio of each
cross section over the fixed-order NLO result, with the band given by
the envelope of all scale variations as in the lowest panel.
Figure taken from Ref.~\protect\cite{Rauch:2016upa}.
}
\label{fig:vbfww_scale_m4l}
\end{figure}
In the following, we will consider the VBF production process
$pp\rightarrow e^+\nu_e\mu^-\bar{\nu}_\mu jj$ to study shower
uncertainties~\cite{Rauch:2016upa}. All results have been generated
with Herwig~7~\cite{Bahr:2008pv,Bellm:2015jjp} using
\VBFNLO~\cite{Arnold:2008rz,Baglio:2014uba,VBFNLO}. Cuts are imposed as
follows
\begin{align}
p_{T,j} &> 30 \GeV \,, & |y_j| &< 4.5 \,, \nonumber\\
p_{T,\ell} &> 20 \GeV \,, & |y_{\ell}| &< 2.5 \,, \nonumber\\
m_{j1,j2} &> 600 \GeV \,, & |y_{j1}-y_{j2}| &> 3.6 \,,
\label{eq:vbfww_cuts}
\end{align}
where partons are clustered into jets with the anti-$k_T$
algorithm~\cite{Catani:1993hr,Ellis:1993tq} using an $R$ separation
parameter of $0.4$. The central value for the factorization,
renormalization and shower-starting scale is taken as the transverse
momentum of the leading jet, $\mu_0 = p_{T,j1}$.

In \fig{fig:vbfww_scale_m4l} we show the invariant mass distribution of
the four final-state leptons. This variable is expected to be rather
insensitive to parton-shower effects. The upper panel shows 
differential cross sections for various combinations of fixed-order
result and parton shower. The black line denotes the fixed-order NLO QCD
result without any parton shower attached, the two blue lines employ the
dipole shower, where the brighter line shows the combination with the LO
result and the darker one with the NLO result, and the red line depicts
the NLO plus angular-ordered shower result. The matching is performed
using the MC@NLO-type, subtractive scheme in both cases. In all cases,
the distribution clearly shows the Higgs peak at 125~GeV, followed by
the continuum production of two on-shell $W$ bosons starting around the
$2M_W$ threshold. The second panel depicts the ratio of the different
cross sections over the fixed-order NLO result, given by the solid lines.
The shaded area describes the scale variation band, which we will
explain afterwards. As one can see from the figure, the fixed-order NLO
QCD curve yields the largest cross section, and all parton-shower
results are smaller by a factor which is approximately constant over the
invariant mass. The reason for this is an effect we have already seen
when discussing the NNLO QCD corrections. The parton shower generates
additional splittings of final-state partons. If these are wide-angle
and hard enough, they will not be clustered into the same jet as the
original parton, but form a separate jet. Hence, the energy of the two
leading jets gets decreased and its invariant mass becomes smaller. Thus
this VBF cut removes a fraction of the events and the cross section
becomes smaller. For the LO curve, the larger strong coupling constant
increases the parton splitting probability and so the drop in cross
section is stronger than for the other results. The four lower panels
show scale uncertainty bands, individually normalized to the central
results. Varied are, from top to bottom, the factorization scale
$\mu_F$, the renormalization scale $\mu_R$, the starting scale of the
shower $\mu_Q$, and all three at the same time, labeled
$\mu_{\text{tot}}$. For the individual scale variations, these are
probed by changing each scale by a factor $2^{\pm1}$ from the central
scale $\mu_0$. For the joint case, all scales are allowed to vary
separately, but the ratio of any two scales must be within the range
$[0.5;2]$. So if any of the scales is varied upwards, none of the others
can fluctuate downwards and vice versa. The result of varying all scales
in this way is also depicted as the band shown in the ratio panel.
Both factorization and renormalization scale are varied consistently
in the hard process and the parton shower, changing all occurrences
simultaneously. 
We see from the figure that the scale variation behavior is dominated by
the fixed-order part. The total LO variation is dominated by the effect
of changing the factorization scale and leads to an uncertainty band of
about 10\%, while the renormalization scale enters through the shower
only and yields small effects. The dependence on the shower scale is
rather small in all cases. From the ratio plot we also see that the
bands of the fixed-order NLO result, the NLO result matched with the
dipole shower and the LO plus dipole shower are all non-overlapping.
Simple scale variation thus tends to underestimate these migration
effects.

\begin{figure}
\begin{center}
\includegraphics[width=0.85\textwidth]{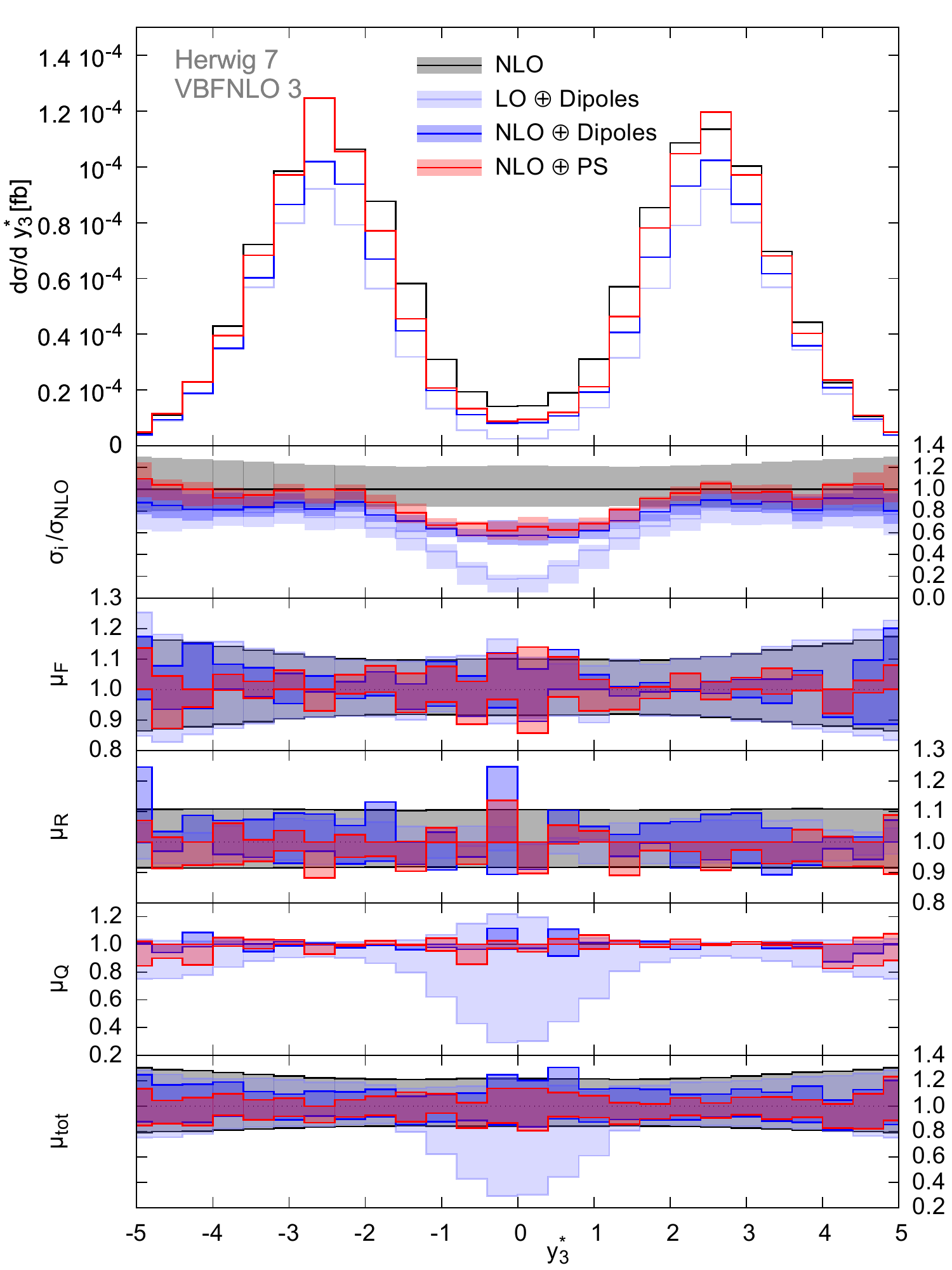} 
\end{center}
\caption{
Rapidity distribution of the third jet relative to the two tagging jets
in $e^+\nu_e\mu^-\bar{\nu}_\mu jj$ production via VBF. 
Curves and quantities plotted are as in \fig{fig:vbfww_scale_m4l}.
Figure taken from Ref.~\protect\cite{Rauch:2016upa}.
}
\label{fig:vbfww_scale_y3star}
\end{figure}
\begin{figure}
\begin{center}
\includegraphics[width=0.85\textwidth]{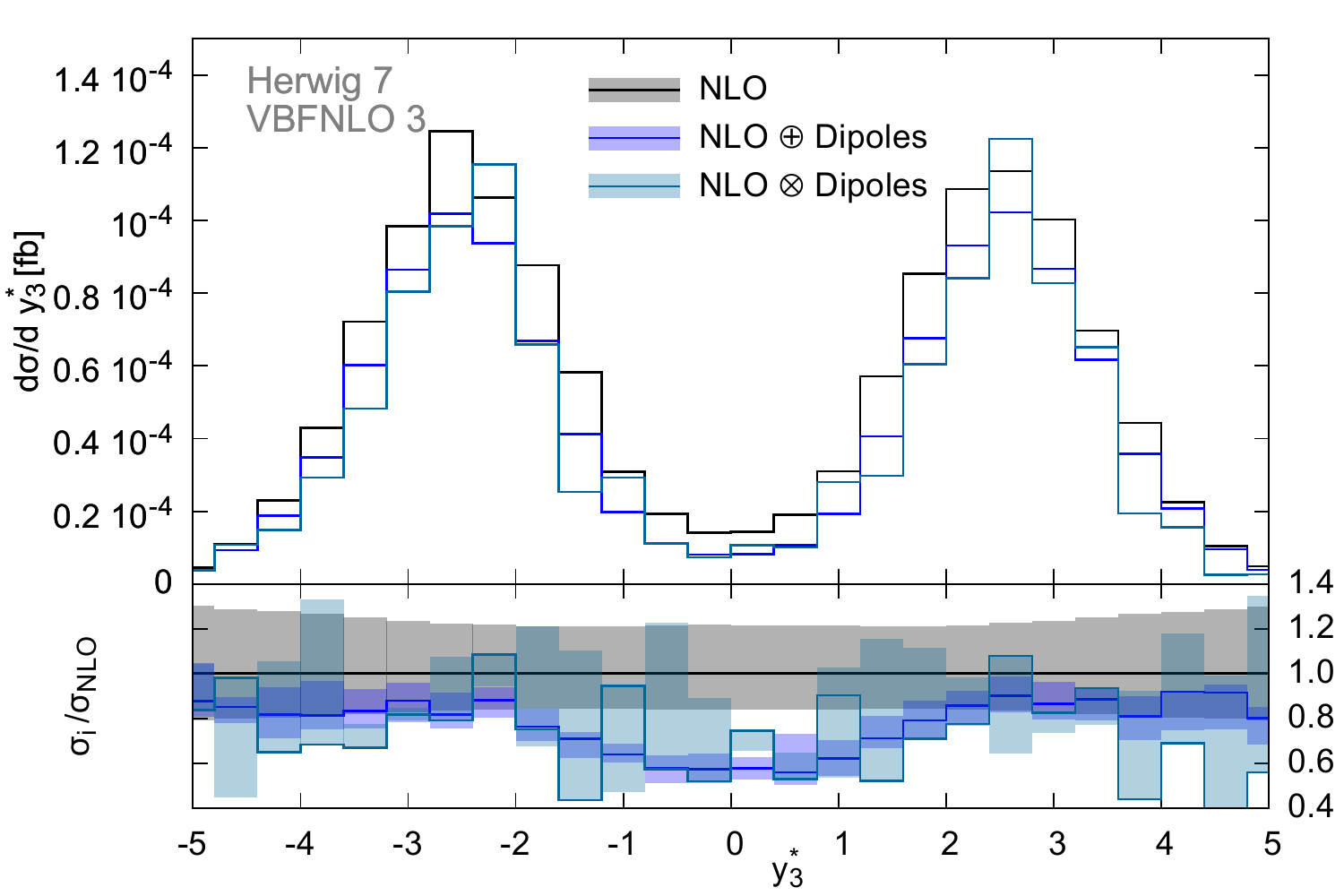} 
\end{center}
\caption{
Rapidity distribution of the third jet relative to the two tagging jets
in $e^+\nu_e\mu^-\bar{\nu}_\mu jj$ production via VBF. 
The upper panel shows the differential cross section for the fixed-order
NLO calculation and combined with the dipole shower using subtractive
MC@NLO-type (``$\oplus$''), and multiplicative POWHEG-type
(``$\otimes$'') matching. The lower panel depicts the ratio of the
differential cross sections over the NLO one. The bands denote the scale
uncertainty band of the respective cross section, obtained by varying
the factorization, renormalization and shower-scale uncertainty by
a factor $2^{\pm1}$ around the central scale $\mu_0$.
Figure taken from Ref.~\protect\cite{Rauch:2016upa}.
}
\label{fig:vbfww_scale_y3star_powheg}
\end{figure}
In \fig{fig:vbfww_scale_y3star} we present results for the variable
$y_3^*$, the rapidity of the third jet relative to the averaged
rapidities of the two tagging jets, which has been defined in
\eq{eq:ystar}. This variable plays an important role for VBF processes,
as a veto on additional jets in the central region allows to reduce
QCD-induced background processes. 
The differential cross section in the upper panel of the figure shows
the same picture we have already seen during the discussion of the
POWHEG-BOX results. At large absolute values of the observable, the
results of the fixed-order calculation and the parton-shower ones agree
well. In the central region, the situation looks different. Here much
less radiation than at NLO is predicted. This is particularly striking
for the LO plus dipole shower curve, where the differential cross
section in the two central bins is only 20\% of the NLO number. In this
case the third jet originates purely from partons generated by the
shower, and radiation is in general too soft to produce sufficient jet
activity. An indication for problematic behavior can be seen in the
large uncertainties when varying the shower scale, though the resulting
band is not wide enough to reach the NLO-matched results. There,
corrections from the hard matrix element increase the cross section and
stabilize the prediction. The uncertainty from varying the shower scale
now becomes small and no longer shows an increase in the central
region, but is rather flat over the whole range. No relevant difference
between the results using the dipole shower and those with the
angular-ordered shower can be observed.

In \fig{fig:vbfww_scale_y3star_powheg} we compare the effects of the two
matching schemes on $y_3^*$. The cyan curve shows results using
multiplicative, POWHEG-type matching, The blue curve using subtractive,
MC@NLO-type matching and the black fixed-order NLO curve are the same as
in \fig{fig:vbfww_scale_y3star}. As parton shower we use in both cases
the dipole shower. The upper panel shows the differential cross section,
and the lower one the ratio over the fixed-order NLO result. The scale
variation bands there are obtained by varying the factorization,
renormalization and, in case of the matched cross sections, the shower
scale jointly by a factor $2^{\pm 1}$. We observe that there is in
general a good agreement between the results of the two matching
schemes. For the integrated cross section, the POWHEG result is closer
to the NLO result, which originated mostly from the region where the
rapidity of the third jet is close to one of the two tagging jets, \ie
the peak region of the distribution. For the phenomenologically
interesting central region, where the third jet has $y_3^*=0$,
the suppression of the cross section seen for MC@NLO-type matching is
equally present for POWHEG-type matching. Barring statistical
fluctuations from finite Monte-Carlo statistics, the two results are
almost on top of each other. The uncertainty band of the POWHEG result
is thereby a bit larger. In total, the parton-shower prediction for this
observable is insensitive to the details of the matching procedure or
the exact shower algorithm. Therefore, it can serve as a useful tool to
reduce the contribution of background processes.

\section{Anomalous Coupling Effects}
\label{chap:anomcoupl}

Having discussed the SM predictions for VBF in the previous chapters, we
now turn to new-physics effects appearing in this process class which preserve
the underlying symmetries of the SM. VBF processes contain triple (TGCs) and
quartic gauge couplings (QGCs). So they are an ideal tool to study these
couplings and look for possible deviations from the SM prediction.

\subsection{Effective Field Theories}

A convenient tool to study new-physics effects at high energies in a
model-independent way are effective field theories (EFTs). They are
based on the following expansion of the EFT Lagrangian
\begin{equation}
\La_\text{EFT} = \La_\text{SM} + \sum_{d>4} \sum_i
\frac{f^{(d)}_i}{\Lambda^{d-4}} \obs^{(d)}_i \,,
\label{eq:eftmaster}
\end{equation}
where $\Lambda$ is the typical scale of new physics. 
$f^{(d)}_i$ are the dimensionless coupling-strength coefficients which are
typically of $\Order{1}$. If the underlying theory is known or
presumed to be loop-induced, it can be convenient to pull out an
additional explicit loop factor like $\frac1{16\pi^2}$ in the expansion.
The operators $\obs^{(d)}_i$ are invariant under the symmetries
of the SM. Additionally, different combinations of operators can be
related and lead to the same effects on physical observables. To find a
minimal set, one uses the equations of motions for the fields and the
fact that adding a total derivative to the Lagrangian does not change
the theoretical predictions..

When $\Lambda$ is larger than the energy scales $E$ of our process, higher
orders in the expansion are suppressed by additional powers of
$\frac{E}{\Lambda}$ and it is sufficient to take the leading, lowest-order
term of the expansion. Once this truncation has been performed, only a
finite number of operators contributes and the theory becomes
predictive. It is however important to keep in mind that after
truncation the expansion is valid only below the scale of new physics.

The operators are constructed out of the following building blocks:
\begin{itemize}
\item Higgs doublet field $\Phi$,
\item covariant derivative $D^\mu$, which reduces to $\partial^\mu$ for singlet fields,
\item field strength tensors $G^{a,\mu\nu}$, $W^{i,\mu\nu}$, $B^{\mu\nu}$,
\item fermion fields $\psi$.
\end{itemize}
In lowest dimension, $d=5$, only one operator
exists~\cite{Weinberg:1979sa}, which generates a Majorana mass term for
neutrinos and violates lepton-number conservation. In general, all
operators with an odd number of dimensions involve fermion fields and
lead to lepton or baryon number violation. Therefore, for the
electroweak gauge-boson interactions discussed here, we need to consider
even dimensions only, starting with
$d=6$~\cite{Buchmuller:1985jz,Hagiwara:1993qt,Hagiwara:1993ck,Grzadkowski:2010es}.

Before turning to the operators, we first introduce the notation and
definitions which will be used in the following.
The Higgs doublet field $\Phi$ in the unitary gauge is given by 
\begin{equation}
\Phi = 
\begin{pmatrix}
0 \\
\frac1{\sqrt{2}} ( v + H )
\end{pmatrix} \,.
\end{equation}
The covariant derivative acting on it is defined as
\begin{equation}
D_\mu = \partial_\mu + i g \frac{\sigma^j}2 W_\mu^j + i g' \frac{Y}2
B_\mu \,,
\end{equation}
where $\frac{\sigma^j}2$ are the $SU(2)$ generators with the Pauli
matrices $\sigma^j$ and $Y$ is the hypercharge of the field on which the
derivative acts. For the operators discussed here, this will always be
the Higgs field with $Y=1$. $g$ and $g'$ denote the $SU(2)$ and $U(1)$
gauge couplings, which are related to the electromagnetic coupling
$e = g s_w = g' c_w$ via sine $s_w$ and cosine $c_w$ of the weak mixing
angle, respectively.

The modified field strength tensors
\begin{align}
\widehat{W}^{\mu\nu} &= i g \frac{\sigma^j}2 W^{j,\mu\nu} 
 = i g \frac{\sigma^j}2 \left( 
 \partial^\mu W^{j,\nu} - \partial^\nu W^{j,\mu} - g \epsilon^{jkl}
W^{k,\mu} W^{l,\nu} \right) \,, \nonumber\\
\widehat{B}^{\mu\nu} &= i g' \frac12 B^{\mu\nu} 
 = i g' \frac12 \left( 
 \partial^\mu B^{\nu} - \partial^\nu B^{\mu} \right) 
\label{eq:modfieldstrength}
\end{align}
contain an additional prefactor such that~\cite{Hagiwara:1993qt}
\begin{equation}
\left[ D^\mu , D^\nu \right] = \widehat{W}^{\mu\nu} + \widehat{B}^{\mu\nu}
\end{equation}
and they are treated on a more equal footing with the covariant
derivative.

The modified dual field-strength tensors are given by 
\begin{align}
\widetilde{W}^{\mu\nu} &= \frac12 \epsilon^{\mu\nu\rho\sigma} \widehat{W}_{\rho\sigma} \,, &
\widetilde{B}^{\mu\nu} &= \frac12 \epsilon^{\mu\nu\rho\sigma} \widehat{B}_{\rho\sigma} \,. 
\end{align}

Using these definitions, one can construct the following independent
CP-conserving $d=6$ operators according to
Refs.~\cite{Hagiwara:1993qt,Hagiwara:1993ck,Degrande:2013rea,VBFNLO},
\begin{align}
\obs_{WWW} &= \Tr\left[ {\widehat{W}^\mu}{}_\nu {\widehat{W}^\nu}{}_\rho {\widehat{W}^\rho}{}_\mu \right] \,, \nonumber\\
\obs_{W} &= \left(D_\mu\Phi\right)^\dagger \widehat{W}^{\mu\nu} \left(D_\nu\Phi\right) \,, \nonumber\\
\obs_{B} &= \left(D_\mu\Phi\right)^\dagger \widehat{B}^{\mu\nu} \left(D_\nu\Phi\right) \,, \nonumber\\
\obs_{WW} &= \Phi^{\dagger} \widehat{W}_{\mu\nu} \widehat{W}^{\mu\nu} \Phi \,, \nonumber\\
\obs_{BB} &= \Phi^{\dagger} \widehat{B}_{\mu\nu} \widehat{B}^{\mu\nu} \Phi \,, \nonumber\\
\obs_{\phi,2} &= \partial_\mu\left(\Phi^\dagger\Phi\right) \partial^\mu\left(\Phi^\dagger\Phi\right) \,.
\label{eq:d6cpeven}
\end{align}
The last operator $\obs_{\phi,2}$ contains only terms involving Higgs
bosons. This includes a term $v^2 (\partial_\mu H)
(\partial^\mu H)$, which gives a contribution to the kinetic term of the
Higgs field. This must be absorbed by a redefinition of the Higgs boson
field, thus changing all couplings involving Higgs bosons.

CP-violating operators can be defined analogously by replacing one
field-strength tensor by its dual
\begin{align}
\obs_{\widetilde{W}WW} &= 
 \Tr\left[ {\widetilde{W}^\mu}{}_\nu {\widehat{W}^\nu}{}_\rho
{\widehat{W}^\rho}{}_\mu \right] \,, \nonumber\\
\obs_{\widetilde{W}} &= \left(D_\mu\Phi\right)^\dagger \widetilde{W}^{\mu\nu}
\left(D_\nu\Phi\right) \,, \nonumber\\
\obs_{\widetilde{B}} &= \left(D_\mu\Phi\right)^\dagger \widetilde{B}^{\mu\nu}
\left(D_\nu\Phi\right) \,, \nonumber\\
\obs_{\widetilde{W}W} &= \Phi^{\dagger} \widetilde{W}_{\mu\nu} \widehat{W}^{\mu\nu}
\Phi \,, \nonumber\\
\obs_{\widetilde{B}B} &= \Phi^{\dagger} \widetilde{B}_{\mu\nu} \widehat{B}^{\mu\nu}
\Phi  \,,
\label{eq:d6cpodd}
\end{align}
which contribute to the triple and quartic vertices of electroweak
bosons. Actually, only four of these five operators are linearly
independent, as the relation~\cite{Arnold:2011wj,Degrande:2013rea}
\begin{equation}
\obs_{\widetilde{W}} + \frac12 \obs_{\widetilde{W}W} = 
\obs_{\widetilde{B}} + \frac12 \obs_{\widetilde{B}B} 
\end{equation}
allows to eliminate one of them.


As mentioned before, the choice of independent operators is not unique.
Additional operators can be constructed, which are also of dimension 6
and invariant under the SM gauge groups. An example for this is the
operator 
\begin{align}
\obs_{\phi W} &= \Phi^{\dagger}\Phi \Tr\left[\widehat{W}_{\mu\nu} \widehat{W}^{\mu\nu} \right] 
\end{align}
of Ref.~\cite{Degrande:2013rea}, which is related to $\obs_{WW}=\frac12 \obs_{\phi W}$.

\begin{table}
\footnotesize{
\renewcommand{\arraystretch}{1.2}
\setlength{\tabcolsep}{5.5pt}
 \begin{tabular}{@{}lccccccccccc@{}}
			& $\obs_{WWW}$ & $\obs_{W}$& $\obs_{B}$ & $\obs_{WW}$ & $\obs_{BB}$ & $\obs_{\phi,2}$& $\obs_{\widetilde{W}WW}$ & $\obs_{\widetilde{W}}$ & $\obs_{\widetilde{B}}$ & $\obs_{\widetilde{W}W}$ & $\obs_{\widetilde{B}B}$ \\\hline
$WWZ$			&X	&X	&X	&	&	&	&X	&X	&X	&	&	\\
$WW\gamma$		&X	&X	&X	&	&	&	&X	&X	&X	&	&	\\
$HWW$			&	&X	&	&X	&	&X	&	&X	&	&X	&	\\
$HZZ$			&	&X	&X	&X	&X	&X	&	&X	&X	&X	&X	\\
$HZ\gamma$		&	&X	&X	&X	&X	&(X)	&	&X	&X	&X	&X	\\
$H\gamma\gamma$		&	&	&	&X	&X	&(X)	&	&	&	&X	&X	\\
$WWWW$			&X	&X	&	&	&	&	&X	&	&	&	&	\\
$WWZZ$			&X	&X	&	&	&	&	&X	&	&	&	&	\\
$WWZ\gamma$		&X	&X	&	&	&	&	&X	&	&	&	&	\\
$WW\gamma\gamma$	&X	&	&	&	&	&	&X	&	&	&	&	\\\hline
 \end{tabular}}
\caption{Vertices induced by each operator are marked with X in the
corresponding column. Vertices which are not relevant for three and four
gauge boson amplitudes have been omitted. Table adopted from
Ref.~\protect\cite{Degrande:2013rea}.}
\label{tab:eft6_overview}
\end{table}
An overview how the different operators affect the vertices relevant for
VBF production processes is given in \tab{tab:eft6_overview}. 
For the operator $\obs_{\phi,2}$ the two entries for $HZ\gamma$ and
$H\gamma\gamma$ are marked in brackets, as this operator does not induce
an additional tree-level contribution as in all other cases, but
modifies the loop-induced SM contribution through an additional factor
on the $HWW$ and $Hf\bar{f}$ couplings. 
One can see that all operators modify three-boson vertices. Therefore,
these will also contribute to diboson production processes, which in
general show a higher sensitivity due to their larger cross sections,
and limits from there need to be taken into account.

As one can see from \tab{tab:eft6_overview}, all dimension-6 operators
which introduce modifications to the QGCs also change
the TGCs, and one would expect to see deviations from
the SM predictions there first. If we want to consider models where
only the quartic couplings are changed, we need to increase the dimension
by two and study EFT operators of dimension~8. Such a scenario could 
be realized with new-physics bosons coupling to the SM gauge
bosons. Then contributions to the effective low-energy QGCs contain
tree-level diagrams with an exchange of the new
particles~\cite{Arzt:1994gp}, while for the TGCs only loop-induced diagrams
are possible, which are suppressed by an additional loop factor.

One can also see that dimension-8 operators are required for
pure modifications of the QGCs from the following
argument~\cite{Degrande:2013rea}. The gauge bosons can either originate
from a field-strength tensor or from the covariant derivative acting on
the Higgs field. Both expression are dimension-2 terms, and the 
gauge field is accompanied by a partial derivative $\partial_\mu$ or the
vacuum expectation value $v$, respectively. Thus we need four of them
for the full operator, ending up with dimension~8. 

This also yields a convenient classification mechanism for the
dimension-8 operators modifying QGCs by sorting them by the number of
covariant derivatives and field-strength tensors. As the number of open
Lorentz indices is one and two, respectively, three different
possibilities exist~\cite{Eboli:2006wa,Eboli:2016kko}. Operators can
either contain only covariant derivatives, called scalar, 
\begin{align}
\obs_{S,0} &= \left[ (D_\mu \Phi)^\dagger D_\nu \Phi \right] \times 
      \left[ ( D^\mu \Phi)^\dagger D^\nu \Phi \right] \,, \nonumber\\
\obs_{S,1} &= \left[ ( D_\mu \Phi )^\dagger D^\mu \Phi  \right] \times 
      \left[ ( D_\nu \Phi )^\dagger D^\nu \Phi \right] \,, \nonumber\\
\obs_{S,2} &= \left[ ( D_\mu \Phi )^\dagger D_\nu \Phi  \right] \times 
      \left[ ( D^\nu \Phi )^\dagger D^\mu \Phi \right] \,,
\label{eq:obsd8s}
\end{align}
only field-strength tensors, called tensor,
\begin{align}
\obs_{T,0} &= \Tr\left[ \widehat{W}_{\mu\nu} \widehat{W}^{\mu\nu} \right] \times 
      \Tr\left[ \widehat{W}_{\alpha\beta} \widehat{W}^{\alpha\beta} \right] \,, \nonumber\\
\obs_{T,1} &= \Tr\left[ \widehat{W}_{\alpha\nu} \widehat{W}^{\mu\beta} \right] 
      \times \Tr\left[ \widehat{W}_{\mu\beta} \widehat{W}^{\alpha\nu} \right] \,, \nonumber\\
\obs_{T,2} &= \Tr\left[ \widehat{W}_{\alpha\mu} \widehat{W}^{\mu\beta} \right] 
      \times   \Tr\left[ \widehat{W}_{\beta\nu} \widehat{W}^{\nu\alpha} \right] \,, \nonumber\\
\obs_{T,5} &= \Tr\left[ \widehat{W}_{\mu\nu} \widehat{W}^{\mu\nu} \right]\times 
      \widehat{B}_{\alpha\beta} \widehat{B}^{\alpha\beta} \,, \nonumber\\
\obs_{T,6} &= \Tr\left[ \widehat{W}_{\alpha\nu} \widehat{W}^{\mu\beta} \right] 
      \times \widehat{B}_{\mu\beta} \widehat{B}^{\alpha\nu} \,, \nonumber\\
\obs_{T,7} &= \Tr\left[ \widehat{W}_{\alpha\mu} \widehat{W}^{\mu\beta} \right] 
      \times \widehat{B}_{\beta\nu} \widehat{B}^{\nu\alpha} \,, \nonumber\\
\obs_{T,8} &= \widehat{B}_{\mu\nu} \widehat{B}^{\mu\nu} \widehat{B}_{\alpha\beta} \widehat{B}^{\alpha\beta} \,, \nonumber\\
\obs_{T,9} &= \widehat{B}_{\alpha\mu} \widehat{B}^{\mu\beta} \widehat{B}_{\beta\nu}
\widehat{B}^{\nu\alpha} \,,
\label{eq:obsd8t}
\end{align}
or two of them each, called mixed,
\begin{align}
\obs_{M,0} &= \Tr\left[ \widehat{W}_{\mu\nu} \widehat{W}^{\mu\nu} \right] \times 
      \left[ ( D_\beta \Phi )^\dagger D^\beta \Phi \right] \,, \nonumber\\
\obs_{M,1} &= \Tr\left[ \widehat{W}_{\mu\nu} \widehat{W}^{\nu\beta} \right] \times  
      \left[ ( D_\beta \Phi )^\dagger D^\mu \Phi \right] \,, \nonumber\\
\obs_{M,2} &= \left[ \widehat{B}_{\mu\nu} \widehat{B}^{\mu\nu} \right ] \times  
      \left [ ( D_\beta \Phi )^\dagger D^\beta \Phi \right ] \,, \nonumber\\
\obs_{M,3} &= \left[ \widehat{B}_{\mu\nu} \widehat{B}^{\nu\beta} \right ] \times  
      \left [ ( D_\beta \Phi )^\dagger D^\mu \Phi \right ] \,, \nonumber\\
\obs_{M,4} &= \left[ ( D_\mu \Phi )^\dagger \widehat{W}_{\beta\nu} 
      D^\mu \Phi  \right] \times \widehat{B}^{\beta\nu} \,, \nonumber\\
\obs_{M,5} &= \left[ ( D_\mu \Phi )^\dagger \widehat{W}_{\beta\nu} 
      D^\nu \Phi  \right] \times \widehat{B}^{\beta\mu} \,, \nonumber\\
\obs_{M,7} &= \left[ ( D_\mu \Phi )^\dagger \widehat{W}_{\beta\nu} 
      \widehat{W}^{\beta\mu} D^\nu \Phi  \right] \,.
\label{eq:obsd8m}
\end{align}

\begin{table}
\centering
\footnotesize{
\renewcommand{\arraystretch}{1.2}
\begin{tabular}{@{}lcccccc@{}} \hline
 &
\begin{tabular}{@{}c@{}}
$\obs_{S,0}$,\\
$\obs_{S,1}$,\\
$\obs_{S,2}$
\end{tabular}&
\begin{tabular}{@{}c@{}}
$\obs_{M,0}$,\\
$\obs_{M,1}$,\\
$\obs_{M,7}$
\end{tabular}&
\begin{tabular}{@{}c@{}}
$\obs_{M,2}$,\\
$\obs_{M,3}$,\\
$\obs_{M,4}$,\\
$\obs_{M,5}$
\end{tabular}&
\begin{tabular}{@{}c@{}}
$\obs_{T,0}$,\\
$\obs_{T,1}$,\\
$\obs_{T,2}$
\end{tabular}&
\begin{tabular}{@{}c@{}}
$\obs_{T,5}$,\\
$\obs_{T,6}$,\\
$\obs_{T,7}$
\end{tabular}&
\begin{tabular}{@{}c@{}}
$\obs_{T,8}$,\\
$\obs_{T,9}$
\end{tabular}\\\hline
 $WWWW$ &  X &  X &    & X   &    &    \\
 $WWZZ$ & X & X & X & X & X &  \\
 $ZZZZ$ & X & X & X & X & X & X \\
 $WWZ\gamma$ &   & X & X & X & X &  \\
 $WW\gamma\gamma$ &   & X & X & X & X &  \\
 $ZZZ\gamma$ &   & X & X & X & X & X \\
 $ZZ\gamma\gamma$ &   & X & X & X & X & X \\
 $Z\gamma\gamma\gamma$ &   &   &   & X  & X  & X  \\
 $\gamma\gamma\gamma\gamma$&  &   &   &X   &X   &X   \\\hline
\end{tabular}
}
\caption{Quartic vertices modified by each dimension-8 
operator are marked with $X$. Table adopted from
Ref.~\protect\cite{Degrande:2013rea}.}
\label{tab:eft8_overview}
\end{table}

In Ref.~\cite{Eboli:2006wa}, two additional operators,
\begin{align}
\obs_{T,3} &= \Tr\left[ \widehat{W}_{\alpha\mu} \widehat{W}^{\mu\beta}
\widehat{W}^{\nu\alpha} \right] \times \widehat{B}_{\beta\nu} \,,\nonumber\\
\obs_{T,4} &= \Tr\left[ \widehat{W}_{\alpha\mu}
   \widehat{W}^{\alpha\mu}  \widehat{W}^{\beta\nu} \right] \times \widehat{B}_{\beta\nu} \,,
\end{align}
have been defined, which however vanish identically. For $\obs_{T,3}$,
the trace is symmetric under permutations of indices $\beta$ and $\nu$,
while the field-strength tensor $\widehat{B}_{\beta\nu}$ is
anti-symmetric, and for $\obs_{T,4}$ the trace itself vanishes.
The operator 
\begin{equation}
\obs_{M,6} = \left[ ( D_\mu \Phi )^\dagger \widehat{W}_{\beta\nu} 
      \widehat{W}^{\beta\nu} D^\mu \Phi \right] \,,
\end{equation}
which has also been introduced in Ref.~\cite{Eboli:2006wa}, is
equivalent to $\obs_{M,0}$ and the relation $\obs_{M,6} = \frac12
\obs_{M,0}$ holds~\cite{ThesisSekulla}.

\begin{figure}
\begin{center}
\begin{tabular}{rl}
\raisebox{-0.5\height}{\includegraphics[width=0.3\textwidth]{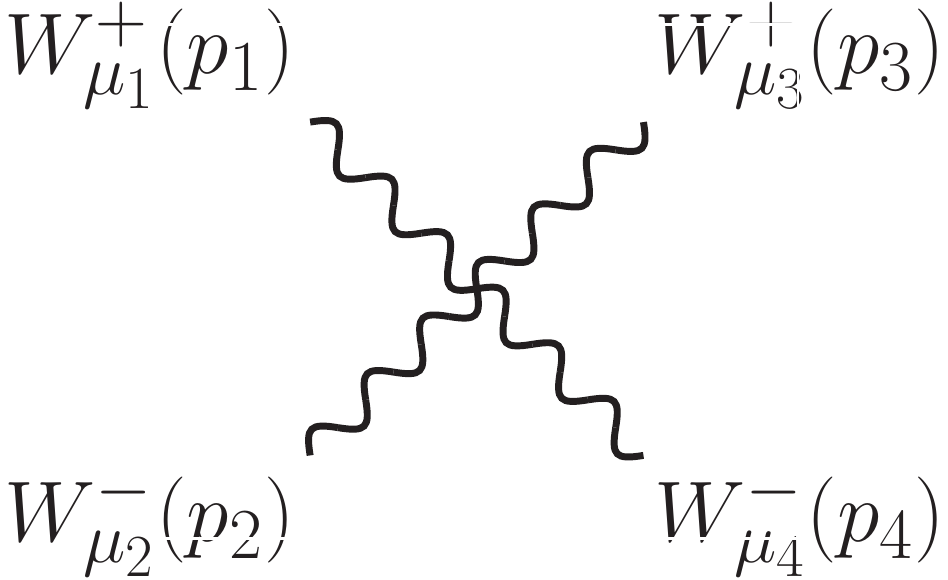}} &
\parbox{0.6\textwidth}{\begin{align*}
i \,2 M_W^4 \biggl[ 
 &\quad 2\,\frac{f_{S,0}}{\Lambda^4} \cdot g_{\mu_1\mu_3} g_{\mu_2\mu_4} \\
 &+\frac{f_{S,1}+f_{S,2}}{\Lambda^4} \cdot \bigl(g_{\mu_1\mu_2} g_{\mu_3\mu_4}
     + g_{\mu_1\mu_4} g_{\mu_2\mu_3} \bigr)
\biggr]
\end{align*}} \\
\raisebox{-0.5\height}{\includegraphics[width=0.3\textwidth]{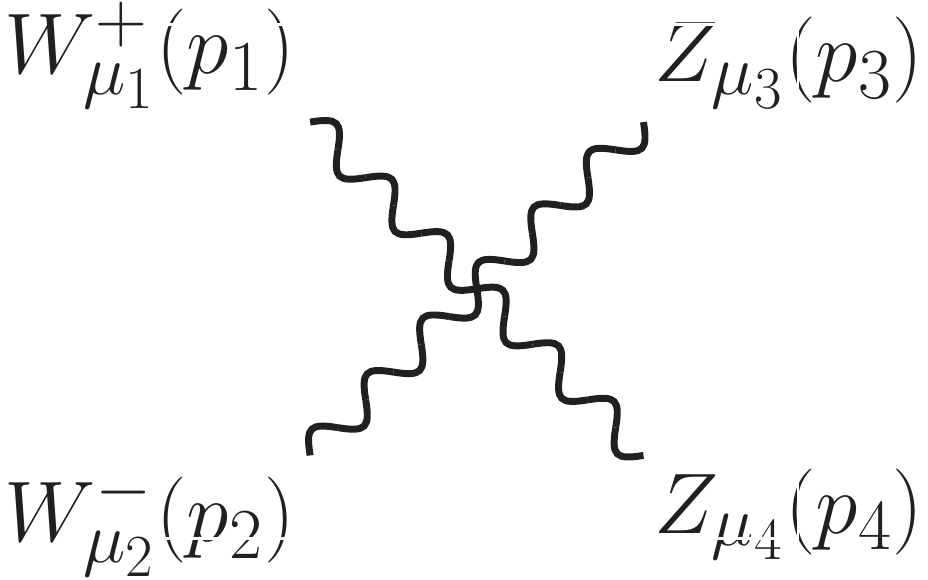}} & 
\parbox{0.6\textwidth}{\begin{align*}
i M_W^2 M_Z^2 \biggl[ 
 &\quad\frac{f_{S,0}+f_{S,2}}{\Lambda^4} \cdot 
    \bigl(g_{\mu_1\mu_3} g_{\mu_2\mu_4} + g_{\mu_1\mu_4} g_{\mu_2\mu_3} \bigr) \\
 &+2\,\frac{f_{S,1}}{\Lambda^4} \cdot g_{\mu_1\mu_2} g_{\mu_3\mu_4}
\biggr]
\end{align*}} \\
\raisebox{-0.5\height}{\includegraphics[width=0.3\textwidth]{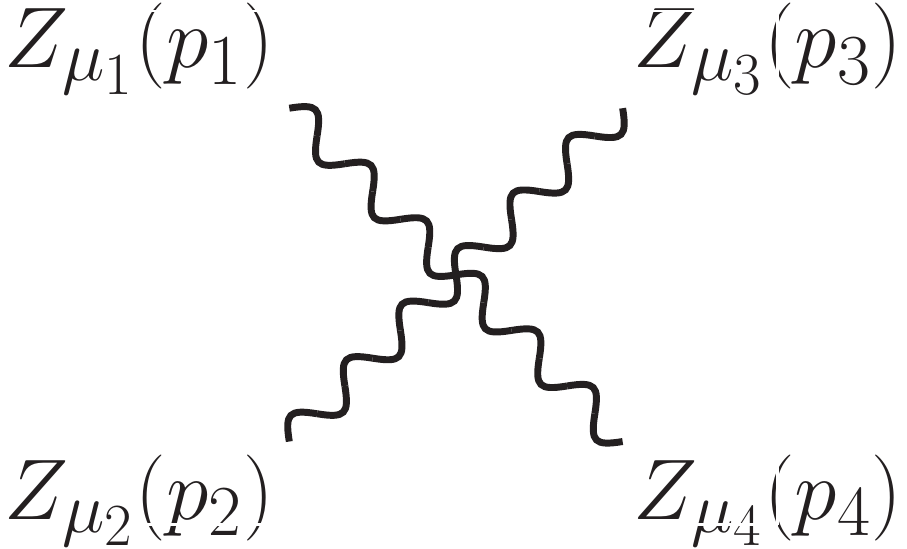}} &
\parbox{0.6\textwidth}{\begin{align*}
i \,2 M_Z^4  
 &\frac{f_{S,0}+f_{S,1}+f_{S,2}}{\Lambda^4} \\
 &\quad \cdot \bigl( g_{\mu_1\mu_2} g_{\mu_3\mu_4} 
 + g_{\mu_1\mu_3} g_{\mu_2\mu_4}
     + g_{\mu_1\mu_4} g_{\mu_2\mu_3} \bigr)
\end{align*}} 
\end{tabular}
\end{center}
\caption{Feynman rules for quartic gauge boson vertices originating from the
operators $\obs_{S,0}$, $\obs_{S,1}$ and $\obs_{S,2}$.
}
\label{fig:fm_LS012}
\end{figure}
The scalar case contains an additional operator
$\obs_{S,2}$~\cite{ThesisMax,ZeppenfeldPrivate,Eboli:2016kko} compared to
Ref.~\cite{Eboli:2006wa}. This operator cannot be constructed out of the
other ones, but is a new, distinct possibility. 
This can be easily seen if we first consider the following building
block appearing in the scalar dimension-8 operators
\begin{align}
\left[ (D_\mu \Phi)^\dagger D_\nu \Phi \right] 
&= \frac12 \left( \partial_\mu H \right) \left( \partial_\nu H \right)
+ M_W^2 W^-_\mu W^+_\nu \left(1+\frac{H}{v}\right)^2
+ \frac{M_Z^2}2 Z_\mu Z_\nu \left(1+\frac{H}{v}\right)^2 \nonumber\\
&\quad+ \frac{i M_Z}2 \bigl( Z_\mu \left(\partial_\nu H\right) 
  - Z_\nu \left(\partial_\mu H\right) \bigl) \left(1+\frac{H}{v}\right) \,.
\label{eq:dmuphidnuphi}
\end{align}
Therefore, for the operators one obtains
\begin{align}
\obs_{S,0}\Bigr|_{\text{4 gauge bosons}} &= 
\quad M_W^4 \ W^- {\cdot} W^- \ W^+ {\cdot} W^+ \nonumber\\
&\quad
+ M_W^2 M_Z^2 \ W^- {\cdot} Z \ W^+ {\cdot} Z \nonumber\\
&\quad
+ \frac14 M_Z^4 \ Z {\cdot} Z \ Z {\cdot} Z \,,
\nonumber\\
\obs_{S,1}\Bigr|_{\text{4 gauge bosons}} &= 
\quad M_W^4 \ W^- {\cdot} W^+ \ W^- {\cdot} W^+ \nonumber\\
&\quad
+ M_W^2 M_Z^2 \ W^- {\cdot} W^+ \ Z {\cdot} Z \nonumber\\
&\quad
+ \frac14 M_Z^4 \ Z {\cdot} Z \ Z {\cdot} Z \,,
\nonumber\\
\obs_{S,2}\Bigr|_{\text{4 gauge bosons}} &= 
\quad M_W^4 \ W^- {\cdot} W^+ \ W^- {\cdot} W^+ \nonumber\\
&\quad
+ M_W^2 M_Z^2 \ W^- {\cdot} Z \ W^+ {\cdot} Z \nonumber\\
&\quad
+ \frac14 M_Z^4 \ Z {\cdot} Z \ Z {\cdot} Z \,,
\label{eq:S0vsS2}
\end{align}
where we have restricted ourselves now to only those terms which lead to
an interaction of four gauge bosons, as only these are relevant for the
discussion on VBF processes later. The corresponding Feynman rules are
depicted in \fig{fig:fm_LS012}.
From these expressions one sees immediately that there is no linear
combination of $\obs_{S,0}$ and $\obs_{S,1}$ which would yield the
correct form for $\obs_{S,2}$ for all three vertices. We will come back
to this issue when discussing relations between different
parametrizations below. 

In \tab{tab:eft8_overview}, taken from Ref.~\cite{Degrande:2013rea},
we show again which QGCs get modified by the different operators. One
can see in particular that the operators $\obs_{T,8}$ and $\obs_{T,9}$
change the couplings between neutral gauge bosons only.

\subsubsection{Non-linear Realization}

The approach described above is based on a power counting in terms of
the canonical dimension of the operators. An alternative possibility is
given by the electroweak chiral Lagrangian, modeled in analogy to the
chiral Lagrangian of QCD~\cite{Weinberg:1978kz}. Here the expansion
parameter is the chiral dimension of the
operator~\cite{Urech:1994hd,Knecht:1999ag,Nyffeler:1999ap,Hirn:2004ze,Hirn:2005fr},
which is 0 for bosons and 1 for derivatives, couplings and fermion
bilinears, and is equivalent to a loop
expansion~\cite{Buchalla:2013eza}. This approach has been originally
formulated for scenarios without Higgs
bosons~\cite{Appelquist:1980vg,Longhitano:1980iz,Longhitano:1980tm,Dawson:1990cc,Appelquist:1993ka,Dobado:1995qy,Dobado:1995ze,Dobado:1997jx,Kilian:2003pc,Belyaev:1998ih,Buchalla:2012qq}
and later extended to include the possibility of a light Higgs
boson~\cite{Feruglio:1992wf,Bagger:1993zf,Koulovassilopoulos:1993pw,Burgess:1999ha,Wang:2006im,Grinstein:2007iv,Alonso:2012px,Buchalla:2013rka,Brivio:2013pma}.
The chiral Lagrangian is often also called non-linear EFT, as the Higgs
and Goldstone fields appear in the operators in a non-linear way, while
for the operators with higher canonical dimension discussed first the
appearance through the fields $\Phi$ is linear.

For quartic gauge boson interactions, two operators are
relevant~\cite{Alboteanu:2008my}:
\begin{align}
\La_4 &= \alpha_4 \left( \Tr\left[ V_\mu V_\nu \right] \right)^2 \,,
\nonumber\\
\La_5 &= \alpha_5 \left( \Tr\left[ V_\mu V^\mu \right] \right)^2 \,,
\label{eq:La45}
\end{align}
where $\alpha_4$ and $\alpha_5$ denote the dimensionless coefficients of
the two operators, respectively. 
Additionally, the following definitions are used 
\begin{align}
V_\mu &= \Sigma \left( D_\mu \Sigma \right)^\dagger = - \left( D_\mu
\Sigma \right) \Sigma^\dagger \,, \nonumber \\
D_\mu\Sigma &= \partial_\mu \Sigma + i g \frac{\sigma^a}2 W^a_\mu \Sigma
- i g' \Sigma B_\mu \frac{\sigma^3}2 \,, \nonumber \\
\Sigma &= \exp\left(-\frac{i}{v} \sigma^a w^a \right) 
\stackrel{\text{unitary gauge}}{=} 1 \,.
\end{align}
The fields $w^a$ denote the triplet of Goldstone bosons, which vanish
in the unitary gauge we are going to employ, and $\sigma^a$ are the
Pauli matrices as before.

Inserting all definitions one obtains
\begin{equation}
\Tr[V_\mu V_\nu] = -\frac{g^2}2 \left( W^+_\mu
W^-_\nu + W^-_\mu W^+_\nu \right) -\frac{g^2}{2c_w^2}
Z_\mu Z_\nu
\end{equation}
and hence
\begin{align}
\La_4 = \alpha_4 \Bigl(
&\quad 8 \frac{M_W^4}{v^4} \left( W^- {\cdot} W^- \ W^+ {\cdot} W^+ + W^- {\cdot} W^+ \ W^- {\cdot} W^+ \right) \nonumber\\
&
+ 16 \frac{M_W^2 M_Z^2}{v^4} \ W^- {\cdot} Z \ W^+ {\cdot} Z \nonumber\\
&
+ 4 \frac{M_Z^4}{v^4} \ Z {\cdot} Z \ Z {\cdot} Z \Bigr) \,,
\nonumber\\
\La_5 = \alpha_5 \Bigl(
&\quad 16 \frac{M_W^4}{v^4} W^- {\cdot} W^+ \ W^- {\cdot} W^+ \nonumber\\
&
+ 16 \frac{M_W^2 M_Z^2}{v^4} \ W^- {\cdot} W^+ \ Z {\cdot} Z \nonumber\\
&
+ 4 \frac{M_Z^4}{v^4} \ Z {\cdot} Z \ Z {\cdot} Z \Bigr) \,.
\label{eq:L4L5}
\end{align}

With the discovery of a light Higgs-like state at the
LHC~\cite{Aad:2012tfa,Chatrchyan:2012xdj,ATLAS-CONF-2015-044}, this
should also be reflected in the operators of the electroweak chiral
Lagrangian. Therefore, in Ref.~\cite{Kilian:2014zja}, a new set has been
defined in analogy to the dimension-8 operators given in
\eq{eq:obsd8s}
\begin{align}
\La_{S,0} &= F_{S,0} \Tr\left[ (D_\mu \hat{H})^\dagger D_\nu \hat{H} \right] \times 
      \Tr\left[ ( D^\mu \hat{H})^\dagger D^\nu \hat{H} \right] \,, \nonumber\\
\La_{S,1} &= F_{S,1} \Tr\left[ ( D_\mu \hat{H} )^\dagger D^\mu \hat{H}  \right] \times 
      \Tr\left[ ( D_\nu \hat{H} )^\dagger D^\nu \hat{H} \right] \,.
\end{align}
Thereby, $\hat{H}$ is a $2\times2$ Hermitian matrix defined as
\begin{equation}
\hat{H} = \frac12 
\begin{pmatrix}
v+H-iw^3 & -i (w^1-iw^2) \\
-i (w^1+iw^2) & v+H+iw^3
\end{pmatrix}
\stackrel{\text{unitary gauge}}{=} \frac{v+H}2
\begin{pmatrix}
1 & 0 \\
0 & 1
\end{pmatrix} \,,
\end{equation}
where $H$ is the physical Higgs boson and $w^i$ are again the Goldstone
bosons. The covariant derivative acting on $\hat{H}$ is given by
\begin{equation}
D_\mu \hat{H} = \partial_\mu \hat{H} - i g \frac{\sigma^a}2 W^a_\mu
\hat{H} + i g' B_\mu \frac{\sigma^3}2 \hat{H} \,.
\end{equation}
The coefficients $F_{S,0}$ and $F_{S,1}$ are dimensionful with a mass
dimension of $-4$.

Inserting the definitions yields
\begin{align}
\Tr[(D_\mu \hat{H})^\dagger D_\nu \hat{H}]
&= \frac12 \left( \partial_\mu H \right) \left( \partial_\nu H \right) + \frac{g^2 v^2}8 \left( W^+_\mu W^-_\nu + W^-_\mu W^+_\nu \right) \left(1+\frac{H}{v}\right)^2 \nonumber\\
&\quad+\frac{g^2 v^2}{8c_w^2} Z_\mu Z_\nu \left(1+\frac{H}{v}\right)^2 \nonumber\\
&= \frac12 \left( \partial_\mu H \right) \left( \partial_\nu H \right)
+ \frac{M_W^2}2 \left( W^+_\mu W^-_\nu + W^-_\mu W^+_\nu \right) \left(1+\frac{H}{v}\right)^2 \nonumber\\
&\quad+ \frac{M_Z^2}2 Z_\mu Z_\nu \left(1+\frac{H}{v}\right)^2 \,,
\label{eq:trdmuhdnuh}
\end{align}
which is similar to the one obtained in \eq{eq:dmuphidnuphi}, but contains an
extra symmetrization of the Lorentz indices. This becomes visible when looking
at the $W$ term and also the mixed $Z (\partial H)$ terms vanish for this
reason.

For the quartic gauge boson vertices, we get
\begin{align}
\La_{S,0}\Bigr|_{\text{4 gauge bosons}} = F_{S,0} \Bigl(
&\quad \frac{M_W^4}2 \left( W^- {\cdot} W^- \ W^+ {\cdot} W^+ + W^- {\cdot} W^+ \ W^- {\cdot} W^+ \right) \nonumber\\
&
+ M_W^2 M_Z^2 \ W^- {\cdot} Z \ W^+ {\cdot} Z \nonumber\\
&
+ \frac{M_Z^4}4 \ Z {\cdot} Z \ Z {\cdot} Z \Bigr) \,,
\nonumber\\
\La_{S,1}\Bigr|_{\text{4 gauge bosons}} = F_{S,1} \Bigl(
&\quad M_W^4 W^- {\cdot} W^+ \ W^- {\cdot} W^+ \nonumber\\
&
+ M_W^2 M_Z^2 \ W^- {\cdot} W^+ \ Z {\cdot} Z \nonumber\\
&
+ \frac{M_Z^4}4 \ Z {\cdot} Z \ Z {\cdot} Z \Bigr) \,.
\label{eq:LS0S1}
\end{align}
A comparison of \eq{eq:L4L5} and~\eq{eq:LS0S1} shows that the functional
form of the two sets is the same when considering quartic gauge-boson
vertices only. This allows to define relations between the operator
coefficients which lead to the same theoretical predictions. These
connections are discussed in the following subsection.

\subsubsection{Relations between Definitions}

As mentioned above, two representations exist to define the operators of
an EFT expansion, and within these approaches again different choices of
equivalent operators are possible. The question is then what the
relations between the different sets are, and which of them are
equivalent. Equivalence hereby means that when setting the operator
coefficients according to the relation, all physics observables produce
the same results at LO for each set. This is true for example if the
sets lead to the same Feynman rules for all vertices deviating from
their SM values.

For practical applications often a weaker condition is sufficient,
namely that the two operator sets agree for a certain subset of Feynman
rules. An example would be experimental studies of a specific process.
To compare the experimental measurements with theory predictions, Monte
Carlo events with anomalous couplings switched on need to be generated.
This can be a rather time-consuming processes, depending on the detail
level of the event simulation. If one wants to quote the results not
only for the operator set with which the study has performed, but also
other parametrizations, it is helpful if the event simulation does not
have to be repeated for the new set. Instead one can use relations,
which allow to simply re-interpret the already derived results. For such
a task it is sufficient if the relations only hold for those vertices
which appear in the Feynman diagrams of the considered process. Other
vertices are allowed to have different conversion rules without spoiling
the results.

In the following, if no specific vertices or vertex classes are
indicated, the relations hold for all and the corresponding operator
sets are equivalent, otherwise this is only true for the quoted subset.

For the dimension-6 operators in \eqs{eq:d6cpeven,eq:d6cpodd}, an
alternative parametrization commonly used for LEP results
exists~\cite{Hagiwara:1986vm}. The modified terms in the Lagrangian take
the form
\begin{align}
\La_{\text{LEP}} &= \sum_{V=\gamma,Z} -i g_{WWV} 
  \left( 
   g_1^V ( W_{\mu\nu}^+W^{-\mu} - W^{+\mu}W_{\mu\nu}^-) V^\nu
   +\kappa_V W_\mu^+W_\nu^-V^{\mu\nu}
  \right.\nonumber\\
&\quad\left.
  +\frac{\lambda_V}{M_W^2} W_\mu^{\nu+}W_\nu^{-\rho}V_\rho^{\mu}
  +ig_4^V W_\mu^+W^-_\nu(\partial^\mu V^\nu+\partial^\nu V^\mu)
  \right.\nonumber\\
&\quad\left.
  -ig_5^V\epsilon^{\mu\nu\rho\sigma}(W_\mu^+\partial_\rho W^-_\nu-\partial_\rho W_\mu^+W^-_\nu)V_\sigma
  +\tilde{\kappa}_VW_\mu^+W_\nu^-\tilde{V}^{\mu\nu}
  \right.\nonumber\\
&\quad\left.
  +\frac{\tilde{\lambda}_V}{M_W^2}W_\mu^{\nu+}W_\nu^{-\rho}\tilde{V}_\rho^{\mu}
\right) \,,
\label{eq:LanomLEP}
\end{align}
where $V_{\mu\nu} = \partial_\mu V_\nu - \partial_\nu V_\mu$ for $V \in
W^\pm$, $Z$, $\gamma$ and the coupling constants are $g_{WW\gamma}=-e$
and $g_{WWZ}=-e\frac{c_w}{s_w}$.
This parametrization is expressed in terms of the fields after
electroweak symmetry breaking and in general does not conserve $SU(2)_L$
symmetry. This choice is motivated by the fact that the maximal LEP energy
of about $200 \GeV$ is still below the electroweak symmetry breaking
scale, while at the LHC, energies beyond this scale are probed.
Electromagnetic gauge invariance imposes $g_1^\gamma =1$ and
$g_4^\gamma=g_5^\gamma = 0$, so there are five independent operators
which conserve both $C$ and $P$, $g_1^Z, \kappa_\gamma, \kappa_Z,
\lambda_\gamma, \lambda_Z$, and six which violate $C$ and/or $P$,
$g_4^Z, g_5^Z, \tilde{\kappa}_\gamma, \tilde{\kappa}_Z,
\tilde{\lambda}_\gamma, \tilde{\lambda}_Z$.

These parameters can be related to the dimension-6 operators for the
anomalous triple gauge couplings,
obtaining~\cite{Hagiwara:1993ck,Wudka:1994ny,Degrande:2013rea}
\begin{align}
g_1^Z & = 1+f_W\frac{M_Z^2}{2\Lambda^2} \,, \nonumber\\
\kappa_\gamma & = 1+(f_W+f_B)\frac{M_W^2}{2\Lambda^2} \,, \nonumber\\
\kappa_Z & =
1+(f_W-f_B\frac{s_w^2}{c_w^2})\frac{M_W^2}{2\Lambda^2} \,, \nonumber\\
\lambda_\gamma & = \lambda_Z =
f_{WWW}\frac{3g^2M_W^2}{2\Lambda^2} \,, \nonumber\\
g_4^Z &= g_5^Z=0 \,, \nonumber\\
\tilde{\lambda}_\gamma & = \tilde{\lambda}_Z =
f_{\tilde{W}WW}\frac{3g^2M_W^2}{2\Lambda^2} \,, \nonumber\\
\tilde{\kappa}_\gamma & = 
f_{\tilde{W}}\frac{M_W^2}{2\Lambda^2} \,, \nonumber\\
\tilde{\kappa}_Z & = 
-f_{\tilde{W}}\frac{s_w^2}{c_w^2}\frac{M_W^2}{2\Lambda^2} \,.
\end{align}
From the first three lines, one can directly deduce the relation~\cite{Hagiwara:1993ck}
\begin{equation}
\Delta g_1^Z=\Delta \kappa_Z + \frac{s_w^2}{c_w^2} \Delta \kappa_\gamma
\end{equation}
with $\Delta g_1^Z = g_1^Z - 1$, $\Delta \kappa_{\gamma,Z} =
\kappa_{\gamma,Z} - 1$, and from the last two lines
\begin{equation}
\tilde{\kappa}_Z = - \frac{s_w^2}{c_w^2} \tilde{\kappa}_\gamma \,.
\end{equation}
The Lagrangian of \eq{eq:LanomLEP} does not contain any quartic gauge
boson vertices in contrast to the dimension-6 operators, so the
relations above do not extend beyond triple gauge couplings. Also this
Lagrangian is therefore not invariant under $SU(2)_L$ gauge
transformations.

For anomalous couplings of the two quartic vertices $W^+W^-\gamma\gamma$
and $ZZ\gamma\gamma$ two further operators have been defined for
LEP~\cite{Amsler:2008zzb,Stirling:1999ek}
\begin{align}
\La_0 &= -\frac{e^2}{16\Lambda^2} a_0 F_{\mu\nu} F^{\mu\nu}
W^{a,\alpha} W^a_{\alpha} \,, \nonumber\\
\La_c &= -\frac{e^2}{16\Lambda^2} a_c F_{\mu\alpha} F^{\mu\beta}
W^{a,\alpha} W^a_{\beta} \,,
\end{align}
with 
\begin{align}
F^{\mu\nu} &= \partial^\mu A^\nu - \partial^\nu A^\mu \,, \nonumber\\
W^a_\mu &= \left( \frac1{\sqrt{2}} (W^+_\mu + W^-_\mu),
\frac{i}{\sqrt{2}} (W^+_\mu - W^-_\mu), \frac{Z_\mu}{c_w} \right) \,,
\end{align}
where $A_\mu$ denotes the photon field.
Vertex-specific relations only can be derived relating them to the mixed
dimension-8 operators~\cite{Degrande:2013rea}
\begin{align}
\frac{a_0}{\Lambda^2} &= g^2 v^2 \left( 
\frac{f_{M,0}}{\Lambda^4} +\frac12 \frac{f_{M,2}}{\Lambda^4}
\pm \frac{f_{M,4}}{\Lambda^4} 
\right) \,, \nonumber\\
\frac{a_c}{\Lambda^2} &= g^2 v^2 \left( 
- \frac{f_{M,1}}{\Lambda^4} -\frac12 \frac{f_{M,3}}{\Lambda^4}
\pm \frac12 \frac{f_{M,5}}{\Lambda^4} +\frac12 \frac{f_{M,7}}{\Lambda^4}
\right) \,,
\end{align}
where the upper sign holds for the $W^+W^-\gamma\gamma$ and the lower
sign for the $ZZ\gamma\gamma$ vertex. As in total eight dimension-8
operators contribute to the two vertices, only certain linear
combinations are related to $a_0$ or $a_c$ and can possibly be
determined from a measurement of those.

As last item we will compare the anomalous contributions to the quartic
gauge-boson vertices, which due to 
historical reasons has lead to some confusion. The situation is
complicated by the fact that there are two different conventions
regarding the exact definition of the modified field-strength tensors
$\widehat{W}^{\mu\nu}$ and $\widehat{B}^{\mu\nu}$ for the dimension-8
operators.
In this article, we follow the same definition,
\eq{eq:modfieldstrength}, which is commonly used for the dimension-6
operators, namely that they are multiplied with the respective coupling
strength $g$ and $g'$. This convention has also been used in
Ref.~\cite{Degrande:2013rea} and the \VBFNLO implementation. 
Refs.~\cite{Eboli:2006wa,Eboli:2016kko}, where the dimension-8 operators have first
been defined, instead uses 
\begin{align}
\widehat{W}^{\text{EGM},\mu\nu} &= \frac{\sigma^j}2 W^{j,\mu\nu} 
 = \frac{\sigma^j}2 \left( 
 \partial^\mu W^{j,\nu} - \partial^\nu W^{j,\mu} - g \epsilon^{jkl}
W^{k,\mu} W^{l,\nu} \right) \,, \nonumber\\
\widehat{B}^{\text{EGM},\mu\nu} &= B^{\mu\nu} 
 = \left( \partial^\mu B^{\nu} - \partial^\nu B^{\mu} \right) \,,
\end{align}
such that
\begin{align}
\widehat{W}^{\text{EGM},\mu\nu} &= \frac1{ig} \widehat{W}^{\mu\nu} \,,
&
\widehat{B}^{\text{EGM},\mu\nu} &= \frac2{ig'} \widehat{B}^{\mu\nu} \,,
\end{align}
where here and in the following all variables following this definition
are marked with the superscript $\text{EGM}$. This convention is
employed in the FeynRules~\cite{Christensen:2008py,Alloul:2013bka} model
file~\cite{FeynRulesModelD8}, which in turn is the common implementation
for use in \program{MadGraph5\_aMC@NLO}~\cite{Alwall:2014hca}.

This change in the modified field-strength tensors then leads to similar
changes for the coefficients of the tensor and mixed operators,
\eqs{eq:obsd8t,eq:obsd8m}, leading to the
relations~\cite{Degrande:2013rea},
\begin{alignat}{2}
 f_{M,0,1}   &= - \frac{1}{g^2}      & \cdot & f_{M,0,1}^\text{EGM} \,, \nonumber\\
 f_{M,2,3}   &= - \frac{4}{g'^2}     & \cdot & f_{M,2,3}^\text{EGM} \,, \nonumber\\
 f_{M,4,5}   &= - \frac{2}{g g'}     & \cdot & f_{M,4,5}^\text{EGM} \,, \nonumber\\
 f_{M,7}     &= - \frac{1}{g^2}      & \cdot & f_{M,7}^\text{EGM}   \,, \nonumber\\
 f_{T,0,1,2} &=   \frac{1}{g^4}      & \cdot & f_{T,0,1,2}^\text{EGM} \,, \nonumber\\
 f_{T,5,6,7} &=   \frac{4}{g^2 g'^2} & \cdot & f_{T,5,6,7}^\text{EGM} \,, \nonumber\\
 f_{T,8,9}   &=   \frac{16}{g'^4}    & \cdot & f_{T,8,9}^\text{EGM} \,,
\end{alignat}
while the scalar operators, \eq{eq:obsd8s}, stay identical.

As can be directly seen from comparing \eq{eq:S0vsS2} with
\eqs{eq:L4L5,eq:LS0S1}, there are also relations between the scalar
dimension-8 operators and the operators from the electroweak chiral
Lagrangian, the latter both without and with a Higgs boson.

For the chiral Lagrangian with Higgs boson, there is actually a full
equivalence to the scalar dimension-8 operators. This can be seen most
easily by comparing \eqs{eq:dmuphidnuphi,eq:trdmuhdnuh}, from which
follows
\begin{equation}
\Tr[(D_\mu \hat{H})^\dagger D_\nu \hat{H}] = \frac12 \Bigl(
\left[ (D_\mu \Phi)^\dagger D_\nu \Phi \right] 
+ 
\left[ (D_\nu \Phi)^\dagger D_\mu \Phi \right] 
\Bigr) \,.
\end{equation}
Hence for the operator coefficients we obtain
\begin{align}
F_{S,0} &= \frac{f_{S,0}+f_{S,2}}{\Lambda^4} \,, & 
f_{S,0} &= f_{S,2} \,, \nonumber\\
F_{S,1} &= \frac{f_{S,1}}{\Lambda^4} \,.
\end{align}

For the non-linear formulation without Higgs boson, a relation can
obviously be valid only for the quartic gauge-boson couplings, as no
vertices with Higgs bosons are modified by $\La_4$ and $\La_5$, which
are however present in the dimension-8 operators. In this case we find
\begin{align}
\alpha_4 &= \frac{v^4}{16} F_{S,0} = \frac{v^4}{16} \frac{f_{S,0}+f_{S,2}}{\Lambda^4} \,, 
& f_{S,0} &= f_{S,2} \,, \nonumber\\
\alpha_5 &= \frac{v^4}{16} F_{S,1} = \frac{v^4}{16} \frac{f_{S,1}}{\Lambda^4} \,,
\label{eq:linvsnonlin}
\end{align}
where simultaneously also the correspondence to the with-Higgs case is
shown. Parameter scenarios with $f_{S,0} \ne f_{S,2}$ cannot be modeled
in either chiral Lagrangian with the set of operators given so far. For
this the inclusion of an additional operator~\cite{Longhitano:1980tm}
\begin{equation}
\La_6 = \alpha_6 \Tr\left[ V_\mu V_\nu \right] 
\Tr\left[ T V^\mu \right] \Tr\left[ T V^\nu \right]
\end{equation}
with $T=\Sigma \sigma_3 \Sigma^\dagger$ is necessary.

From these relations we see the importance of the new operator
$\obs_{S,2}$, which allows us to write the relations in the same way for
all three quartic vertices with massive gauge bosons. If, as in the
original set, this operator is not present, only vertex-specific
conversion rules exist. This possibility arises only due to the fact
that each of the three vertices contains at least two identical
particles. The resulting symmetrization when generating Feynman rules
from the Lagrangian means that for each of the $WWWW$ and $WWZZ$
vertices only two different Lorentz structures exist instead of the
naively expected three. Therefore, two operators with different Lorentz
structures are sufficient to generate all possibilities when considering
a single vertex only. For the $ZZZZ$ vertex only one Lorentz structure
exists for the scalar operators as all particles are identical. 
The vertex-specific relations without $\obs_{S,2}$
read~\cite{Degrande:2013rea}
\begin{align}
&\text{$WWWW$-vertex:\quad} & 
  \alpha_4 &= \frac{v^4}{8} \frac{f_{S,0}}{\Lambda^4} \,, &
  \alpha_4 + 2\alpha_5 &= \frac{v^4}{8} \frac{f_{S,1}}{\Lambda^4} \,,
  \nonumber\\
&\text{$WWZZ$-vertex:\quad} & 
  \alpha_4 &= \frac{v^4}{16} \frac{f_{S,0}}{\Lambda^4} \,, &
  \alpha_5 &= \frac{v^4}{16} \frac{f_{S,1}}{\Lambda^4} \,,
  \nonumber\\
&\text{$ZZZZ$-vertex:\quad} & 
  \alpha_4&+\alpha_5 = \frac{v^4}{16}
    \frac{f_{S,0}+f_{S,1}}{\Lambda^4}\,.
  \hspace*{-10em} 
\label{eq:linvsnonlinspec}
\end{align}
These relations should however be considered a kludge only, and if
possible the non-vertex-specific ones in \eq{eq:linvsnonlin} including
the operator $\obs_{S,2}$ used instead. For some processes this is even
mandatory. An example is the VBF-production process of $pp\rightarrow
W^+W^-jj$. For certain parton combinations like $ud \rightarrow ud
W^+W^-$ both $WWWW$- and $WWZZ$-vertices appear in the same subprocess.
In that case, the two operator sets can be related only according to
\eq{eq:linvsnonlin}.

\subsection{Unitarity}
\label{sec:anomunitarity}

\begin{figure}
\begin{center}
\includegraphics[width=0.7\textwidth]{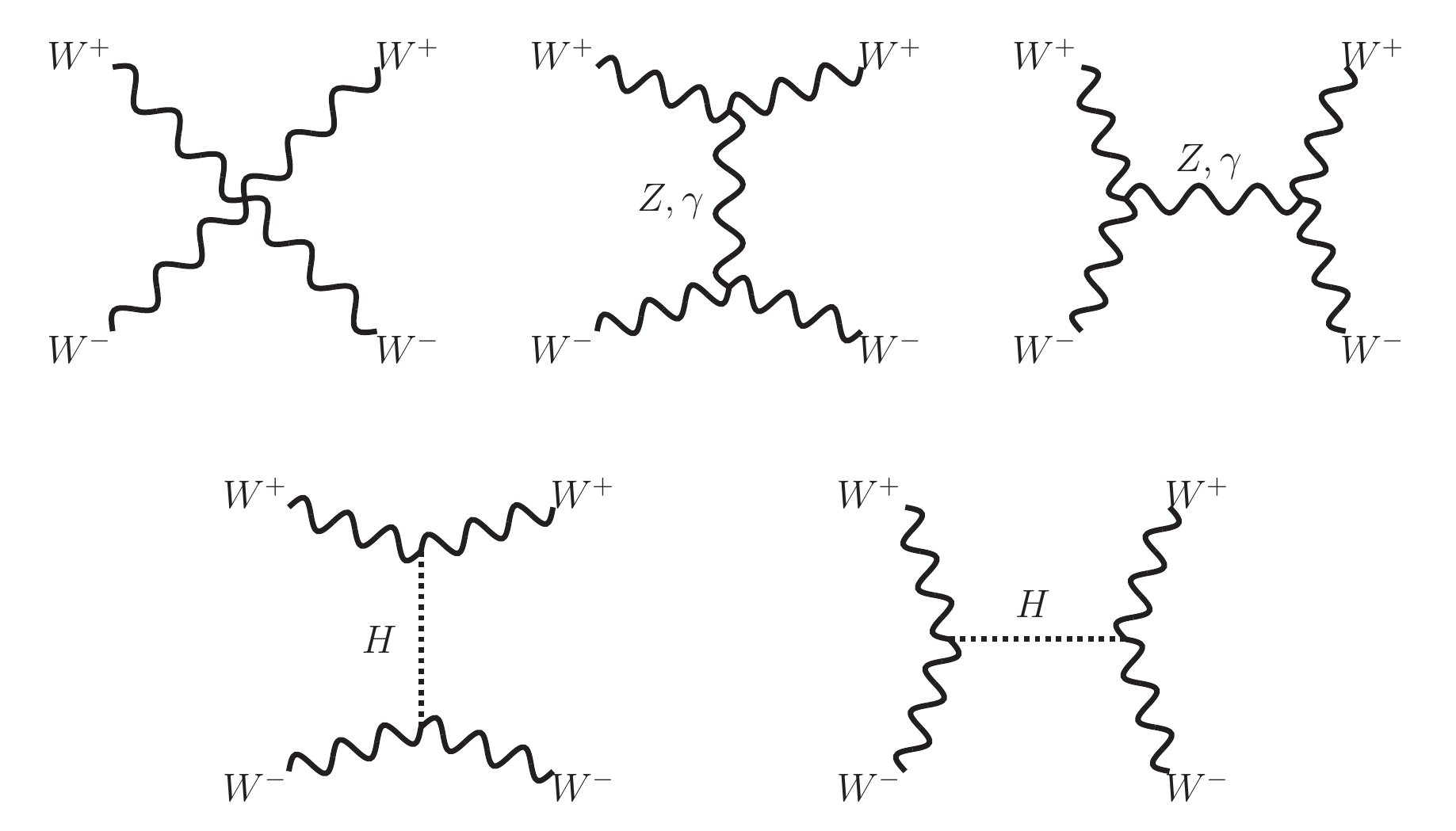}
\end{center}
\caption{Feynman diagrams contributing to $WW$ scattering.
}
\label{fig:fm_wwscatter}
\end{figure}
The high-energy behavior of differential cross sections in vector-boson
scattering is the result of a delicate cancellation between the different
diagrams. For simplicity, let us consider only $2\rightarrow2$
scattering of two on-shell longitudinal $W$ bosons, $W^+_L W^-_L
\rightarrow W^+_L W^-_L$. The contributing Feynman diagrams are
depicted in \fig{fig:fm_wwscatter}.
In the high-energy region, where $E\gg M_W$, the
longitudinal polarization vector of each $W$ can be approximated by
\begin{equation}
\epsilon_L^\mu = \frac{p^\mu}{M_W} + \Order{\frac{M_W}{E}} \,.
\end{equation}
Considering the diagram with the quartic gauge-boson vertex only, its
matrix element rises with $\frac{s^2}{M_W^4}$. The same behavior happens
also for the two diagrams with a $Z$/$\gamma$ $s$- and $t$-channel exchange,
with the same factor with an opposite sign such that the two terms
cancel and the remaining dependence is $\frac{s}{M_W^2}$. This in turn
is canceled by the two Higgs exchange diagrams depicted in the lower row
of \fig{fig:fm_wwscatter}. Thus the leading energy behavior of the full
matrix element is constant in energy~\cite{Lee:1977eg}.

Switching on anomalous couplings, this cancellation is in general
violated. For example, the dimension-8 operators discussed before can
modify the quartic vertex only, while leaving all others at their
respective SM values. The corresponding rise in the amplitude at high
energies means that at large enough values, unitarity of the scattering
matrix gets violated. Then the probability interpretation does not hold
any longer, as scattering must happen more often than the actual flux of
incoming particles allows. This is of course an unphysical situation and
needs to be avoided. The underlying reason is the breakdown of the
expansion in \eq{eq:eftmaster}, where neglected higher-order terms are
not sufficiently suppressed any longer and need to be taken into account
as well.

\subsubsection{Unitarity Bounds}

First, we need to determine where unitarity becomes violated. A
convenient tool for this task is partial wave analysis~\cite{Jacob:1959at}. 
The starting point is unitarity of the $S$ matrix
\begin{align}
\Id &= S^\dagger S = \left(\Id+iT\right)^\dagger \left(\Id+iT\right) \nonumber\\
    &= \Id + T^\dagger T + i (T-T^\dagger) \nonumber\\
\Rightarrow -i (T-T^\dagger) &= T^\dagger T 
\label{eq:Sunitarity}
\end{align}
with the matrix $T=\frac1{i}(S-\Id)$. This expression is now
placed between two two-body states with momenta $p_1$, $p_2$ and
helicities $\lambda_1$, $\lambda_2$, and $k_1$, $k_2$, $\kappa_1$,
$\kappa_2$, respectively,
\begin{multline}
-i \bigl(
\bra{k_1 k_2 \kappa_1 \kappa_2} (T-T^\dagger) \ket{p_1 p_2 \lambda_1 \lambda_2}
\bigr) \\
= \sum_{n} \bra{k_1 k_2 \kappa_1 \kappa_2}
T^\dagger \ket{n} \bra{n} T \ket{p_1 p_2 \lambda_1 \lambda_2}  
\,,
\label{eq:Sunitstate}
\end{multline}
where on the right-hand side we have inserted a full set of states $\Id
= \sum_{n} \ket{n}\bra{n}$.

For simplicity, we choose a specific Lorentz frame and take the two
incoming momenta aligned along the $z$ axis in their center-of-mass
frame, $\vec{p}_1 = - \vec{p}_2$, so $T$ depends only on the
center-of-mass energy $E$.  Then for a $2\rightarrow2$ scattering
process $AB\rightarrow CD$, one obtains~\cite{Jacob:1959at}
\begin{multline}
\bra{p_3 p_4 \lambda_3 \lambda_4} T \ket{p_1 p_2 \lambda_1 \lambda_2} \\
\begin{aligned}
&= (2\pi)^4 \delta^{(4)}(p_1+p_2-p_3-p_4) 
\bra{\theta\phi \lambda_3 \lambda_4} T(E) \ket{0 0 \lambda_1 \lambda_2} \\
&= (2\pi)^4 \delta^{(4)}(p_1+p_2-p_3-p_4) \\
&\qquad \cdot
\M\bigl(A(p_1,\lambda_1),B(p_2,\lambda_2)\rightarrow
C(p_3,\lambda_3),D(p_4,\lambda_4)\bigr) \,,
\end{aligned}
\end{multline}
where the momenta have been replaced by the scattering angles $\theta$
and $\phi$, with the normalization 
\begin{equation}
\braket{p_1' p_2' \lambda_1' \lambda_2'}{p_1 p_2 \lambda_1 \lambda_2} 
= 2 E_1 2 E_2 (2\pi)^6
\delta^{(3)}(\vec{p}_1{}'-\vec{p}_1)\delta^{(3)}(\vec{p}_2{}'-\vec{p}_2)\delta_{\lambda_1'\lambda_1}\delta_{\lambda_2'\lambda_2}
\,.
\end{equation}
$\M$ denotes the usual matrix element of the $2\rightarrow2$ scattering
process derived from Feynman rules.

The matrix element can now be decomposed into partial waves
\begin{multline}
\bra{\theta\phi \lambda_3 \lambda_4} T(E) \ket{0 0 \lambda_1 \lambda_2}
\\
= 16\pi \sum_J (2J+1) e^{i(\lambda_{12}-\lambda_{34})\phi}
d^J_{\lambda_{12}\lambda_{34}}(\theta) \
\bra{\lambda_3 \lambda_4} T^J(E) \ket{\lambda_1 \lambda_2}
\label{eq:partwavedecomp}
\end{multline}
with $\lambda_{12}=\lambda_1-\lambda_2$, $\lambda_{34}=\lambda_3-\lambda_4$ and the
Wigner $d$-functions $d^J_{\lambda\kappa}(\theta)$ (see \eg
Ref.~\cite{Agashe:2014kda} for tabulated values). For any initial or
final state with identical particles an additional factor of $\sqrt{2}$
must be inserted on the right-hand side of \eq{eq:partwavedecomp} to
account for the correct normalization of the wave function.

The Wigner $d$-functions obey a completeness relation
\begin{equation}
\int_{-1}^1 \di \cos\theta \ d^J_{\lambda\kappa}(\theta)
d^{J'}_{\lambda\kappa}(\theta)
= \frac2{2J+1} \delta_{JJ'} \,,
\label{eq:wignerorth}
\end{equation}
which will be exploited in the next step.

Multiplying \eq{eq:partwavedecomp} with $\frac1{32\pi} \int_{-1}^1 \di \cos\theta
\ d^{J'}_{\lambda_{12}\lambda_{34}}(\theta)$, we can project out a specific partial wave
\begin{align}
a^J_{\lambda_{12}\lambda_{34}} &\equiv e^{i(\lambda_{12}-\lambda_{34})\phi} \bra{\lambda_3 \lambda_4}
T^J(E) \ket{\lambda_1 \lambda_2} \nonumber\\
&= \frac1{32\pi} \int_{-1}^1 \di \cos\theta \ d^{J}_{\lambda_{12}\lambda_{34}}(\theta)
\bra{\theta\phi \lambda_3 \lambda_4} T(E) \ket{0 0 \lambda_1 \lambda_2} 
\nonumber\\
&= \frac1{32\pi} \int_{-1}^1 \di \cos\theta \ d^{J}_{\lambda_{12}\lambda_{34}}(\theta)
\M\bigl(A(p_1,\lambda_1),B(p_2,\lambda_2) \rightarrow
C(p_3,\lambda_3),D(p_4,\lambda_4)\bigr) \,.
\label{eq:partwavedecompr}
\end{align}
The factor for identical particles in the initial or final
state becomes $\frac1{\sqrt{2}}$ on the right-hand side for each pair.

Going back to \eq{eq:Sunitstate}, we now apply this to the special case
of identical initial and final states, which yields
\begin{multline}
-i \bigl(
\bra{p_1 p_2 \lambda_1 \lambda_2} (T-T^\dagger) \ket{p_1 p_2 \lambda_1 \lambda_2}
\bigr) \\
\begin{aligned}
&= \sum_{n} \bra{p_1 p_2 \lambda_1 \lambda_2}
T^\dagger \ket{n} \bra{n} T \ket{p_1 p_2 \lambda_1 \lambda_2} \\
&= \sum_{n} \Bigl| \bra{n} T \ket{p_1 p_2 \lambda_1 \lambda_2} \Bigr|^2
\,.
\end{aligned}
\end{multline}

Furthermore, we restrict ourselves to two-particle states $\ket{n_2}$ in
the sum over states, and write down the helicity sum and the momentum
integration explicitly. As non-negative terms on the right-hand side
are dropped, the equation becomes an inequality. Leaving out the global
factor for energy-momentum conservation on both sides, one gets
\begin{multline}
2 \Im \Bigl( \M\bigl(A(p_1,\lambda_1),B(p_2,\lambda_2) \rightarrow
A(p_1,\lambda_1),B(p_2,\lambda_2)\bigr) \Bigr)\\
\begin{aligned}
&\ge \sum_{CD} \sum_{\lambda_3,\lambda_4} \int \frac{\di^3 p_3}{(2\pi)^3 2 E_3}
\frac{\di^3 p_4}{(2\pi)^3 2 E_4} 
(2\pi)^4 \delta^{(4)}(p_1+p_2-p_3-p_4) \\
&\qquad \cdot \M^{*}\bigl(A(p_1,\lambda_1),B(p_2,\lambda_2) \rightarrow
C(p_3,\lambda_3),D(p_4,\lambda_4)\bigr) \\
&\qquad \cdot \M    \bigl(A(p_1,\lambda_1),B(p_2,\lambda_2) \rightarrow
C(p_3,\lambda_3),D(p_4,\lambda_4)\bigr) \,. 
\end{aligned}
\label{eq:unitcondsame}
\end{multline}

We then obtain after inserting the partial-wave expansion
\begin{multline}
16\pi \sum_J (2J+1)\, d^J_{\lambda_{12}\lambda_{12}}(0) \, 2\, \Im\bigl(
a^J_{\lambda_{12}\lambda_{12}}(AB\rightarrow AB)\bigr) \\
\begin{aligned}
&\ge \sum_{CD} \sum_{\lambda_3,\lambda_4} \int\di\Omega\ \frac1{16\pi^2}
\frac{\abs{\vec{p}_3}}{E} \left(16\pi\right)^2\, \sum_{J',J} (2J'+1)(2J+1) \\
&\qquad \cdot
\bigl(d^{J'}_{\lambda_{12}\lambda_{34}}
a^{J'}_{\lambda_{12}\lambda_{34}}(AB\rightarrow CD)\bigr)^{*} \ 
d^J_{\lambda_{12}\lambda_{34}}
a^J_{\lambda_{12}\lambda_{34}}(AB\rightarrow CD) \\[1ex]
&\ge 16\pi \sum_J (2J+1) \sum_{CD}\sum_{\lambda_3,\lambda_4} \sum_{J'} (2J'+1) \\
&\qquad\cdot
a^{J'*}_{\lambda_{12}\lambda_{34}}(AB\rightarrow CD) 
a^J_{\lambda_{12}\lambda_{34}}(AB\rightarrow CD)
\int_{-1}^1\di\cos\theta\ d^{J'}_{\lambda_{12}\lambda_{34}}
d^{J}_{\lambda_{12}\lambda_{34}} \\
&\ge 16\pi \sum_J (2J+1)\, 2\sum_{CD}\sum_{\lambda_3,\lambda_4}
\abs{a^J_{\lambda_{12}\lambda_{34}}(AB\rightarrow CD)}^2 \,.
\end{aligned}
\end{multline}
Thereby, we have simplified the phase-space integral by applying
$\frac{\abs{\vec{p}_3}}{E} \le \frac12$, used that the Wigner
$d$-functions are real, and exploited the orthogonality condition given
in \eq{eq:wignerorth}. As angular momentum is conserved, the above
equation must hold for each $J$ separately. The sum on the right-hand
side can be split into a part with same initial and final state and same
helicity difference, $AB=CD$ and
$\lambda_3-\lambda_4=\lambda_{12}\equiv\lambda$, a sum over same states
but different helicities, and a sum over different states. The first
part might be associated with a multiplicity factor, depending on how
often such a combination appears in the sum.  This factor can simply be
dropped without affecting the inequality.  Noting that
$d^J_{\lambda\lambda}(0)=1$, one then obtains for the partial-wave
coefficients
\begin{gather}
\begin{aligned}
\Bigl(\Re\,a^J_{\lambda\lambda}(AB\rightarrow AB)\Bigr)^2 
+ \Bigl(\Im\,a^J_{\lambda\lambda}(AB\rightarrow AB)\Bigr)^2
- \Im\,a^J_{\lambda\lambda}(AB\rightarrow AB)& \\
\quad + \sum_{\kappa\ne\lambda}
  \abs{a^J_{\lambda\kappa}(AB\rightarrow AB)}^2
+ \sum_{CD \ne AB}\sum_{\kappa}
  \abs{a^J_{\lambda\kappa}(AB\rightarrow CD)}^2
&\le 0 \,, 
\end{aligned}\nonumber\\[1ex]
\begin{aligned}
\Bigl(\Re\,a^J_{\lambda\lambda}(AB\rightarrow AB)\Bigr)^2 
+ \Bigl(\Im\,a^J_{\lambda\lambda}(AB\rightarrow AB) - \frac12
\Bigr)^2& \\
\quad + \sum_{\kappa\ne\lambda}
  \abs{a^J_{\lambda\kappa}(AB\rightarrow AB)}^2
+ \sum_{CD \ne AB}\sum_{\kappa}
  \abs{a^J_{\lambda\kappa}(AB\rightarrow CD)}^2
&\le \Bigl(\frac12\Bigr)^2 \,.
\end{aligned}
\label{eq:unitcircle}
\end{gather}
From the last line it becomes clear that for
$a^J_{\lambda\lambda}(AB\rightarrow AB)$, with all others set to zero,
this inequality describes a circle around $(0,\frac12)$ with radius
$\frac12$, the so-called Argand circle. 
For unitarity considerations based on tree-level matrix elements, which
will be used in the following, this condition is slightly too
restrictive. Here we assume that small imaginary parts will be
generated by higher-order corrections, and the inequality above is
applied to the real part of the partial-wave coefficient only.
For all other partial waves the inequality restricts them to be inside a
circle with radius $\frac12$ around the origin.
Therefore, the unitarity condition for each partial wave of a
$2\rightarrow2$ matrix element reads 
\begin{equation} 
\abs{\Re\,a^J_{\lambda\kappa}} \le \frac12 \,.
\label{eq:unitaritycond}
\end{equation}
For simplicity, we have restricted ourselves to a specific initial and
final state with \eg fixed helicities. Instead, we could have used
normalized linear combinations of states, for which the derivation is
valid as well, and \eq{eq:unitaritycond} also holds for those.

Studies of unitarity-violation bounds from anomalous couplings have been
performed in
Refs.~\cite{Baur:1987mt,Gounaris:1993fh,Gounaris:1994cm,Gounaris:1995ed,Degrande:2012wf,Corbett:2014ora}.
A tool to calculate this tree-level unitarity bound for anomalous
quartic gauge couplings in $2\rightarrow2$ vector-boson scattering
processes is available in Ref.~\cite{VBFNLOFF}.

For all processes with higher final-state multiplicities than 2, which
includes all VBF processes, one would in principle need an extension of
the above method. Up to now no such calculation is known in the
literature. As the unitarity violation happens only in the vector boson
scattering part, one extrapolates from the known results. Thereby, the
virtual $t$-channel vector bosons together with the attached quark lines
are replaced by corresponding on-shell counterparts. Also the
two final-state bosons are placed on-shell. Their invariant mass is kept
fixed and determines the relevant energy scale. The resulting
$2\rightarrow2$ vector-boson scattering process can now be analyzed
according to \eq{eq:unitaritycond}, and determines whether unitarity is
preserved or not.

\subsubsection{Unitarity Restoration}

Having determined the condition for conservation of unitarity, the next
question is what to do in case unitarity is violated. Here, one needs
not to worry about arbitrarily high energies. Only energy scales smaller
than the center-of-mass of the respective collider are an issue.  For
higher energies, one can always postulate that some unitarity-restoring
mechanism kicks in between the largest reachable energy and the energy
scale where unitarity violation starts.

One of the simplest approaches is to ignore the parameter space above
which unitarity becomes violated. For variables which are directly
related to the center-of-mass energy this can be easily achieved by not
analyzing any data from this region. Difficulties may arise for
observables where the unitarity-violation effect is spread over the
whole range. Extracting information from those then becomes problematic.
Another drawback is that one still expects some deviation from the SM
prediction even in the unitarity-violating phase-space regions, but with
a smaller contribution. Not using this at all potentially throws away
information.

\paragraph{Form Factor}

Therefore, a possible refinement consists of damping the effect of
anomalous couplings such that the unitarity bound is always respected.
This damping can be achieved by multiplying the amplitude with an ad-hoc
energy-dependent factor, a so-called form factor.
In fact, the method described before can be seen as a special case of
this, with a step function switching off anomalous couplings at the
violation threshold. The shape of the form factor is in principle
arbitrary, but the required properties give some guidance. The
magnitude of the partial wave depends only on the center-of-mass energy
$E$ of the $2\rightarrow2$ scattering process, so this should hold for
the form factor as well. Where unitarity is not an issue, at small
energies, the original amplitude should be mostly unaltered, \ie the
factor close to $1$. At very large energies, the form factor should
counterbalance the growth of the amplitude with energy, \ie if the
amplitude shows a leading $E^N$ behavior, the form factor should be
proportional to $E^{-n}$ with $n \ge N$.

A common choice is a dipole form factor~\cite{Baur:1988qt}, defined
as~\cite{Arnold:2008rz,Baglio:2014uba,VBFNLO}
\begin{equation}
F(E) = \left(1+\frac{E^2}{\Lambda^2_{\text{FF}}}\right)^{-p} \,,
\label{eq:dipoleff}
\end{equation}
where $\Lambda_{\text{FF}}$ parametrizes the characteristic scale where
form-factor effects become relevant and $p$ is the damping exponent. The
latter must be chosen large enough that the amplitude growth is
balanced, \ie the minimal allowed value is $p=2$ for anomalous QGCs.
Larger values are possible as well, then the anomalous couplings will be
further damped at large energies.

For anomalous Higgs couplings in VBF-$H$ production, also the following
form factors have been used commonly~\cite{Figy:2004pt,Hankele:2006ma}
\begin{align}
F_1 &= \left(1-\frac{q_1^2}{\Lambda^2_{\text{FF}}}\right)^{-1} \cdot 
\left(1-\frac{q_2^2}{\Lambda^2_{\text{FF}}}\right)^{-1} \,, \nonumber\\
F_2 &= -2 \, \Lambda^2_{\text{FF}} \  
C_0\left(q_1^2,q_2^2,(q_1+q_2)^2,\Lambda^2_{\text{FF}},\Lambda^2_{\text{FF}},\Lambda^2_{\text{FF}}\right)
\,,
\label{eq:dipoleffvertex}
\end{align}
where $q_1$ and $q_2$ denote the $t$-channel momentum transfer on each
side of the VBF process, and $C_0$ is the usual scalar three-point
function~\cite{Passarino:1978jh}. Note that $q_1^2$ and $q_2^2$ become
large negative numbers when the momentum transfer increases. The use of
$C_0$ is inspired by the fact that the origin of such an anomalous
coupling could be a loop of heavy particles which leads to an effective
vertex when integrated out.

The form factor tool~\cite{VBFNLOFF} already mentioned above can also
calculate the necessary scale $\Lambda^2_{\text{FF}}$ of a dipole form
factor \eq{eq:dipoleff} for anomalous QGCs. The input is the maximum
energy of the collider $\sqrt{s}$, the exponent $p$, and of course the
anomalous couplings themselves. From this, the largest value of
$\Lambda^2_{\text{FF}}$ is determined which still fulfills the unitarity
condition \eq{eq:unitaritycond} up to a center-of-mass energy of
$\sqrt{s}$.

\paragraph{$K$-Matrix}

Depending on the exact shape of the form factor, the damping at larger
energies can exceed the amount required by the unitarity condition. The
effect of anomalous couplings is maximized if we can find a form factor
such that it exactly fulfills the unitarity bound. There are several
methods which aim to construct such a factor, like the 
inverse amplitude method~\cite{Truong:1988zp,Truong:1991gv,Dobado:1989qm,Dobado:1996ps,Oller:1997ng,Dobado:1999xb,GomezNicola:2001as,Espriu:2014jya},
N/D unitarization~\cite{Truong:1991gv,Truong:1991ab}, 
Pad\'e
unitarization~\cite{Pade1892,Basdevant:1969rw,Basdevant:1969sz,Dicus:1990ew}, 
the $K$-matrix~\cite{Heitler1,Heitler:1947mca,Schwinger:1948yk,Berger:1991uj,Gupta:1993tq,Chanowitz:1999se,Alboteanu:2008my,Kilian:2014zja}
or the closely related $T$-matrix~\cite{Kilian:2014zja,Kilian:2015opv}.
For anomalous QGCs, the $K$-matrix approach has been used most, so we
will shortly summarize it, following the original presentation in
Ref.~\cite{Alboteanu:2008my}. It has been worked out for the chiral
Lagrangian approach without Higgs, using the Lagrangian terms $\La_4$
and $\La_5$ defined in \eq{eq:La45}. The translation to dimension-8
operators is easily done by applying the relations of
\eq{eq:linvsnonlin}. Isospin conservation thereby requires us to set
$f_{S,0}=f_{S,2}$.

The amplitude for the process $W^+W^- \rightarrow ZZ$ is dominated by
scattering of longitudinal bosons at high center-of-mass energies. There
the polarization vectors can be approximated by $\epsilon^\mu_L(p) =
\frac{p^\mu}{M}$.
This yields for the leading energy dependence of the amplitude, using
the Feynman rule shown in \fig{fig:fm_LS012},
\begin{equation}
A(W^+W^- \rightarrow ZZ) = A(s,t,u) 
= C_{02} (t^2+u^2) + C_{1} s^2
\label{eq:AmpWWZZ}
\end{equation}
with
\begin{align}
C_{02} &= \frac14 \frac{f_{S,0}+f_{S,2}}{\Lambda^4} \,, & f_{S,0}&=f_{S,2} \nonumber\\
C_{1}  &= \frac12 \frac{f_{S,1}}{\Lambda^4} \,,
\end{align}
and the momenta replaced by the Mandelstam variables $s$, $t$ and $u$.
The SM part of the amplitude has been left out, as its leading term is
constant in energy and does not need any unitarization on its own.

All other combinations of gauge bosons can be
expressed by this master amplitude using the relations
\begin{align}
A(W^+Z \rightarrow W^+Z) &= A(t,s,u) \,, \nonumber\\
A(W^+W^- \rightarrow W^+W^-) &= A(s,t,u)+A(t,s,u) \,, \nonumber\\
A(W^+W^+ \rightarrow W^+W^+) &= A(t,s,u)+A(u,s,t) \,, \nonumber\\
A(ZZ \rightarrow ZZ) &= A(s,t,u)+A(t,s,u)+A(u,s,t) \,.
\label{eq:amplrel}
\end{align}

To study unitarity, we need the eigenamplitudes both in spin and
isospin. The gauge bosons are isospin eigenstates
with~\cite{Ballestrero:2011pe,ThesisMax}
\begin{align}
\ket{W^\pm} &= \mp \ket{1;\pm1} \,, &
\ket{Z} &= \ket{1;0} \,,
\end{align}
which can be combined into two-gauge-boson product states, using the
Condon-Shortley sign convention,
\begin{align}
\ket{W^\pm W^\pm} &= \ket{2;\pm2} \,, \nonumber \\
\ket{W^\pm Z} &= \mp \frac1{\sqrt{2}} \ket{2;\pm1} - \frac1{\sqrt{2}} \ket{1;\pm1} \,, \nonumber \\
\ket{W^\pm W^\mp} &= - \frac1{\sqrt{6}} \ket{2;0} \mp \frac1{\sqrt{2}} \ket{1;0}
- \frac1{\sqrt{3}} \ket{0;0} \,, \nonumber\\
\ket{ZZ} &= \frac2{\sqrt{6}} \ket{2;0} - \frac1{\sqrt{3}} \ket{0;0}
\,.
\end{align}

Inserting these relations into \eq{eq:amplrel} and solving for the
isospin eigenamplitudes yields
\begin{align}
A_0(s,t,u) &= 3 A(s,t,u) + A(t,s,u) + A(u,s,t) \nonumber\\
 &= C_{02} (2s^2+4t^2+4u^2) 
  + C_{1}  (3s^2+t^2+u^2) \,, \nonumber\\
A_1(s,t,u) &= A(t,s,u) - A(u,s,t) \nonumber\\
 &= C_{02} (u^2-t^2) 
  + C_{1}  (t^2-u^2) \,, \nonumber\\
A_2(s,t,u) &= A(t,s,u) + A(u,s,t) \nonumber\\
 &= C_{02} (2s^2+t^2+u^2) 
  + C_{1}  (t^2+u^2) \,. 
\label{eq:amplisospin}
\end{align}

These are then decomposed into partial waves using
\eq{eq:partwavedecompr}
\begin{align}
a^{IJ}(s) &= \frac12 \frac1{32\pi} \int_{-1}^1 \di\cos\theta\ A_I(s,t,u)
d^J_{00}(\theta) \nonumber\\
 &= \frac1{32\pi} \int_{-s}^0 \frac{\di t}{s} A_I(s,t,u) P_J(s,t,u)
\,,
\label{eq:amplspinisospin}
\end{align}
where we have replaced the integration over the angle with the
integration over the Mandelstam variable. When neglecting the
gauge-boson masses, their relation is $t=-\frac{s}2(1-\cos\theta)$ and
from $u=-s-t$. The Wigner $d$-functions with two zero indices are the
ordinary Legendre polynomials $P_J$. The extra factor $\frac12$ is due
to the wave-function normalization, as the eigenamplitudes have
identical states for both initial and final state.  The non-vanishing
coefficients are
\begin{alignat}{4}
32\pi\,a^{00}(s) &=
   & \frac{14}{3} & \, C_{02} \, s^2 
\ +& \frac{11}{3} & \, C_{1}  \, s^2        
  \,, \nonumber\\
32\pi\,a^{02}(s) &=
   & \frac{4}{15} & \, C_{02} \, s^2 
\ +& \frac{1}{15} & \, C_{1}  \, s^2        
  \,, \nonumber\\
32\pi\,a^{11}(s) &=
   & \frac{1}{3} & \, C_{02} \, s^2  
\ -& \frac{1}{3} & \, C_{1}  \, s^2          
  \,, \nonumber\\
32\pi\,a^{20}(s) &=
   & \frac{8}{3} & \, C_{02} \, s^2  
\ +& \frac{2}{3} & \, C_{1}  \, s^2          
  \,, \nonumber\\
32\pi\,a^{22}(s) &=
   & \frac{1}{15} & \, C_{02} \, s^2  
\ +& \frac{1}{15} & \, C_{1}  \, s^2          
  \,\phantom{,}
\end{alignat}
using
\begin{align}
P_0(s,t,u) &= 1 \,, & 
P_2(s,t,u) &= -2+3\frac{t^2+u^2}{s^2} \,, \nonumber\\ 
P_1(s,t,u) &= \frac{t-u}{s} = \frac{u^2-t^2}{s^2} \,, &
P_3(s,t,u) &= 9\frac{t^2-u^2}{s^2}+10\frac{t^3-u^3}{s^3} \,, 
\end{align}
\ie all coefficients with $J=3$ and higher vanish. As $s$, $t$ and $u$
are not independent of each other, there is no unique way to write the
Legendre polynomials as functions of them. The expressions shown here are
the form we will need later when back-substituting.

\begin{figure}
\begin{center}
\includegraphics[width=0.7\textwidth]{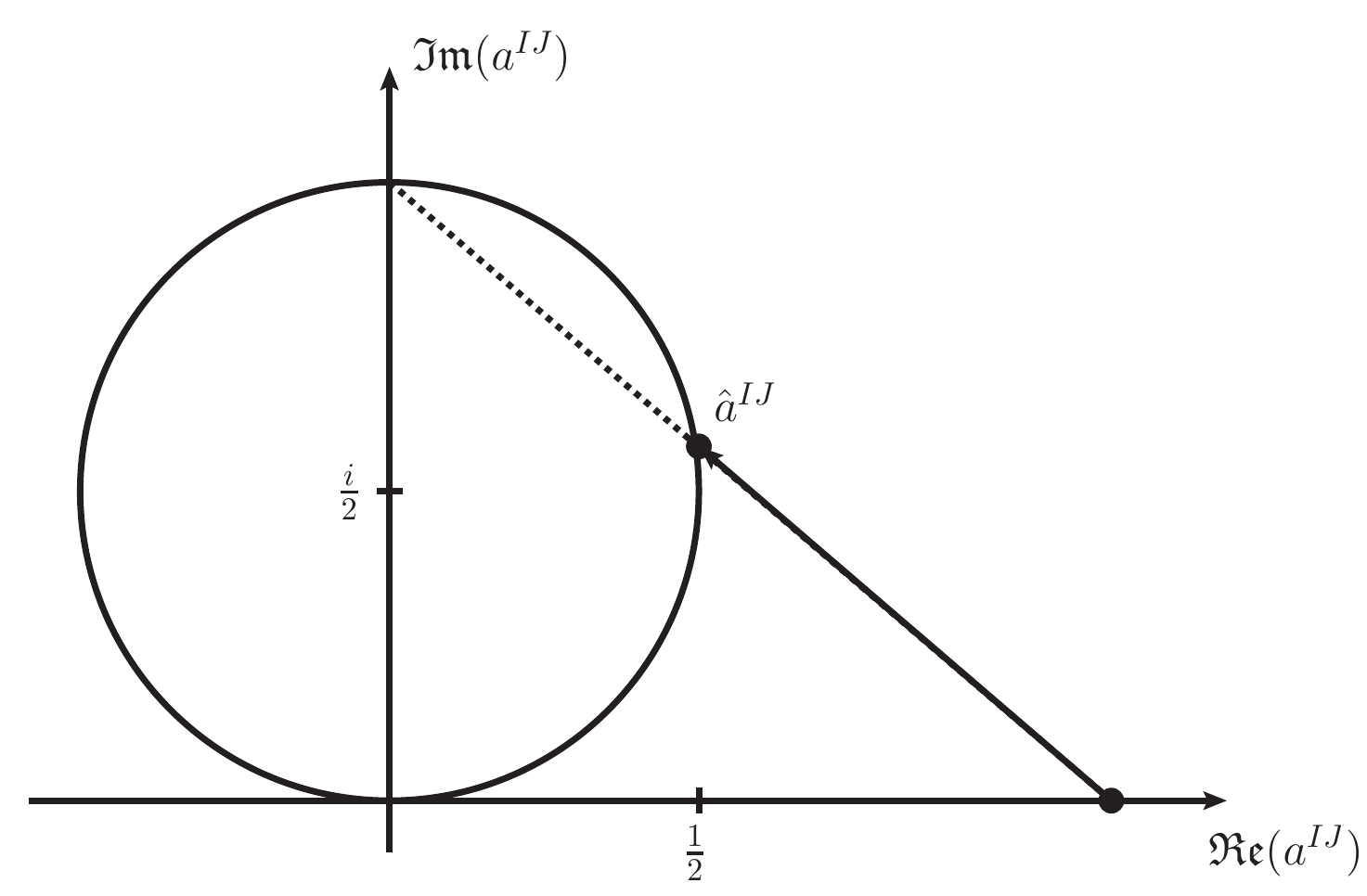}
\end{center}
\caption{Graphical representation of the Argand circle and the
$K$-matrix unitarization method, an inverse stereographic projection,
for the partial wave $a^{IJ}$ leading to the unitarized partial wave
$\hat{a}^{IJ}$.}
\label{fig:argandK}
\end{figure}
The partial waves computed at tree-level are real. To project this onto
the Argand circle, we draw a straight line from $a^{IJ}$ to the top of
the circle and determine the intersection between the line and the
circle. This yields for the unitarized partial wave 
\begin{equation}
\hat{a}^{IJ} = \frac{a^{IJ}}{1-i a^{IJ}} \,.
\end{equation}
A graphical representation of this procedure is drawn in
\fig{fig:argandK}. When $a^{IJ}$ goes to infinity, the unitarized
partial wave reaches the top of the circle, $\hat{a}^{IJ}=i$. As this
position is associated with the crossing of a resonance (compare with a
Breit-Wigner propagator, which becomes purely imaginary at the resonance
peak), the $K$-matrix scheme formally corresponds to placing a resonance
at infinity, which also has infinite width~\cite{Alboteanu:2008my}.
For later it is useful to split the unitarized partial wave into the
original expression and a counter-term containing the correction,
\begin{equation}
\Delta a^{IJ} = \hat{a}^{IJ} - a^{IJ} 
= \frac{i \left(a^{IJ}\right)^2}{1-ia^{IJ}} \,,
\label{eq:partwavect}
\end{equation}
yielding explicitly
\begin{align}
\Delta A^{00} \equiv 32\pi \Delta a^{00} &= 
  - \frac{( 14 C_{02} + 11 C_1 )^2}%
  {  3 \bigl( ( 14 C_{02} + 11 C_1 ) s^2 +  96\pi i \bigr)}
  s^4 \,, \nonumber\\
\Delta A^{02} \equiv 32\pi \Delta a^{02} &= 
  - \frac{(  4 C_{02} +    C_1 )^2}%
  { 15 \bigl( (  4 C_{02} +    C_1 ) s^2 + 480\pi i \bigr)}
  s^4 \,, \nonumber\\
\Delta A^{11} \equiv 32\pi \Delta a^{11} &= 
  - \frac{(    C_{02} -    C_1 )^2}%
  {  3 \bigl( (    C_{02} -    C_1 ) s^2 +  96\pi i \bigr)}
  s^4 \,, \nonumber\\
\Delta A^{20} \equiv 32\pi \Delta a^{20} &= 
  - \frac{(  8 C_{02} +  2 C_1 )^2}%
  {  3 \bigl( (  8 C_{02} +  2 C_1 ) s^2 +  96\pi i \bigr)}
  s^4 \,, \nonumber\\
\Delta A^{22} \equiv 32\pi \Delta a^{22} &= 
  - \frac{(    C_{02} +    C_1 )^2}%
  { 15 \bigl( (    C_{02} +    C_1 ) s^2 + 480\pi i \bigr)}
  s^4 \,.
\label{eq:partwavectexp}
\end{align}

This needs to be translated back to the physical amplitudes by inverting
the relations \eq{eq:amplspinisospin}, \ie using \eq{eq:partwavedecomp},
\begin{equation}
\Delta A_I(s,t,u) = \sum_J (2J+1) P_J(s,t,u) 32\pi\Delta a^{IJ}(s) \,,
\end{equation}
yielding explicitly
\begin{align}
\Delta A_0(s,t,u) &= 
  \frac{\Delta A^{00}-10 \Delta A^{02}}{s^2} s^2 + \frac{15 \Delta A^{02}}{s^2} (t^2+u^2)
  \,, \nonumber\\
\Delta A_1(s,t,u) &= 
  \frac{3 \Delta A^{11}}{s^2} (u^2-t^2)
  \,, \nonumber\\
\Delta A_2(s,t,u) &= 
  \frac{\Delta A^{20}-10 \Delta A^{22}}{s^2} s^2 + \frac{15 \Delta A^{22}}{s^2} (t^2+u^2)
  \,. 
\end{align}
This is then put into relations derived from inverting
\eq{eq:amplisospin} and inserted into \eq{eq:amplrel} to yield for the
unitarized on-shell scattering
amplitudes~\cite{Alboteanu:2008my,ThesisMax}
\begin{multline}
\widehat{A}(W^+W^-\rightarrow ZZ) \\
\shoveleft\qquad
= A(W^+W^-\rightarrow ZZ) + \frac13 \Delta A_0(s,t,u) - \frac13 \Delta A_2(s,t,u) \\
\shoveleft\qquad
\begin{alignedat}[b]{2}
&=& (t^2+u^2) & \cdot \Biggl( C_{02} + 
     \frac{5 \bigl(\Delta A^{02}-\Delta A^{22}\bigr)}{s^2} \Biggr) \\
&& + s^2      & \cdot \Biggl( C_{1} + 
     \frac{\bigl(\Delta A^{00}-\Delta A^{20}\bigr) 
        - 10 \bigl(\Delta A^{02}-\Delta A^{22}\bigr)}{3 s^2} \Biggr) 
 \,, 
\end{alignedat}\hfill
\label{eq:unitampWWZZ}
\end{multline}
\begin{multline}
\widehat{A}(W^+Z\rightarrow W^+Z) \\
\shoveleft\qquad
= A(W^+Z\rightarrow W^+Z) + \frac12 \Delta A_1(s,t,u) + \frac12 \Delta A_2(s,t,u) \\
\shoveleft\qquad
\begin{alignedat}[b]{2}
&=&  s^2 & \cdot \Biggl( C_{02} + 
     \frac{\Delta A^{20}-10\Delta A^{22}}{2 s^2} \Biggr) \\
&& + u^2 & \cdot \Biggl( C_{02} + 
     \frac{15 \Delta A^{22} + 3 \Delta A^{11}}{2 s^2} \Biggr) \\
&& + t^2 & \cdot \Biggl( C_{1} + 
     \frac{15 \Delta A^{22} - 3 \Delta A^{11}}{2 s^2} \Biggr) 
 \,,
\end{alignedat}\hfill
\label{eq:unitampWZWZ}
\end{multline}
\begin{multline}
\widehat{A}(W^+W^-\rightarrow W^+W^-) \\
\shoveleft\qquad
= A(W^+W^-\rightarrow W^+W^-) + \frac13 \Delta A_0(s,t,u) + \frac12 \Delta A_1(s,t,u) + \frac16 \Delta A_2(s,t,u) \\
\shoveleft\qquad
\begin{alignedat}[b]{2}
&=&  s^2 & \cdot \Biggl( C_{02} + C_{1} \\ 
&&& \phantom{\cdot \Biggl( \ } + \frac{\bigl(2 \Delta A^{00} + \Delta A^{20}\bigr) 
           - 10 \bigl(2 \Delta A^{02} + \Delta A^{22}\bigr) }{6 s^2} \Biggr) \\
&& + t^2 & \cdot \Biggl( C_{02} + C_{1} +
     \frac{10 \Delta A^{02} - 3 \Delta A^{11} + 5 \Delta A^{22}}{2 s^2} \Biggr) \\
&& + u^2 & \cdot \Biggl( 2 \cdot C_{02} + 
     \frac{10 \Delta A^{02} + 3 \Delta A^{11} + 5 \Delta A^{22}}{2 s^2} \Biggr) 
 \,,
\end{alignedat}\hfill
\label{eq:unitampWWWW}
\end{multline}
\begin{multline}
\widehat{A}(W^+W^+\rightarrow W^+W^+) \\
\shoveleft\qquad
= A(W^+W^+\rightarrow W^+W^+) + \Delta A_2(s,t,u) \\
\shoveleft\qquad
\begin{alignedat}[b]{2}
&=&  s^2 & \cdot \Biggl( 2 \cdot C_{02} + 
     \frac{\Delta A^{20} -10 \Delta A^{22}}{s^2} \Biggr) \\
&& + (t^2+u^2) & \cdot \Biggl( C_{02} + C_{1} +
     \frac{15 \Delta A^{22}}{s^2} \Biggr) 
 \,,
\end{alignedat}\hfill
\label{eq:unitampWWss}
\end{multline}
\begin{multline}
\widehat{A}(ZZ\rightarrow ZZ) \\
\shoveleft\qquad
= A(ZZ\rightarrow ZZ) + \frac13 \Delta A_0(s,t,u) + \frac23 \Delta A_2(s,t,u) \\
\shoveleft\qquad
\begin{alignedat}[b]{2}
&=&  s^2 & \cdot \Biggl( 2 \cdot C_{02} + C_{1} \\ 
&&& \phantom{\cdot \Biggl( \ } + \frac{\bigl( \Delta A^{00} + 2 \Delta A^{20} \bigr) 
       -10 \bigl(\Delta A^{02} + 2 \Delta A^{22}\bigr)}{3 s^2} \Biggr) \\
&& + (t^2+u^2) & \cdot \Biggl( 2 \cdot C_{02} + C_{1} +
     \frac{5 \bigl(\Delta A^{02} + 2 \Delta A^{22}\bigr)}{s^2} \Biggr) 
 \,.
\end{alignedat}\hfill
\label{eq:unitampZZZZ}
\end{multline}
As unitarization of the partial wave and hence of the $K$-matrix depends
on the center-of-mass energy $\sqrt{s}$ only, crossing symmetry between
the different amplitudes becomes broken, as is obvious from the explicit
form of the amplitudes. 

\begin{figure}
\begin{center}
\includegraphics[width=0.6\textwidth]{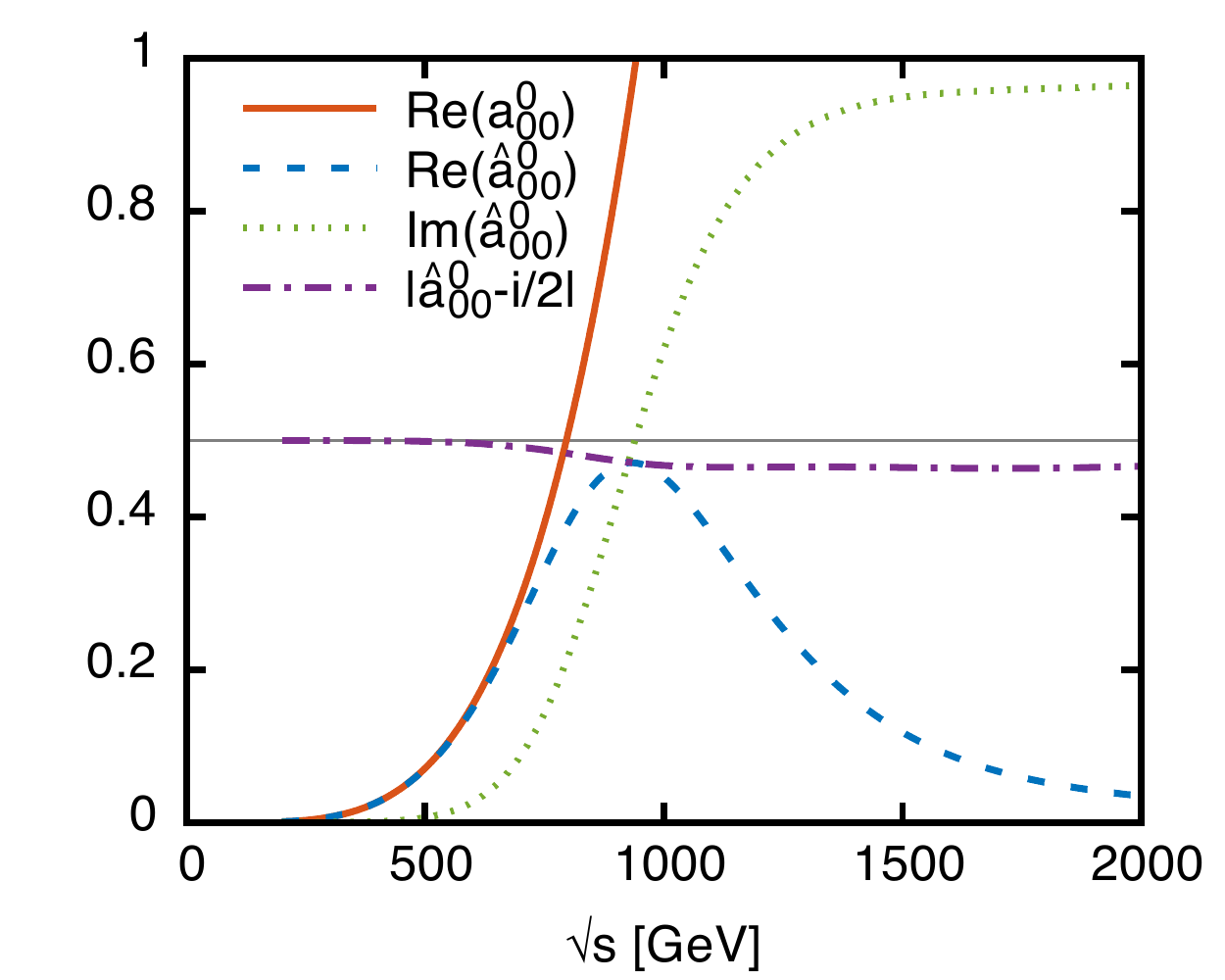} 
\end{center}
\caption{Comparison of the $J=0$ partial wave $a^0_{00}$ induced by the
anomalous coupling $\frac{f_{S,1}}{\Lambda^4}=400 \text{ TeV}^{-4}$ with
its $K$-matrix-unitarized version $\hat{a}^0_{00}$ for the process
$W^+W^+\rightarrow W^+W^+$. The SM contribution to the amplitude has
been turned off. 
}
\label{fig:pwave_reim}
\end{figure}
\begin{figure}
\begin{center}
\includegraphics[width=0.6\textwidth]{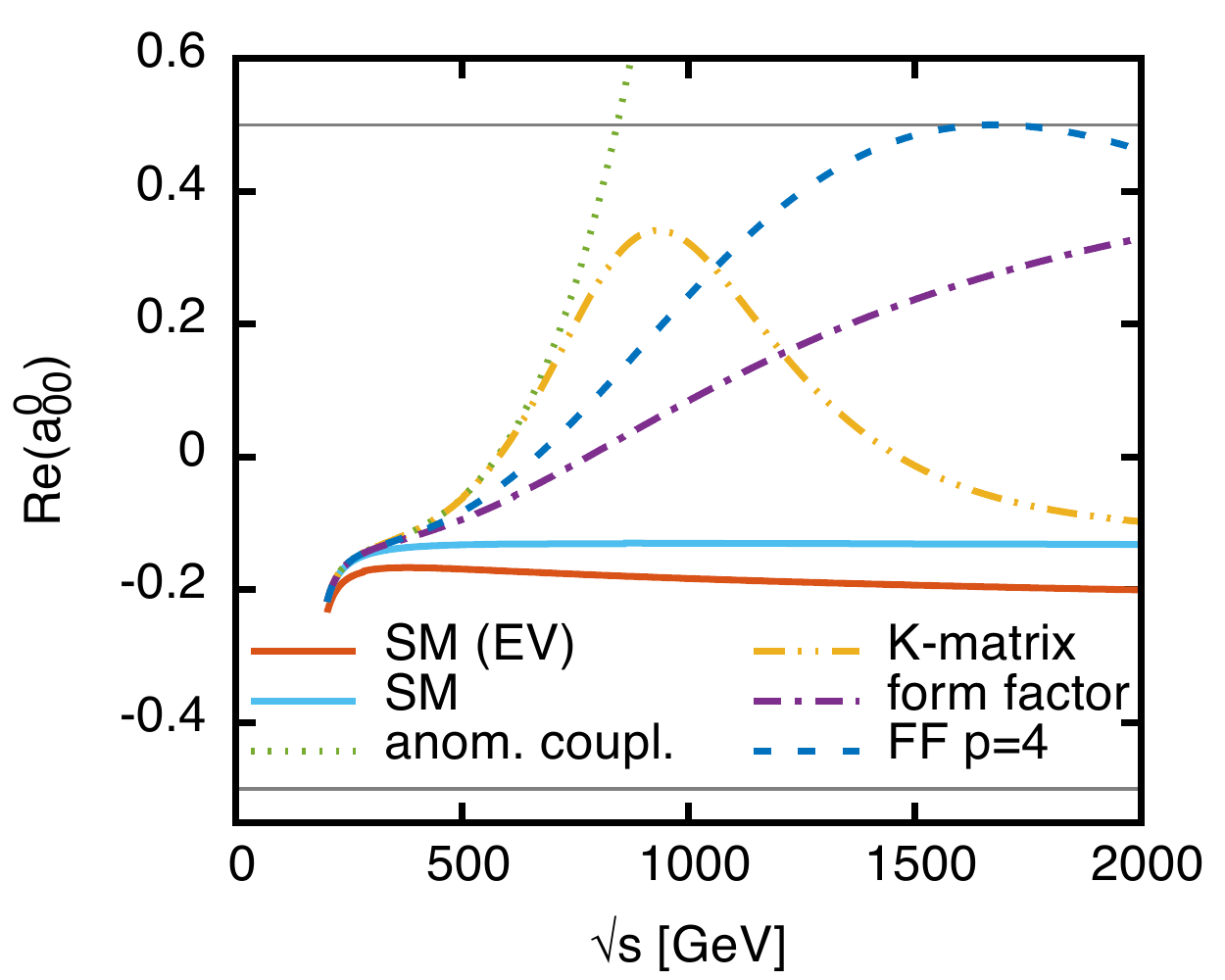} 
\end{center}
\caption{Zeroth partial wave of the $W^+W^+\rightarrow W^+W^+$ amplitude
for the SM and with anomalous coupling $\frac{f_{S,1}}{\Lambda^4}=400
\text{ TeV}^{-4}$, either unmodified or including different
unitarization prescriptions. All curves show longitudinal boson
scattering except the one labeled ``SM (EV)'', which is the most
restrictive linear combination and dominated by scattering of transverse
bosons. Additionally the unitarity bound at $\pm 0.5$ is plotted.
}
\label{fig:pwave_comparison}
\end{figure}
Figs.~\ref{fig:pwave_reim} and~\ref{fig:pwave_comparison} show the
effect of unitarization on partial waves. Both plots have been generated
using code from the \VBFNLO form factor tool~\cite{VBFNLOFF}.
In \fig{fig:pwave_reim} we apply $K$-matrix unitarization to the process
$W^+W^+\rightarrow W^+W^+$. As anomalous coupling,
$\frac{f_{S,1}}{\Lambda^4}=400 \text{ TeV}^{-4}$ is used and the SM
contribution to the amplitude switched off. This value is compatible
with the current experimental bounds from run-I of the
LHC~\cite{Aad:2014zda} and testable with additional data from run-II.
We compare the $J=0$ partial wave before and after applying $K$-matrix
unitarization. This requires us to set $\lambda_1=\lambda_2$ and
$\lambda_3=\lambda_4$, but leaves the individual values of the initial
and final-state polarizations still open. As the anomalous coupling
operator $\obs_{S,1}$ generates interactions between longitudinal bosons,
it is the $\lambda_i=0$ combination which we need to consider here. We
see that the unmodified partial wave~$a^0_{00}$, which is real, quickly
rises with the center-of-mass energy $\sqrt{s}$ and exceeds the
unitarity bound \eq{eq:unitaritycond} at $\sqrt{s}=795\text{ GeV}$.
Applying $K$-matrix unitarization, the unitarized partial
wave~$\hat{a}^0_{00}$ becomes complex. The figure shows its real and
imaginary part as well as its distance from the center of the Argand
circle. These three curves exhibit exactly the behavior one expects from
the discussion beforehand.  When the partial wave approaches the
unitarity limit, the projection onto the circle induces an imaginary
part and cuts off the real part at $0.5$. For large values the projected
partial wave approaches imaginary unity. Due to the projection onto the
circle, the distance should be constant and have a value of $0.5$. The
small deviation one observes is due to the fact that in the derivation
of the $K$-matrix we have taken the longitudinal polarization vectors in
the high-energy limit and neglected terms proportional to the
gauge-boson mass, whose contribution we see here.
Fig.~\ref{fig:pwave_comparison} shows a comparison of SM, pure anomalous
couplings, and different unitarization methods for the zeroth partial
wave. As in the previous plot, we study the process $W^+W^+\rightarrow
W^+W^+$ with anomalous coupling $\frac{f_{S,1}}{\Lambda^4}=400 \text{
TeV}^{-4}$. To obtain the most restrictive bound in the SM, we compute
all possible polarization combinations and then diagonalize the
resulting $3\times3$ matrix, which yields the curve labeled "SM (EV)",
where EV stands for ``largest eigenvalue''.
Its largest contribution comes from the scattering of transverse gauge
bosons, where all bosons have the same polarization. All other curves
show the scattering of longitudinal gauge bosons, which is the dominant
source for interactions induced by the anomalous coupling. 
For the SM part, there is actually an additional complication due to
virtual photon exchange in the $t$-channel. As photons are massless,
they mediate long-range interactions, which lead to a Coulomb
singularity when the scattering angle $\theta$ approaches zero. In the
full VBF process, the initial-state gauge bosons are virtual themselves
and no singularity appears. For the partial-wave analysis, which
includes an integral over the scattering angle, we exclude this region
by introducing a lower cutoff for the angle. This comes at the cost of
making the partial-wave coefficient dependent on its value. Thereby, the
full process can give us some guidance at which momentum transfer the
contribution to the cross section becomes negligible.  Converted into a
scattering angle for the $2\rightarrow2$ process of $W$-boson
scattering, this corresponds to about one degree, which we use as the
minimal angle. We have also studied the effects of varying this cutoff.
Lowering it to $0.01$ degrees enlarges the absolute value of the SM
partial-wave coefficients by approximately $0.01$. Thus there is only a
mild dependence on the actual value.

Switching the anomalous coupling on, the corresponding partial wave,
which now includes the SM contribution, crosses the unitarity bound at
$\sqrt{s}=834\text{ GeV}$. All unitarization methods restrict the real
part of the partial wave to be inside the unitarity limits. The
$K$-matrix method does not even reach it completely due to the effect of
the added SM contribution. For the dipole form factor, \eq{eq:dipoleff},
the minimal exponent to counterbalance the rise induced by the anomalous
coupling is $p=2$. In this case one needs to set
$\Lambda_{\text{FF}}=832\text{ GeV}$ to exactly touch the unitarity
bound at high energies. In the figure only the rise is visible, the
limit would be fulfilled only at the highest checked energy, which is
the collider center-of-mass energy taken as 13~TeV. Taking higher
exponents is possible as well. Therefore the figure also shows a second
form factor curve with $p=4$, for which the maximal scale is
$\Lambda_{\text{FF},p=4}=1667\text{ GeV}$. This setting yields a larger
contribution at smaller energies, and the unitarity bound is touched at
the same value. For higher energies, the larger damping lets the partial
wave decrease again and eventually it approaches the SM-only curve for
very high energies.

To make use of the unitarization in vector-boson scattering, the
expressions in eqs.~\ref{eq:unitampWWZZ} to~\ref{eq:unitampZZZZ} need to
be generalized to the off-shell case as the final step.
Taking the amplitude for $W^+Z\rightarrow W^+Z$, 
\begin{equation}
A(W^+Z\rightarrow W^+Z) = C_{02} (s^2+u^2) + C_{1} t^2
\label{eq:AmpWZWZ}
\end{equation}
as an example, the procedure is as follows. First, one compares the
corresponding coefficients in front of the original and the unitarized
amplitude, \ie 
\begin{alignat*}{2}
s^2: \quad&\quad C_{02} \quad&\quad \rightarrow \quad&\quad C_{02} + 
     \frac{\Delta A^{20}-10\Delta A^{22}}{2 s^2} \,, \\
t^2: \quad&\quad C_{1} \quad&\quad \rightarrow \quad&\quad C_{1} + 
     \frac{15 \Delta A^{22} - 3 \Delta A^{11}}{2 s^2} \,, \\
u^2: \quad&\quad C_{02} \quad&\quad \rightarrow \quad&\quad C_{02} + 
     \frac{15 \Delta A^{22} + 3 \Delta A^{11}}{2 s^2} \,.
\end{alignat*}
Then we identify the squares of the Mandelstam variables with the
corresponding factors of $g_{\mu\nu}g_{\rho\sigma}$ in the Feynman rules
in \fig{fig:fm_LS012}. Then we obtain 
\begin{center}
\begin{tabular}{rl}
\raisebox{-0.5\height}{\includegraphics[width=0.35\textwidth]{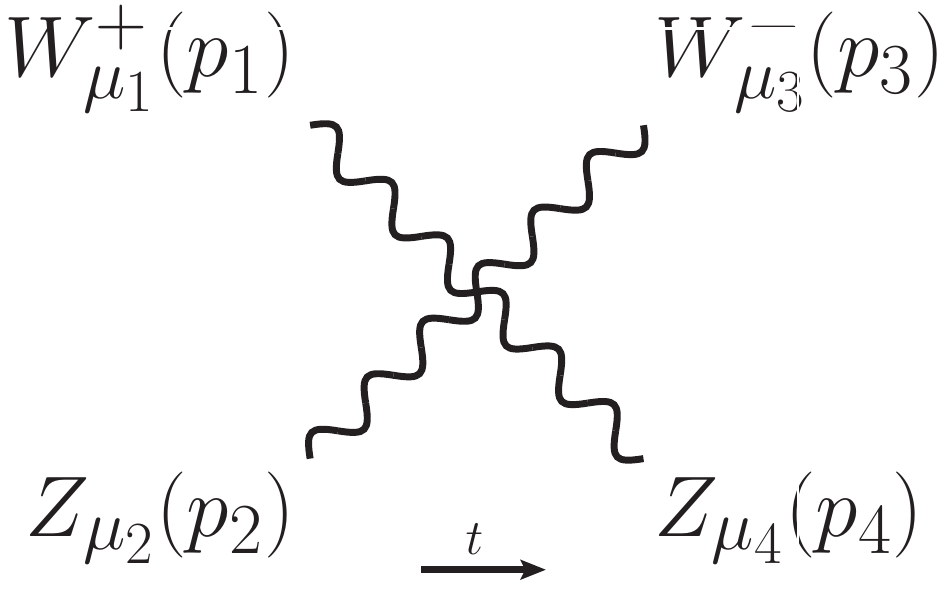}} & 
\parbox{0.6\textwidth}{\begin{align*}
i & M_W^2 M_Z^2 \\
&\biggl[ \quad \Bigl( \frac{f_{S,0}+f_{S,2}}{\Lambda^4} 
   + 2\,\frac{\Delta A^{20}-10\Delta A^{22}}{s^2}
   \Bigr) && g_{\mu_1\mu_2} g_{\mu_3\mu_4} \\
 &+     \Bigl( 2\,\frac{f_{S,1}}{\Lambda^4} 
   + 2\,\frac{15 \Delta A^{22} - 3 \Delta A^{11}}{s^2}
   \Bigr) && g_{\mu_1\mu_3} g_{\mu_2\mu_4} \\
 &+     \Bigl( \frac{f_{S,0}+f_{S,2}}{\Lambda^4} 
   + 2\,\frac{15 \Delta A^{22} + 3 \Delta A^{11}}{s^2}
   \Bigr) && g_{\mu_1\mu_4} g_{\mu_2\mu_3} 
\biggr] \,,
\end{align*}} \\
\end{tabular} 
\end{center}
and analogously for all other vertices.
As for the amplitudes, the resulting Feynman rules are no longer
invariant under crossing initial and final state. Therefore, the 
flow of time has been indicated by a small arrow. This and the
corresponding Feynman rules for the other vertices can now be used for
our calculation of VBF processes to give $K$-matrix unitarized cross
sections.

\subsection{Cross Section Results}
\label{sec:anomcs}

As last part of the discussion on anomalous couplings, we apply the
previously obtained results to VBF processes and show distributions of
differential cross sections. 

\begin{figure}
\begin{center}
\includegraphics[width=0.45\textwidth]{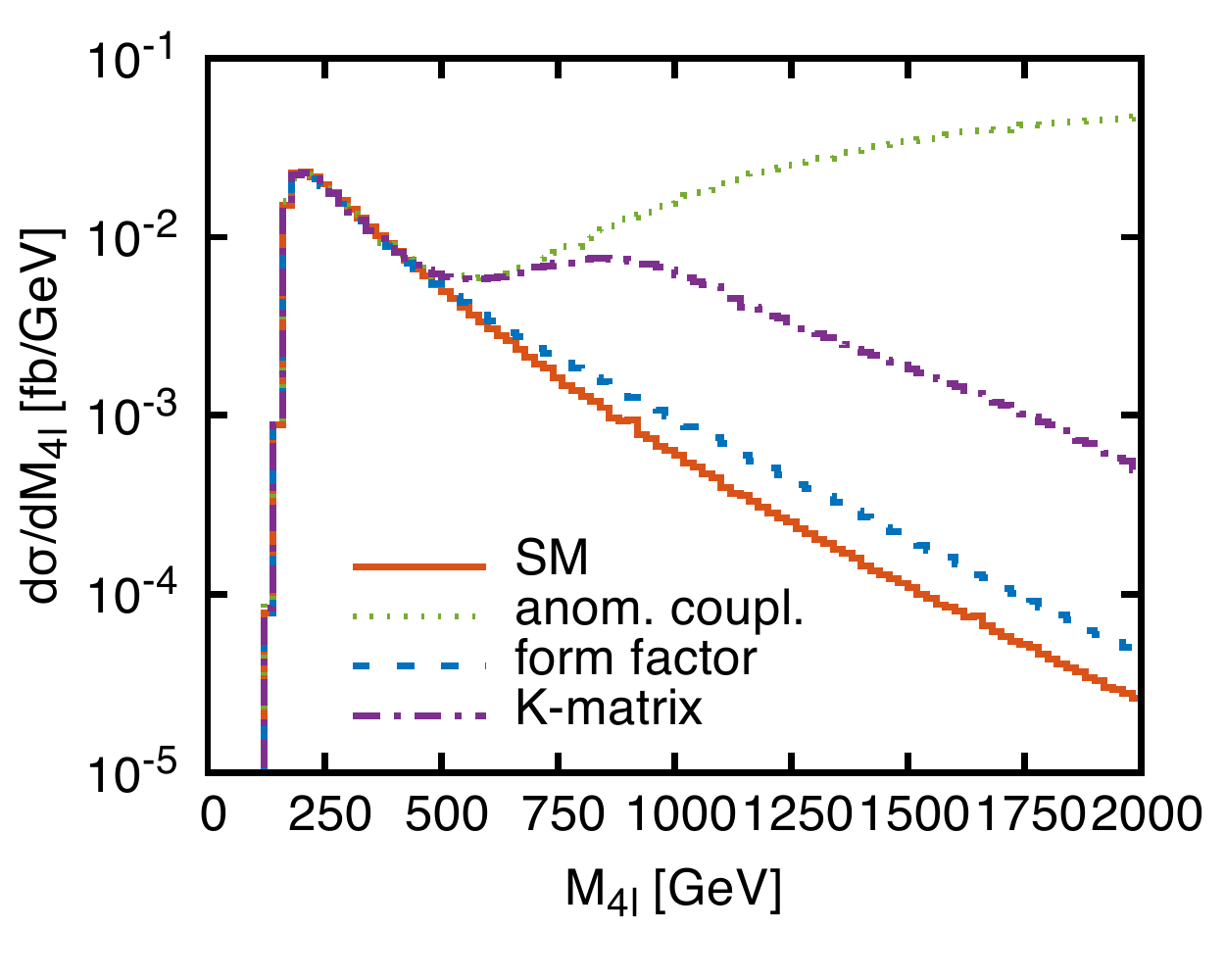} \quad
\includegraphics[width=0.45\textwidth]{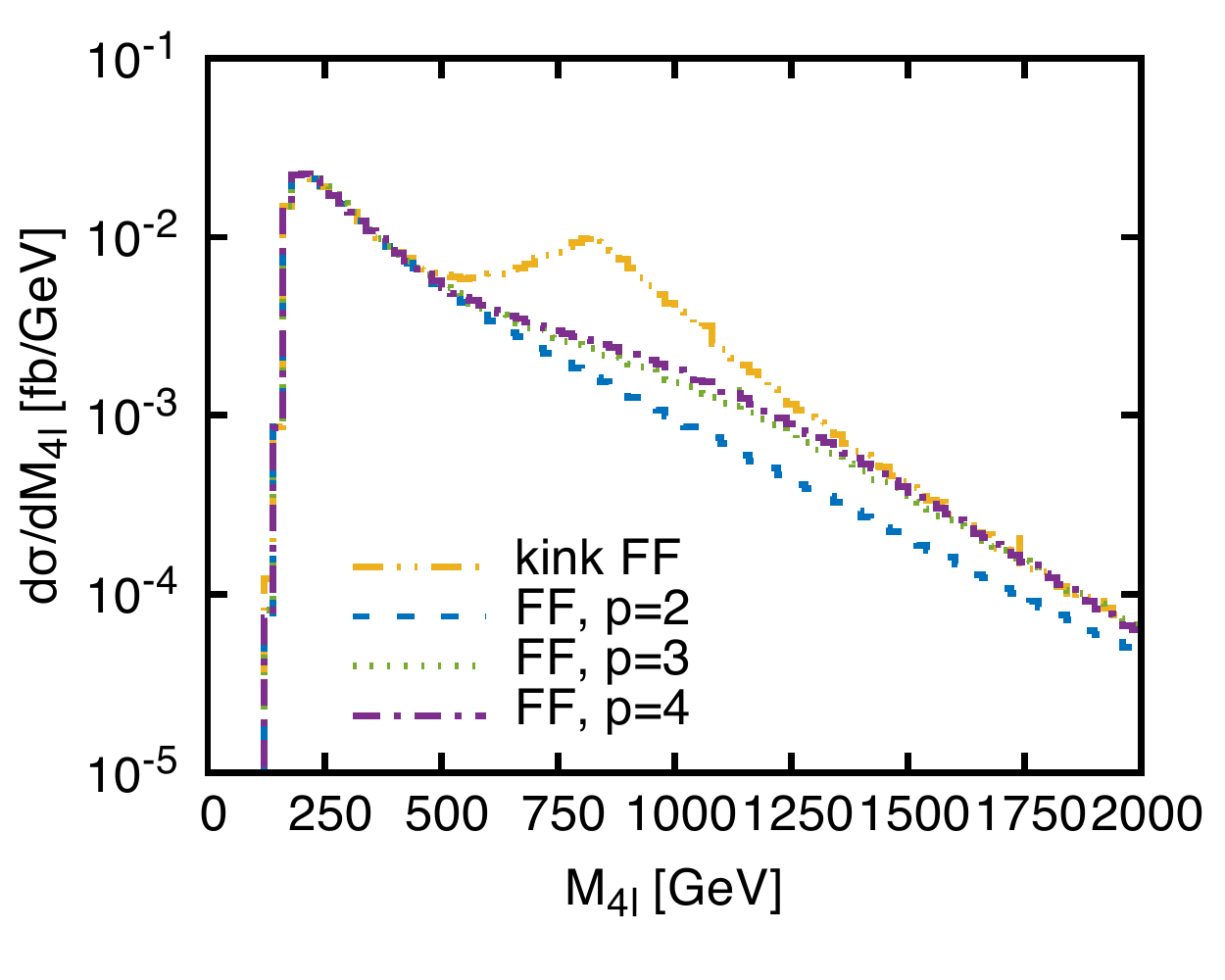} 
\end{center}
\caption{
Invariant mass distribution of the four final-state leptons for the VBF
production process $pp\rightarrow e^+\nu_e \mu^+\nu_\mu jj$ at NLO QCD
for the LHC with a center-of-mass energy of 13~TeV. \textit{Left:}
Comparison of the SM contribution, with switching on the anomalous QGC
$\frac{f_{S,1}}{\Lambda^4}=400 \text{ TeV}^{-4}$, and including
unitarization either via form factor, \eq{eq:dipoleff}, or using the
$K$-matrix method. \textit{Right:} Different choices of the exponent of
the dipole form factor with corresponding scales and a kink function
form factor clipping the amplitude at the unitarity boundary.
}
\label{fig:anomdist_m4l}
\end{figure}
The distribution of vector-boson scattering processes most directly
connected to anomalous QGCs is of course the invariant mass of the four
final-state leptons. Fig.~\ref{fig:anomdist_m4l} shows this distribution
at NLO QCD accuracy for VBF-$W^+W^+jj$ production with leptonic decays
for the LHC at a center-of-mass energy of 13~TeV. The electroweak
parameters are set as defined in \eq{eq:EWpara} and the cuts chosen
according to \eqs{eq:generalcuts,eq:vbfcuts}. As anomalous QGC we take
again $\frac{f_{S,1}}{\Lambda^4}=400 \text{ TeV}^{-4}$. The curves have
been generated with \VBFNLO~\cite{Arnold:2008rz,Baglio:2014uba,VBFNLO}.
The left panel compares the differential cross sections of the SM, with
the anomalous coupling switched on, and with two unitarization methods,
namely $K$-matrix and a dipole form factor, \eq{eq:dipoleff}, with
exponent $p=2$ and scale $\Lambda_{\text{FF}}=832\text{ GeV}$.
Visible deviations from the SM result induced by anomalous couplings set
in at an invariant mass of about 400~GeV. After this, the curve with
the unmodified anomalous coupling quickly rises and shows a clear
unitarity-violating behavior. At the upper end of the plot at 2~TeV, the
differential cross section is larger than the SM result by three orders
of magnitude. Its leading $s^4$ behavior still leads to a rise, as
it compensates the $\frac1{s}$ factor in the flux factor and the PDF
suppression. The $K$-matrix curve also shows a strong rise where the
anomalous-coupling effect sets in, but then the rise is damped by the
projection onto the Argand circle. Eventually the contribution falls in
parallel with the SM contribution, with an approximately constant factor
between them.  Unitarizing the amplitude with a dipole form factor, the
difference rises gradually with growing invariant mass. The whole
picture is actually similar to what we have seen already in
\fig{fig:pwave_comparison} for the partial wave analysis. Both anomalous
couplings and $K$-matrix show a rise, then the $K$-matrix gets cut off
at the unitarity bound, while the form factor part grows slowly but
steadily. The right panel shows different choices for the form factor
settings. Besides the curve with $p=2$ that was already part of the left
panel, also the ones with $p=3$ and $\Lambda_{\text{FF},p=3}=1342\text{
GeV}$ as well as $p=4$ and $\Lambda_{\text{FF},p=4}=1667\text{ GeV}$ are
plotted. Higher exponents combined with the maximally allowed value for
$\Lambda_{\text{FF}}$ yield a larger contribution at smaller
invariant masses. When going to higher values, the additional damping
starts to set in and reduces the differential cross section again
towards the SM curve, thus crossing the lines with lower exponents.
Additionally we define a kink-function like form factor as 
\begin{equation}
F_{\text{kink}}(E) = 
\begin{cases}
1 & \text{for $E \le \Lambda_{\text{FF,kink}}$} \,, \\
\left(\frac{\Lambda_{\text{FF,kink}}}{E}\right)^4 & \text{for $E > \Lambda_{\text{FF,kink}}$} \,,
\end{cases}
\end{equation}
with $\Lambda_{\text{FF,kink}} = 834\GeV$ chosen as the scale where
unitarity violation happens. A partial wave analysis of the on-shell
scattering process shows that this leads to an unmodified rise up to the
form factor scale and then leaves the partial wave constant up to tiny
effects of $\Order{\frac{M}{E}}$. Thus, it stays at the unitarity bound
and maximizes the contribution, similar to the $K$-matrix projection.
This is also visible in the differential cross section plot in
\fig{fig:anomdist_m4l}. Up to the form factor scale, the curve follows
exactly the one for the unmodified anomalous coupling. There, a kink
is visible and the distribution falls again. The dropoff is steeper than
for the $K$-matrix, as there the decomposition into eigenstates and
moving into the complex plane allows for a larger contribution.
The various curves also illustrate that depending on the unitarization
method chosen, the largest deviation from the SM happens in different
phase-space regions, and all need to be considered when comparing
with experimental data.

\begin{figure}
\begin{center}
\includegraphics[width=0.45\textwidth]{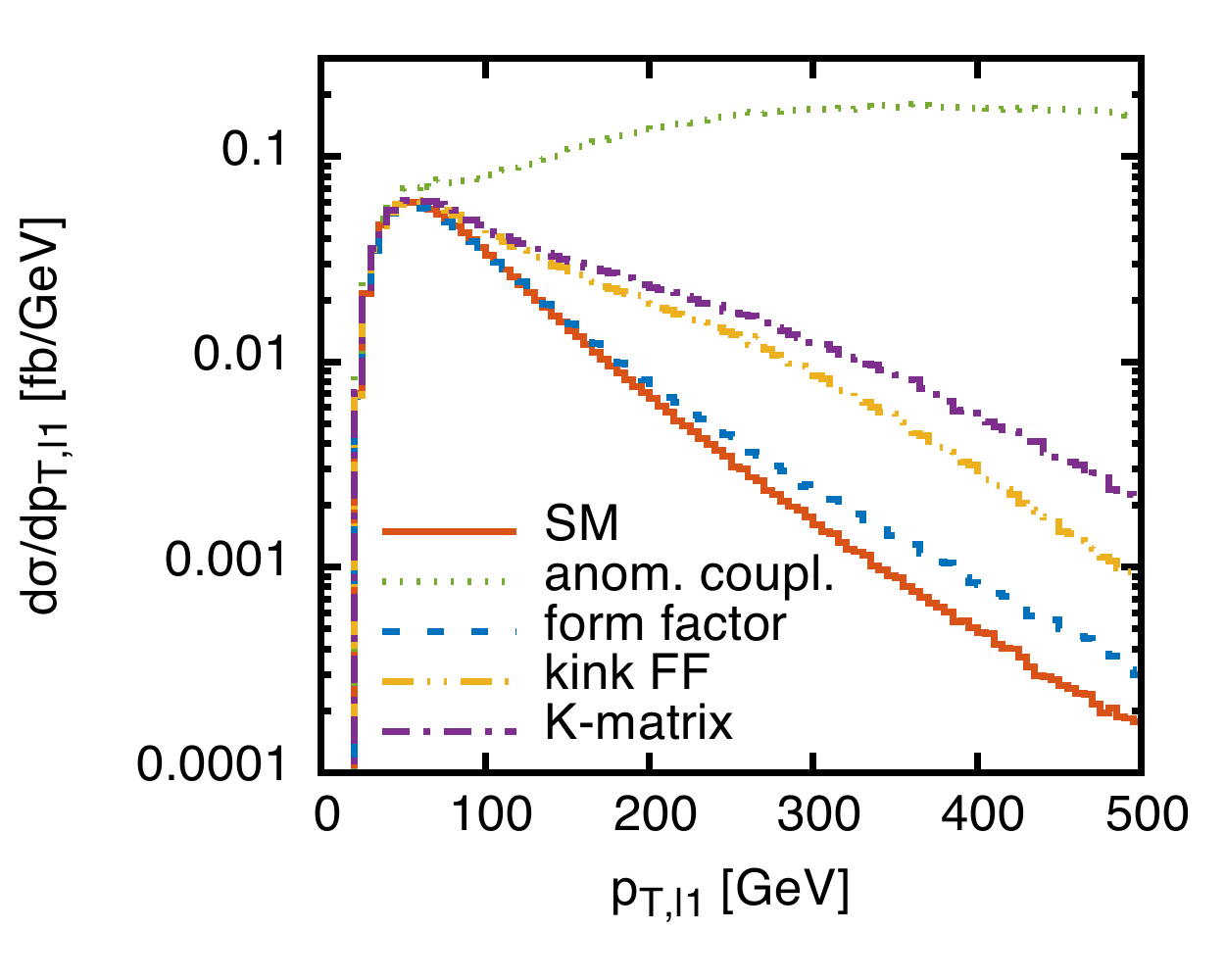} \quad
\includegraphics[width=0.45\textwidth]{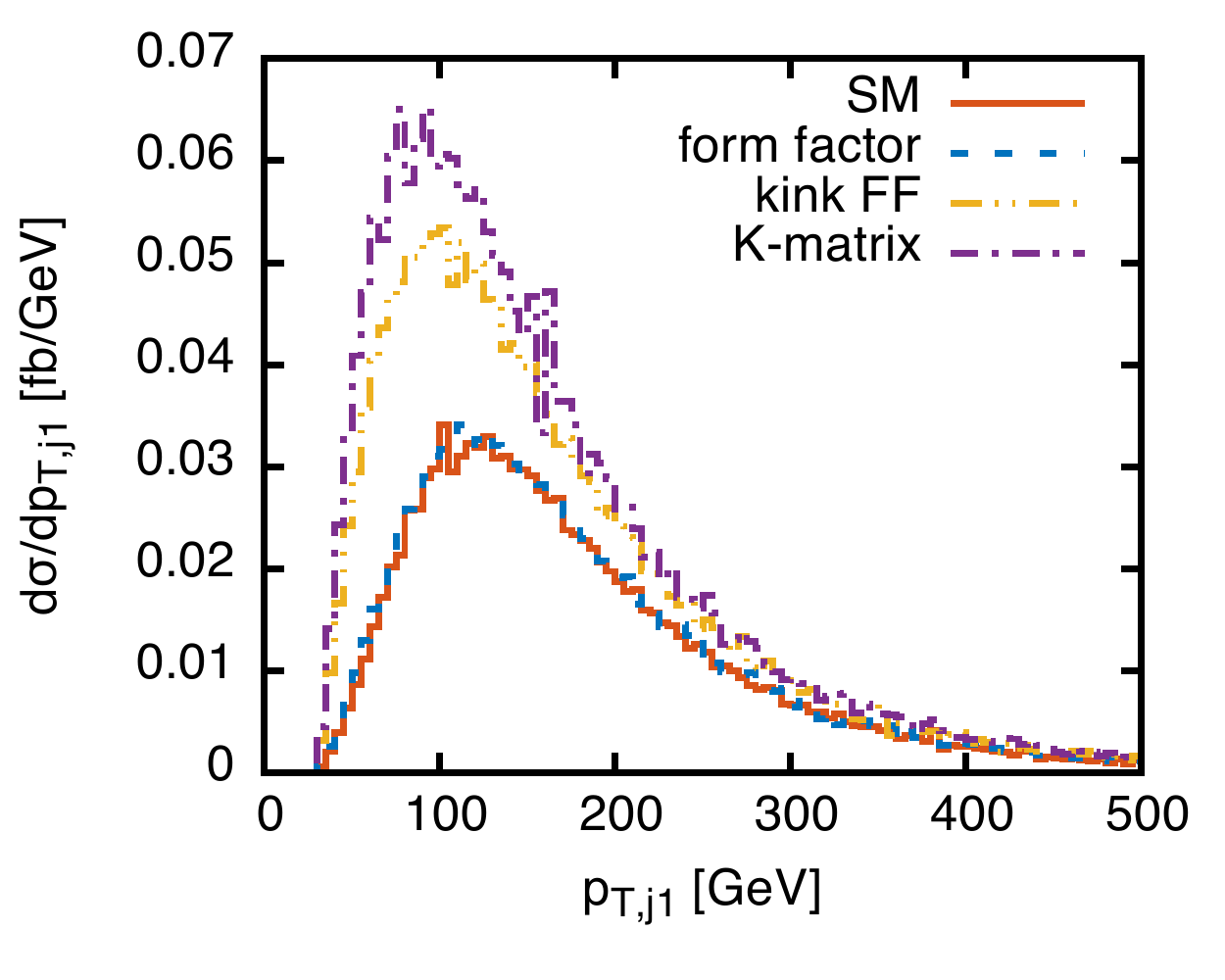} 
\end{center}
\caption{
Transverse momentum distribution of the leading lepton (\textit{left})
and the leading jet (\textit{right}) for the VBF production process
$pp\rightarrow e^+\nu_e \mu^+\nu_\mu jj$ at NLO QCD for the LHC with a
center-of-mass energy of 13~TeV. 
}
\label{fig:anomdist_pTlj1}
\end{figure}
The left panel of \fig{fig:anomdist_pTlj1} shows the distribution of the
transverse momentum of the leading lepton, \ie the one with the larger
transverse momentum. The behavior of the different curves is similar to
the previous plot. With anomalous couplings switched on, the curve
exceeds the one obtained from applying the $K$-matrix or a kink form
factor basically everywhere. This means that the unitarity-violating
contributions get spread out over the whole range of the distribution.
On the right-hand side of \fig{fig:anomdist_pTlj1} we finally present
the transverse momentum distribution of the leading jet. This observable
is not directly connected to the four-lepton invariant mass, and so the
main effect is a global rise of the cross section when compared to the
SM case. Small shape changes do occur however. This is best visible when
looking at the respective positions of the cross-section peak. The
larger the contribution from anomalous couplings, the more it shifts
towards slightly smaller values. This effect originates from the chosen 
anomalous coupling operator $\obs_{S,1}$, which enhances the fraction of
longitudinally polarized gauge bosons. These lead to smaller transverse
momenta of the jets than transverse polarized ones. If for example the
operator $\obs_{T,0}$ is used instead, the position of the peak remains
almost unchanged.

\section{Experimental Status}
\label{chap:experiment}

Having discussed the theoretical backgrounds of vector-boson fusion and
vector-boson scattering so far, we finally turn to an overview of the
current experimental results and future prospects. For a recent overview
on the experimental status and prospects see also
Refs.~\cite{Szleper:2014xxa,Green:2016trm}.

\subsection{LHC Run-I Results}

The VBF production process studied most so far is production
of a Higgs boson plus two jets. The presence of this mode is well
established with a significance of 5.4 standard deviations (4.7
expected) in the combination of ATLAS and CMS
results~\cite{ATLAS-CONF-2015-044}. The Higgs decay modes which enter
this result are $\gamma\gamma$, $\tau\tau$, $WW$ and $ZZ$. 
The ratio of the measured cross section over the SM prediction is
determined as $\mu_\text{VBF}=1.18^{+0.25}_{-0.23}$. Hence, there is
good agreement between theory and experiment. Additionally, the VBF-$H$
production process is an ideal place to study Higgs decays into
invisible particles, which lead to a signature of missing transverse
momentum in the detector~\cite{Eboli:2000ze}. Current data places an
upper limit of 58\% (44\% expected) on the invisible branching fraction
at 95\% CL~\cite{Chatrchyan:2014tja}.

\begin{figure}
\begin{center}
\includegraphics[width=0.5\textwidth]{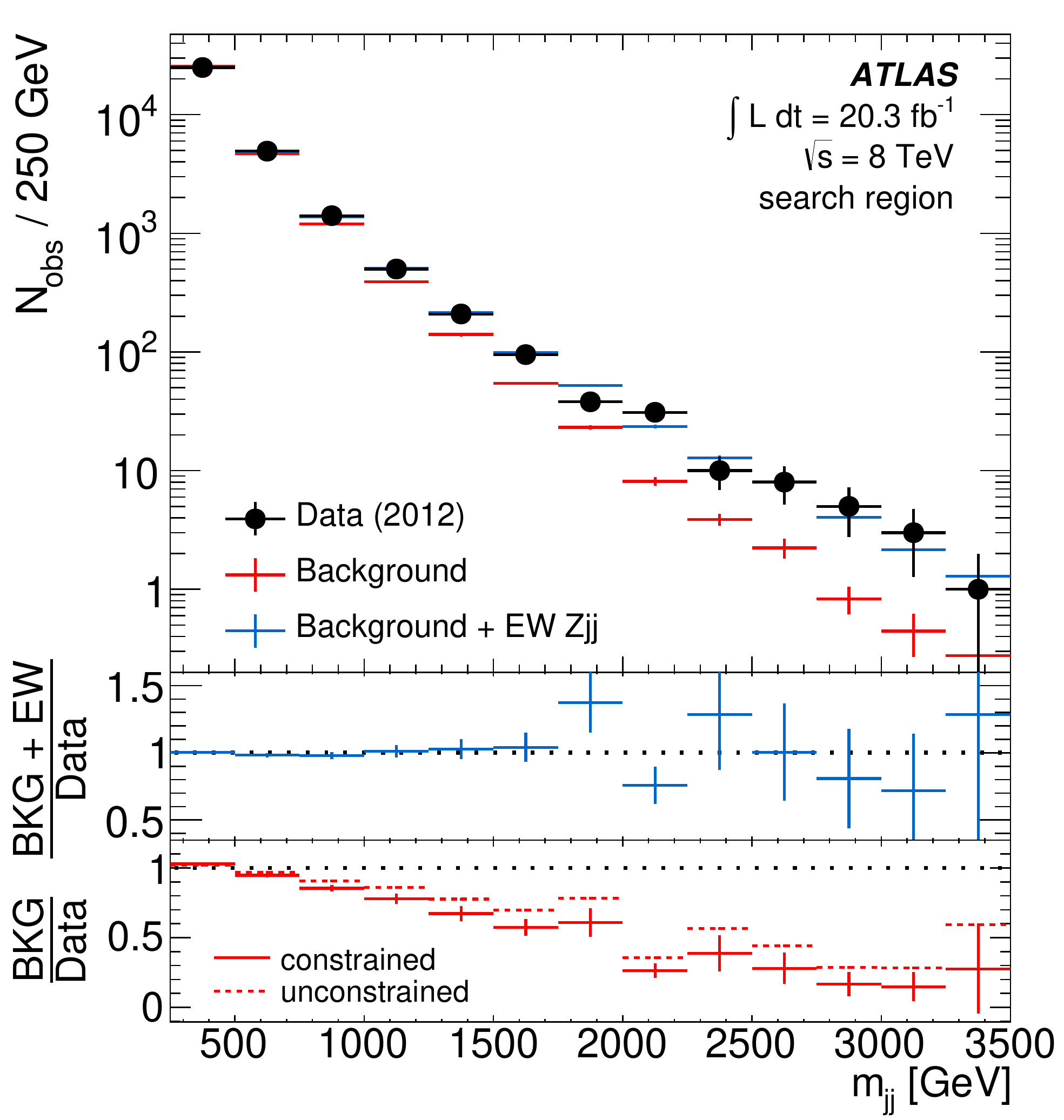}
\end{center}
\caption{Invariant-mass distribution of the two leading jets in $Zjj$
production. The label ``EW'' denotes the VBF production mechanism,
and ``BKG'' all other ones. 
Figure taken from Ref.~\protect\cite{Aad:2014dta}.
}
\label{fig:ATLAS_Zjj}
\end{figure}
In the following we will focus on the production of electroweak gauge
bosons. VBF-production of a single $Z$ boson in association with two
jets has been studied by ATLAS~\cite{Aad:2014dta} and
CMS~\cite{Chatrchyan:2013jya,Khachatryan:2014dea}. The final-state
signature of two opposite-sign, same-flavor leptons plus two jets is
dominated by the QCD-induced process. As VBF-specific cut, a lower bound
on the invariant mass of the two leading jets of 250~GeV has been
imposed. With this, the predicted fraction of VBF-induced events is
still only 4.0\%, while 94.7\% are QCD-induced, and the rest originates
from semi-leptonic diboson production. Raising this bound to 1~TeV
increases the VBF fraction to 12\% only. Therefore, a shape analysis has
been employed in Ref.~\cite{Aad:2014dta}. The obtained invariant-mass
distribution of the two leading jets is depicted in \fig{fig:ATLAS_Zjj},
where the label ``EW'' refers to the prediction of the VBF production
process and ``BKG'' to the rest, which is dominated by QCD-induced
production. One clearly sees the typical VBF behavior, which predicts
comparably more cross section in the very high-mass region and becomes
the dominant source of events. The two lower panels of the figure show
the ratio of the theory prediction to the data\footnote{The two labels
in the lower panel refer to the background template fit.}. One already
sees by eye that the background-only hypothesis does not reproduce the
measured data, while there is good agreement when including the
electroweak production process. A statistical analysis shows that the
significance exceeds the observation level of five standard deviations,
thus marking the first observation of a vector-boson fusion process.
More recently, also VBF-$W$ production has been studied by
CMS~\cite{Khachatryan:2016qkk} and a first measurement of the cross
section performed, which agrees well with the SM expectation.

\begin{figure}
\begin{center}
\raisebox{-0.5\height}{\includegraphics[width=0.48\textwidth]{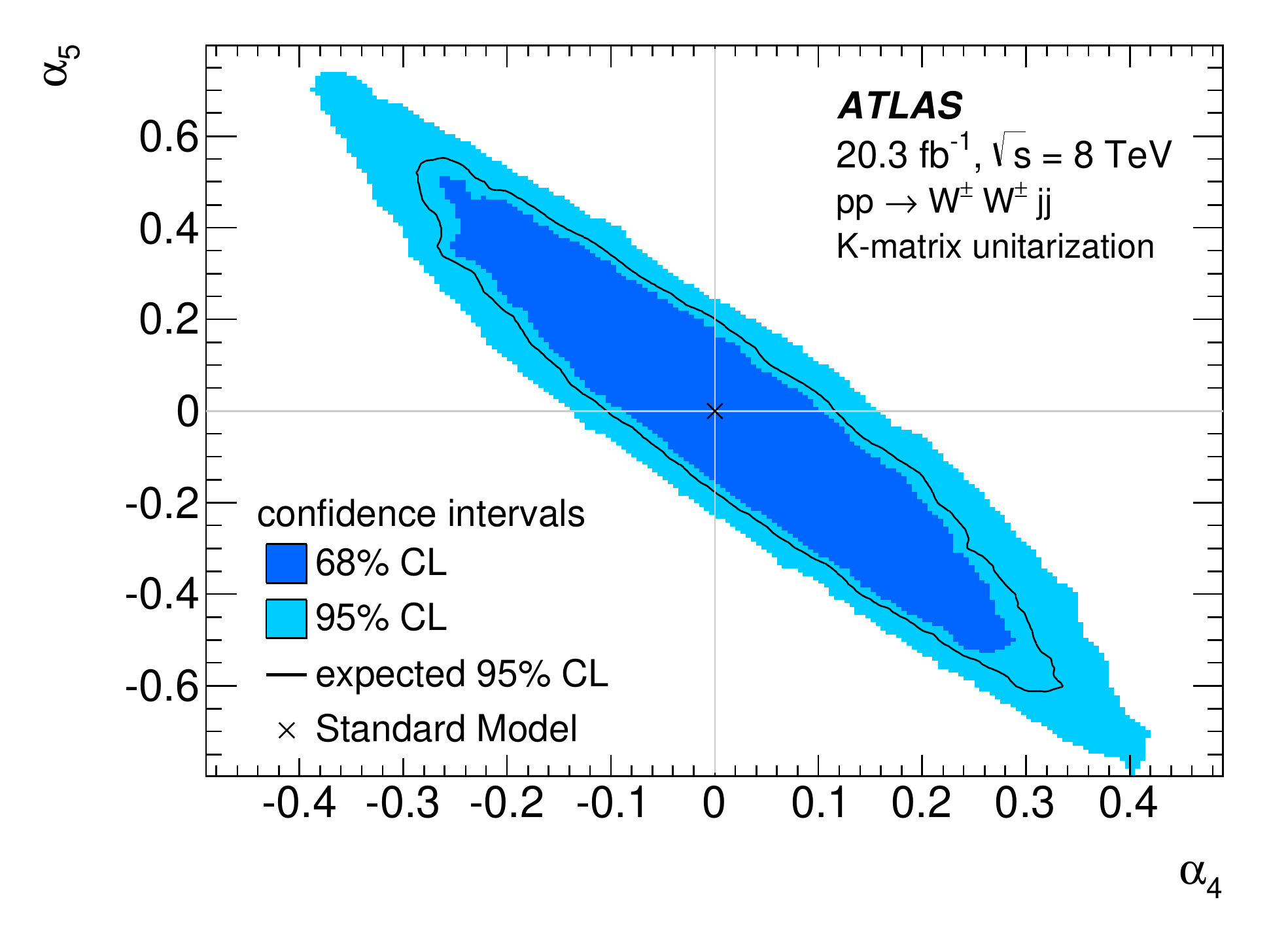}}\quad
\raisebox{-0.5\height}{\includegraphics[width=0.43\textwidth]{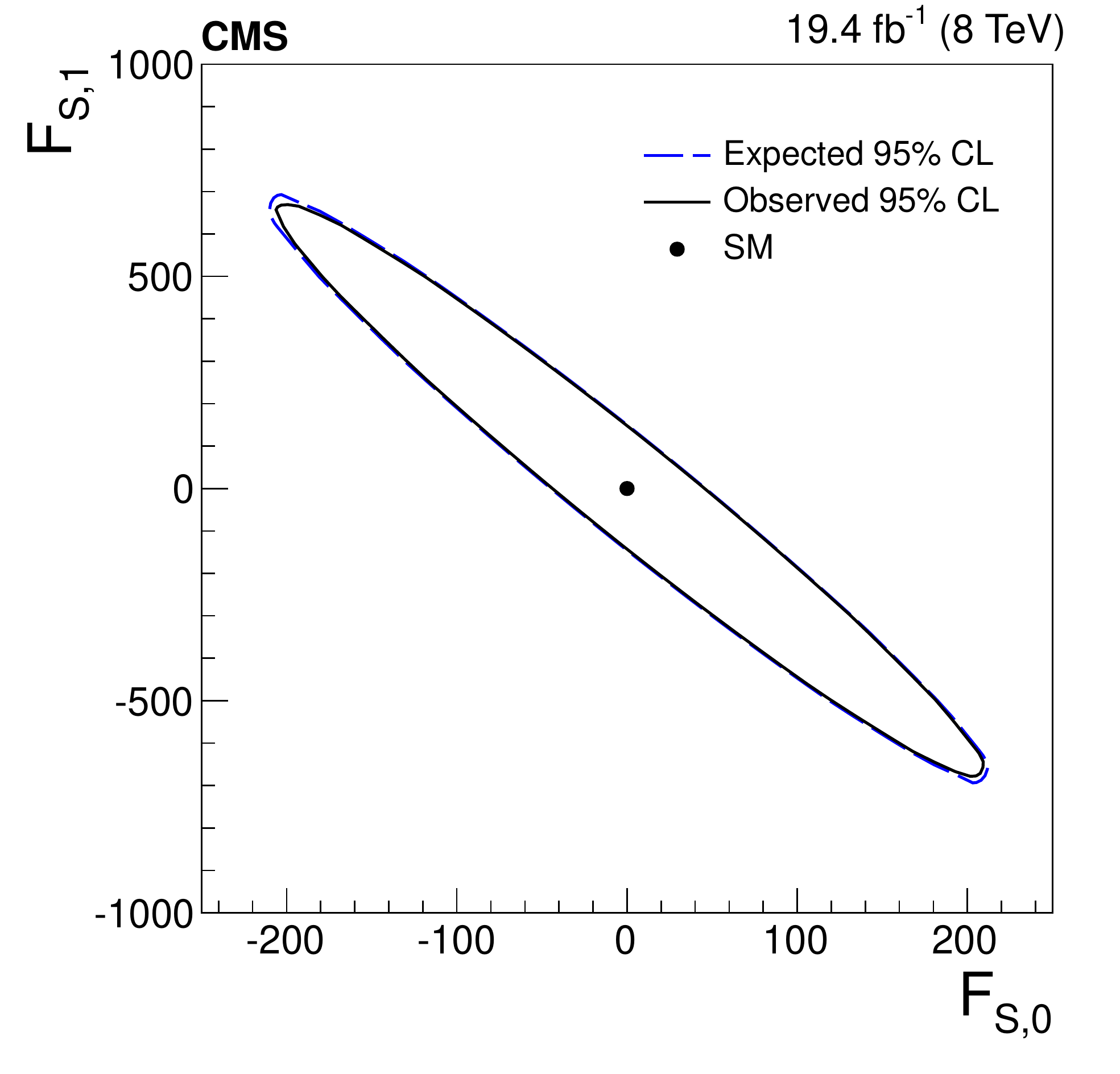}}
\end{center}
\caption{
Limits on anomalous quartic gauge-boson vertices determined from
same-sign $WW$ production plus two jets.
\textit{Left}: 
ATLAS result using the operator set of the chiral Lagrangian,
\eq{eq:La45}, and $K$-matrix unitarization. 
Figure taken from Ref.~\protect\cite{Aad:2014zda}.
\textit{Right}: 
CMS result for the dimension-8 operators of \eq{eq:obsd8s}, using
capital letters for the coefficients, setting $\Lambda=1\text{ TeV}$ and
not applying any unitarization procedure. 
Figure taken from Ref.~\protect\cite{Khachatryan:2014sta}.
The relation between the two operator sets is given by
\eq{eq:linvsnonlinspec}.
}
\label{fig:ATLASCMS_WWjj_anom}
\end{figure}
Moving on to vector-boson scattering, production of two leptonically
decaying same-sign $W$ bosons plus two jets has been studied both by
ATLAS~\cite{Aad:2014zda} and CMS~\cite{Khachatryan:2014sta}. This final
state is advantageous, as the corresponding QCD-induced process cannot
be produced from initial-state gluons at leading order. Thus the
fraction of electroweak events is larger than for other boson
combinations. Positive evidence (3.6~standard deviations) for
electroweak $W^+W^+jj$ production is reported by ATLAS, while CMS
observes a $1.9~\sigma$ excess. Both results are then used to obtain
limits on anomalous QGCs, shown in \fig{fig:ATLASCMS_WWjj_anom}. ATLAS
thereby uses the operator set of the chiral Lagrangian, \eq{eq:La45},
and applies $K$-matrix unitarization, whereas CMS extracts limits on
the operators $\obs_{S,0}$ and $\obs_{S,1}$, \eq{eq:obsd8s}, and does not
apply any unitarization procedure. The relation between the two sets of
operators is given by the $WWWW$-vertex entry of \eq{eq:linvsnonlinspec}.
As an example, the parameter point $F_{S,0}=200 \TeV^{-4}$, $F_{S,1}=-600
\TeV^{-4}$, 
located towards the lower right end of the CMS 95\%~CL region, translates
into $\alpha_4\simeq 0.092$, $\alpha_5\simeq -0.184$. This is well inside
the 68\%~CL region of the ATLAS plot, despite observing the process with
greater significance. The difference is due to the lack of unitarization
method in the CMS result, so large event counts are predicted at high
invariant masses, which are not present in the data and therefore lead
to stronger limits. The limits, up to which the results
respect the unitarity bound without the need for a specific unitarization
method, are given as $F_{S,0} = 0.016 \TeV^{-4}$ and $F_{S,1} = 0.050
\TeV^{-4}$, when the other coefficient is set to
zero~\cite{Khachatryan:2014sta}.

Other vector-boson scattering processes studied so far are 
$WZjj$ production~\cite{Aad:2016ett} and semi-leptonic $WVjj$
production~\cite{Aaboud:2016uuk} in ATLAS, and 
$W\gamma jj$ production~\cite{CMS-PAS-SMP-14-011} and $Z\gamma jj$
production~\cite{CMS-PAS-SMP-14-018} in CMS. Evidence for electroweak
production has been found in the last process.

\subsection{Future Prospects}

Both experiments have also studied expectations on vector-boson
scattering and limits an anomalous QGCs from future running of the
LHC~\cite{ATL-PHYS-PUB-2012-005,ATL-PHYS-PUB-2013-006,CMS-PAS-FTR-13-006}.
These have been performed for a center-of-mass energy of 14~TeV and
integrated luminosities of up to $3000\text{ fb}^{-1}$. The actual
LHC run-II center-of-mass energy of 13~TeV is slightly lower, but these
are not expected to change the general picture. 

All VBF processes greatly benefit from the higher center-of-mass energy.
As we have seen in \tabs{tab:nlocs8,tab:nlocs13}, the cross section
increases by about a factor 3 to 4 when going from 8 to 13~TeV. Then not
only the same-sign $W$ channel can be observed, but also the other
combinations of massive bosons, $W^+W^-$, $WZ$ and
$ZZ$~\cite{ATL-PHYS-PUB-2012-005,ATL-PHYS-PUB-2013-006,CMS-PAS-FTR-13-006}.
Given no relevant deviation from the SM prediction is observed, the
various dimension-8 operators are expected to be constrained to values
below 0.2 to 1~TeV$^{-4}$ at 95\%~CL. Deviations from operators with a
size of several TeV$^{-4}$ can even lead to a discovery when $3000\text{
fb}^{-1}$ of luminosity have been
collected~\cite{ATL-PHYS-PUB-2012-005,ATL-PHYS-PUB-2013-006,CMS-PAS-FTR-13-006}.

\section{Summary}
\label{chap:conclusion} 

Vector-boson fusion and vector-boson scattering are important production
processes at the LHC. From the theoretical side, they are characterized
by quark--(anti-)quark scattering through $t$-channel exchange of an
electroweak gauge boson. As there is no color connection between the two
quark lines, a very characteristic feature emerges. The two final-state
quarks appear as so-called tagging jets in the forward region of the
detectors.  Additional jet radiation in the region between the two
tagging jets is strongly suppressed. This distinguishes them from the
QCD-induced production mode of the same final state, which shows
enhanced jet activity in the central region.  We have also discussed the
impact of interference effects between the two.  These become negligible
when imposing VBF cuts, which require a large invariant mass and
rapidity separation of the two tagging jets.

Higher-order corrections play an important role to provide accurate
predictions. The size of the NLO QCD corrections for VBF processes is
moderate, but they help to stabilize the predictions under variations of
the renormalization and factorization scales. For these, taking the
momentum transfer of the exchanged bosons proves to be an advantageous
choice. We have derived explicit expressions for the vertex corrections.
These are both UV-divergent, which is cured by renormalization, and
IR-divergent, which is solved by adding diagrams with real emission of a
gluon. Corrections beyond NLO QCD are available for VBF-Higgs production
only. Both NLO EW and NNLO QCD contributions lead to a reduction of the
cross section. In the latter case the change is outside the uncertainty
bands given by a scale variation of the NLO QCD cross section.

Recent theoretical progress has been the matching of NLO QCD
calculations with parton showers. We have first given an overview of the
general properties, before discussing issues specific to VBF processes.
The distribution of jets in the central region has been an open question
from the combination of LO cross sections and parton showers, as
predictions of various programs have differed significantly. Using NLO
QCD results, this has now stabilized and a rather low jet activity is
expected, making this observable a good tool to distinguish VBF from
background processes. We have also discussed the impact of using
different matching schemes and showers, which agree well with each
other. 

The appearance of triple and quartic gauge couplings makes VBF processes
an important tool to measure these couplings and probe possible
deviations from the SM prediction. They are conveniently parametrized
using effective field theories. We have given an overview of the most
common operator schemes used and presented the relations among them. When
their coefficients become large, unitarity of the $S$ matrix can be
violated. Through partial-wave expansion we have derived the relevant
bound on scattering amplitudes. Unitarity can be restored for example
through the use of form factors or an inverse stereographic projection
back onto the unitarity circle, the $K$-matrix method. For the latter we
have presented explicit expressions using the scalar dimension-8
operators.

First evidence for some VBF production processes has already been
gathered during the run-I phase of the LHC. As no relevant deviations
from the SM prediction have been observed, this also allows to place
constraints on the anomalous coupling contributions. The higher
center-of-mass energy of subsequent LHC runs will strongly benefit the
VBF processes. All combinations of electroweak gauge boson pairs will be
observable with the final expected luminosity. Stringent constraints can
be placed on new-physics contributions to the quartic gauge couplings. A
further enhancement of the precision could be reached when increasing
the center-of-mass energy. Current proposals with energies up to 100~TeV
allow for an ultimate test of the SM.

\begin{acknowledgments}
It is a pleasure to thank Dieter Zeppenfeld for many valuable
discussions, helpful comments on the manuscript as well as his
encouragement to write this report. 
I am grateful to Johannes Bellm, Paco Campanario, Bastian Feigl, Stefan
Gieseke, Simon Plätzer and Franziska Schissler for helpful discussions
on the topics presented here. 
I would also like to thank Graeme Nail, Robin Roth and Marco Sekulla for
a careful reading of the manuscript.  
\end{acknowledgments}

\bibliographystyle{apsrev4-1}
\bibliography{papers}

\end{document}